\documentclass[review]{elsarticle}

\usepackage{graphicx}

\usepackage[T1]{fontenc}
\usepackage{dsfont}               
\usepackage{mathrsfs}             
\usepackage{slashed}              
\usepackage{amsmath}
\usepackage{amssymb}
\usepackage{amsbsy}
\usepackage{amsfonts}
\usepackage{subfigure}
\usepackage{sidecap}
\usepackage{bm}
\usepackage{pifont}
\newcommand{\xmark}{\ding{55}}%

\numberwithin{equation}{section}
\numberwithin{table}{section}
\numberwithin{figure}{section}

\journal{Progress in Particle and Nuclear Physics}

\usepackage{titlesec}
\usepackage{sectsty}
\titleformat{\section}{\normalfont\Large\bfseries}{\thesection}{1em}{}
\titleformat{\subsection}{\normalfont\large\bfseries}{\thesubsection}{1em}{}
\titleformat{\subsubsection}{\normalfont\normalsize\bfseries}{\thesubsubsection}{1em}{}
\usepackage[colorlinks,citecolor=blue,linktoc=all,linkcolor=cyan]{hyperref}

\begin{document}

\begin{frontmatter}

\title{Neutrinos and nucleosynthesis of elements}


\author[UWroc]{Tobias Fischer}
\author[CUG]{Gang Guo}
\author[GSI,TUD]{Karlheinz Langanke}
\author[GSI,TUD]{Gabriel Mart\'inez-Pinedo}
\author[UMN]{Yong-Zhong Qian}
\author[ASIOP,ASIAA,NCTS]{Meng-Ru Wu\corref{mycorrespondingauthor}}
\cortext[mycorrespondingauthor]{Corresponding author}
\ead{mwu@gate.sinica.edu.tw}

\address[UWroc]{Institute of Theoretical Physics, University of Wroclaw, 50-204 Wroclaw, Poland}
\address[CUG]{School of Mathematics and Physics, China University of Geosciences, Wuhan, 430074, China}
\address[GSI]{GSI Helmholtzzentrum f\"ur Schwerioneneforschung, 64291 Darmstadt, Germany}
\address[TUD]{Technische Universit\"at Darmstadt, 64289 Darmstadt, Germany}
\address[UMN]{School of Physics and Astronomy, University of Minnesota, Minneapolis, Minnesota 55455, U.S.A.}
\address[ASIOP]{Institute of Physics, Academia Sinica, Taipei 11529, Taiwan}
\address[ASIAA]{Institute of Astronomy and Astrophysics, Academia Sinica, Taipei 10617, Taiwan}
\address[NCTS]{Physics Division, National Center for Theoretical Sciences, Taipei 106, Taiwan}

\begin{abstract}
Neutrinos are known to play important roles in many astrophysical scenarios from the early period of the big bang to current stellar evolution being a unique messenger of the fusion reactions occurring in the center of our sun. In particular, neutrinos are crucial in determining the dynamics and the composition evolution in explosive events such as core-collapse supernovae and the merger of two neutron stars. 
In this paper, we review the current understanding of supernovae and binary neutron star mergers by focusing on the role of neutrinos therein. 
Several recent improvements on the theoretical modeling of neutrino interaction rates in nuclear matter as well as their impact on the heavy element nucleosynthesis in the supernova neutrino-driven wind are discussed, including the neutrino-nucleon opacity at the mean field level taking into account the relativistic kinematics of nucleons, the effect due to the nucleon-nucleon correlation, and the nucleon-nucleon bremsstrahlung. 
We also review the framework used to compute the neutrino-nucleus interactions and the up-to-date yield prediction for isotopes from neutrino nucleosynthesis occurring in the outer envelope of the supernova progenitor star during the explosion. Here improved predictions of energy spectra of supernova neutrinos of all flavors have had significant impact on the nucleosynthesis yields. 
Rapid progresses in modeling the flavor oscillations of neutrinos in these environments, including several novel mechanisms for collective neutrino oscillations and their potential impacts on various nucleosynthesis processes are summarized. 
\end{abstract}

\begin{keyword}
Core-collapse supernova\sep neutron star merger \sep neutrino \sep nucleosynthesis

\end{keyword}

\end{frontmatter}

\newpage

\thispagestyle{empty}
\tableofcontents

\newpage
\section{Introduction}\label{sec:intro}
Finding the exact origins of nature's different elements and their isotopes has been a long quest since the early 20th century. 
The seminal papers in 1957 from Margaret Burbidge, Geoffrey Burbidge, William A. Fowler, and Fred Hoyle (B2FH) and independently from Alistair Cameron established the foundation that stars and the associated astrophysical explosions are the main sites that produced nearly all different nuclei after the Big Bang nucleosynthesis during the cosmic evolution~\cite{B2FH,Cameron:1957}. 
Several major nucleosynthesis processes and the general requirements in nuclear physics and astrophysics conditions, including the hydrogen burning, helium burning, $\alpha$ process, slow ($s$-) and rapid ($r$-) neutron capture processes, as well as $p$- or $\gamma$-processes, were discussed in \cite{B2FH,Cameron:1957}. 
Tremendous advances were made since then to unravel the complexities in astrophysics and nuclear physics that are involved in these nucleosynthetic processes; see e.g., reviews \cite{Kappeler:2011xx,Rauscher.Dauphas.ea:2013,Pignatari:2016xx,Cowan:2019pkx,Arcones:2022jer}. 
Among these, the charged particle reactions (CPR), in which case the Coulomb barrier needs to be overcome, are essential for the understanding of the big bang nucleosynthesis and the evolution of stars from the main sequence to later stages. However, they cannot account for the formation of elements significantly heavier than iron, such as barium, europium as well as the actinides such as uranium and plutonium, which requires the capture of neutrons by the seed nuclei formed via CPR ($r$-process) or by existing nuclei inside stars ($s$-process).
Meanwhile, other nucleosynthetic processes (e.g., neutrino- ($\nu$-) process~\cite{Woosley90,Heger05a}, the $\nu p$ process~\cite{Froehlich06b,Pruet06}, $i$-process~\cite{Cowan:1977xx,Johns2016xx}) were found to potentially be responsible for producing certain isotopes whose traces are present in stellar observations or in the solar system, together with $p$- or $\gamma$-process~\cite{Rauscher.Dauphas.ea:2013,Pignatari:2016xx}. High-energy collisions with cosmic rays are a production source of
the light elements beryllium, boron and also lithium~\cite{Prantzos:2012}. 

Explosive astrophysical events such as a variety of core-collapse supernovae (CCSNe) and the merger of two neutron stars (NS) or a neutron star with a black hole (BH) have long been considered as sites where a number of the above nucleosythetic processes can take place.  
In particular, the involved short dynamical timescale of $\mathcal{O}(1)$~s as well as the associated neutron-rich, hot, and dense environments provide ideal conditions for the $r$-process to occur. 
Moreover, the extreme conditions of density and temperature allow the production of neutrinos. 
These neutrinos not only affect or even determine the dynamical evolution of the systems, but also play essential roles in shaping the formation of nuclei in nucleosynthesis processes associated with these events, including the $r$-process as well as the $\nu$- and $\nu p$ processes. 
Consequently, knowledge of how neutrinos interact with nuclei and nuclear matter and how flavor oscillations of neutrinos happen at relevant astrophysical conditions are needed to robustly model these events and their nucleosynthesis yields. 
The resulting observational signatures, which include the direct ones: neutrinos, electromagnetic, and gravitational waves (GWs), and the indirect ones: the fossil records of elemental abundances in stars as well as the presence of the short-lived radioactive nuclei in the Milky Way or nearby galaxies and/or their traces in meteorites and deep-sea crusts, can then be combined altogether to advance our understanding.

In this article, we review the recent progress in elucidating the role of neutrinos on the nucleosynthesis of elements. 
In Sec.~\ref{sec:astro}, we introduce the two main astrophysical sites, CCSNe and binary NS mergers (BNSMs) and discuss the general evolution of these systems and the role of neutrinos therein. 
In Sec.~\ref{sec:third}, we review recent work on neutrino interaction with 
matter and nuclei as well as their impact on CCSNe nucleosynthesis.
We then discuss recent findings of flavor oscillations of neutrinos in SNe and in BNSMs, and their implications for  different nucleosynthesis processes in Sec.~\ref{sec:osc}. 
We conclude in Sec.~\ref{sec:sum}.

\newpage
\section{Astrophysical sites}\label{sec:astro}

The canonical sites where neutrinos are copiously produced and affect 
the evolution of the system as well as nucleosynthesis therein are the CCSNe and BNSMs. 
Both sites involve conditions of high temperature $T\gtrsim \mathcal{O}(10)$~MeV and mass density $\rho\gtrsim\mathcal{O}(10^{14})$~g~cm$^{-3}$.
Under these conditions, weak interactions that allow neutrino productions can proceed with time scale much shorter than $\mathcal{O}(1)$~s, which allows massive production of neutrinos in all flavors ($\nu_e$, $\nu_\mu$, $\nu_\tau$ and their antineutrinos).
The density of the environment can be so high that even neutrinos are ``trapped'' for certain duration and only diffuse out slowly.
Consequently, the interaction of neutrinos with matter around the nuclear saturation density sets the evolution of the system.
Moreover, neutrino interaction with nucleons and nuclei at these sites can determine or change the composition in the matter ejected, and therefore are critical in shaping the nucleosynthesis outcome of these events. 
In the following sections, we will review the current understanding of CCSNe and BNSMs, with particular focus on the role of neutrinos.

\subsection{Core-collapse supernovae}
\label{sub:first}
\subsubsection{Supernova progenitors}
Progenitors of CCSNe are massive stars with a zero-age main sequence (ZAMS) mass heavier than about 8--10~$M_\odot$.
Such stars spent about several 
million years on the main sequence burning the hydrogen burning stage, 
followed by the core helium burning that lasts about several hundred thousand years.
In contrast to low- and intermediate-mass stars, these massive stars further enter the advanced nuclear burning stages, starting with core carbon burning that lasts for several hundred years. 
Details about the more advanced nuclear burning stages, e.g., oxygen and neon burning, which can last for several months up to years, strongly depend on the stellar evolutionary track determined by the ZAMS mass and metallicity (see also Ref.~\cite{Phillips1999phst.book}).
Those with ZAMS mass ranging between 8--10~M$_\odot$, which develop stellar cores composed of mostly oxygen, neon and magnesium (c.f. Ref.~\cite{Nomoto87,Jones13,Limongi2023arXiv231200107L} and references therein) 
can undergo core collapse triggered by electron captures on neon and magnesium nuclides and result in the so-called electron-capture supernovae. 
Heavier stars can further enter the silicon burning phase, which lasts only for a day or less, and develop iron cores before core collapse happens.

The time spans of different nuclear burning stages are mostly dictated by the nuclear burning timescales involved, and are typically much longer than the cooling as well as the dynamical timescales. 
Thus, hydrostatic simulations of the evolution of such stars in spherically symmetric (1D) models have been employed for several decades \cite{LimongiChieffi2000ApJS129,Woosley2002RvMP74,Heger03,UmedaNomoto2008ApJ673,Eggenberger2008Ap&SS316} to provide a variety of progenitors as the initial conditions for modeling SN explosions. 
However, it is also well known that the stellar structure can be convectively unstable, featuring nuclear burning regions with nearly constant entropy per baryon, resulting in composition mixing. 
This important multi-dimensional phenomenon has been typically implemented into the stellar evolution models through the mixing length theory~\cite{Bohm-Vitense:1958,Kippenhahn:2013}.  
Other important multi-dimensional effects such as the stellar rotation, magnetic fields, and mass loss have also been studied but remain as the major sources of uncertainty \cite{Heger2000ApJ528,UmedaNomoto2003Natur422,Hirschi2004A&A425,Iwamoto2005Sci309,Tominaga2007ApJ660,Hirschi2012A&A546A,Langer2012A&A542,ChieffiLimongi2013ApJ764}.
Among those, the mass loss rate determines the core evolution at the onset of core carbon burning and hence has a direct impact on the later SN evolution. Low mass loss rates result in a compact carbon core and lead to a high density silicon and sulfur layer, which, in turn, will power a high mass accretion rate onto the proto-neutron star formed later during the SN evolution powering high neutrino luminosities. This aspect of CCSN phenomenology has been analyzed in terms of the compactness parameter as a useful, though ad hoc, diagnostic tool~\cite{O'Connor11}. 
For stars residing in close binary system, the binary interactions can also affect the associated SN progenitor properties and have been under intense investigations~\cite{Langer:2012jz,Yoon:2017nuh,Schneider:2020vvh,Laplace:2021vre,Kinugawa:2023hdo,Wang:2024dwq}. 

Another interesting aspect is that the nuclear burning timescale during the final oxygen and in particular the silicon burning phase becomes much closer to the hydrodynamical timescale. 
The former can be approximated as
\begin{equation}
\tau \propto 0.52~{\rm days}~\left(\frac{Q}{2\times 10^{17}~{\rm erg~g}^{-1}} \right)
\left(\frac{M_{\rm core}}{1.5~{\rm M}_\odot}\right)\left(\frac{3.5\times L_\odot}{L}\right)~,
\end{equation}
where $Q$ measures the energy output for the ``schematic'' process, $^{28}{\rm Si} \,+ \, ^{28}{\rm Si} \longrightarrow \,^{56}{\rm Ni}$, $L =3.5\ L_\odot$ is a typical value for the silicon-burning luminosity found in stellar evolution calculations and 1.5~$M_\odot$ is a typical core mass.  
Recently, multi-dimensional simulations of the silicon burning stage have been investigated  (for a recent review, see  Ref.~\cite{KupkaMuthsam:2017} and references therein).
Simulations for a short epoch of the evolution during the oxygen as well as the silicon-sulfur burning stages~\cite{Arnett:1998,Couch2015ApJ,BMuller16,Janka2020ApJ,Takashi2021MNRAS,FieldsCouch2021ApJ921} demonstrated that the implemented initial perturbations will grow to large scale sizes, which locally reduces the density in the silicon-sulfur layer and hence leading to a lower ram pressure at the later SN shock evolution. 
However, even though all these models show a direct impact on the later SN dynamics, favoring the SN shock revival, the choice of the initial perturbations, mimicking the presence of convection at start from a spherically symmetric stellar model, are still under debate.

During the final burning phases, in particular the silicon burning, the Fermi energy of electrons reaches values large enough to trigger electron capture on nuclei so that the core develops a neutron excess. 
Neutrino emission from thermal processes and from nuclear electron captures become the main cooling mechanisms at these stages, lowering the entropy of the stellar core and causing the stellar core to deleptonize. 
The processes involved give rise to the production of low energy, sub-MeV to few MeV neutrinos.

\subsubsection{Stellar core collapse} 
The aforementioned neutrino losses  
cause the continuous deleptonization of the stellar core, which keeps the central entropy per baryon at a low value of few $k_{\rm B}$. 
The associated decrease of the electron abundance, $Y_e\equiv n_e/n_b$ where $n_e$ and $n_b$ are the number density of electrons and baryons, results in the loss of pressure support, as the partly degenerate electron gas provided the dominant contribution to the pressure and hence stability of the stellar core. 
The subsequent adiabatic contraction of the stellar core cause the continuous rise of the central temperature and density, 
such that the nuclear composition is dominated by iron-group nuclei which are in nuclear statistical equilibrium at $T\gtrsim 6$~GK $(0.5~{\rm MeV})$.
At these high temperatures, in addition to neutrino losses from electron capture on nuclei, a second process contributes to destabilize the core, namely the photodisintegration of heavy nuclei. 
The photodisintegration leads to the presence of free nucleons and alpha nuclei. 
Since electron captures on free protons are significantly faster than electron captures on heavy iron group nuclei \cite{juoda}, this might accelerate the deleptonization of the contracting stellar core.
However, modern supernova simulations indicate that the abundance of free protons stays low during the collapse~\cite{langanke03}.

\begin{figure}[htp]
\centering
\includegraphics[width=0.46\columnwidth]{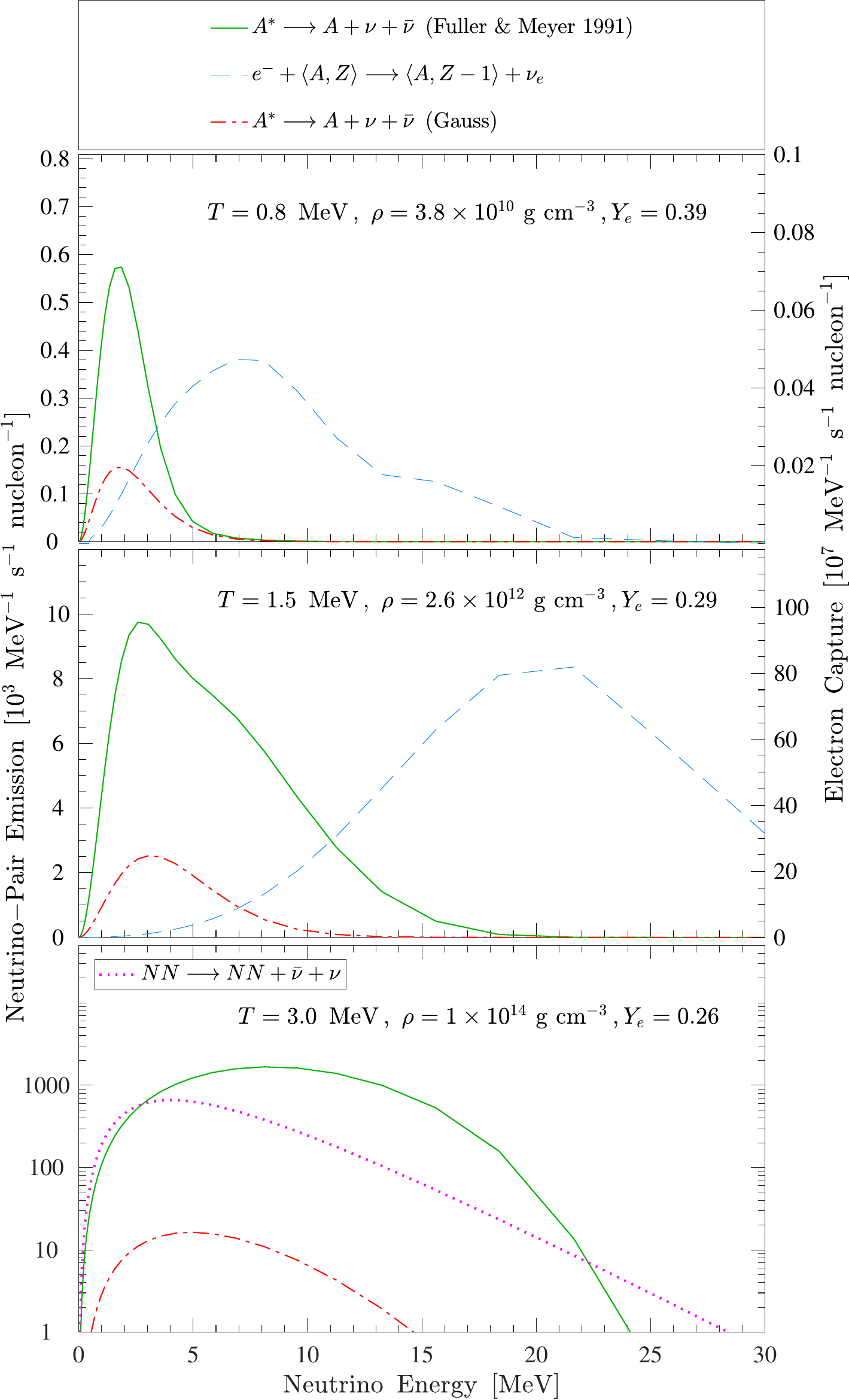}
\caption{(Figure reproduced from Ref.~\cite{Fischer13}.) Neutrino emission spectra from  electron captures on heavy nuclei~\cite{juoda} (blue dashed lines) and from de-excitations of heavy-nuclei~\cite{Fischer13}, comparing two different approximated treatments of the nuclear de-excitation function, from Ref.~\cite{Fuller:1991}  (green solid line) and assuming a Gaussian distribution (red dash-dotted line), at low and intermediate temperatures and densities (top and middle panels) and at high temperatures/densities (bottom panel). In the bottom panel the $\nu\bar\nu$-emission rate from the nucleon-nucleon ($NN$) bremsstrahlung is shown (magenta dotted line) instead of nuclear electron captures.  Note that the scales (right axis) for $\nu_e$ from electron captures are different from the pair production spectrum (left axis). All spectra shown here are for a single neutrino species. }
\label{fig:vspectra}
\end{figure}

Nuclear de-excitation by neutrino pair emission has been proposed as an additional channel of neutrino losses in Ref.~\cite{Fuller:1991}. 
This idea was revisited in Ref.~\cite{Fischer13} and applied to stellar core collapse simulations, implementing the rates of Ref.~\cite{Fuller:1991} 
and a simplistic rate based on Gaussian strength functions. 
The latter also takes into account the first forbidden transitions.
Fig.~\ref{fig:vspectra} compares these neutrino pair emission rates with electron captures on nuclei at low and intermediate temperatures and densities (top and middle panels) as well as to the nucleon--nucleon bremsstrahlung at high density (bottom panel). 
The de-excitation process greatly enhances the neutrino luminosities of  
$\bar\nu_e$ as well as the heavy lepton flavor neutrinos $\nu_{\mu/\tau}$ and their antineutrinos  
during the stellar collapse phase \cite{Fischer13}, for which the only source otherwise is the electron-position annihilation. 
However, the entire collapse phase is still dominated by nuclear electron captures, whose rates are significantly higher (as shown in Fig.~\ref{fig:vspectra}), and hence the electron neutrino luminosity is larger by about 4 orders of magnitude. 

\subsubsection{Supernova core-bounce and post-bounce evolution}\label{sec:SNpostbounce}
The stellar core collapse halts due to nuclear repulsion when the core density exceeds the saturation density at $n_{\rm sat}\simeq 0.15~{\rm fm}^{-3}~(\rho_{\rm sat}\simeq 2.4\times 10^{14}~{\rm g}~{\rm cm}^{-3})$.
During the collapse, a phase transition  similar to the gas-liquid transition, from inhomogeneous matter -- conditions at which heavy nuclei and unbound nucleons coexist -- to homogeneous nuclear matter is completed around $n_{\rm sat}$. 
While the precise density value of this Mott transition is still an active subject of research,  common approaches modeling this transition include the first-order phase transition with the Maxwell constructions~\cite{LSEOS}, the geometric excluded volume approach~\cite{Typel10,HS,Fischer2016EPJA} and recently the development of an effective model taking into account medium modified nuclear binding energies, consistent with quantum statistical calculations for light nuclear clusters based on scattering phase shifts~\cite{Fischer20c}. 

During the collapse, the inner core contracts subsonically and becomes disconnected from the supersonic outer core at the sonic point.  
Consequently, when the nuclear repulsion halts the collapse of the inner core, the outer part 
is still falling inwards. 
This leads to the formation of a stagnation front, which first develops into a sound wave and later, when reaching the sonic point, becomes a hydrodynamic shock wave. This phenomenon is known as the core bounce. 
Following the core bounce, mass accretion continues with a rate reaching several $M_\odot$ per second for $\mathcal{O}(10)$~ms and results in the formation of a compact central object known as the proto-neutron star (PNS)\footnote{The exact mass accretion rate and duration can sensitively depend on progenitor mass and metallicity.}. 

\begin{figure}[htp]
\centering
\includegraphics[width=0.6\columnwidth,angle=-90]{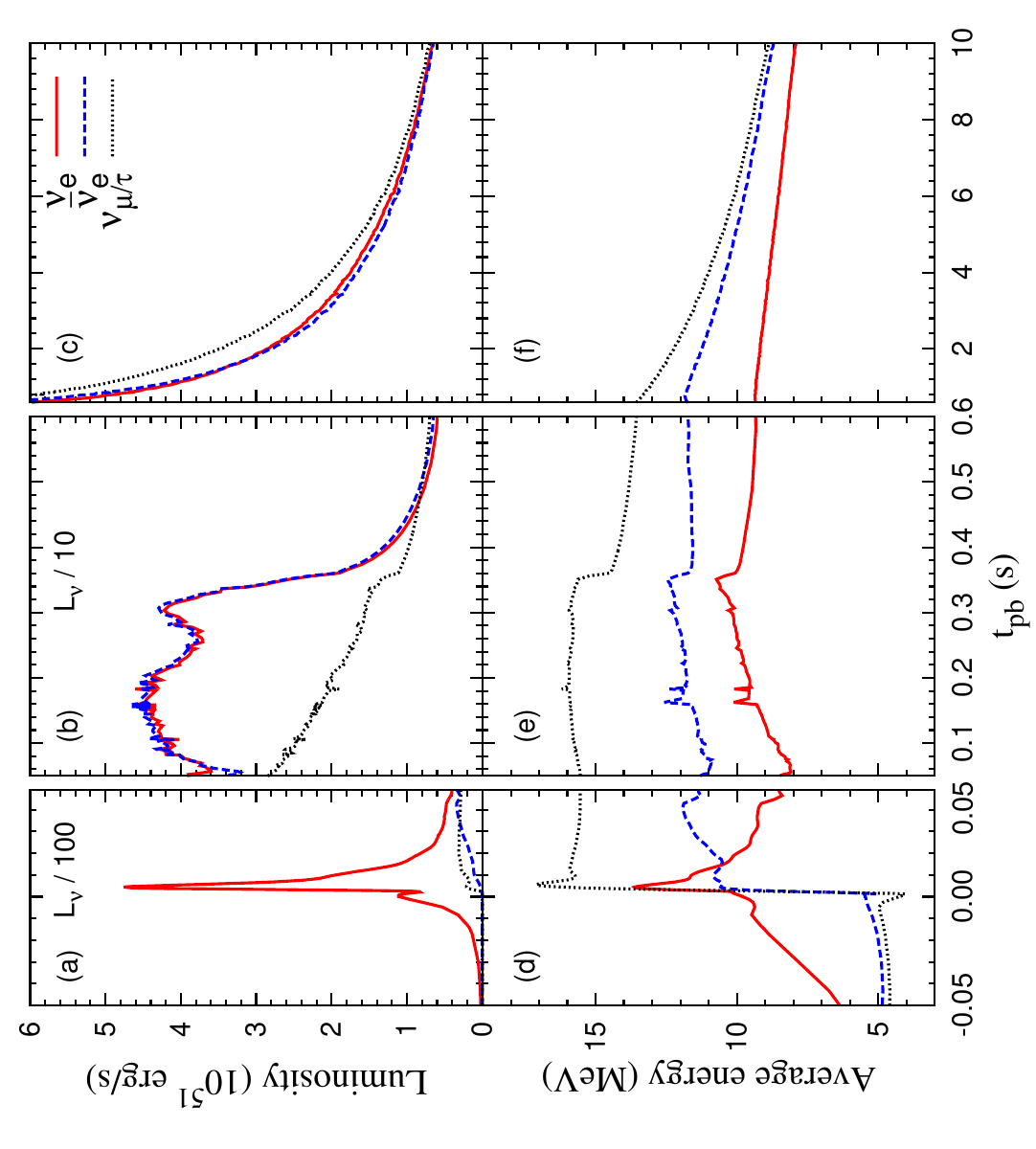}
\caption{(Figure reprinted from Ref.~\cite{Wu:2014kaa}.) Evolution of neutrino luminosities and average energies as a function of the post-bounce time, $t_{\rm pb}$ obtained in a spherically symmetric SN simulation with an 18~$M_\odot$ progenitor in Ref.~\cite{Fischer09}.}
\label{fig:neutrinos}
\end{figure}

The neutrinos emitted during the collapse can no longer escape the core freely when the density reaches $\rho \sim 10^{12}$~g~cm$^{-3}$, corresponding to temperatures of a few MeV, and become trapped by the nuclear absorption processes as well as their coherent scatterings with nuclei. 
This effect of neutrino trapping is visible as the $\nu_e$ luminosity slightly dips right before the core bounce shown in Fig.~\ref{fig:neutrinos}. 
In general, there is a variety of weak processes operating to produce neutrinos of different flavors, in addition to those associated with nuclear electron captures and nuclear de-excitations. 
In particular, the neutrino pair reactions, dominantly the electron-positron annihilation and at higher baryon density also the nucleon-nucleon bremsstrahlung, produce heavy lepton flavor neutrinos and antineutrinos. 
The $\bar\nu_e$ luminosity rises  
slightly slower than $\nu_{\mu/\tau}$ 
due to the final state blocking of $\nu_e$, which suppresses their production via the pair processes. 
Notice that muonic weak reactions, including the charged current and purely leptonic processes, already give rise to the non-negligible abundance of $\mu^-$ and $\mu^+$ at core bounce, once high energy $\nu_\mu$ and $\bar\nu_\mu$ ($\gtrsim 50-100$~MeV) are present. 
However, their abundances remain small on the order of $\sim 10^{-4}$ and do not affect the dynamical evolution at this stage~\cite{Fischer20d}. 

The bounce shock expands rapidly outwards, suffering energy loss of about 8~MeV per baryon from the dissociation of infalling heavy iron-group nuclei into nucleons.
Furthermore, when propagating across the neutrinospheres, a large number of electron neutrinos due to electron captures on protons made by photo-dissociation are released producing the so-called $\nu_e$ burst, as shown in Fig.~\ref{fig:neutrinos} which lasts about 20--50~ms. 
As a result, the electron fraction drops rapidly in that region from initially about $Y_e\simeq 0.3$ to $Y_e\simeq 0.15$ after the launch of the $\nu_e$ burst, which is often referred to as deleptonization or neutronization burst. 
These energy losses cause the dynamic shock wave to turn into an accretion shock front, which stalls at around $\mathcal{O}(100)$~km about $\sim 50$~ms post bounce. 
The later evolution is determined by mass accretion from the still infalling outer layers of the progenitor stars, as well as neutrino heating and cooling of the material located in the post shock layer, all of which would occur in the presence of convection and the possible existence of (magneto) hydrodynamic instabilities~\cite{Janka12}. 

\begin{figure}[thb]
\centering
\includegraphics[width=0.475\columnwidth]{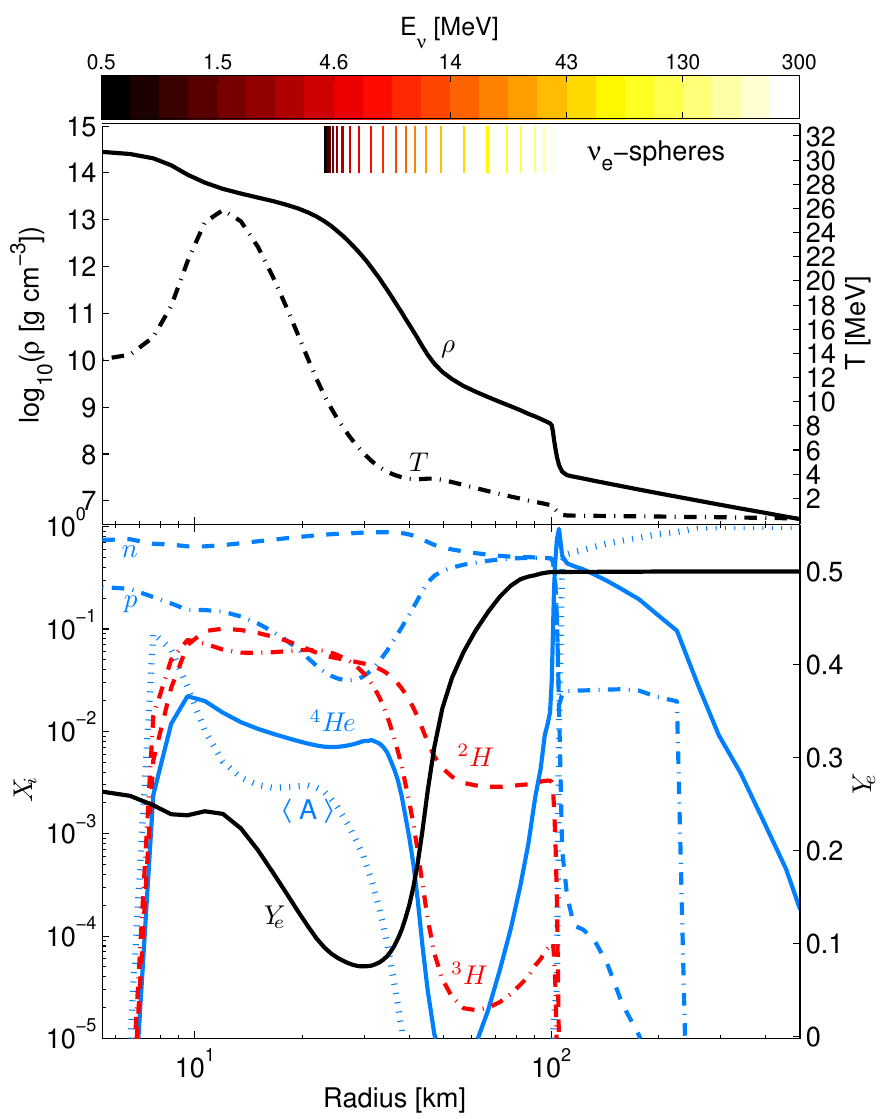}
\caption{{\em Top panel:} Radial profile of rest mass density $\rho$ (solid line) and temperature $T$ (dash-dotted line), showing in addition the neutrino energy dependent neutrinospheres at the top for $E_{\nu_e}=0.5-300$~MeV (color coded). {\em Bottom panel:} mass fractions $X_i$ of the most abundant nuclear species (blue and red lines) and the electron abundance $Y_e$ (black solid line). (Figure reprinted from Ref.~\cite{Fischer14})}
\label{fig:acc}
\end{figure}

The complex phenomenon of neutrino decoupling during this phase is illustrated in Fig.~\ref{fig:acc}, where the radial profile of mass density and temperature taken from a spherically symmetric simulation based on Boltzmann neutrino transport at about 300~ms post bounce are shown. 
Due to the energy dependence of the various neutrino reaction cross sections, neutrinos of different energies decouple at different radii, shown by the vertical lines in Fig.~\ref{fig:acc} for $\nu_e$ as an example, defined as where the corresponding optical depths equal to $2/3$~\cite{Fischer12}.

Low energy $\nu_e$, of about 0.5--1.5~MeV, decouple deep inside the PNS, at high densities and temperatures on the order of few times $10^{13}$~g~cm$^{-3}$ and $T\simeq 8-12$~MeV, while high energy $\nu_e$, of about 200--300~MeV, decouple only behind the accretion shock (i.e. the sharp density gradient located at about 100~km in Fig.~\ref{fig:acc}), at low densities and temperatures, on the order of $\rho\simeq 10^{8}-10^{9}$~g~cm$^{-3}$ and $T\simeq 1-3$~MeV.
Outside the shock wave, neutrinos are freely streaming and at the PNS interiors neutrinos are trapped.

During the post bounce evolution, neutrinos of all flavors are produced by various weak reactions (summarized in Table~\ref{tab:nu-reactions} below). 
However, since the post-shock layers have temperatures above 1~MeV,
the abundance of heavy nuclei and hence processes involving them are strongly suppressed. 
Furthermore, the dissociated nuclei give rise to free baryons through accretion from the infalling outer layers of the progenitor, which accumulate at the PNS surface, forming a thick layer shown in Fig.~\ref{fig:acc} behind the shock around the region of neutrino decoupling. 
There, the electronic charged current processes (1)--(3) in Table~\ref{tab:nu-reactions} dominate the production of $\nu_e$ and $\bar\nu_e$, while both charged current and the neutrino-nucleon scattering contribute to their decoupling~\citep{Raffelt01}. 
Their luminosities can be approximated as~\cite{Fischer09},
\begin{equation}
L_{\nu_e} \sim 1.5\times 10^{52}~\left(\frac{M}{1.5~{\rm M}_\odot}\right) \left(\frac{80~{\rm km}}{R}\right)\left(\frac{\dot{M}}{0.03~{\rm M}_\odot/{\rm s}}\right)~{\rm erg}~{\rm s}^{-1},
\label{eq:Lnu_mdot}
\end{equation}
for typical values of the PNS mass $M$ and radius $R$ as well as for a mass accretion rate $\dot M$. 
In contrast, heavy lepton flavor neutrinos are dominantly produced by pair processes, they decouple at generally higher temperatures and densities. Hence, their luminosity is determined by diffusion. 

\subsubsection{Neutrino-driven supernova explosions}\label{sec:n-driven-sn}
\begin{figure}[htb]
\centering
\includegraphics[width=0.75\columnwidth]{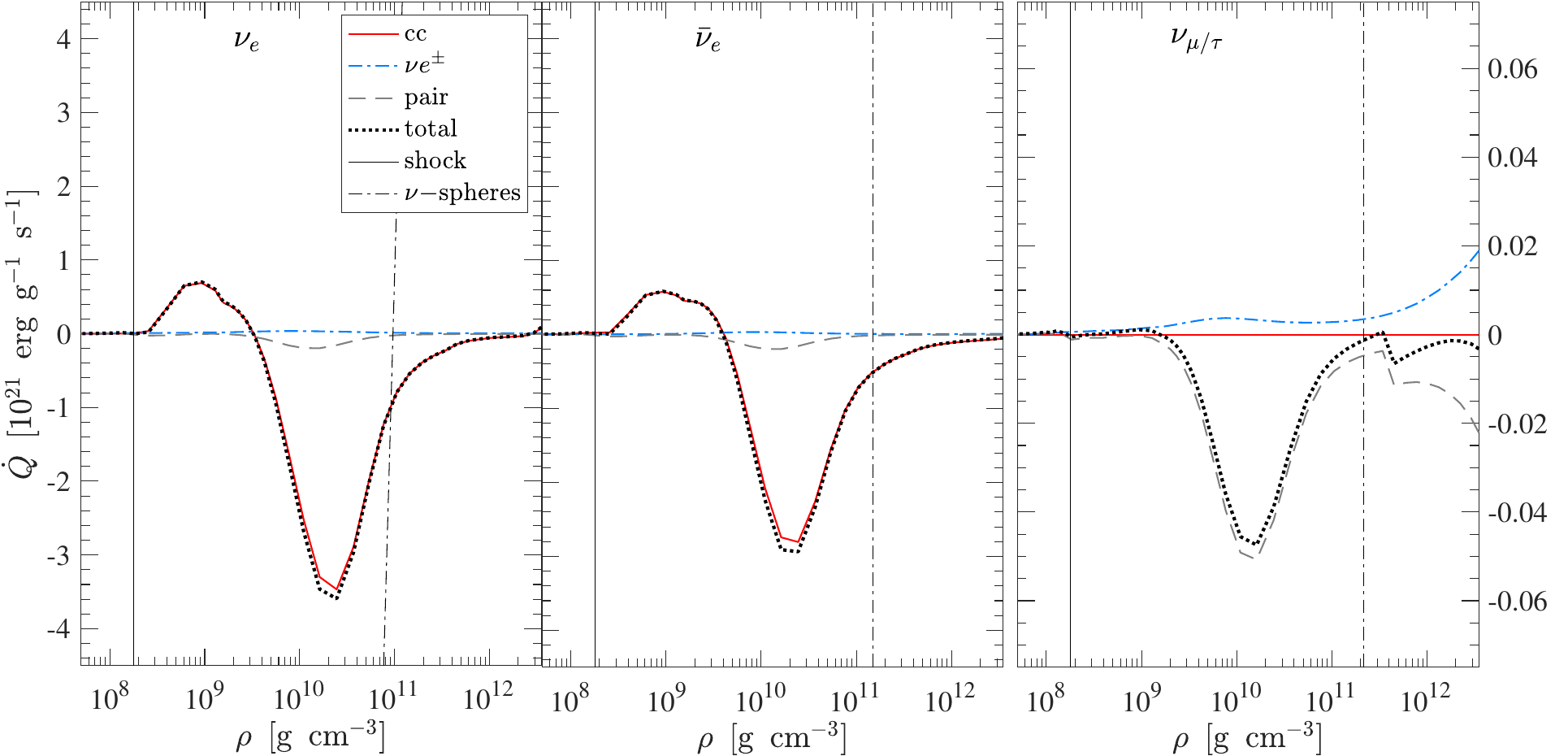}
\caption{Neutrino heating ($\dot{Q}>0$) and cooling rates ($\dot{Q}<0$) for all flavors, separated into charged-current  reactions, denoted as ``cc'' (red solid  lines), neutrino-electron scattering $\nu e^\pm$ (blue dash-dotted  lines) and pair processes (gray dashed lines), corresponding to a typical post bounce situation at about 250~ms. Thin vertical lines mark the locations of the shock (solid lines) and the corresponding neutrino spheres (dash-dotted lines). (CCSN simulation data from Ref.~\cite{Fischer20d})}
\label{fig:acc_heating}
\end{figure}

During the phase of shock stalling, the energy-dependent decoupling of neutrinos described above results in the development of regions between the shock and the PNS surface where net heating (with the specific heating rate $\dot Q>0$) and net cooling ($\dot Q<0$) from various reactions are obtained. 
Fig.~\ref{fig:acc_heating} shows a snapshot of $\dot Q$ as a function of the rest mass density ($\rho$) from the charged current interactions, neutrino electron/positron scattering ($\nu e^\pm$) and pair processes, for $\nu_e$, $\bar\nu_e$ and $\nu_{\mu/\tau}$ as representative for the heavy lepton flavors, taken from the 1D CCSN simulations of Ref.~\cite{Fischer20d}. 
These panels show that generally the $\nu_e$ and $\bar\nu_e$ charged current reactions dominate the cooling and heating in this region. 
Since the electron and positron capture rates on nucleons decrease much faster with declining density (and therefore temperature and radius) than the $\nu_e$ and $\bar\nu_e$ capture rates (see e.g., Ref.~\cite{Janka12}), net cooling (heating) develops at higher (lower) densities.  
As a result, there exists a gain radius (in 1D simulations) above (below) which the net $\dot Q>0$ ($\dot Q<0$) from all reactions is achieved for neutrinos to deposit (take) energy into (from) the medium. 

The heating of material in the post shock region by neutrinos results in the continuous rise of the entropy per baryon. 
This energy deposition is the source of the shock revival, turning the accretion front back into a dynamic shock wave which expands to increasingly larger radii. 
This concept is the physical foundation for the neutrino-driven SN explosion scenario, first reported in Refs.~\cite{Colgate66,Bethe85}. 
If successful, this determines the onset of the CCSN explosion, leaving behind the nascent hot and lepton rich PNS as remnant. 

In spherically symmetric CCSN simulations, self-consistent neutrino-driven explosions could be obtained only for the low-mass progenitors with the ZAMS masses of about 8--10~$M_\odot$~\cite{kitaura06}.  
With these low-mass progenitors, the shock revival due to neutrino heating occurs shortly after shock stalling, on the order of less than 100~ms after bounce, due to a sharp density gradient separating the stellar core from the low-density envelope. 
The resulting SN explosions generally have low explosion energies in 1D or in multi-dimensional simulations, about $10^{50}$~erg with ejected amount of nickel possibly ranging between $\mathcal{O}(10^{-3}-10^{-1})$~$M_\odot$~\cite{Wanajo09,Wanajo.Mueller.ea:2018,Burrows:2019rtd,Zha22}. 
However, quantitative and important differences between multi-dimensional simulations and spherically symmetric cases exist, particularly concerning the explosive nucleosynthesis. 
In multi-dimensional simulations, it may allow for the existence of neutron-rich convective lumps which are only mildly exposed to neutrinos that are absent in the corresponding 1D simulations~\cite{Wanajo11}. 
The explosive nucleosynthesis in these inner layers have recently been studied based on two or three dimensional CCSN simulations \cite{Wanajo.Mueller.ea:2018,Wang:2023vkk}, with the production of elements up to zirconium including possibly some $p$-nuclei such as $^{92}$Mo. 
However, the final answer about the contribution of innermost CCSN ejecta to the nucleosynthesis will require 3-dimensional simulations with accurate neutrino transport, given the significantly different dynamics observed in 2D and 3D models \cite{Bollig.Yadav.ea:2021}.

Neutrino-driven explosions of more massive iron-core progenitors require multi-dimensional simulations, where the neutrino heating efficiency increases in the presence of convection and the development of hydrodynamic instabilities, such as the standing accretion shock instability (SASI) \cite{Fogli09}, exposing material in the heating region to neutrino heating more efficiently. 
The success of multidimensional SN simulations has been demonstrated in the past decades 
for neutrino-driven explosion models. 
These neutrino radiation hydrodynamics models typically implement different approximated but sophisticated Boltzmann neutrino transport schemes to capture the important role of neutrino heating and cooling. 
In the following, we briefly describe (in alphabetical order) the assumptions employed in several state-of-the-art multi-dimensional SN simulation codes developed by different research groups.
\begin{itemize}
\item The {\tt 3DnSNe} three-dimensional, magnetohydrodynamics simulation code of Ref.~\cite{Matsumoto2024MNRAS528} is based on the spherical coordinates. It employs the Isotropic Diffusion Source Approximation (IDSA) of Ref.~\cite{Liebendorfer09}, which treats separately trapped and freely streaming neutrinos at the level of spectral distributions, carefully matched via a source term. The module of 3DnSNe implements a complete set of neutrino-matter interaction rates~\cite{Kotake18}. The intermediate computational costs of the IDSA enable detailed qualitative comparison of the spatial dimensionality dependence of the SN simulations~\cite{Takiwaki14}.
\item The {\tt 3DGRMHD} fully general relativistic three-dimensional neutrino-radiation magneto-hydrodynamics model in Cartesian coordinates~\cite{KurodaT16}. 
It is based on a spectral two-moment M1 approach, where both the neutrino radiation energy density (zeroth angular moment) and the momentum density (first angular moment) are evolved, with M1 tensor closure for the second and third moments of the neutrino radiation fields~\cite{Shibata11}.
{\tt 3DGRMHD} enabled the systematic quantification of the impact of magnetic fields, in combination with stellar rotation, in neutrino-driven SN explosions~\cite{KurodaT20,Shibagaki2023arXiv230905161S}. 

\item The neutrino transport of the {\tt CHIMERA} Oak Ridge code extends the spherically symmetric multi-group flux limited diffusion scheme of Ref.~\cite{Bruenn85} in multiple spatial dimensions (see Ref.~\cite{Bruenn20} and references therein, where also the details about the flux limiter are provided). 
{\tt CHIMERA} has a flexible hydrodynamics module that solves the Newtonian equations, w/o approximate general relativistic monopole component, in Cartesian, polar or spherical coordinates, where as the transport module implements the ray-by-ray+ treatment and hence requires spherical coordinates. 
A representative set of {\tt CHIMERA} axially symmetric and three-dimensional simulations can be found in Refs.~\cite{Bruenn16,lentz15}.
\item The neutrino transport of the {\tt FLASH-M1} open-source code 
implements the {\tt $\nu$GR1D} energy-dependent M1 scheme~\cite{O'Connor14}, via an analytic closure using the Eddington factor, that interpolates between the optically thick and optically thin regimes.
\item The {\tt FORNAX} Princeton simulation code 
solves multi-group two-moment, velocity-dependent transport equations keeping terms up to order $\mathcal{O}(v/c)$ in co-moving frame in spherical coordinates, with analytic M1 tensor closure for the second and third radiation moments~\cite{Skinner19}. 
The neutrino transport has been compared quantitatively with the ray-by-ray+ treatment in core-collapse SN simulations in axial symmetry~\cite{Skinner16}.
{\tt FORNAX} yields a comprehensive catalogue of self-consistent axially symmetric and three-dimensional core-collapse SN explosions (see Refs.~\cite{Burrows2021Natur,Vartanyan2022MNRAS510,Vartanyan2023MNRAS526} and references therein).
\item The {\tt VERTEX} Garching simulation code 
implements the ray-by-ray+ approach~\cite{Buras06b}, featuring a two-moment neutrino transport scheme in spatial angular bins with minimal coupling. The system of equations is closed by a variable Eddington factor obtained by solving model Boltzmann equations iteratively given hydrodynamic profiles along each radial ray and being coupled locally to the spherical solution of that ray.
Besides this advantage over other multi-group transport schemes, implementing an approximate angular dependence, the ray-by-ray+ approach neglects transverse fluxes and lateral transport, however, non-radial advection of neutrinos as well as the non-radial neutrino pressure is coupled to the radiation hydrodynamics equations.
Axially symmetric and three-dimensional {\tt VERTEX} simulation results of SN explosions for a variety of non-rotating massive stars can be found in Ref.~\cite{Summa16,Bollig21}, whereas the role of rotation has been studied in Ref.~\cite{Summa18}.
The {\tt VERTEX} ray-by-ray+ results are compared with full multi-dimensional, multi-group neutrino transport, based on an analytical closure scheme of the moments equations, finding quantitative good agreement in 3D simulations besides natural stochastic variations known to occur due to multi-dimensional phenomena in turbulent environments \cite{Glas2019ApJ873}. \\
\end{itemize}
We note that in the above models, {\tt CHIMERA}, {\tt FLASH-M1}, {\tt FORNAX} and {\tt VERTEX} implement general relativistic contributions into the transport schemes in an approximate fashion through an effective GR potential~\cite{Marek06}. 
Also noted is that a qualitative comparison of the aforementioned transport schemes in 1D SN simulations, including in addition results of {\tt AGILE-BOLTZTRAN}, which is based on general relativistic neutrino radiation hydrodynamics in 1D and implements Boltzmann neutrino transport~\cite{Mezzacappa93a,Mezzacappa93b,Mezzacappa93c,Liebendorfer04}, has been presented in Ref.~\cite{O'Connor18_GlobalComparison}.
Moreover, implementation of full Boltzmann neutrino transport in axial symmetry has been developed by Nakamura~et~al.~\cite{Nagakura14,Nagakura17}. 
Although at present it is applied only to hydrostatic background and under special relativity due to the high computational cost~\cite{Nagakura18}, 
it enables the quantitative comparison with the aforementioned approximate multi-dimensional neutrino transport schemes~\cite{Richers:2017}. 

A general caveat that all present successful multi-dimensional neutrino-driven CCSN explosion models face is the obtained low explosion energies, more precisely the diagnostic explosion energy (see e.g., definition in Sec.~2.4 of Ref.~\cite{Fischer10}) on the order of a few times $10^{50}$~erg.
This is somewhat in tension with the observed SN explosions for which canonically explosion energies on the order of a few times $10^{51}$~erg are deduced \cite{Nomoto06}. 
This deficit might be related to the insufficient simulation duration that current SN simulations can consider
due to computational limitations; c.f. Refs.~\cite{Bollig.Yadav.ea:2021,Wang:2023vkk}, or alternatively the lack of some incompletely understood input physics in the models; see, e.g., Sec.~\ref{sec:osc}.
A possible boost of the explosion energy, in agreement with observations, could also be obtained when considering alternative explosion mechanism, which will be discussed below.

Weak processes involving muons have to be considered once the conditions of temperature and density inside the PNS are such that a substantial amount of $\nu_\mu$ have energies in excess of the rest mass of muons, $m_\mu=105.66$~MeV, and/or the chemical potential of the electrons exceeds this value.  
The involved processes not only include the muonic charged current processes and (anti)neutrino-muon scattering but also the set of purely leptonic reactions, such as the muon decay and its inverse as well as lepton flavor exchange and conversion processes (see Table 1 in Ref.~\cite{Fischer20d}). 
Simulations including this set of muonic weak reactions into the collision integral of the neutrino transport equation show that muons are produced from the absorption of high energy muon-(anti)neutrinos on (protons)neutrons, shortly before bounce. 
During the post-bounce evolution, the presence of (anti)muons replaces (positrons)electrons, which has a softening effect but leaves a small impact on the evolution of the PNS, featuring slightly higher central rest mass density and a smaller radius. 
However, it can result in an enhancement of the neutrino heating efficiency in the gain region in multi-dimensional simulations, enabling neutrino-driven shock revival in some cases \cite{bollig17}. 

If the shock revival fails, then mass accretion onto the PNS will continue until the maximum mass, given by the nuclear equation of state (EoS), is reached and the PNS collapses into a BH, swallowing eventually the entire progenitor star. 
The failed SN explosion scenario has long been studied in the context of neutrino emission, featuring a sudden termination of neutrino emission at the moment of BH formation~\cite{Sumiyoshi07,Fischer09,O'Connor11}. 
However, the actual moment of neutrino termination is defined when the neutrinospheres fell through the event horizon (more precisely the apparent horizon), which has recently been demonstrated in Ref.~\cite{Rahman22}. 

There are additional proposals for alternative SN mechanisms. 
If a rotating collapsing star boosts an extremely strong magnetic field, the magnetic pressure along the rotational axis supports the shock expansion leading to a jet-like explosion in what is called a magneto-rotational SN~\cite{Bisnovatyi-Kogan70,LeBlancWilson70,Walder12,Wada2015,Grunhut17,Winteler12}. 
Another alternative scenario invokes a first-order phase transition of hadronic matter to a quark-gluon plasma at high densities~\cite{Sagert09}.
Related to this transition, a second shock wave is produced moving with ultra-relativistic velocities triggering the explosion~\cite{Fischer11,Fischer18,Zha20,KurodaT21,Jakobus:2022ucs,KhosraviLargani2023arXiv230412316K}. 
For heavy progenitor stars with rotations, which lead to BH formation on collapse, jets may be powered by accretion disk and result in energetic explosions named collapsars as a plausible mechanism for Type Ic SNe associated with long gamma-ray bursts~\cite{Woosley:1993wj,MacFadyen:1998vz,Woosley:2006fn, Kumar:2014upa}.
Other interesting possibilities such as the jittering jets explosion mechanism~\cite{Soker2010,Soker:2023mbr} or the collapse induced thermonuclear explosions~\cite{Kushnir:2015mca} have also been proposed. 
Notably, these alternative scenarios may result in different nucleosynthesis yields from those obtained in neutrino-driven SNe and are potential sites of $r$-process nucleosynthesis~\cite{Winteler12,Bugli21,Nishimura.Sawai.ea:2017,Reichert23,fischer20b,Siegel:2017nub}. 
We note here that the nucleosynthesis yields in magneto-rotational SNe are very sensitive to the magnitude of the magnetic-field enhancement during collapse due to the magneto-rotational instability. High magnetic fields favor a fast ejection of material that is barely exposed to neutrinos and allow for the production of heavy $r$-process elements~\cite{Obergaulinger.Reichert:2023}.

\subsubsection{Proto-neutron star deleptonization and the neutrino-driven wind}
\label{subsec:pnsevol}

In the case of a successful shock revival and the subsequent continuous shock expansion to increasingly larger radii, the mass accretion onto the central remnant PNS eventually ceases. 
Neutrino emissions from the nascent PNS lead to the subsequent PNS cooling and deleptonization for $\mathcal{O}(10)$~s, which is consistent with the detection of $\bar\nu_e$ from SN1987A \cite{hirata87}. 

\begin{figure}[htp]
\centering
\includegraphics[width=0.8\columnwidth]{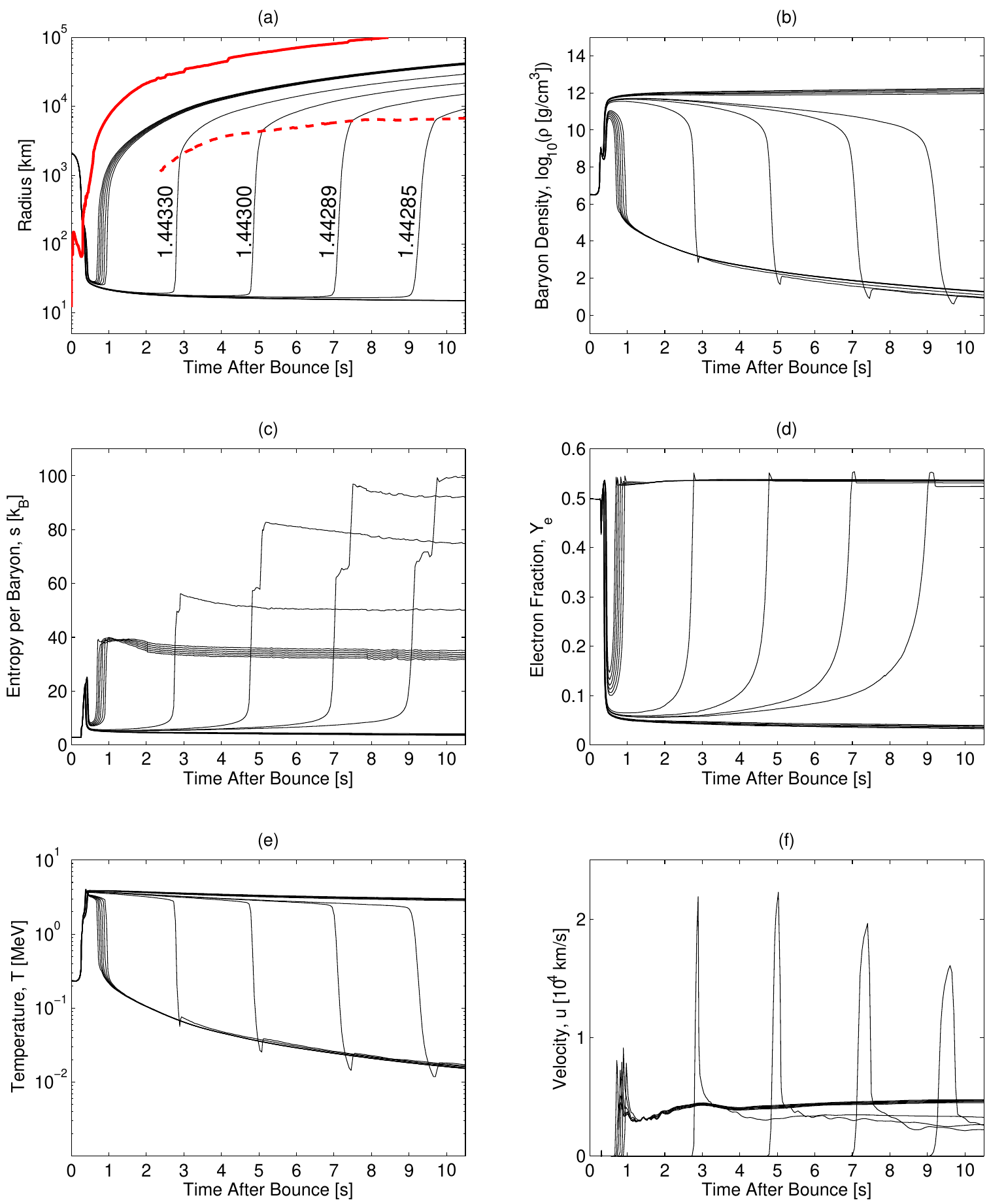}
\caption{(Figure reprinted from Ref.~\cite{Fischer10}) Evolution of dynamic and thermodynamic properties for a selection of mass tracers during the neutrino-driven wind phase of a parametrized neutrino-driven CCSN explosion launched from 11.2~$M_\odot$ progenitor. 
Credit: Fischer et.~al., A\&A, 517, A80, 2010~\cite{Fischer10}, reproduced with permission $\copyright$ ESO.}
\label{fig:NDW-tracers}
\end{figure}

An important aspect in this phase is the energy deposition of the decoupling neutrinos in the surface layer of the PNS, which drives a continuous low-mass outflow, known as the neutrino-driven wind~\cite{Duncan86,Woosley94,Witti94,Qian96,Thompson01b}. 
Roughly speaking, it requires about 7--10 (anti)neutrino absorption processes for (anti)neutrinos with average energies between 10--15~MeV, to overcome the gravitational binding energy of a baryon of about 104~MeV located at the surface of a PNS with typical mass of 1.5~$M_\odot$. 
The neutrino-driven wind can also expand with supersonic velocities and collide with the slower moving material ejected by the SN explosion.
If this happens, a reverse shock known as the neutrino-driven wind termination shock forms and can affect the thermodynamical properties in the wind~\cite{Janka95,Burrows95,Arcones07}. 
Fig.~\ref{fig:NDW-tracers} shows the time evolution of several selected mass tracers of the neutrino-driven wind\footnote{These are typically selected in a post-processing fashion in spherically symmetric models, while they have to be predefined in multi-dimensional simulations.} from an 1D model~\cite{Fischer10} to illustrate the general wind properties, including the dynamical timescale, shown here via the velocities in panel~(f), the entropy of the expanding wind in panel~(c), the $Y_e$ in panel~(d), together with the radial evolution in panel~(a) as well as the corresponding density and temperature in panel~(b) and (e). The solid and dashed lines in panel~(a) mark the locations of the bounce shock, which expands continuously to increasingly larger radii, and the reverse shock, respectively.

The early studies of the neutrino-driven wind based on neutrino radiation hydrodynamics~\cite{Woosley94,Witti94} simulations or by the steady-state wind models~\cite{Qian96,Thompson01b} suggested that heavy-element nucleosynthesis via the $r$-process may be obtained due to the moderate neutron richness of $Y_e\sim 0.38-0.5$ and high entropy per nucleon $\sim 150-400$~$k_B$ in the wind. 
However, more recent long term 1D simulations carried out to more than 10 seconds post bounce with accurate Boltzmann neutrino transport~\cite{Huedepohl10,Fischer10} as well as multi-dimensional simulations lasting for several seconds with approximated neutrino transport scheme~\cite{Stockinger:2020hse,Bollig21,Witt21,Wang:2023vkk} found that the physical conditions in the wind do not enable the $r$-process nucleosynthesis.
Instead, the composition in the wind may be proton-rich [$Y_e\gtrsim 0.5$; see e.g., Fig.~\ref{fig:NDW-tracers}(d)] and may allow the $\nu p$ process~\cite{Froehlich06b,Pruet06} to occur, which will be discussed further in Sec.~\ref{subsec:ndw_nucl}.

The modeling of neutrino decoupling also plays a key role in determining the wind properties and the associated nucleosynthesis conditions. 
During the PNS deleptonization, the PNS contracts due to the neutrino cooling, which leads to a steeper density profile around the PNS surface (see e.g., Fig.~1 in Ref.~\cite{Fischer12}).
In comparison to the post-bounce mass accretion phase prior to the shock revival, shown in the top panel of Fig~\ref{fig:acc}, the energy-dependent neutrinosphere locations shift to higher densities around $10^{12}-10^{13}$~g~cm$^{-3}$. 
The subsequent evolution of the neutrino luminosities and average energies are shown in Fig.~\ref{fig:neutrinos}, during the PNS deleptonization phase up to about 10~s. 
It shows the continuously decreasing fluxes and average energies for all flavors as neutrinos decouple at higher densities (lower temperatures) at later times~\cite{Fischer12}. 
Comparing the earlier accretion phase to the later deleptonization phase, an important change is that a scattering atmosphere, which existed for heavy lepton flavor neutrinos at earlier phase~\cite{Raffelt01,Keil03}, also develops for $\nu_e$ and $\bar\nu_e$.  
This is because the increasingly dominant role of (nearly) elastic neutrino--nucleon scattering over the inelastic charged current processes, as the latter is further suppressed by final state blocking~\cite{Fischer12}. 
Since neutrino--nucleon scattering is flavor blind, the neutrino spectra and fluxes of all flavors become increasingly similar during the PNS deleptonization, as shown in Fig.~\ref{fig:neutrinos}. 
Although this general trend is robust and insensitive to the detailed treatment of neutrino interactions with matter around the neutrinosphere, precise prediction of neutrino luminosities and average energies are needed to accurately model the nucleosynthesis condition and yields in the neutrino-driven wind, as will be the focuses of Secs.~\ref{sec:CC} and~\ref{subsec:ndw_nucl}.

Another important point to note here is related to the multidimensional nature of the SN explosion. 
Recent multidimensional simulations carried over to several seconds post the core bounce found the mass accretion can persist for seconds after the onset of explosion, especially for explosions that are more aspherical with heavier progenitors, which 
might alter the appearance of a spherically symmetric neutrino-driven wind~\cite{Stockinger:2020hse,Bollig21,Nagakura:2021MNRAS,Witt21,Wang:2023vkk} assumed in 1D models.
Consequently, the clean separation of the explosion and PNS cooling phase obtained in 1D models, which is reflected in the rapid drop of the neutrino fluxes shown in Fig.~\ref{fig:neutrinos} becomes no longer possible in those cases, and the relation between the thermodynamical condition of the wind and the remnant property can be affected~\cite{Wang:2023vkk}. 
On the other hand, for light progenitors in the ZAMS mass range of 8--10~M$_\odot$ and depending on metallicity, where explosions are not too aspherical, spherical neutrino-driven wind solutions remain a relatively good approximation to results obtained in multidimensional simulations~\cite{Janka08,Sandoval:2021hnk,Nagakura:2021MNRAS500,Vartanyan2023MNRAS526}. 

The asymmetric nature of the SN explosions may further leave imprints on the morphology and chemical composition distribution in SN remnants.
Attempts to model the very late-time evolution of SN explosions until the shock break-out or even the remnant evolution in multi-dimensions have been performed with various assumptions~\cite{Wongwathanarat:2014yda,Orlando:2016jxx,Wongwathanarat:2016jvy,Utrobin:2018mjr,Muller:2018gok,Orlando:2019vdf,Ono:2019zhr,Jerkstrand:2020hlf,Orlando:2020igr,Sandoval:2021hnk,Witt21}. 
These studies revealed potentially intriguing dependence of observables of lightcurves and nuclear $\gamma$ lines on the progenitor model and structure, as well on the interaction of SN shock with the circumstellar medium, relevant for the observation of Cassiopeia~A and SN1987A.

\subsubsection{Presupernova and supernova neutrino detection}\label{sec:nu-det}

As mentioned above, massive stars are profuse sources of neutrinos even before they become SNe. During later stages of their evolution, their central temperature and density increase dramatically, and $\nu\bar\nu$ pair production by photo-neutrino emission, plasmon decay, and $e^\pm$ pair annihilation become the dominant mechanism of energy loss (see e.g., \cite{Itoh96,Guo16}). Likewise, $\nu_e$ and $\bar\nu_e$ production by weak nuclear processes 
become more and more significant as such stars evolve. These neutrinos not only provide cooling that drives the evolution of massive stars, but also serve as potential signatures of their evolution and advance warning for the ensuing SNe. With the capability of detecting low-energy $\bar\nu_e$ of several MeV by the gadolinium-loaded Super-Kamiokande (SK) detector \cite{SK:2022} and by the next generation of detectors such as the Jiangmen Underground Neutrino Observatory (JUNO) \cite{juno}, there is growing interest in detecting these presupernova neutrinos \cite{Kato20}. It is plausible to detect presupernova $\bar\nu_e$ emitted during the late oxygen burning or the silicon burning stages a few days before the explosion of a massive star within a few kpc \cite{Odrzywolek:2004a,Odrzywolek:2004b,Kutschera:2009,Odrzywolek:2010,kato15,asakura16,kato17,Patton:2017}. At a relatively close distance of $222^{+48}_{-34}$~pc \cite{distance}, Betelgeuse with a mass of $20_{-3}^{+5}\,M_\odot$ \cite{mass16} would be a candidate star. However, its current evolutionary status is uncertain. It is most likely undergoing core Helium burning according to Refs.~\cite{Joyce2020,Wheeler2023}, but it may already be in the late stage of core carbon burning as suggested by a recent study \cite{Saio2023}. In the latter case, Betelgeuse would be a promising source of presupernova neutrinos for detection by JUNO.

The main presupernova neutrino signals in JUNO result from $\bar\nu_e+p\to n+e^+$ (inverse $\beta$-decay, IBD) and $\nu + e^-\to\nu +e^-$ (electron scattering, ES).
Considering neutrinos produced by $e^\pm$ pair annihilation from a star at 222~pc during the last day prior to its core collapse and assuming the normal ordering (NO) of neutrino masses, 
Ref.~\cite{Guo19} estimated about 123.8, 243.5, 416.0 and 497.1 IBD events with neutrino energy of 1.8--4 MeV for progenitor models of 12, 15, 20, and 25 $M_\odot$, respectively.
The corresponding numbers of ES events with recoil electron kinetic energy of 0.8--2.5 MeV are 117.2, 212.9, 380.9, and 479.8, respectively.
The ratio of the expected number of IBD events for the inverted ordering (IO) to that for the NO is $N_{\rm IBD}^{\rm IO}/N_{\rm IBD}^{\rm NO}\approx 0.29$, and
the corresponding ratio for the ES events is $N_{\rm ES}^{\rm IO}/N_{\rm ES}^{\rm NO}\approx 1.23$ \cite{Guo19}. The high sensitivity of the IBD events to the yet-unknown neutrino mass ordering
is due to the Mikheyev-Smirnov-Wolfenstein (MSW) effect \cite{MSW78,MSW85}, which governs the flavor transformation during neutrino propagation through the stellar interior.
Note that while for each progenitor model, the number of events of each kind has substantial uncertainties (within a factor of $\sim2$) due to uncertainties in treatment of stellar evolution, 
the ratios $N_{\rm IBD}^{\rm NO}/N_{\rm ES}^{\rm NO}$, $N_{\rm IBD}^{\rm IO}/N_{\rm ES}^{\rm IO}$, $N_{\rm IBD}^{\rm IO}/N_{\rm IBD}^{\rm NO}$, and $N_{\rm ES}^{\rm IO}/N_{\rm ES}^{\rm NO}$ 
are nearly independent of the stellar models. 
Because $N_{\rm IBD}^{\rm NO}/N_{\rm ES}^{\rm NO}\approx 4.2N_{\rm IBD}^{\rm IO}/N_{\rm ES}^{\rm IO}$, it is possible to distinguish the neutrino mass ordering not only without knowing the exact mass of the SN progenitor but also allowing for significant uncertainties in the predicted numbers of IBD and ES events.
An important issue to consider is the backgrounds. For IBD events, the largest backgrounds come from reactor $\bar\nu_e$ and geo-$\bar\nu_e$, which contribute about 15.7 and 1.1 events, respectively, within the same time and energy windows as for the signals. These backgrounds will be well measured and are below the expected number of IBD events for a star like Betelgeuse (17--$25\,M_\odot$) even considering the uncertainties in stellar models. Unlike the IBD events, which can be identified by coincident detection of the positron and the neutron produced in the reaction, ES events only cause single hits in JUNO and suffer from much higher backgrounds of different origins. The dominant background, due to $\beta^+$ decay
of the cosmogenic $^{11}$C produced by $(\gamma, n)$ spallation following a muon shower, is at the level of a few $10^4$ events per day \cite{juno}. 
It is feasible to determine the neutrino mass ordering from the nearly model-independent ratio of the IBD and ES event numbers as discussed above when 
this background is suppressed by a factor of $\sim 2.5$--10 depending on the mass of the stellar source \cite{Guo19}. 
Such background reduction might be achieved by three-fold coincident detection of the muon initiating the shower, the neutron from the spallation production of $^{11}$C, and the positron from the $^{11}$C decay \cite{PhysRevC.71.055805}. 

Compared to presupernova neutrinos, neutrinos from the ensuing SN have much higher average energies and orders of magnitude higher luminosities. To date only neutrinos from a single event, SN 1987A at a distance of $\approx 50$~kpc, have been detected \cite{PhysRevLett.58.1490, PhysRevD.38.448,PhysRevLett.58.1494,ALEXEYEV1988209}. The approximately twenty neutrino events observed have been extensively studied to understand SN neutrino emission and neutrino properties (see e.g., \cite{arnett1989} for a review of earlier works and e.g., \cite{Loredo:2001rx} and \cite{Costantini_2007} for detailed analysis methodologies). 
While parametric models of neutrino emission were commonly used to extract several parameters from the relatively sparse SN 1987A data for comparison with the results from numerical SN models, it is more straightforward to compare the SN models directly with the data. Specifically, Bayesian techniques can be used to rank the SN models in the latter approach. Such analyses were carried out in Refs.~\cite{Olsen21,Olsen22a} using three baseline models from the Garching group \cite{Garching} with progenitors of 9.6, 20, and $27\,M_\odot$. These models are representative of current 1D SN simulations and cover a range of neutrino emission. Variation of the overall flux normalization and simple scenarios of neutrino oscillations were also explored. With the data on the 11 events observed in a time span of $\approx 12$~s in the Kamiokande II detector, it was found that models with a brief accretion phase of neutrino emission are the most favored \cite{Olsen21,Olsen22a}. This result is not affected by varying the overall flux normalization or considering neutrino oscillations. In addition, $p$-value tests of the predicted event distributions in energy and arrival time showed no evidence that any of the best-fit models are incompatible with the data \cite{Olsen21,Olsen22a}. Another study \cite{librc2023} focused on the first $\sim 0.5$--1.5~s of neutrino emission and examined a number of 1D, 2D, and 3D simulations. Given that the exact model for SN 1987A is not known and that even the state-of-the-art simulations need further improvements, perhaps it is not surprising that those simulations examined by Ref.~\cite{librc2023} generally disagree with the data. Yet another study \cite{fiorillo2023supernova} showed that the relatively short cooling timescales of $\sim5$--9~s resulting from current treatment of convection in the PNS are not
easily compatible with the duration of the SN 1987A neutrino emission except for rare fluctuations in the sparse data. Based on the above discussion, while some discrimination among the neutrino emission models could be provided by the SN 1987A data, no definitive conclusions can be drawn.

More useful will be a future neutrino signal from a Galactic SN, which is expected to produce a greater number of neutrino events in current and planned detectors (e.g., \cite{Scholberg}). 
Focusing on an early phase of neutrino emission spanning 500 ms, which is closely related to the explosion mechanism, the authors of Ref.~\cite{Hyper-Kamiokande:2021frf} performed a detailed study of the feasibility of distinguishing between five SN simulations (1D \cite{totani98,nakazato13b,couch20}, 2D \cite{vartanyan19}, 3D \cite{tamborra14}) using hypothetical neutrino data in the planned Hyper-Kamiokande (HK) detector (e.g., \cite{Hyper-Kamiokande:2018ofw}). 
Taking into account flavor oscillations due to the MSW effect, they performed detailed reconstruction of 100 and 300 simulated events, which were dominated by IBD contributions (ES and charged-current reactions of $\nu_e$ and $\bar\nu_e$ on $^{16}$O were also included). They concluded that all the five simulations could be distinguished with 300 events, which are expected from a SN at a distance of $\sim 60$--100 kpc based on these simulations. Further analyses similar to the above were carried out in Ref.~\cite{Migenda:2019xbm} to study the feasibility of distinguishing between four simulations (1D \cite{nakazato13b}) that differ specifically in the mass or initial metallicity of the progenitor. 
A complementary study was carried out in Ref.~\cite{Olsen22b} using seven 1D SN simulations from the Garching group \cite{Garching}, none of which were considered in Refs.~\cite{Hyper-Kamiokande:2021frf,Migenda:2019xbm}. Each of these simulations covers neutrino emission for at least 9~s and therefore, includes both the accretion and PNS cooling phases. 
These simulations correspond to progenitor masses of 9.6--$27\,M_\odot$, and predict $\sim 2500$--7150 ($\sim 3550$--9940) IBD events in a detector similar to SK (an idealized SK with perfect detection efficiency, perfect energy resolution, and no background) for a SN at 10 kpc.
It was found that for each of the three neutrino oscillation scenarios considered (no oscillations or standard MSW effect with the NO or IO), all the neutrino emission models can be distinguished from each other with the IBD signal in a detector similar to SK from a SN at a known distance up to $\sim 25$ kpc or at an unknown distance up to $\sim 10$ kpc. In addition, provided that the emission model is known, for example, from observations of the SN progenitor, the three oscillation scenarios can be distinguished from each other with an idealized SK
and a SN at a known distance of 10 kpc. Based on the analyses of Ref.~\cite{Olsen22b},
it should be quite possible to distinguish between 1D SN neutrino emission models using the IBD signal in HK from a future Galactic SN, at least when the same neutrino oscillation scenario is assumed for all the models.

\subsection{Binary neutron star mergers}\label{sub:second}

\begin{figure}[htb]
\centering
\includegraphics[width=\columnwidth]{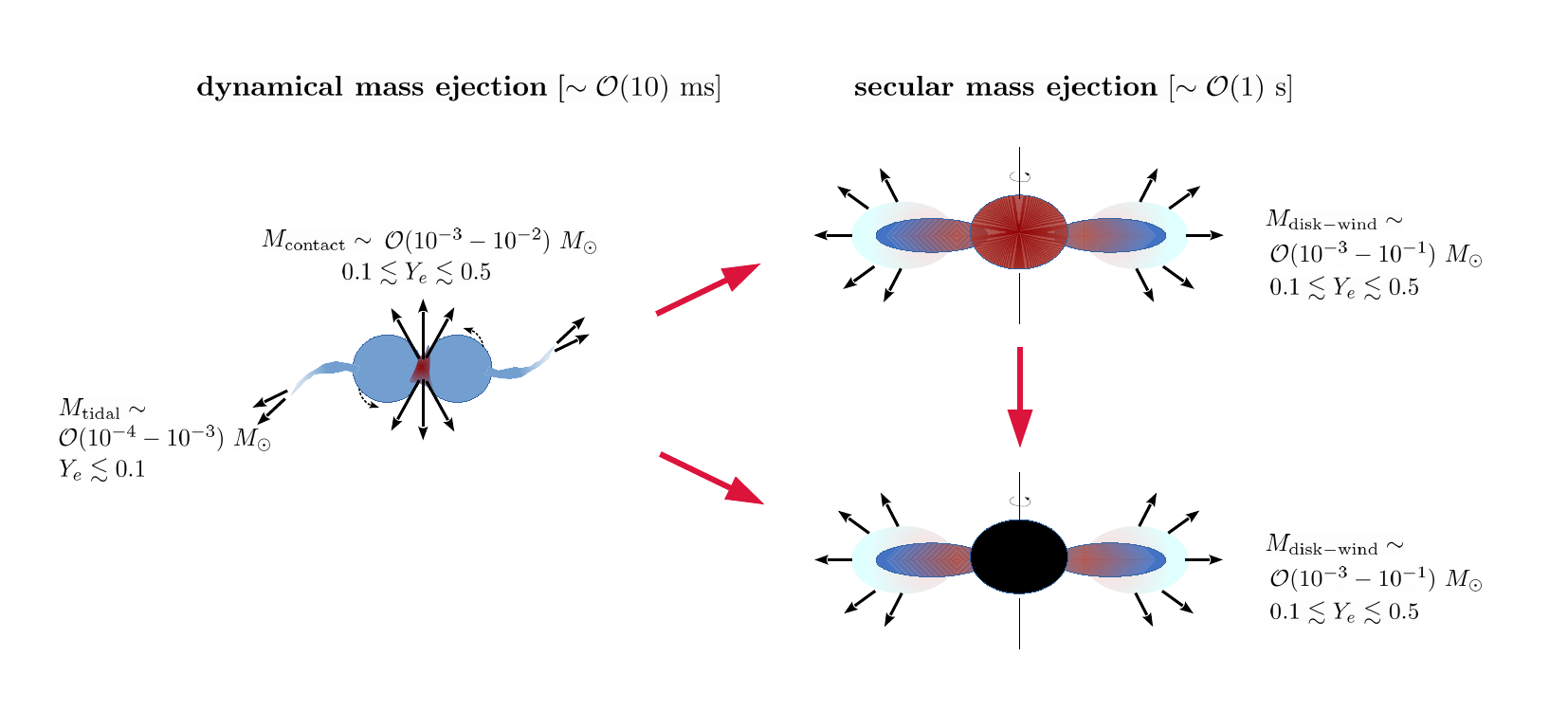}
\caption{Schematic plot showing the major mass ejection phases of BNSMs.
Left: Dynamical mass ejection during the merging of two NSs from tidal disruption and from the contact interface.
Right: Secular mass ejection from the accretion disk during the remnant phase, where the central object can be a HMNS (upper panel) or BH (lower panel).
Red arrows indicate the possible evolution paths while black arrows indicate the (angular) velocity of the object(s) or ejecta.
Neutrinos play important roles in both phases in affecting the $Y_e$ of the ejecta (see text for details).
}
\label{fig:BNSM_schematic}
\end{figure}

The mergers of two NSs or one NS with one BH were proposed to be promising sites of heavy element production through the $r$-process nucleosynthesis several decades ago~\cite{Lattimier1974,Symbalisty1982,Eichler1989}.
With the advances in numerical simulations, it became increasingly clear that they may even be the most dominating sites of $r$-process~\cite{Freiburghaus1999,Rosswog:1998hy,Goriely:2011vg,Fernandez:2013tya,Just:2014fka,Wanajo2014} and can produce unique electromagnetic emissions powered by the energy generation from the decay of unstable $r$-process nuclei, called kilonovae~\cite{Metzger10,Barnes:2013wka,Tanaka:2013ana}.
This association was further confirmed by the detection of kilonovae associated with short gamma-ray bursts~\cite{Tanvir:2013pia,Yang:2015pha,Jin:2016pnm} as well as the seminal discovery of GW170817 and the associated kilonova AT2017gfo directly linked to the BNSM~\cite{LIGOScientific:2017vwq,LIGOScientific:2017ync}. 
Meanwhile, studies also revealed that neutrinos can play substantial roles in the evolution of the BNSM remnant and in determining the nucleosynthesis yields in the associated outflows. 
In this section, we briefly outline the general picture of BNSM evolution and focus on recent findings related to neutrinos therein. 
Readers interested in detailed reviews on BNSMs, their GWs and kilonovae can refer to Refs.~\cite{Cowan:2019pkx,Shibata:2019wef,Metzger:2019zeh,Baiotti:2019sew,Nakar:2019fza}. 

\subsubsection{Merger and dynamical ejecta}
NS binaries (NSBs) can  originate from binary star systems that remain gravitationally bound after undergoing two SN explosions or from dynamic captures in dense stellar regions.
For an isolated NSB, its orbital separation shrinks over time due to the loss of angular momentum and energy by the emission of GWs. 
The typical inspiral timescale from an initial separation of $\sim\mathcal{O}(1)$~A.U. to their eventual merge is $\sim \mathcal{O}(100)$~Myr, depending on the system's initial eccentricity.  
For most systems, the orbits become quasi-circular before mergers occur. 

\begin{figure}[htb]
\centering
\includegraphics[width=0.45\columnwidth]{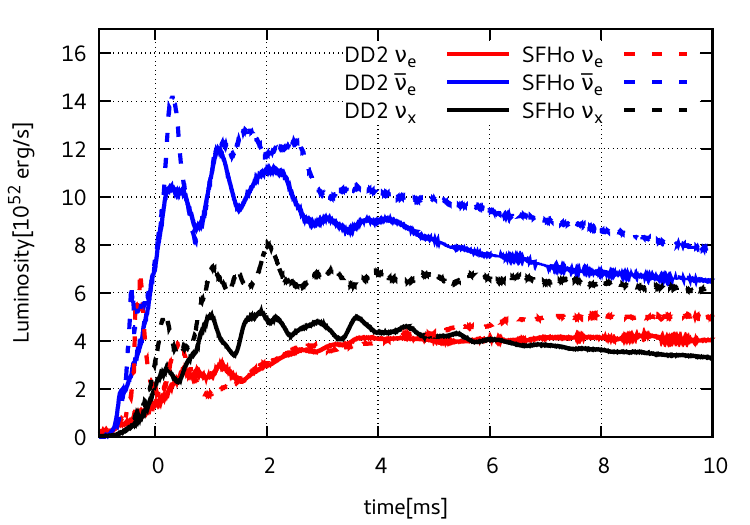}
\includegraphics[width=0.42\columnwidth]{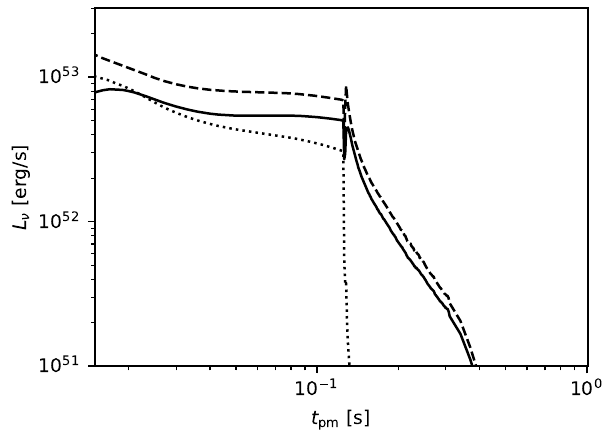}
\caption{Left panel (adapted from Ref.~\cite{George:2020veu}): The energy luminosity of different neutrino species during the early post-merger time of BNSMs from simulations done in \cite{Ardevol-Pulpillo:2018btx} with two NSs of $1.35$~$M_\odot$ each and two nuclear EoSs -- DD2~\cite{Typel10,HS,Hempel12} and SFHo~\cite{HS,SFH}. 
Right panel (courtesy of O. Just): The energy luminosity of $\nu_e$ (solid), $\bar\nu_e$ (dashed), and $\nu_x$ (dotted) during BNSM remnant evolution phase from the model sym-n1-a6 of Ref.~\cite{Just:2023wtj}.}
\label{fig:merger_nu_lumi}
\end{figure}

Shortly before the merger, the strong gravitational forces between the two NSs lead to the tidal disruption of materials at the NS surfaces.  
A part of these tidally disrupted material can become unbound and get ejected from the system while some part of them fall back to form an accretion disk (see Fig.~\ref{fig:BNSM_schematic}). 
The ejected mass is typically $\sim \mathcal{O}(10^{-4}-10^{-3})$~$M_\odot$ and distributed close to the merger plane.  
At merger, the violent collision of the two NSs (with relative velocity of $\sim 0.3$~c) leads to shock formation and can also efficiently eject materials from the contact interface of similar or even larger amount (see also Fig.~\ref{fig:BNSM_schematic}). 
Moreover, the temperature of the remnant rises significantly due to the collision, up to $\sim \mathcal{O}(50)$~MeV.  
Consequently, the heated matter seeks to reach a new thermodynamical equilibrium, during which the weak interactions (electron and positron captures) work to raise $Y_e$ and produce $\nu_e$ and $\bar\nu_e$ while $\nu_x$ (denoting $\nu_\mu$, $\nu_\tau$ and their antineutrinos collectively) are also generated via pair processes.   
The resulting total neutrino luminosity can reach $L_{\nu,\rm{tot}}\sim\mathcal{O}(10^{53})$~erg~$s^{-1}$, with $L_{\bar\nu_e}\gtrsim L_{\nu_e}\sim L_{\nu_x}$ (see e.g., the left panel of Fig.~\ref{fig:merger_nu_lumi} and Ref.~\cite{Cusinato:2021zin} for a compilation of neutrino emission property from BNSMs).
The dominant emission of $\bar\nu_e$ over other species is expected since the remnant must protonize for some period to reach the newly established equilibrium. 
This fact also has important consequence for the flavor conversion of neutrinos, which will be addressed in Sec.~\ref{sec:osc}. 
Consequently, charged-current neutrino absorption on nucleons further raise the $Y_e$ of the dynamical ejecta, particularly for the component closer to the polar direction (perpendicular to the merger plane)~\cite{Radice:2018pdn,Fujibayashi:2020dvr,Kullmann:2021gvo}.

The resulting $Y_e$ distribution obtained in recent simulations that included nucleonic weak interactions in dynamical ejecta typically ranges in between $0.05\lesssim Y_e\lesssim 0.5$ 
with an average value $\langle Y_e\rangle\sim 0.2-0.3$~\cite{Radice:2018pdn,Fujibayashi:2020dvr,Kullmann:2021gvo}.
This is in sharp contrast to the low $Y_e\lesssim 0.1$ obtained in earlier work without including weak interactions~\cite{Rosswog:1998hy,Goriely:2011vg}.
As a result of the broad $Y_e$ distribution, the $r$-process in the dynamical ejecta can produce a wide range of nuclei from the first peak up to the third peak and actinides~\cite{Wanajo2014,Radice:2018pdn,Fujibayashi:2020dvr,Kullmann:2021gvo}. 
The exact $Y_e$ distribution and the nucleosynthesis yield for a given set of binary parameter still depend on a number of factors that need to be further clarified, including the treatment of neutrino transport~\cite{Foucart:2022bth}, the EoS (that affects the post-merger temperature and the amount of ejecta)~\cite{Janka:2022krt}, as well as the flavor oscillations of neutrinos (see Sec.~\ref{sec:oscnucleo_bnsm}). 
Moreover, potential impact due to the existence of muon~\cite{Loffredo:2022prq} and/or pions~\cite{Fore:2019wib,Vijayan:2023qrt} in BNSMs remain to be clarified. 

Similar to studies that included the possible occurrence of hadron-quark phase transition in CCSNe (see the end of Sec.~\ref{sec:n-driven-sn}), recent works also found that a phase transition happening early in the post-merger phase can leave interesting imprints on the emitted GW signals~\cite{Most:2018eaw,Bauswein:2018bma,Prakash:2021wpz}.
These works seem to suggest that it does not significantly affect the amount of dynamical ejecta and the associated $r$-process nucleosynthesis~\cite{Bauswein:2019skm,Prakash:2021wpz}.

For BH-NS mergers, it requires a relatively low mass BH with $M_{\rm BH}\lesssim 10$~$M_\odot$ with relative high BH spin with dimensionless spin parameter $\chi\gtrsim 0.5$ to tidally disrupt the NS before it is being swallowed by the BH~\cite{Foucart:2020ats,Kyutoku:2021icp}. 
When these conditions are satisfied, mass ejection and $r$-process nucleosynthesis in both the dynamical ejecta as well as the post-merger secular winds can occur.  
Although the fraction of BH-NS mergers that contribute to the $r$-process production remains largely uncertain, it will be better constrained by the ongoing and future GW telescopes. 
For the dynamical ejecta of BH-NS mergers, neutrinos do not play a significant role for clear reason that there is no central neutrino emitting source. 
However, at the post merger phase, neutrino emission from the BH-NS post-merger remnants could have similar impact as in that of BNSM remnant case that will be discussed below.

\subsubsection{Secular evolution of remnant and its outflow}
After the early phase of dynamical mass ejection, the remnant of a BNSM settles into a nearly axisymmetric system consisting of a hypermassive NS (HMNS), or a BH in the case of prompt collapse, surrounded by an accretion disk/torus whose mass is $\sim 10^{-3}-10^{-1}$~$M_\odot$. 
The lifetime of the HMNS can range from $\mathcal{O}(10-100)$~ms, depending on the system's initial mass before collapsing to a BH or further settles into a supermassive NS (SMNS) after the differential rotation has been removed~\cite{Fujibayashi:2022ftg,Just:2023wtj}.  
Both the HMNS and the accretion disk are also strong neutrino emitters for $\mathcal{O}(100)$~ms; see e.g., the right panel of Fig.~\ref{fig:merger_nu_lumi}.
Earlier studies of the long term evolution of BH--torus systems revealed that
the viscous heating inside the torus can inflate the disk when neutrino emission becomes inefficient at the time scale of $\mathcal{O}(1)$~s. 
The viscous heating, together with the further energy generation from the associated nuclear recombination of $^4$He, is capable of unbinding $\sim 20-50\%$ of the disk mass within a few seconds. 
Thus, the amount of disk outflow can dominate that of the dynamical ejecta and is the primary candidate that accounts for the observed kilonova associated with GW170817~\cite{Metzger:2019zeh}. 

Although hydrodynamic and magneto-hydrodynamic simulations indicated that the detailed disk outflow properties such as their velocity and the mass outflow rate can sensitively depend on the initial magnetic field configuration inside the disk~\cite{Fahlman:2022jkh},
the predicted $Y_e$ distribution of the ejecta appears to be robust. 
For cases where the central remnant consists of a BH, the resulting $Y_e$ in disk outflow ranges between $0.1\lesssim Y_e\lesssim 0.5$, with an average value $\langle Y_e\rangle\sim 0.2-0.3$, similar to the dynamical ejecta.
The density and temperature condition inside the torus (where neutrinos are trapped initially) allow weak interactions to set the torus $Y_e$ close to values given by reaction equilibrium~\cite{Fernandez:2013tya,Just:2022flt}. 
Neutrino absorption on nucleons plays an important role for a small amount of the early ejecta along the polar direction, which is exposed to larger amount of neutrino flux~\cite{Fernandez:2013tya,Just:2014fka}. 
These neutrino-driven/aided ejecta mainly contribute to the high $Y_e$ portion.
For the majority of the outflow, neutrino absorption does not affect much their $Y_e$ evolution. 

\begin{figure}[htb]
\centering
\includegraphics[width=0.5\columnwidth]{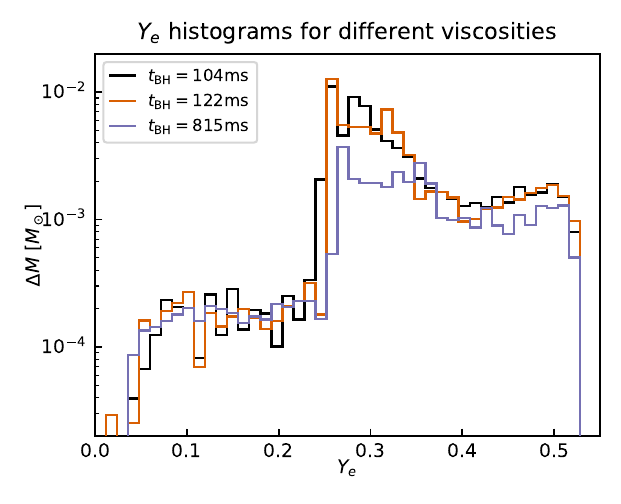}
\caption{The $Y_e$ distribution of disk outflow from the BNSM remnant, which initially consists of a HMNS that later collapses to a BH taken from Ref.~\cite{Just:2023wtj} (courtesy of O. Just). 
Different colors represent post-merger models adopting different viscosity parameters, which result in different times for the BH formation. }
\label{fig:ye_tau_hmns}
\end{figure}

However, the role of neutrino absorption on the remnant outflow may be different for cases where a HMNS exists for a substantial amount of time after the merger. 
This is mainly because before the HMNS collapses to a BH, its neutrino emission can dominate that
from the disk and can thus strongly affect the outflow $Y_e$~\cite{Perego:2014fma}.
Ref.~\cite{Lippuner:2017bfm} first investigated how the lifetime of the HMNS and the associated neutrino emission impact the outflow and found that the presence of the HMNS can lead to much higher averaged $Y_e$ values than cases where BHs form promptly. 
Subsequent work further confirmed that the secular ejecta from a remnant with delayed formation of the BH, i.e., having a HMNS whose lifetime is $\sim \mathcal{O}(100-1000)$~ms indeed have $Y_e\gtrsim 0.25$ (see Fig.~\ref{fig:ye_tau_hmns}) and produce much reduced amount of lanthanides than cases with prompt formation of BH~\cite{Fujibayashi:2022ftg,Kiuchi:2022nin,Just:2023wtj}. 
In particular, the high $Y_e$ conditions in the outflow are obtained not only during the phase before the formation of the BH, but also in the epoch after the BH formation, since the thermodynamic conditions inside the disk are altered during the phase where a HMNS still exists. 

Besides the impact on disk cooling and the nucleosynthesis in the ejecta, neutrinos can also play non-negligible roles in jet-launching in post-merger environments responsible for short gamma-ray bursts. 
Although studies suggest that the energy deposition from neutrino interaction and annihilation is likely not enough to launch a relativistic jet~\cite{Just:2015dba}, the inclusion of neutrino processes can affect the lifetime of the HMNS as well as help clear out the baryon loading in the vicinity of the BH, thereby affecting the time of jet-launching powered mainly by the Blandford–Znajek mechanism~\cite{Sun:2022vri}.

\newpage
\section{Neutrino interactions and nucleosynthesis}
\label{sec:third}

As neutrinos play crucial roles in various phases of CCSNe and BNSMs, accurate modeling of the production and decoupling of neutrinos are among the central focuses of efforts made over the last couple of decades to improve the accuracy in modeling the evolution of CCSNe and BNSMs as well as their associated nucleosynthesis yields. 
These include the improved treatment of neutrino-nucleon interaction taking into account various modifications in the presence of the nuclear medium, which will be reviewed in Sec.~\ref{sec:neutrino-nucleon}. 
The impact on the nucleosynthesis condition and yields in the neutrino-driven wind will be discussed in Sec.~\ref{subsec:ndw_nucl}.
Beyond neutrino-nucleon interaction, calculations of weak interaction rates involving nuclei are critical in determining the precise evolution during the collapse phase of CCSN. 
Moreover, the profuse amount of SN neutrinos also interact with existing nuclei in the stellar mantle, which can reshape the abundances of certain key isotopes and make important contribution to their abundances at present day through the neutrino nucleosynthesis. 
We briefly review the advances made in evaluating the neutrino-nucleus interaction in Sec.~\ref{Sec:v-nucleus}, and discuss in Sec.~\ref{sec:nu-nucleo} recent progresses in understanding the neutrino nucleosynthesis in CCSNe.

Table~\ref{tab:nu-reactions} lists the most relevant neutrino interactions with leptons, nucleons, and nuclei that are discussed in this review in determining the production and decoupling of neutrinos.  
Note that the neutrino-nucleus interactions discussed in Secs.~\ref{Sec:v-nucleus} and Sec.~\ref{sec:nu-nucleo} are not included in this Table. 

\begin{table}[ht!]
\centering
\caption{
Major neutrino reactions most relevant for modeling the CCSN evolution. References that list the detailed expression of the interaction rates are given.}
\begin{tabular}{rlcc}
\hline
\hline
& & Weak process & References \\
\hline
1 & charged current & $l^- + p \rightleftarrows \nu_l + n $ & \cite{Bruenn85,Reddy98,Guo20a,Fischer20a,Fischer20d} \\ 
2 & charged current & $l^+ + n \rightleftarrows \bar\nu_l + p $ & \cite{Bruenn85,Reddy98,Guo20a,Fischer20a,Fischer20d} \\
3 & charged current -- neutron decay & $n \rightleftarrows p + \bar\nu_e + e^-$ & \cite{Fischer20a} \\
4 & nuclear electron captures & $e^- + (A,Z) \rightleftarrows \nu_e + (A,Z-1) $  & \cite{juoda} \\
5 & nuclear de-excitations & $(A,Z)^* \rightleftarrows (A,Z) + \nu + \bar\nu$ & \cite{Fuller:1991,Fischer13} \\
6 & elastic neutrino--nucleon scattering & $\nu + N \rightleftarrows N + \nu' $ & \cite{Bruenn85,Mezzacappa93a} \\
7 & coherent neutrino--nucleus scattering & $\nu + (A,Z) \rightleftarrows \nu' + (A,Z)$ & \cite{Bruenn85,Mezzacappa93a} \\
8 & inelastic neutrino--lepton scattering & $\nu + l^\pm \rightleftarrows \nu' + l^\pm$ & \cite{Bruenn85,Mezzacappa93c,Guo20a,Fischer20d} \\
9 & electron--positron annihilation & $e^- + e^+ \rightleftarrows  \nu + \bar{\nu}$ & \cite{Bruenn85} \\
10 & nucleon--nucleon bremsstrahlung & $N + N \rightleftarrows  N + N + \nu + \bar{\nu} $ & \cite{hannestad98,Fischer2016AA,Guo:2019cvs} \\
11 & electron neutrino pair annihilation & $\nu_e + \bar\nu_e \rightleftarrows  \nu_{\mu/\tau} + \bar\nu_{\mu/\tau}$ & \cite{Buras02} \\
12 & lepton flavor exchange  & $\nu_e + \mu^- \rightleftarrows  \nu_\mu + e^-$ & \cite{Guo20a,Fischer20d} \\
13 & lepton flavor exchange & $\bar\nu_e + \mu^+ \rightleftarrows  \bar\nu_\mu + e^+$ & \cite{Guo20a,Fischer20d} \\
14 & lepton flavor conversion & $\nu_e + e^+ \rightleftarrows  \nu_\mu + \mu^+$ & \cite{Guo20a,Fischer20d} \\
15 & lepton flavor conversion & $\bar\nu_e + e^- \rightleftarrows  \bar\nu_\mu + \mu^-$ & \cite{Guo20a,Fischer20d} \\
16 & muon decay & $\mu^- \rightleftarrows  e^- + \nu_\mu + \bar\nu_e$ & \cite{Guo20a} \\
\hline
\end{tabular}
\\
Note: $\nu=\{\nu_e,\bar{\nu}_e,\nu_{\mu/\tau},\bar{\nu}_{\mu/\tau}\}$, $N=\{n,p\}$ and $l=\{e,\mu\}$ unless stated otherwise\\
\label{tab:nu-reactions}
\end{table}

\subsection{Neutrino-nucleon reactions in SN matter}
\label{sec:neutrino-nucleon}

The neutrino-nucleon interaction is governed by the current-current interaction from the standard model of particle physics. In vacuum, the neutrino-nucleon reactions across all known energy scales have been studied, considering the full kinematics, the complete weak interaction terms as well as different degrees of freedom 
of constituents relevant at different energy scales \cite{Formaggio:2012cpf}. The situation is very different for neutrino interactions with the hot and dense nuclear matter, where nuclear medium effects could play an important role and alter the reaction rates significantly. However, for strongly interacting nuclear system, a complete and consistent treatment of the full hadronic interaction terms, the relativistic kinematics, and more importantly, the nuclear medium effects, is too involved and theoretically impossible (see e.g., \cite{Burrows:2002jv,Burrows:2004vq} for reviews). In practice, various approximations have been adopted.

\subsubsection{Neutrino-nucleon opacity at mean field level}
\label{sec:CC} 

\begin{figure}[htb]
\centering
\includegraphics[width=0.65\columnwidth]{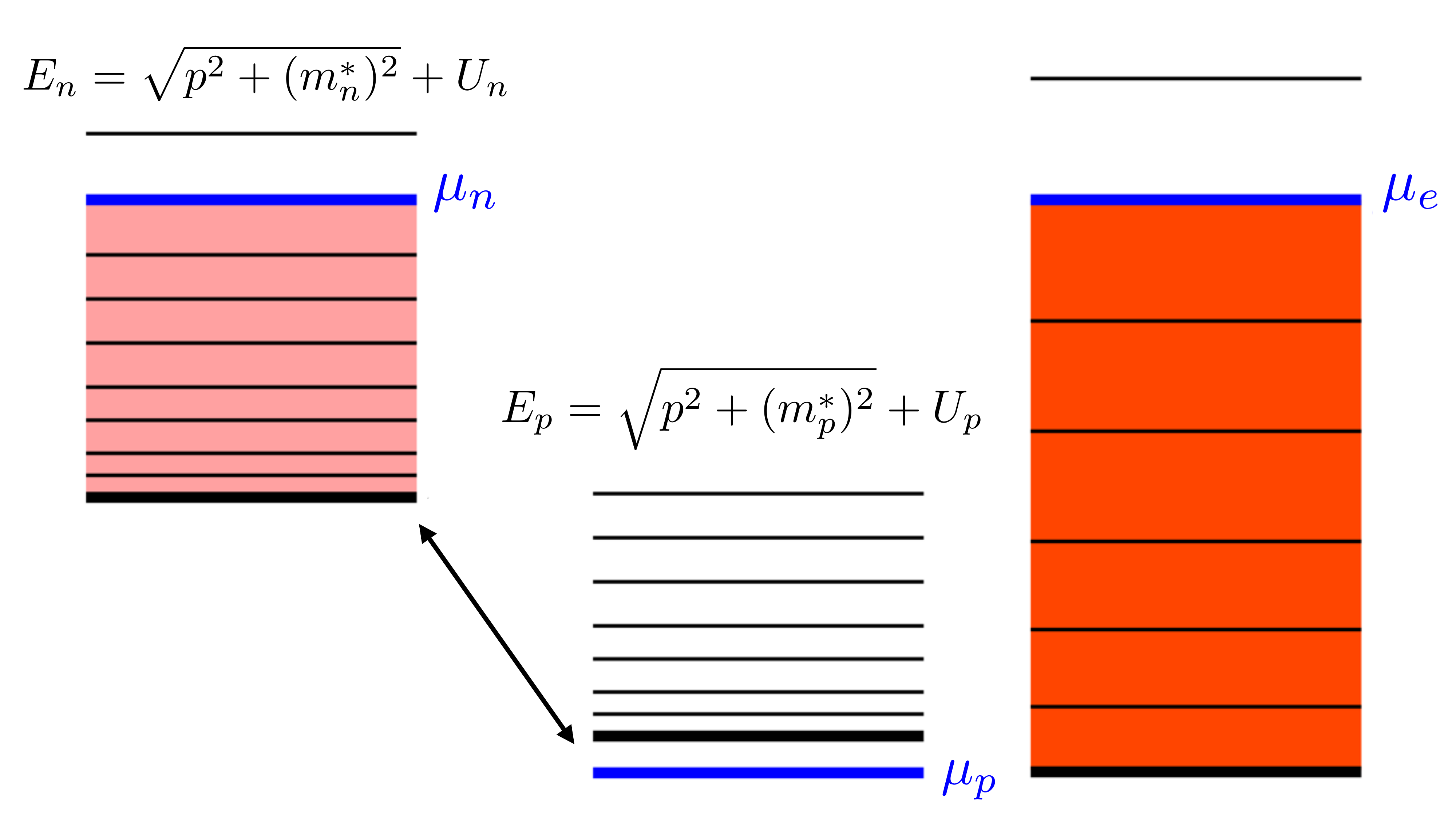}
\caption{Schematic illustration of the neutron-rich situation, given large differences between the chemical potentials of neutrons $\mu_n$ and protons $\mu_p$, typically encountered in astrophysical situations feature hot and dense nuclear matter, where neutrons and protons are treated within the mean-field approximation given by their respective effective masses $m_N^*$ and single-particle potentials $U_N$ both of which enter the particle's dispersion relations $E_N$.}
\label{fig:cc-scheme}
\end{figure}

In the mean field approximation, nucleons in nuclear medium can be viewed as independent quasiparticles, with modified dispersion relations parametrized in terms of effective masses $m_{n,p}^*$
and mean field potentials $U_{n,p}$ as $E_{n,p}=\sqrt{m_{n,p}^{*2}+p_{n,p}^2}+U_{n,p} \equiv E_{n,p}^* + U_{n,p}$. 
The effective masses and potentials can be deduced from relativistic mean field theory,
based on which the EoS of nuclear matter
can be derived \cite{HS,Shen98,Typel10,Shen10,Furusawa11,Shen11a,Shen11b,Hempel12,SFH}.
Under typical neutron rich conditions encountered in CCSNe, the neutron chemical potential ($\mu_n$) exceeds that of protons ($\mu_p$) by a large amount as illustrated in Fig.~\ref{fig:cc-scheme}. 
Since most commonly used nuclear EoS are based on the relativistic mean field approach, relativistic dispersion relations are usually employed. 
The nuclear EoS can also be computed using nonrelativistic formalism along with nonrelativistic parameterizations of nuclear potentials, such as the widely used Lattimer and Swesty (LS) EoS~\cite{LSEOS}. 
Correspondingly, nucleons take nonrelativistic energy-momentum relation as
$E_{\rm{NR}} = p^2/(2m^*) + U_{\rm{NR}}$. Taking the relativistic form, $E \approx E_{\rm{NR}} + m \approx \sqrt{m^{*2} + p^2}+m-m^*+U_{\rm{NR}}$ with $m$ being the bare nucleon mass.
For noninteracting nuclear medium, $m_{n,p}^*$ becomes the bare nucleon masses and $U_{n,p}\to 0$. 

Neglecting the residual correlation between the dressed nucleons at the mean field level, the differential rates for neutrino-nucleon scattering, $\nu(K_1) + N(P_2) \to \nu(K_3) + N(P_4)$, and neutrino absorption on nucleon, $\nu(K_1) + N(P_2) \to l(K_3) + N(P_4)$, can be obtained from the Fermi's golden rule as \cite{Reddy98}
\begin{align}
{d^2\Gamma \over d\omega d\cos\theta_{13}} =  {(1-f_3) k_3 \over 8\pi^2 E_1} \int \frac{d^3\bm{p}_2}{\left(2\pi\right)^3} \int \frac{d^3\bm{p}_4}{\left(2\pi\right)^3}
 \frac{\left|\mathcal{M}\right|^2}{4E_2^*E_4^*} 
  \times  \left(2\pi\right)^4\delta^{(4)}\!\left(K_1+\!P_2-\!K_3-\!P_4\right) f_2 \left(1-f_4\right),
\label{eq:diff_rates_mfa}  
\end{align}
where $\omega$ is the energy transfer to the nuclear medium, $\theta_{13}$ is the angle between the three-momentum of leptons, $\left|\mathcal{M}\right|^2$ is the squared scattering amplitude summed over final state spins and averaged over initial state spins, and $f_{i}$ is the standard Fermi-Dirac distribution function. By integrating over $\omega$ and $\cos\theta_{13}$, the total scattering rate or the inverse mean free path of neutrino as a function of $E_\nu$ can be computed. Note that the same expression shown in Eq.~(\ref{eq:diff_rates_mfa}) can also be derived equivalently within the framework of finite temperature field theory \cite{roberts17}.

\begin{figure}[htp]
\centering
\includegraphics[width=0.45\columnwidth]{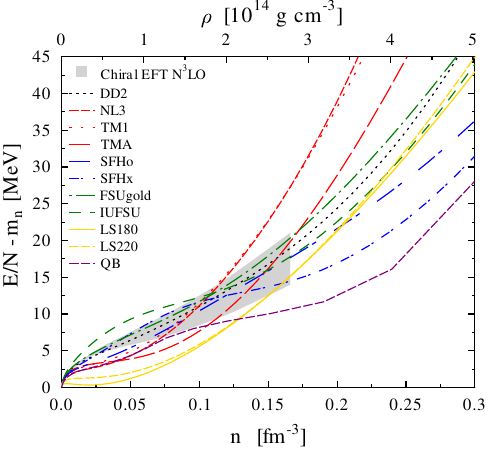}
\includegraphics[width=0.45\columnwidth]{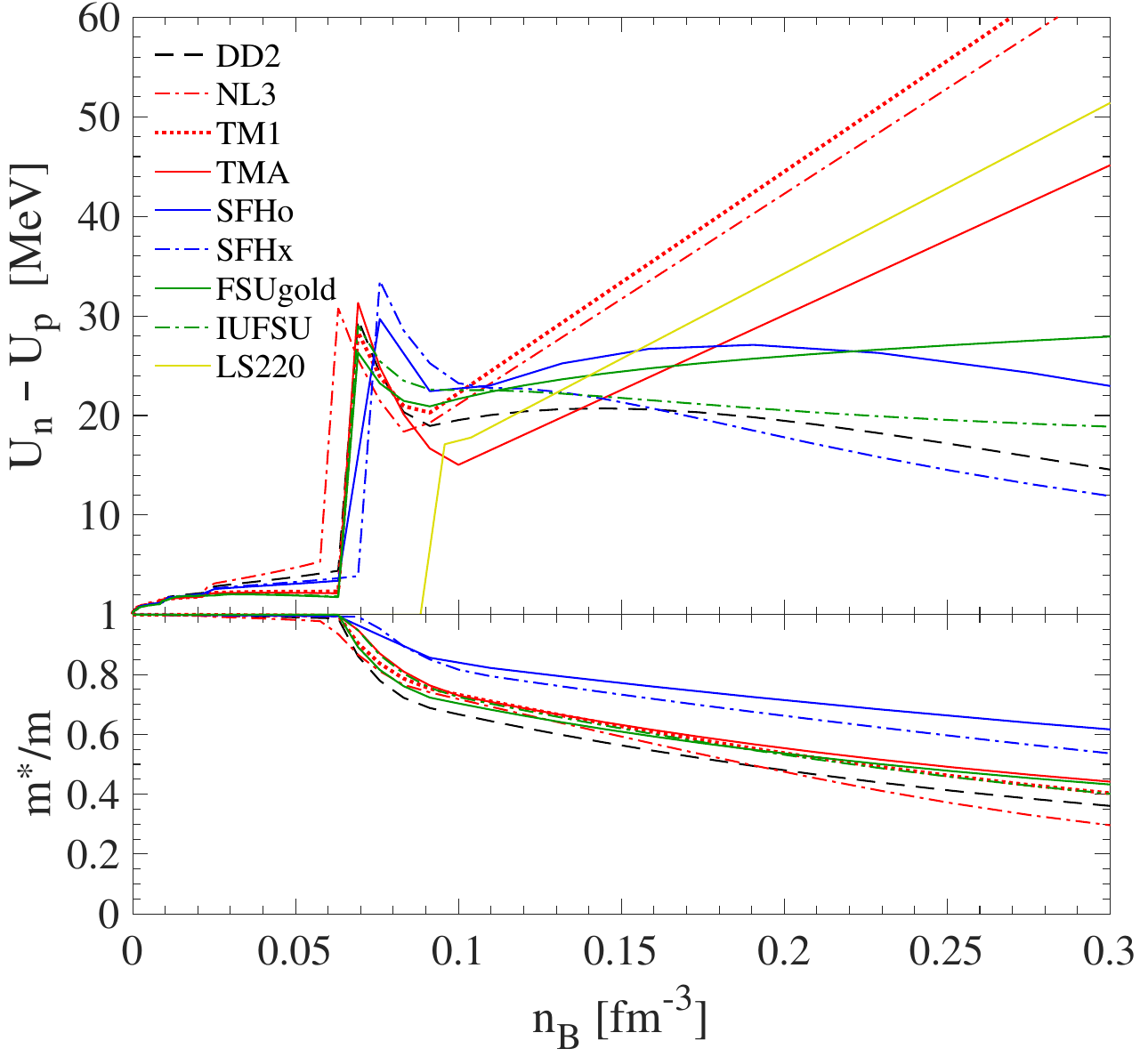}
\caption{{\em Left panel}~(Figure adapted from Ref.~\cite{Fischer17}):~Energy per nucleon for neutron matter from chiral effective field theory interactions computed up to N$^3$LO (gray band), including uncertainty estimates due to the many-body calculation, low-energy constants and the regularization cutoffs in the 3N and 4N forces \cite{Tews13,Krueger13}.  The colored lines show the energy per nucleon for a selection of RMF EoS from the catalogue of Ref.~\cite{Hempel12} as well as two commonly used Skyrme models (LS180, LS220) \cite{LSEOS}, and a quark bag (QB) model EoS \cite{Fischer09,Fischer11}.  
{\em Right panel:}~Mean-field potential difference, $U_n-U_p$, between neutrons and protons and effective masses, $m^*/m$, for the same selection of commonly used nuclear matter EoS, for a fixed electron fraction of $Y_e=0.3$ and a temperature of $T=7$~MeV.
}
\label{fig:enm_unup}
\end{figure}

The nuclear EoS only has a modest impact on neutral current neutrino-nucleon scattering rates at the mean field level, but can affect directly the charged current rates. In what follows, we focus on the charged current channel. Taking into account the effective nucleon masses $m_N^*$ and the mean-field potential $U_N$, both the transition matrix elements and the phase spaces of nucleons in Eq.~\eqref{eq:diff_rates_mfa} are modified. 
The evolution of SN dynamics as well as nucleon abundances depend sensitively on the EoS taken, which are also related to the neutrino-nucleon interaction rates.
Constraints on the EoS at low density, in particular up to nuclear saturation density, are given by chiral effective field theory for pure neutron matter, which is shown in the left panel of Fig.~\ref{fig:enm_unup} (gray band) from Refs.~\cite{Tews13,Krueger13}, in comparison to a representative set of relativistic and non-relativistic nuclear EoSs \cite{LSEOS,HS,Hempel12,SFH} and a quark bag model \cite{Sagert09,Fischer11}, all of which have been commonly used in astrophysical simulations of SNe as well as BNSMs. 
This comparison indicates that there is some tension among most models, although there are a few exceptions that exhibit qualitative agreement.
The EoS at supersaturation densities can be constrained by the observation of massive pulsars of about 2~$M_\odot$ \cite{Antoniadis13,Fonseca:2021}, but the existing calculations still differ greatly.
Recently, it has become possible to further constrain the EoS using the GW data obtained from the first BNSM event GW170817 \cite{LIGOScientific:2017ync,Abbott18,Lattimer18}, through the determination of the tidal deformability which, in turn, provided constraints for the radius of neutron stars of 8.9--13.2~km for intermediate mass neutron stars of 1.2--1.6~$M_\odot$. Some tension exists between these results and those obtained from the NICER (Neutron star Interior Composition ExploreR) NASA X-ray satellite mission, where rather large radii were found from the analysis of two independent groups \cite{NICER_Miller2019,NICER_Watts2019,NICER_Miller2021,NICER_Riley2021}.  

These variations of the EoSs are also reflected in the mean field potentials and the effective masses of nucleons, which become more evident at supersaturation densities, as shown in the right panel of Fig.~\ref{fig:enm_unup}. 
Note, however, the relevant density range for the neutrino decoupling from matter, i.e.~the location of the neutrinospheres, corresponds to densities below 0.1~fm$^{-3}$, where the differences between the different EoSs are somewhat less pronounced due to the low-density constraints, not only from chiral effective field theory but also for the EoS at nuclear saturation density which is confirmed experimentally, such as the nuclear symmetry energy and the corresponding slope. Further details and a detailed discussion comparing these EoSs in simulations of CCSNe as well as the corresponding complete set of nuclear matter properties can be found in Ref.~\cite{Fischer14}. 
It is important to note here that the nuclear symmetry energy, $S(\rho)$, determines the mean field potential difference, since $\mu_n-\mu_p=4\beta S(\rho)\propto U_n-U_p$, with 
the asymmetry parameter $\beta=1-2Y_p$ for a given proton-to-nucleon ratio $Y_p$.
Some EoSs, such as NL3, TM1, TMA, and LS220, lead to a steeply rising $U_n-U_p$ with density, as shown in the right panel of Fig.~\ref{fig:enm_unup}, which feature a stiff symmetry energy and a too high slope parameter compared to that derived from experiments~\cite{Lattimer13}, while other EoSs indicate a negative slope towards high densities. 
It would be interesting, that a priori it is not possible at current stage, 
to determine whether the slope remains positive at all densities or not,
which requires more experimental and observational constraints in the future.

The neutrino-nucleon rate or opacity also depends on how the integral in Eq.~\eqref{eq:diff_rates_mfa} is evaluated using different approximations. Taking nonrelativistic energy-momentum relation for nucleons and keeping only the leading terms of the matrix elements in the heavy nucleon limit (i.e., the terms proportional to $g_V^2$ and $g_A^2$), the differential rates taking into account the thermal motion and the final state blocking can be solved analytically in terms of the density and spin dynamical structure factors \cite{Reddy98}. Taking further the so-called elastic approximation where no energy is transferred to the heavy nucleons, the neutrino-nucleon opacity has a simple analytical form, see Eqs.~(34) and (35) of \cite{Reddy98}. The nucleon recoil and weak magnetism can be included in the elastic opacity approximately \cite{horowitz02}. It was shown that weak magnetism can enhance/suppress
the neutrino/antineutrino rates and affects the
neutrino spectra and SN dynamics~\citep{horowitz02,Kotake18}.    

\begin{figure}[htp]
\centering
\includegraphics[width=0.4\columnwidth]{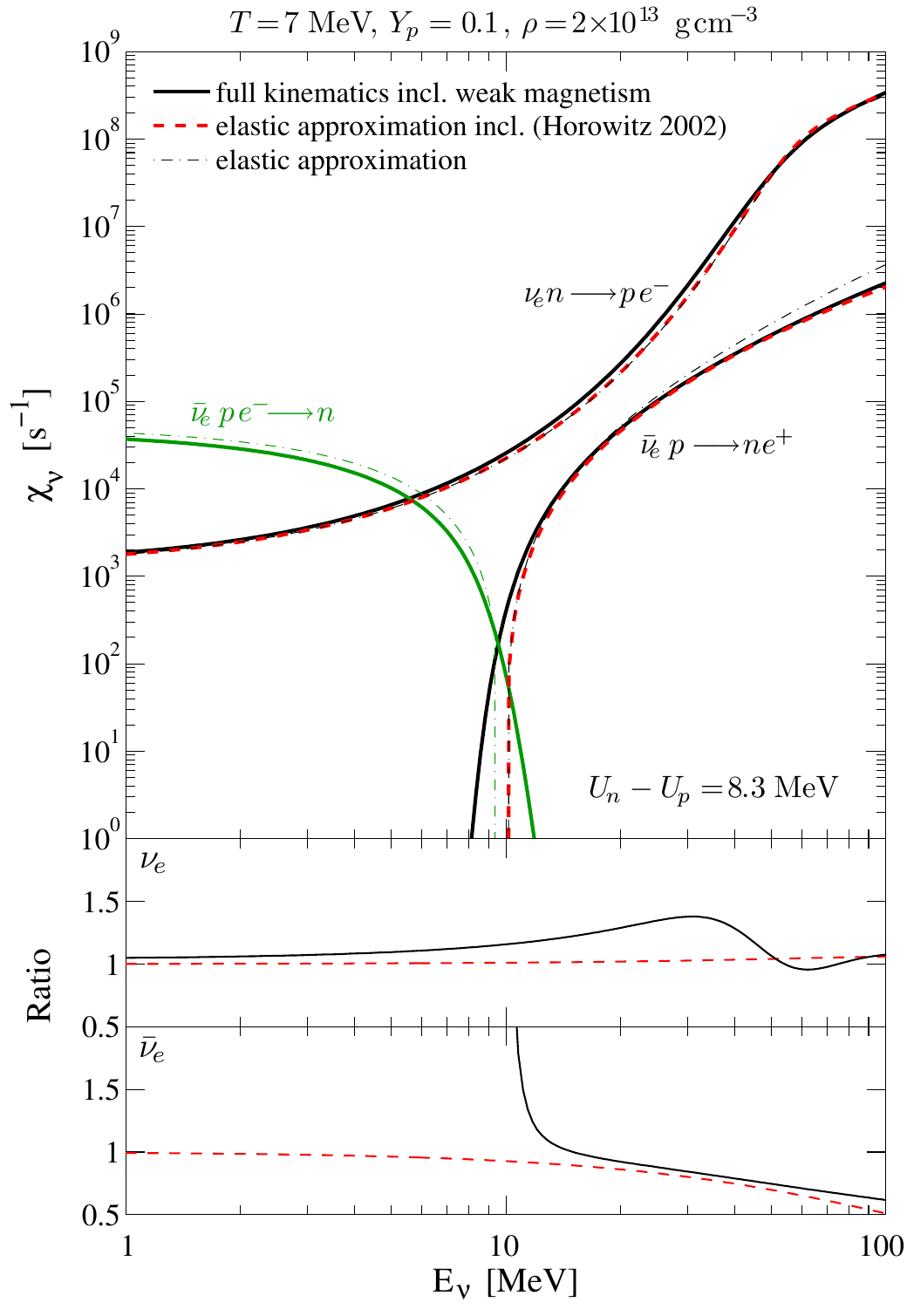}
\includegraphics[width=0.41\columnwidth]{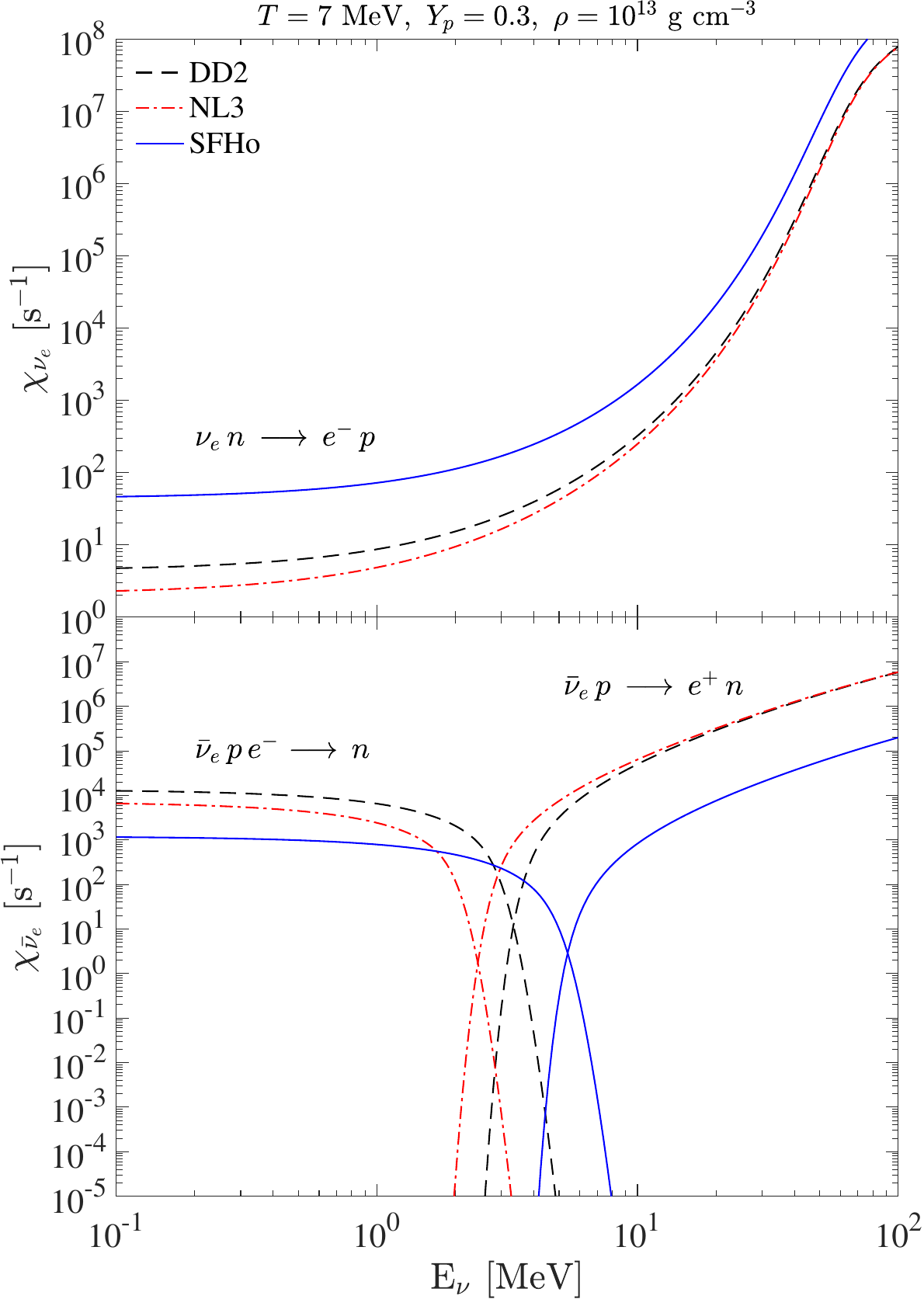}
\caption{Charged current rates for $\nu_e$ and $\bar\nu_e$ at typical conditions. {\em Left panel}~(Figure adapted from Ref.~\cite{Fischer20a}):~Comparing the full kinematics treatment at the mean-field level including weak magnetism (solid lines) with the elastic approximation (dashed lines) of Ref.~\cite{Bruenn85} and including approximately inelastic contributions and weak magnetism corrections (dash-dotted lines) following Horowitz (2002) \cite{horowitz02}. The green lines correspond to the inverse neutron decay. The corresponding ratios of the results with full kinematics and weak magnetism (solid lines) and the results considering the correction factors following Horowitz (2002) \cite{horowitz02} (dashed lines) to those using the elastic approximation without including the weak magnetism are also shown in the lower figures for $\nu_e$ and $\bar\nu_e$, respectively. {\em Right panel:}~Comparing the charged-current rates for selected relativistic mean field EoS, DD2 ($U_n-U_p=2.1$, $m^*/m=0.9861$), NL3 ($U_n-U_p=1.2$, $m^*/m= 0.9902$) and SFHo ($U_n-U_p=4.5$, $m^*/m=0.9855$)
with full kinematics and weak magnetism considered.}
\label{fig:CC_rates}
\end{figure}

A relativistic treatment of both full kinematics (nucleon thermal motion, recoil, and blocking) and weak magnetism for charged current neutrino reactions has been performed~\cite{roberts17,Fischer20}, and its impact on CCSN dynamics, neutrino signals as well nucleosynthesis was explored based on 1D simulation~\cite{Fischer20}. 
Figure~\ref{fig:CC_rates} (left panel) compares the charged-current rates in the full kinematics treatment at the mean field level, including weak magnetism contributions, with the elastic approximation of Ref.~\cite{Bruenn85} and including in addition approximately inelastic contributions and corrections due to weak magnetism following Ref.~\cite{horowitz02}. 
The latter has been commonly used in most CCSN studies. From this comparison it becomes evident that the full kinematics treatment cannot be adequately captured solely by a multiplicative factor varying with neutrino energy given in Ref.~\cite{horowitz02}.
This is particularly relevant in the energy domain when the $\bar\nu_e$ rates drop to zero due to the energy threshold for this process, requiring $E_{e^+}=E_{\bar\nu_e}-(E_n - E_p)>0$. At the mean-field level, this corresponds roughly to $E_{\bar\nu_e} \gtrsim (m_n^*-m_p^*) + (U_n-U_p)$, as is illustrated in Fig.~\ref{fig:CC_rates}, modulo momentum transfer contributions which even further suppress the rates towards lower energies. 
Note that a threshold does not exist for $\nu_e$ since the condition $E_{e^-}=E_{\nu_e}+(E_n - E_p)$ is always fulfilled. Furthermore, it is interesting to note that the inverse neutron decay, which was previously largely neglected, contributes significantly to the $\bar\nu_e$ opacity (green lines in Fig.~\ref{fig:CC_rates}). For this reaction the threshold is given by $E_{e^-}=-E_{\bar\nu_e}+(E_n - E_p)$, i.e. the medium modifications enhance the rate substantially with increasing density. 

The right panel of Fig.~\ref{fig:CC_rates} compares in addition the dependence of the charged-current rates for $\nu_e$ (top panel) and $\bar\nu_e$ (bottom panel) for three selected EoSs: DD2 \cite{Typel10,HS,Hempel12}, NL3 \cite{HS,Hempel12}, and SFHo \cite{HS,SFH}. Since each EoS has different single-particle properties as well as a different density dependence, as was discussed before, there is a certain degeneracy of the charged current weak rates. The largest impact is due to the threshold for the $\bar\nu_e$ absorption on protons as well as for the inverse neutron decay. Small (large) values of the neutron-proton mean field potential differences result in the suppression of the $\bar\nu_e$ absorption rate at low (high) energy, while the opposite is found for the $\bar\nu_e$ opacity from the inverse neutron decay, such that the integrated $\bar\nu_e$ opacity becomes less sensitive to the value of $U_n-U_p$. 
In other words, the symmetry energy impact on the charged current rates entering directly the rate expressions at the mean field level via the mean field potentials, is somewhat less strong than originally anticipated \cite{MartinezPinedo12,MartinezPinedo14}. 
Spectral differences between $\nu_e$ and $\bar\nu_e$ are rather driven by the different abundances of neutrons and protons, which might differ greatly for the EoSs commonly used, at intermediate densities relevant for the neutrino decoupling. This abundance difference arises from the different descriptions of the nuclear medium, particularly concerning the abundances of light and heavy nuclear clusters at nuclear subsaturation densities.
A detailed discussion about this aspect can be found in Refs.~\cite{Fischer17,Fischer20c}. 
It depends sensitively on two aspects: {\em (i)} the description of the nuclear clusters where commonly the modified nuclear statistical equilibrium of Ref.~\cite{HS} is employed based on thousands of nuclear species with tabulated and partly calculated nuclear masses versus the quantum statistical approach, implemented in particular for light clusters including Pauli blocking as well as two- and effective three-body scattering phase shifts, and {\em (ii)} the Mott transition already mentioned in Sec.~\ref{sec:SNpostbounce}. 
Different treatments of the Mott transition give significantly different nucleon abundances in the density region of $\rho\simeq 10^{12}-10^{14}$~g~cm$^{-3}$, and accordingly the charged current rates differ substantially \cite{Fischer20c}.
Nevertheless, definite conclusions cannot be drawn about the impact of neutrino rates using different EoSs on SN dynamics and the neutrino signals by simply looking at the neutrino opacity.
The system is too complex since the underlying equations are highly non-linear. For instance, the different EoSs exhibit  softer or stiffer high-density behaviors, resulting in different SN dynamics, different abundance for each species, as well as different locations of the neutrinospheres, 
all of which might compensate and entangle with the direct impact on the neutrino opacity \cite{Fischer2016EPJA,Fischer17}.
In addition to studies associated with known EoSs,
neutrino-nucleon rates at the mean field level have also been computed using
effective nucleon masses and potentials obtained from virial expansion at low densities \cite{Horowitz:2012us}, and from pseudo-potentials constructed from nucleon scattering phase shifts \cite{Rrapaj:2014yba}, realistic nucleon-nucleon potentials \cite{Shen:2003ih} and those derived from chiral effective field theory \cite{Vidana:2022ket}, and Skyrme interactions \cite{Oertel:2020pcg,Hutauruk:2022bii,Duan:2023amg}.

\begin{figure*}[htbp] 
\centering  
\includegraphics[width=0.49\textwidth]{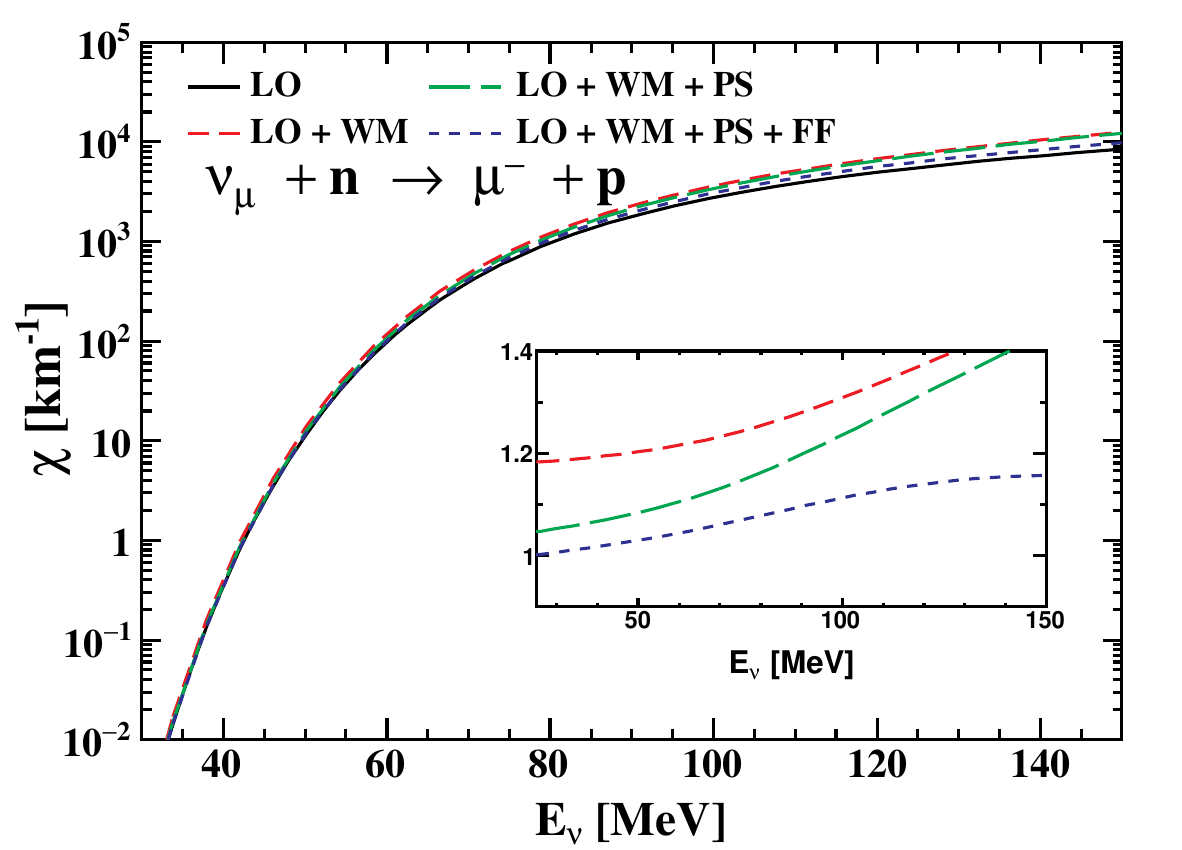}%
\includegraphics[width=0.49\textwidth]{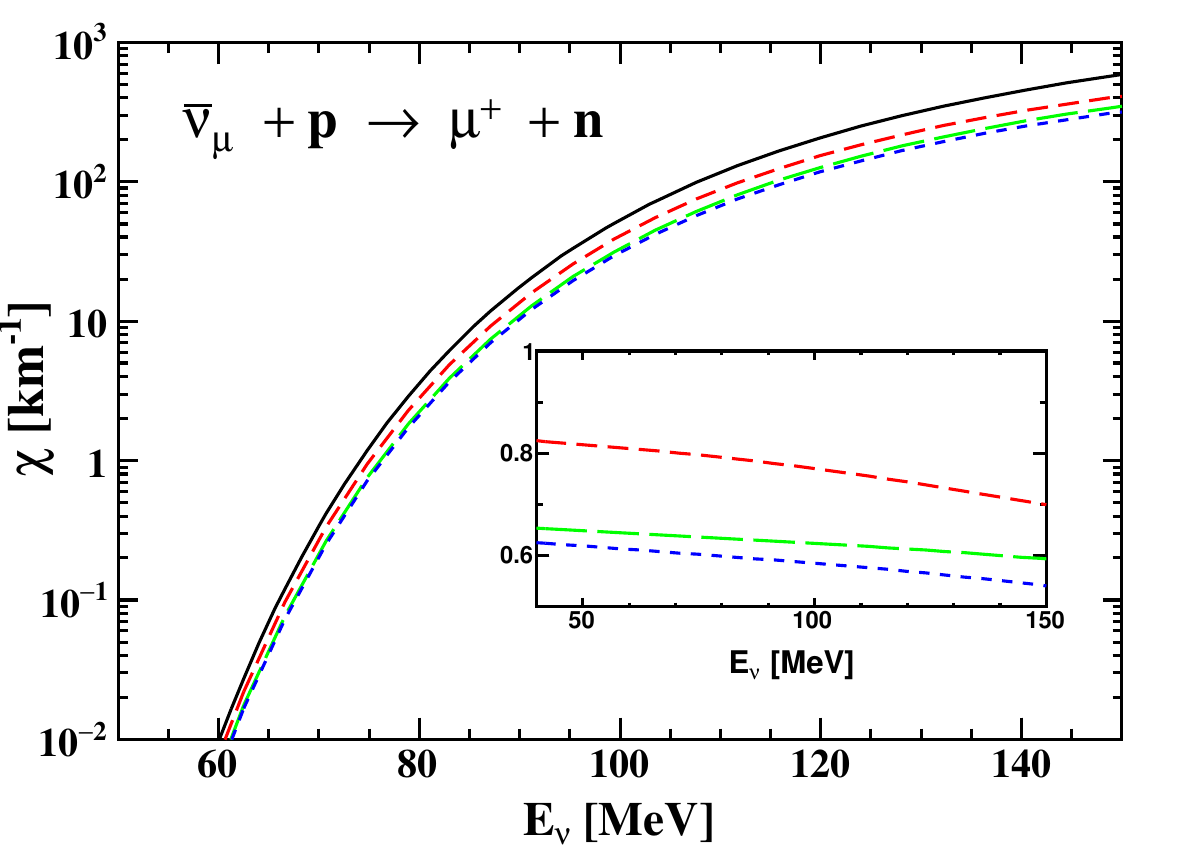}          
\caption{Opacities for $\nu_\mu+n\to \mu^-+p$ and $\bar\nu_\mu+p\to \mu^++n$
and the effects of weak magnetism (WM), pseudoscalar term (PS), and form factors (FF). The relative ratios of the opacities to that considering only the leading-order terms (LO, the black lines), i.e., terms proportional to $g_V^2, g_A^2$, and $g_Vg_A$, are shown in the inset as functions of $E_\nu$. Note that for all the results presented, full kinematics including nucleon recoil and blocking are always considered. The condition is taken from a 2D simulation \cite{bollig17} considering the muonic rates using LS200 at $r\approx 13.6$~km and at 0.4~s post core bounce, with $T=38.3$~MeV and $\rho=10^{14}$~g/cm$^3$. Figure adapted
from Ref.~\cite{Guo20a}.}
\label{fig:wm_pseu_form}                   
\end{figure*}

Due to a large rest mass, the production of $\mu^\pm$ and their role in SN dynamics were largely ignored in the previous literature. Their relevance has recently been explored in 2D simulations \cite{bollig17}, indicating that the presence of $\mu^\pm$  softens the EoS, induces faster core contraction, and thus leads to higher luminosities and averaged energies of neutrinos, thereby stimulating the neutrino-driven explosions. A follow-up study considered the muonization of SN matter in more details and focused on the possible impact on neutrino signals \cite{Fischer20d}. Interestingly, muonization could lead to a burst of $\nu_\mu$ shortly after core bounce when the shock passes the $\nu_\mu$ neutrinosphere \cite{Fischer20d}. Note that the
accumulation of muons is closely related to the charged current reactions of $\nu_\mu$ ($\bar\nu_\mu$), which are produced in large amounts via electron-positron pair annihilation and nucleon-nucleon bremsstrahlung early on before core bounce. To obtain accurate charged current $\nu_\mu$-nucleon rates, Ref.~\cite{Guo20a} included the pseudoscalar term and considered the $q^2$-dependent nucleon form factors in the hadronic current at the mean field level, which are important due to large energy and momentum transfer. Fig. \ref{fig:wm_pseu_form} compares the effects of weak magnetism (WM),
the pseudoscalar term (PS), and the form factors (FF) on the neutrino opacities for
$\nu_\mu+n\to \mu^-+p$ at one selected condition with $T=38.3$ MeV and $\rho=10^{14}$~g/cm$^3$
from a 2D CCSN simulation \cite{bollig17}
using the EoS LS200 \cite{LSEOS}, where
the relevant muonic reactions have been implemented. As shown in the figure, 
weak magnetism enhances (suppresses) the absorption rates of $\nu_\mu$ ($\bar\nu_\mu$) by $\sim$
20\% at $E_\nu=20$ MeV and by 30\%--50\% at $E_\nu=150$ MeV. The interference between the axial-vector term and the pseudoscalar term is always negative. Therefore, inclusion of the pseudoscalar term suppresses the opacities for both $\nu_\mu$ and $\bar\nu_\mu$
absorption on nucleons. Similarly, taking into account the form factor effect leads to reduced couplings for all interaction terms and thus reduces the opacities. We note that the enhancement for $\nu_\mu$ due to weak magnetism is largely canceled by the effects of pseudoscalar term and form factor, while for $\bar\nu_\mu$, all effects contribute negatively and the absorption opacity is significantly quenched compared to the  result
obtained at lowest order.   

\subsubsection{Effects of nucleon-nucleon correlation}\label{subsec:nncorre}
The nucleon-nucleon correlation due to the nuclear force affects neutrino-nucleon opacities significantly. Because of the non-perturbative nature as well as the complexity of the nuclear force, a complete treatment of the nucleon correlations is still lacking. In what follows, we briefly review the methods adopted in the literature to study the correlation effects on neutrino opacities in hot and dense nuclear matter and highlight some progress made during the past few years.

For low energy neutrinos interacting with nonrelativistic nucleons, the reaction rates are dominated by the density (vector) dynamical structure factor, $S_V(\omega, q)$, and the spin (axial) dynamical structure factor, $S_A(\omega, q)$, where $\omega$ and $q$ are the energy and momentum transfer to the nuclear medium \cite{Raffelt:1993ix,Horowitz:2006pj}. The energy exchange $\omega$ is of order ${\rm max}[q\sqrt{T/m_N}, q^2/m_N]$ for neutrino-nucleon scattering, which is much smaller than $q$ for heavy nucleons. When integrating over a range of $\omega$ to obtain the total rate, the dynamical structure factors can be approximately replaced by
$S_{V,A}(q, \omega) \to \delta(\omega) \int d\omega S_{V,A}(\omega, q) \equiv 2\pi\delta(\omega) S_{V,A}(q)$, where $S_{V,A}(q)$ are the static structure factors. This corresponds to the elastic approximation in which energy exchange to the nuclear medium can be neglected. In some approaches discussed below, only the static structure factors can be computed.    

Nucleon correlation effects on neutrino opacities beyond mean field level are often studied based on the random phase approximation (RPA) for both the neutral current and the charged current reactions \cite{Burrows98,Burrows:1998ek,Reddy99}. To account for the screening of coupling between nucleons and the external probe due to particle-hole excitations, an infinite series of ring or bubble diagrams are summed in RPA when calculating the polarization function, whose imaginary part gives the dynamical structure factor from the fluctuation-dissipation theorem. Compared to previous studies using simple and energy-independent interactions in terms of the Landau parameters \cite{Burrows98,Reddy99,Shen:2003ih}, more complete particle-hole interactions derived from 
the Skyrme interactions have been employed to study the RPA correlations \cite{Navarro:1999ij,Hernandez:1999zz,Pastore:2014aia,Dzhioev:2018ovi,Oertel:2020pcg,Pascal:2022qeg,Duan:2023amg}. In addition to those nonrelativistic studies, RPA calculations were also performed within the framework of a relativistic mean field theory \cite{Reddy99,Mornas:2002ji}.
The net effect of RPA correction is to redistribute the strength of the response function to a broad energy region and generally to suppress the neutrino-nucleon opacities \cite{Burrows98,Burrows:1998ek}. Remarkably, nuclear interactions from chiral effective
field theory have recently been employed to study the RPA effects beyond the mean field level \cite{Shin:2023sei}. Different from the studies in \cite{Burrows98,Burrows:1998ek}, Ref.~\cite{Shin:2023sei} found that nucleon correlation enhances the density response and reduces the spin response for neutral current scattering. For charged current reactions, the RPA tends to redistribute the strength to higher energy for neutrino
absorption and lower energy for antineutrino absorption, leading to suppressed $\nu_e$ absorption rates at all relevant energy regions and enhanced (suppressed) $\bar\nu_e$ absorption rates at low (high) energy \cite{Shin:2023sei}. As a consequence, the spectral difference between $\nu_e$ and $\bar\nu_e$ is reduced, which tends to
reduce the neutron richness
in the neutrino-driven wind from the PNS surface.

At low density and/or high temperature, the virial expansion in
terms of the fugacity $z\equiv e^{\mu/T}$, where $\mu$ is the relativistic chemical potential of nucleons, can provide a model-independent description of the EoS of nuclear matter \cite{Horowitz:2005zv,Horowitz:2005nd}, as well as the static neutrino response functions in the long wavelength limit, i.e., $S_{V,A}(q=0)$, for neutrino-nucleon neutral current scattering  \cite{Horowitz:2006pj}. The expansion is up to the second order using the second virial coefficients from the scattering phase shifts and is only valid for $z\lesssim 0.5$. The virial expansion indicates that $S_{V}(q=0)$ is enhanced while $S_{A}(q=0)$ is suppressed due to nucleon correlation, consistent with the findings of \cite{Shin:2023sei}.

For CCSN simulations, the neutrino interactions at the neutrinosphere is highly relevant. The neutrinosphere is typically neutron-rich with densities $\sim 10^{12-13}$~g/cm$^3$ and temperatures $\sim 10$~MeV. With a large neutron-neutron scattering length and short effective range, the neutrinosphere could be approximately modeled as a warm and low density unitary gas of neutrons. For any unitary Fermi gas with infinitely large scattering length and zero effective interaction range, the detail of the interaction is irrelevant due to the absence of length scales and all systems share universal static and dynamic properties (see e.g., \cite{Liu2013}). Unitary gases of cold atoms have been produced and well studied in laboratories. There are also many theoretical calculations for unitary gases, which are, however, not available for the neutron gas. For example, the fourth-order virial coefficients for a unitary gas have been obtained, and can be used to extend the virial results of the static neutrino responses up to $z\approx 1$ \cite{Lin:2017spm}. Based on a lattice formulation, the static structure factors with a finite momentum transfer, $S_{V,A}(q)$, of a unitary gas under SN conditions were also derived \cite{Alexandru:2019gmp}. Similar lattice studies using the leading-order pionless effective field theory were further performed to compute the static structure factors of neutron matter  \cite{Alexandru:2020zti}. 

Note that the full dynamical structure factors $S_{V,A}(\omega, q)$ based on virial expansion or lattice calculation are not yet available. To obtain 
the dynamical structure factors perturbatively, Ref.~\cite{Bedaque:2018wns} considered a
pseudopotential from neutron-neutron phase shifts and computed the contributions to the polarization functions arisen from two-loop diagrams. Including these corrections due to neutron-neutron interaction, the dynamical structure factors for neutron matter are computed up to the second order in $z$. The study presented in \cite{Bedaque:2018wns} should be compared to the one employing nuclear potentials based on chiral effective field theory in \cite{Shin:2023sei}, where the single-particle energies of nucleon are calculated at the Hartree-Fock mean field level and the vertex corrections are resummed in the framework of RPA, both incorporating an infinite number of diagrams.

\subsubsection{Nucleon-nucleon bremsstrahlung} \label{subsec:nnbrems}

\begin{figure}[htp]
\centering
\includegraphics[width=0.35\columnwidth]{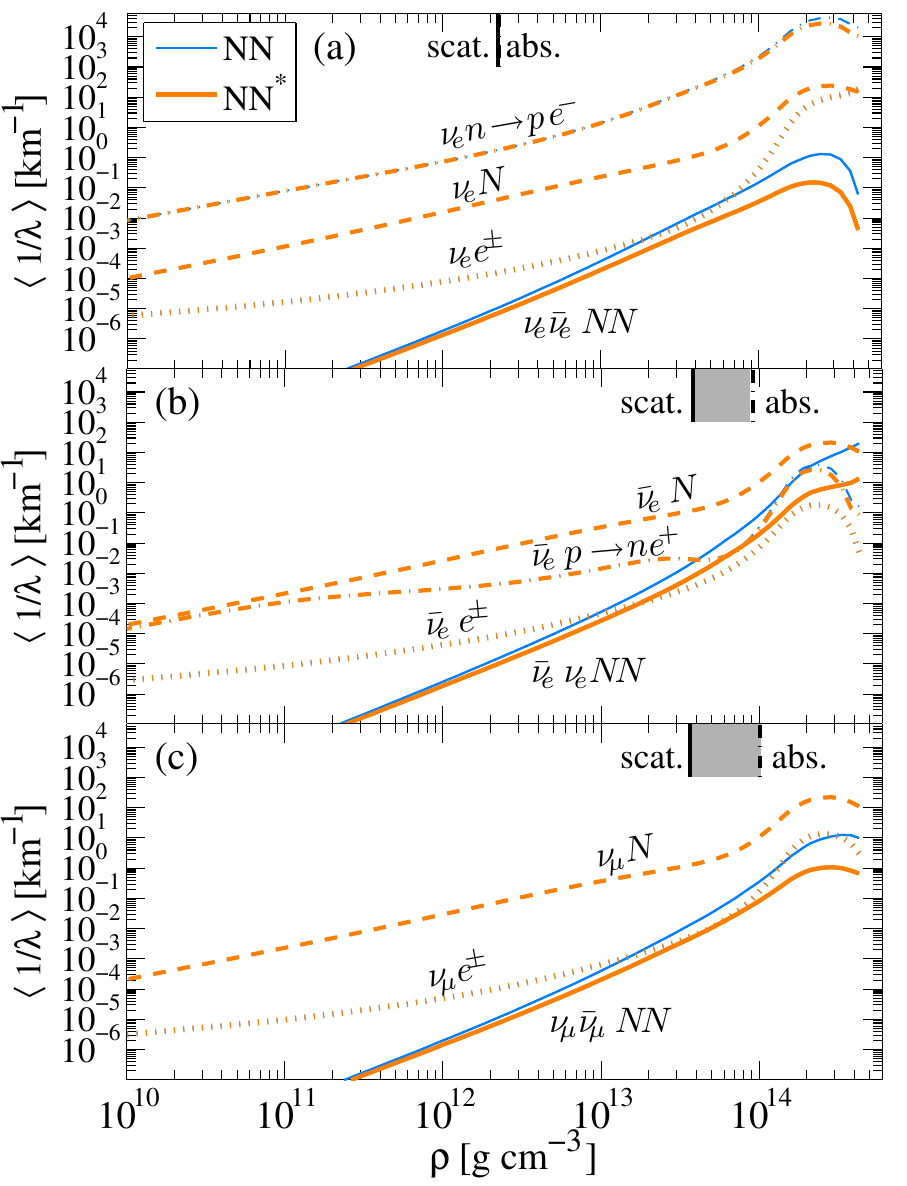}
\caption{Inverse mean free paths for $\nu_e$ in panel~(a), $\bar\nu_e$ in panel~(b) and collectively for all heavy lepton flavors $\nu_\mu$ in panel~(c), for all four selected neutrino interaction channels considered in SN simulations, charged current reactions ($\nu_e n\rightarrow p e^-$, $\bar\nu_e p\rightarrow n e^+$), elastic neutrino nucleon/nucleus scattering ($\nu N$), inelastic neutrino electron/positron scattering ($\nu e^\pm$) and nucleon--nucleon bremsstrahlung ($\nu\bar\nu N N$), for which two different treatments are employed, vacuum one-pion exchange (blue lines) \cite{Friman79,hannestad98} and including approximately corrections of the $\pi NN$ vertex \cite{Fischer2016AA}. Not shown are the other neutrino-pair processes, i.e. electron-positron annihilation and neutrino-pair annihilation. The vertical solid and dash-dotted lines mark the locations of the neutrinospheres of last inelastic and elastic scatterings, respectively. The gray region in between marks the presence of a scattering atmosphere, following the argumentation of Refs.~\cite{Raffelt01,Keil03}. 
Credit: reprinted from Fischer et.~al., A\&A, 593, A103, 2016~\cite{Fischer2016AA}, reproduced with permission $\copyright$ ESO.}
\label{fig:mfp}
\end{figure}

It had long been argued that the neutrino fluxes and spectra at the decoupling are determined dominantly by the charged current rates. This turns out to be the case during the post-bounce mass accretion phase prior to the shock revival and onset of explosion, in which case both charged current rates for $\nu_e$ and $\bar\nu_e$ are comparably large and a scattering atmosphere develops only for the heavy lepton neutrinos, i.e. the spatial separation of neutrinosphere of inelastic scattering at high density and elastic scattering at lower density. The situation changes after the shock revival, when mass accretion ceases and the PNS enters the deleptonization phase. Whereas the neutrinospheres of last inelastic scattering dominate the neutrino decoupling for $\nu_e$, an extended scattering atmosphere develops not only for the heavy lepton neutrino but also for $\bar\nu_e$. This situation is illustrated in Fig.~\ref{fig:mfp} for a typical situation during the early PNS deleptonization phase, showing the inverse mean free paths for all flavors and the four weak reaction channels that are considered in the neutrino transport codes employed for CCSN studies in general, charged current, elastic neutrino nucleon/nucleus scattering, inelastic neutrino electron/positron scattering and neutrino pair processes. The phenomenon that $\nu_e$ also develop a scattering atmosphere at late times, following the argumentation of Refs.~\cite{Raffelt01,Keil03}, was realized in Ref.~\cite{Fischer12}, analyzing simulation results that were based on three-flavor Boltzmann neutrino transport \cite{Fischer10}. This has important consequences for the evolution of the neutrino luminosities $L_\nu$ and average energies $\langle E_\nu \rangle$ that become increasingly similar for all flavors during the PNS deleptonization phase, which is illustrated in Fig.~\ref{fig:delept} (b), as all flavors are becoming dominated by neutrino-nucleon scattering towards late times during the PNS deleptonization, a weak process that is flavor blind. 

As shown in Fig.~\ref{fig:mfp}, nucleon-nucleon ($NN$) bremsstrahlung plays a non-negligible role for inelastic scattering of $\bar\nu_e$, $\nu_x$, and $\bar\nu_x$. Neutrino pair emission from nucleon-nucleon bremsstrahlung is also an important process for the production of $\nu_x$ and $\bar\nu_x$. For the bremsstrahlung rates, two different treatments are shown, vacuum one-pion exchange (NN) \cite{Friman79,hannestad98} and taking into account approximate medium corrections of the $\pi NN$ vertex (NN$^*$) \cite{Fischer2016AA} based on Fermi-liquid theory \cite{Migdal78,Migdal90,Voskresensky01}. It assumes that the nucleons are only slightly excited above the Fermi sea, i.e. all processes occur in the vicinity of the Fermi surface and that the vertices of the processes are dressed by $NN$ correlations. The subsequent medium modification of the one-pion exchange process includes the summation of {\em all} particle-hole loops and a contribution of the residual $s$-wave $\pi NN$ interaction as well as $\pi\pi$-scattering. This involves effectively $\rho$-meson exchange ($t$-channel) and the $\sigma$-meson as correlated $\pi\pi$ ($s$-wave contribution). It results in the suppression of the $NN$ bremsstrahlung opacity for all flavors (see Fig.~\ref{fig:mfp}), in the density domain relevant for CCSNe and PNS deleptonization, up to a few times the nuclear saturation density. At higher densities, higher order corrections become important, due to nucleon--hole (and also $\triangle$--hole) excitations, such that the subsequent $p$-wave pion polarization in nuclear matter leads to the softening of the in-medium pion dispersion relation, which is not captured by the approximate dressing of the vertex functions. Modern nuclear interactions from chiral effective field theory
have been used to study neutrino bremsstrahlung rate based
on the Landau's theory of Fermi liquids \cite{Lykasov:2008yz,Bacca:2008yr,Bacca:2011qd}. To account for
the nonperturbative feature of the nuclear force, NN bremsstrahlung rates were obtained in an independent study based on the $T$-matrix elements extracted from measured phase shifts \cite{bartl}. Using a two-body nuclear potential from chiral effective field theory, Ref.~\cite{Guo:2019cvs} solved the Lippmann-Schwinger equation exactly for the $T$-matrix including non-diagonal and off-shell
contributions, based on which improved bremsstrahlung rates were computed. Compared to the case using one-pion exchange potential, the bremsstrahlung rates from the $T$-matrix are higher at density $\lesssim 10^{12-13}$~g/cm$^3$ and are suppressed at higher densities \cite{Guo:2019cvs}.
Employing the $T$-matrix rates rather than the ones based on one-pion-exchange potential in spherically symmetric simulations has a marginal impact on neutrino luminosities, resulting in a change of approximately $5\%$-$10\%$~\cite{Bartl:2016iok,Betranhandy:2020cdf}. Additionally, it leads to a slight increase of $\sim 0.5$ MeV in the average energies of heavy lepton flavor neutrinos.

Similarly to neutrino bremsstrahlung, the so-called modified Urca processes, such as $\bar\nu_e + e^- + p + N \to n + N$ and $\bar\nu_e + p + N \to e^+ + n + N$ with $N$ a speculator nucleon, and similarly for $\nu_e$, could act as additional opacity sources for low energy neutrinos (as discussed in a recent work \cite{Pascal:2022qeg}). The nucleon-nucleon collision leads to a broadening in the neutrino response functions. Such a collision broadening effect was approximately treated for the charged current neutrino-nucleon reactions (see, e.g., \cite{Roberts12,Pascal:2022qeg}). Note that the effects due to mean field correction and nucleon-nucleon correlation and collision on neutrino opacities, neutrino signals, and the related nucleosynthesis could interfere with one another \cite{Roberts12}. A complete and consistent treatment further taking into account the relativistic kinematics, full hadronic weak interaction terms, and the potential presence of light cluster and nuclear pasta is still lacking.

\subsubsection{Additional degrees of freedom} \label{subsec:addof}

With slightly larger masses compared to $\mu^\pm$, abundant pions can also appear in hot and dense nuclear matter. The thermal pions not only soften the EoS but also contribute to the opacity of low energy muon neutrinos via $\bar\nu_\mu + \mu^- \to \pi^-$ and $\nu_\mu + \pi^- \to \mu^-$ \cite{Fore:2019wib}. In addition to pions, hyperons such as $\Lambda$ and $\Sigma^-$ may be present \cite{Pons:1998mm} and participate in weak reactions with neutrinos (see e.g., \cite{Prakash:1992,Reddy:1996tw,Reddy98,Mornas:2004vt}). 

Even for purely nucleonic matter, the composition vary with density and temperature. Besides the appearance of light clusters at low densities, nucleons can bind into strange configurations and shapes, commonly referred to as nuclear pasta, at low temperatures and  subsaturation densities (see e.g., \cite{Ravenhall:1983,Hashimoto:1984,Williams:1985prf,Oyamatsu:1993zz,Lorenz:1992zz,Sumiyoshi:1995np,Pethick:1995di,KIDO:2000,Watanabe:2000rj} and a recent review \cite{Lopez:2020zne}). 
The presence of nuclear pasta can modify the neutrino scattering rates and thus affect the transport of neutrinos, the cooling of NS, and the related SN neutrino signals \cite{Watanabe:2000zt,Horowitz:2004yf,Horowitz:2004pv,Alloy:2010fk,Alcain:2014cda,Furtado:2015vga,Horowitz:2016fpa,Roggero:2017pag,Schuetrumpf:2019hqe,Lin:2020nxy}.
With typical energies of a few tens MeV, neutrinos produced from SN core have wavelengths comparable to the size of the pasta structures and can scatter coherently on the nucleon cluster with enhanced cross sections \cite{Horowitz:2004yf,Horowitz:2004pv}. Incorporating such an enhanced cross section with the  heterogeneous nuclear pasta in SN simulations, Ref.~\cite{Horowitz:2016fpa} found that the neutrino luminosity increases significantly at late times of 10 s after core collapse. However, Ref.~\cite{Roggero:2017pag} considered the dissolution of nuclei and pasta due to finite temperature effects and concluded that the impact of nuclear pasta on SN neutrino opacities and signals is only modest.

\subsection{Impact of neutrino-nucleon rates on PNS evolution and nucleosynthesis in the $\nu$-driven wind}\label{subsec:ndw_nucl}

\begin{figure}[htp]
\centering
\subfigure[Radial profiles]{
\includegraphics[width=0.41\columnwidth]{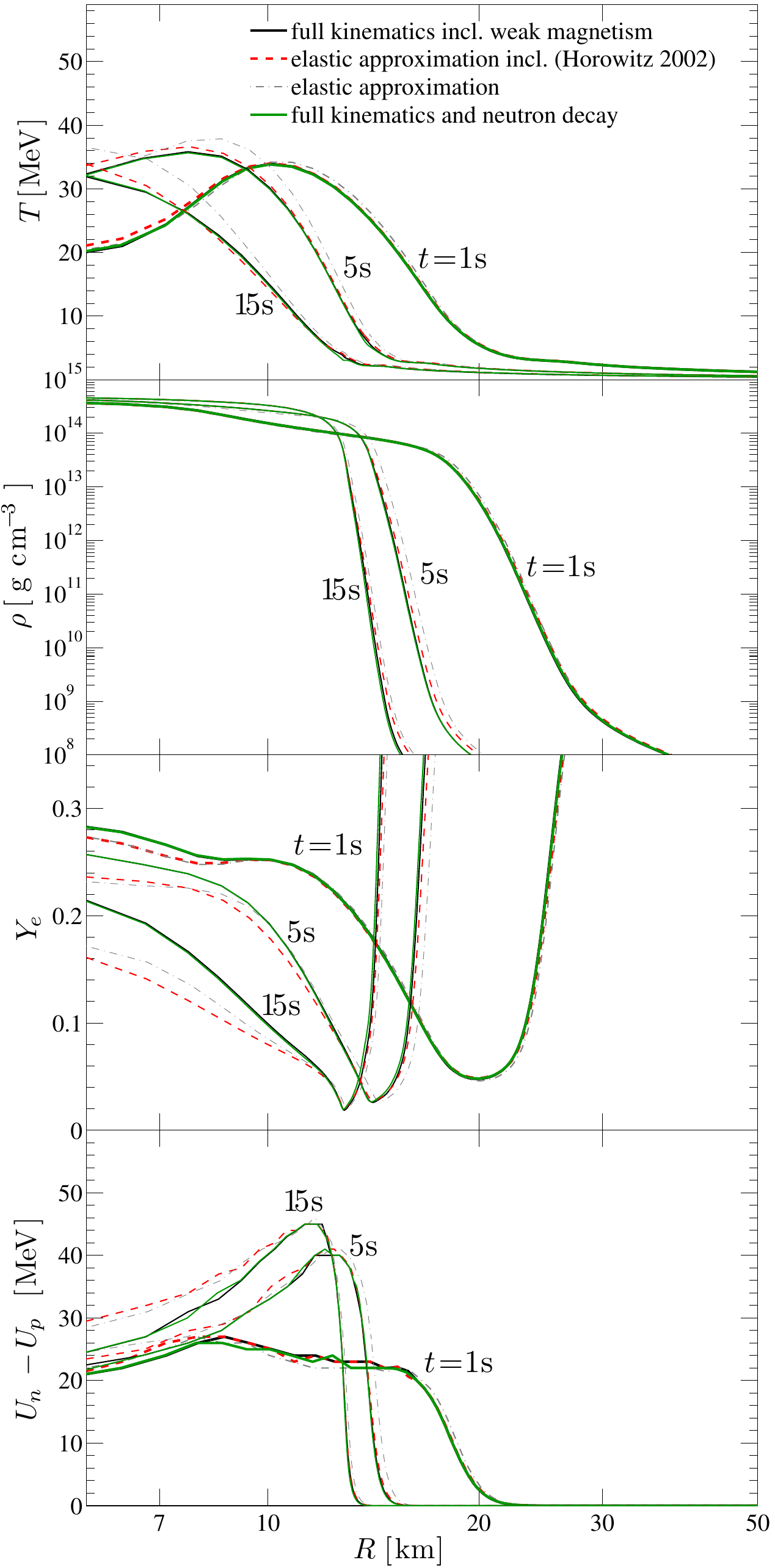}
}
\subfigure[Evolution of neutrino luminosities and average energies]{
\includegraphics[width=0.42\columnwidth]{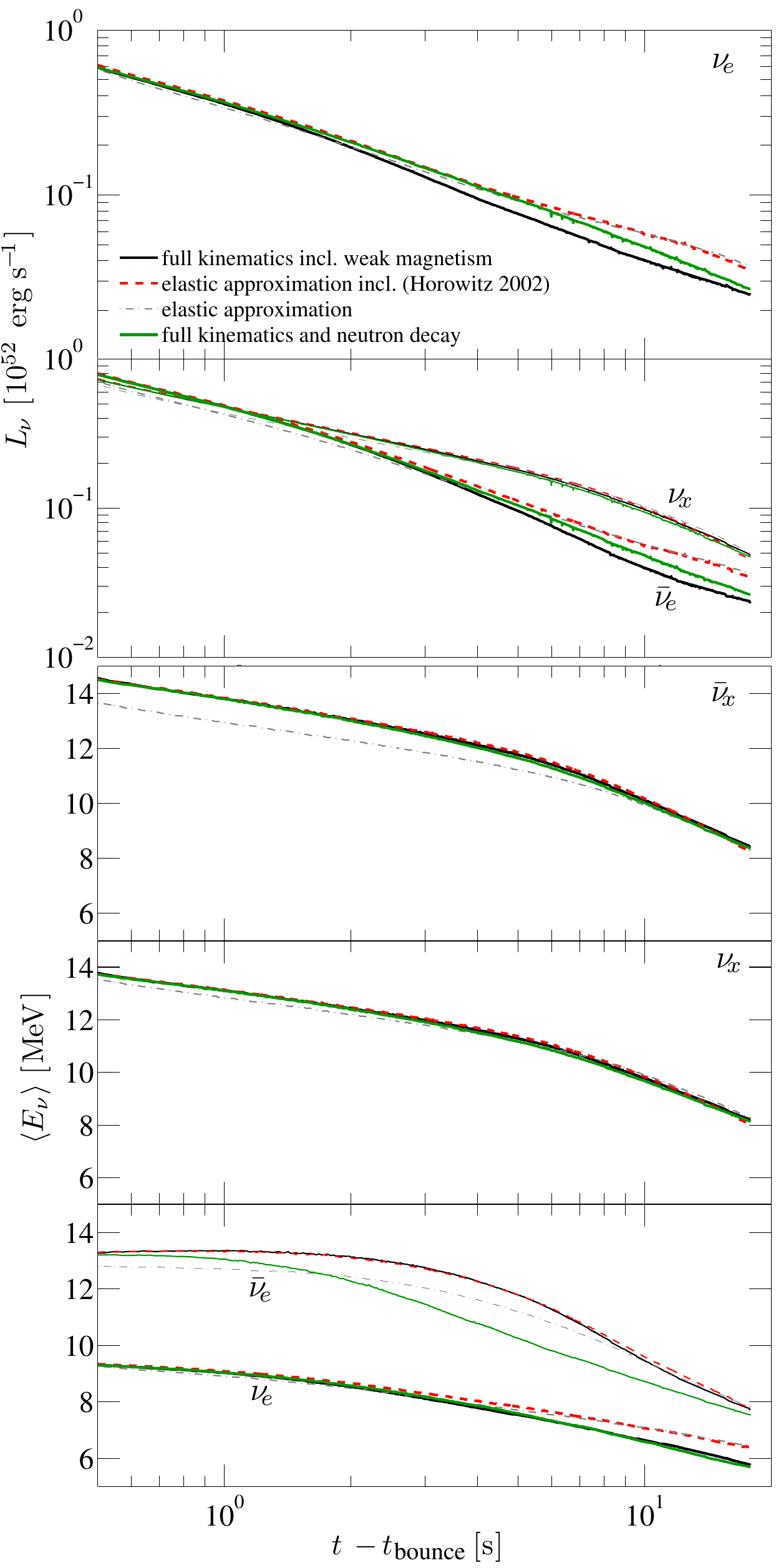}
}
\caption{PNS evolution, comparing four different treatments of the electronic charged-current reactions (see Sec.~\ref{sec:neutrino-nucleon}). Panel~(a):~Radial profiles of selected quantities (from top to bottom), temperature $T$, rest mass density $\rho$, electron fraction $Y_e$ and mean field potential difference between neutrons and protons, $U_n-U_p$, at three selected post-bounce times. Panel~(b):~Evolution of the neutirno luminosities $L_\nu$ and average energies $\langle E_\nu \rangle$, for all flavors ($\nu_x$ and $\bar\nu_x$ denote collectively $\nu_{\mu/\tau}$ and $\bar\nu_{\mu/\tau}$).  (Figures reprinted from Ref.~\cite{Fischer20a})}
\label{fig:delept}
\end{figure}

\begin{figure}[htp]
\centering
\includegraphics[width=0.5\columnwidth]{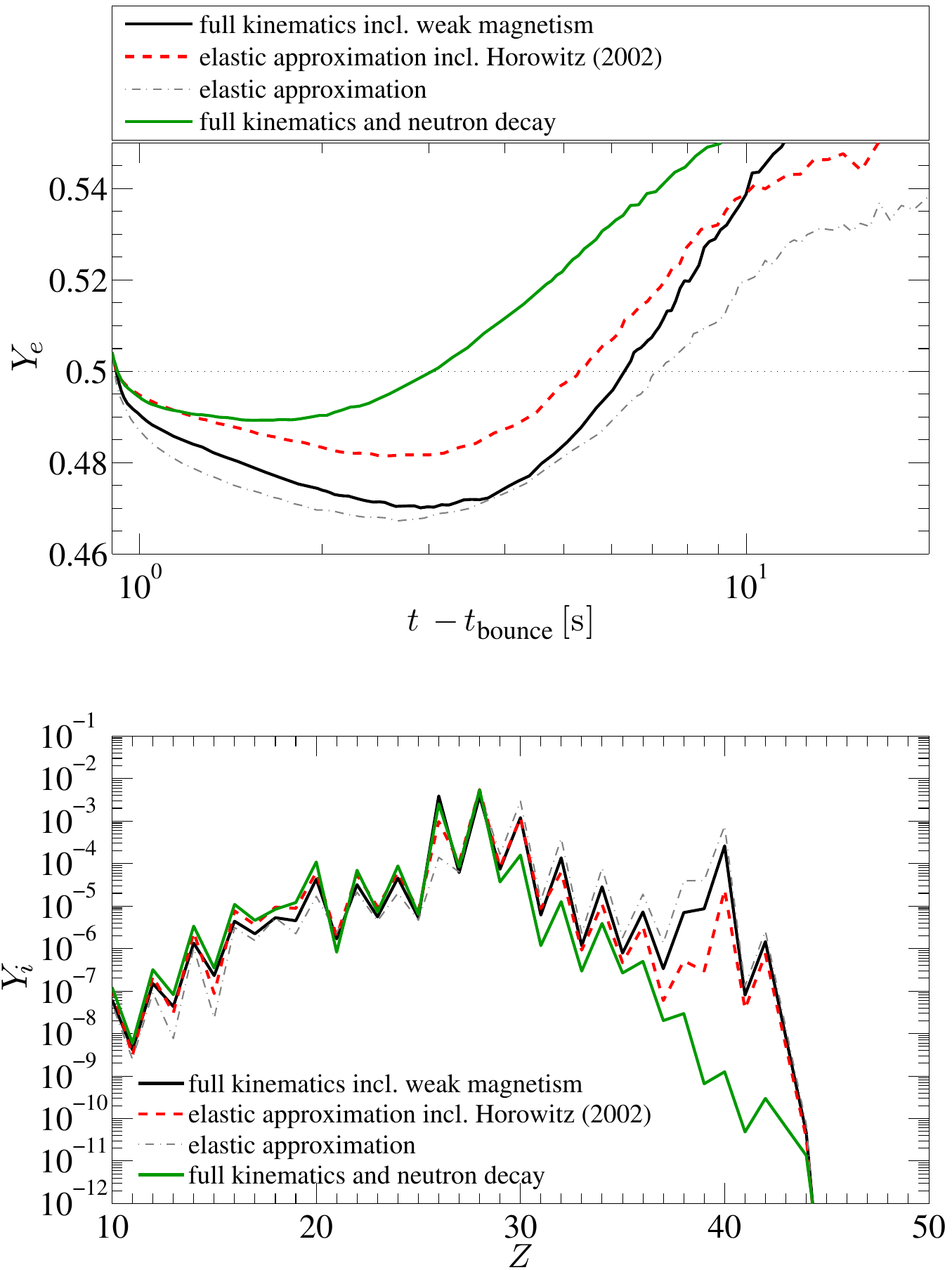}
\caption{Evolution of the neutrino-driven wind $Y_e$ (top panel) and elemental yields from integrated  nucleosynthesis analysis (bottom panel), comparing the different treatments of the charged-current weak rates introduced above. The data are obtained from simulations of the PNS deleptonization within general relativistic neutrino radiation hydrodynamics based on Boltzmann neutrino transport. (Figure reprinted from Ref.~\cite{Fischer20a})}
\label{fig:NDW_ye-abund}
\end{figure}

Accurate neutrino transport and a complete set of weak processes are both key input physics in the simulations of the PNS deleptonization, during which the neutrino-driven wind can be launched as described in Sec.~\ref{subsec:pnsevol}. 
Since these winds initially have high temperatures on the order of few MeV at the PNS surface before being ejected [Fig.~\ref{fig:NDW-tracers}(e)], the nuclear composition are mostly composed of free protons and neutrons in the beginning, and their ratio is sensitively determined by the neutrino luminosities, spectra and their evolution during the PNS deleptonization. 
These nucleons can combine to nuclei once the matter reach colder regions at larger radii. 
The outcome of the nucleosynthesis depends strongly on the proton-to-nucleon ratio $Y_e$, especially whether $Y_e<0.5$ (neutron rich) or $Y_e>0.5$ (proton rich). As the neutrino-nucleon reactions are fast enough, the initial value of $Y_e$ is simply given by \cite{Qian96} 
\begin{equation}
Y_e\approx\frac{\lambda_{\nu_en}}{\lambda_{\nu_en}+\lambda_{\bar\nu_ep}}~,
\label{eq:ye_eq}
\end{equation}
where $\lambda_{\nu_en}$ and $\lambda_{\bar\nu_ep}$ are the rates for $\nu_e$ and $\bar\nu_e$ captures on free nucleons, respectively. Consequently, $Y_e$ is mainly determined by the neutrino luminosities and their spectra, in particular the difference between $\nu_e$ and $\bar\nu_e$.
For example, a larger (smaller) $\langle E_{\bar\nu_e} \rangle - \langle E_{\nu_e} \rangle$ leads to a lower (higher) $Y_e$.
Recent long-term 1D simulations found that the neutrino luminosities and average energies of different flavors in this phase generally decrease with time and become increasingly similar; see e.g., Fig.~\ref{fig:delept}(b) and Refs~\cite{Huedepohl10,Fischer10}.
This is mainly because the neutrino--nucleon scattering, which is flavor blind, becomes increasingly more important than the inelastic charged current processes, as the latter is further suppressed by the final state blocking~\cite{Fischer12}. 
Thus, the resulting $Y_e$ in the wind are found to be around 0.5 (see e.g., Fig.~\ref{fig:NDW-tracers}(d) or Fig.~\ref{fig:NDW_ye-abund}). 
Besides $Y_e$, the wind dynamical timescale and the entropy, which can be related to the high-density EoS, in terms of the mass and radius of the nascent PNS, i.e. their ratio $M/R$ (see e.g., Ref.~\cite{Qian96}), also play important roles in determining the nucleosynthesis outcome. Typically, a more compact PNS that results from a softer high-density EoS features higher entropy in the neutrino-driven wind.

Since the neutron-richness of the neutrino-driven wind is directly affected by the energy spectra of $\nu_e$ and $\bar\nu_e$, which can depend on the evolution history of the PNS and the decoupling of neutrinos of all flavors, in recent years, a number of studies investigated the impact of both the charged-current and neutral-current neutrino nucleon interactions on the PNS evolution and the nucleosynthesis conditions~\cite{MartinezPinedo12,Roberts12,Fischer2016AA,Bartl:2016iok,Pascal:2022qeg}. 
In particular, Refs.~\cite{MartinezPinedo12,Roberts12} found that including the medium modification to the charged-current rates can significantly lower the $Y_e$ to slightly neutron-rich during the early neutrino-driven wind phase. 
More recently, Ref.~\cite{Fischer20a} investigated in detail the improved description including consistent treatment of full kinematics, weak magnetism contributions, as well as the (inverse) neutron decay reactions (see Sec.~\ref{sec:CC}), in accordance with the underlying EoS. 
Figure~\ref{fig:delept} compares radial profiles of selected quantities at three post-bounce times during the PNS deleptonization in the panel (a) as well as the corresponding neutrino luminosities and average energies evolution in the panel (b), for four different treatments of the charged-current weak processes involving the unbound baryon: 
i) the elastic rates without inelastic contributions and corrections due to
weak magnetism;
ii) elastic approximation but taking into account nucleon recoil and weak magnetism correction formulated in Ref.~\cite{horowitz02}; iii) and iv) the fully inelastic rates of Ref.~\cite{Fischer20a} without and with neutron decay (see Sec.~\ref{sec:neutrino-nucleon}).
In particular, it was found that the improved treatment results in higher $Y_e$ values inside the PNS as shown in Fig.~\ref{fig:delept}(a).
This is due to the modified condition of weak equilibrium established as well as the reduction of the $\bar\nu_e$ average energies during the initial deleptonization phase before the impact of the charged-current reactions diminishes, which reduces the deleptonization timescale indicated by the lower central temperatures at smaller radii at intermediate and later times of about $>5~{\rm s}$ post bounce, as shown in Fig.~\ref{fig:delept}(b). 

Moreover, as shown in Fig.~\ref{fig:delept}(b), the inclusion of weak magnetism and recoil mainly suppresses the opacity for $\bar\nu_e$ and thus shifts the neutrinosphere of last inelastic scattering to higher densities and temperatures. Therefore, the average energies of $\bar\nu_e$ will be enhanced. Naively, one would expect a reduced $Y_e$, or an enhanced neutron fraction, due to a larger spectral difference between $\nu_e$ and $\bar\nu_e$. However, this is not the case since weak magnetism also reduces the cross section for $\bar\nu_e$ capture on protons and $\lambda_{\bar\nu_ep}$ in Eq.~\eqref{eq:ye_eq} as well. The reduction in $\lambda_{\bar\nu_ep}$ is primary and the net effect is to enhance $Y_e$ to be closer to 0.5 within a few seconds post core bounce (see e.g., the red and black lines in the upper panel of Fig.~\ref{fig:NDW_ye-abund}). 
Compared to the elastic approach with approximate inclusion of weak magnetism and recoil \cite{horowitz02}, the full kinematics treatment with weak magnetism gives rise to larger charged current opacities for $\nu_e$, and $\bar\nu_e$ to a lesser
extent, which  reduces the $\nu_e$ luminosity more evidently. 
Consequently, the ejected winds turn more neutron rich (see the black line in the upper panel of Fig.~\ref{fig:NDW_ye-abund}). However, when inverse neutron decay is considered as an additional source for $\bar\nu_e$, the spectral difference between $\nu_e$ and $\bar\nu_e$ decreases significantly, turning material slightly neutron rich at the early deleptonization stage within $\sim 3$ s post bounce, and proton rich later.

The bottom panel of Fig.~\ref{fig:NDW_ye-abund} shows
the corresponding nucleosynthesis yields in the neutrino-driven wind with these four different treatments of the charged current neutrino-nucleon rates for $\nu_e$ and $\bar\nu_e$ as described above. As already found in previous studies, the neutrino-driven wind at early times are not neutron rich enough to support an $r$-process, but the light neutron capture nuclei with atomic numbers of $38<Z<42$ could be produced under the slightly neutron-rich conditions encountered in simulations without the inclusion of inverse neutron decay. Taking into account the inverse neutron decay, however, the neutron fraction is reduced and the nucleosynthesis path terminates even before reaching the light neutron capture nuclei.   

In fact, most of the ejected matter in the SN neutrino wind
is turned slightly proton-rich by neutrino interactions, which may allow a nucleosynthesis process called $\nu p$-process \cite{Froehlich06a,Froehlich06b,Pruet06,Wanajo06} to occur.
Upon reaching colder regions, the ejected proton-rich matter freezes-out mainly into $\alpha$ nuclei, with the excess protons left as free particles. 
In principle, heavier nuclei can be synthesized from the freeze-out abundances by subsequent proton captures, interrupted by $\beta^+$ decays. This situation is realized in explosive hydrogen burning in X-ray bursters.
However, in this scenario the mass flow is strongly hampered by the increasing Coulomb barrier of the produced nuclei and by the so-called waiting points.
These are $\alpha$ nuclei ($^{56}$Ni, $^{64}$Ge...) which have relatively long $\beta$ half lives and for which proton captures are strongly hindered by the small or negative proton binding energies of the final nuclei. 
In contrast to X-ray bursts, the free protons in the SN wind are immersed in a strong flux of neutrinos emitted by the PNS. By absorption of $\bar\nu_e$ on ree protons, these can be turned into a fresh supply of free neutrons which can interact with the already produced nuclei. 
In this way the matter flux can overcome the waiting point nuclei by $(n,p)$ reactions and can continue to produce intermediate mass nuclei up to the mass range $A \sim 80-100$ or above. 
The final abundance distribution and the efficiency of $\nu p$ process nucleosynthesis can depend sensitively on several factors, including the $Y_e$ value of the wind, the $\bar\nu_e$ flux, as well as the formation of the wind termination shock~\cite{Wanajo:2010mc,Arcones:2012wj}.
Although recent 1D SN simulations tend to suggest that the $\nu p$ process does not occur very efficiently in producing $p$-nuclei heavier than e.g., $^{94}$Mo, see e.g., Ref.~\cite{Pllumbi:2014saa,Fischer20a} and Fig.~\ref{fig:NDW_ye-abund}, 
other unaccounted effects such as nuclear correlation on the neutrino-nucleon rates \cite{Pascal:2022qeg,Shin:2023sei}, the flavor oscillations of neutrinos, which will be discussed in Sec.~\ref{sec:oscnucleo_snenuwind}, or the multidimensional nature (see below) may still alter this conclusion. 

In principle, the waiting points can also be overcome by absorbing antineutrinos which would change the proton to a neutron inside the nucleus. But the rates are noticeably disfavored compared to the two-step process by $p(\bar{\nu_e},e^+)n$ and $(n,p)$ reactions.  Ref.~\cite{Sieverding18} has explored the influence of the various
charged- and neutral-current neutrino-nucleus reactions on the $\nu p$ process abundances and finds negligible effect. 
Also, during the long-term PNS evolution muons should be considered as well.
In particular, as the deleptonizing and cooling PNS is evolving towards the cold remnant neutron star, the muon abundance increases continuously as the electron and muon chemical potentials become increasingly similar. However, at present, no impact on the neutrino fluxes and the spectra nor their evolution could be identified from the inclusion of muonic weak processes (c.f. the appendix of Ref.~\cite{Fischer21}). 

We note that most of the above discussions are based on 1D simulations, which cannot capture several multidimensional effects that may be important in affecting the neutrino properties and the wind outflow conditions. 
For instance, recent multi-dimensional long term simulations carried over several seconds post bounce generally found that the neutrino-driven wind can be rather aspherical due to the sustained fallback materials, depending on the mass of the progenitor and the asphericity of the explosion~\cite{Stockinger:2020hse,Bollig21,Witt21,Wang:2023vkk}.
In particular, Ref.~\cite{Wang:2023vkk} examined the detailed hydrodynamical and thermodynamical conditions, as well as $Y_e$ in the neutrino-driven wind from 3D simulations with spectral transport of neutrinos. 
They found that the wind properties can resemble those obtained with 1D models for lighter progenitors, but differ more significantly from 1D cases for heavier progenitors. 
However, the resulting nucleosynthesis patterns still seem to be in broad agreement with what is obtained in 1D simulations, as the $Y_e$, entropy, and expansion velocities allow only the production of heavy elements up to $Z\sim40$ and $A\sim 90$.
Moreover, preliminary studies suggested that the convection inside the PNS can possibly alter the decoupling of neutrinos of different flavors and therefore changing the resulting $Y_e$ in the wind~\cite{mirizzi16}, or even generate the gravito-acoustic waves that may strongly affect the wind dynamics and the nucleosynthesis~\cite{Nevins:2023tug}.

Finally, we note that there are exceptions, in which the neutrinos are not primarily responsible for the determination of the nucleosynthesis conditions if the matter is ejected relativistically. Among them are electron-capture SN \cite{kitaura06} and the alternative scenarios briefly mentioned at the end of sec.~\ref{sec:n-driven-sn} -- the magneto-rotational SNe in which a jet forms and quark-hadron phase transition driven SN explosions. Both these latter cases result in the initial ejection of very neutron rich material, while slightly neutron rich pockets are ejected in electron-capture supernovae, followed by the standard neutrino-driven wind on a longer timescale on the order of several seconds, as discussed above.

\subsection{Neutrino-nucleus interaction}
\label{Sec:v-nucleus}
The neutrinos produced in various stages of the SN can interact with nuclei in different ways with potential relevance for the SN dynamics, nucleosynthesis and detection. In this section we will briefly discuss how the interaction between SN neutrinos and nuclei can be described and which consequences they have on these various aspects.   

Neutrino reactions on nuclei can be described in perturbation theory.  
The formalism for the calculation of the various neutrino-nucleus reaction cross sections has been developed
in Ref.~\cite{Connell72}. It is based on a state-by-state evaluation connecting initial and final nuclear states by  multipole operators weighted by an appropriate phase space factor. For the typical low energies and momentum transfers involved, the description reduces to a nuclear structure problem where mainly allowed and first-forbidden transitions are of importance. Furthermore, the energies are too low to allow for charged-current reactions induced by muon and tau neutrinos. These neutrino families contribute only to inelastic neutrino-nucleus scattering which is neutrino flavor blind. 

Fermi transitions, which only contribute to elastic neutrino scattering and to $(\nu_e,e^-)$ reactions, can only connect isobaric analog states, where the transition strength is given by the Fermi sum rule. 
There has been significant experimental and theoretical
progress in determining the Gamow-Teller (GT) strength functions in nuclei. Here one distinguishes between
GT$_-$ and GT$_+$ strengths, depending on whether a neutron is changed into a proton (GT$_-$) or a proton into a neutron (GT$_+$). The former contributes to
($\nu_e,e^-)$ reactions, while the latter to ($\bar\nu_e,e^+$). For inelastic scattering the relevant strength is GT$_0$ where the identity of the proton or neutron struck by the neutrino is conserved.
The GT$_-$ and GT$_+$ strengths can be studied by
intermediate energy charge-exchange reactions like
$(p,n)$, $(^3$He,$t)$ or $(n,p)$, $(d,^2$He), respectively.
Recent reviews on this subject can be found in
\cite{Frekers18,Langanke21}.
As important results these studies revealed that the GT distributions are strongly fragmented in the final nucleus and the total GT strengths are noticeably reduced compared to the expectations from a model of independent particles \cite{Fuller82}. Both observations are related to 
strong correlations among the nucleons. Such correlations are accounted for in the nuclear shell model \cite{Caurier05} and indeed the experimentally determined GT strengths are well reproduced by large-scale shell model calculations \cite{Caurier99,Caurier05,Suzuki22}, if the latter are renormalized by a constant factor
\cite{Brown88,Langanke95,Martinez97}. The GT$_-$ distribution shows a noticeable concentration of the strength in the so-called GT resonance whose centroid is usually also well reproduced within the quasiparticle random phase approximation (QRPA) which, however, does not
describe the fragmentation as well as the shell model
(see for example \cite{Scott14}).

The shell model has been used to calculate the GT distributions for the about 100 mid-mass nuclei in the iron-nickel-region which dominate the core composition in the early phase of the collapse \cite{Caurier99,Langanke00}. These GT distributions have then been used to calculate appropriate electron capture rates, from which the $(\nu_e,e^-$) rates are derived by detailed balance. The relation of electron capture and
$(\nu_e,e^-$) rates have also been exploited in SN simulations at higher densities. However, as the electron and neutrino energies are larger at this stage of the collapse, the rates were derived by QRPA calculations \cite{langanke03,Janka07}, including contributions from forbidden transitions and most importantly using partial occupation numbers as derived from shell model Monte Carlo calculations at finite temperatures accounting for the relevant nuclear correlations \cite{Johnson92,Koonin97}. All rates have been evaluated
taking the finite temperature of the environment into account which is quite important for electron capture at nuclear shell gaps \cite{Dzhioev20,Langanke21,Litvinova21,Dzhioev22,Giraud2022}.
Electron antineutrinos are much less abundant during collapse than electron neutrinos. The ($\bar\nu_e,e^+$) rates are derived from positron capture rates by detailed balance. 

The only neutral-current cross section measured for nuclei heavier than the deuteron is the transition from the ground state to the $T=1$ state at 15.11 MeV in $^{12}$C \cite{Zeitnitz94,Armbruster98}. Hence one has to use indirect experimental information to validate $(\nu,\nu'$) cross sections. Due to the relatively low energies of SN neutrinos, allowed GT$_0$ transitions 
dominate the cross sections. As the GT$_0$ strength is closely related to M1 excitations of nuclei, precision
M1 data obtained by inelastic electron scattering are a useful constraint to the extent that the isoscalar and orbital pieces present in the M1 data can be neglected.
Here three experimental observations help: 1) the isovector component dominates over the isoscalar piece; 2) the orbital and spin M1 responses are usually
experimentally well separated with the major orbital strength, for example the collective scissors mode, residing at significantly lower energies (at about 2-4 MeV) than the major spin excitations (at about 7-9 MeV)
\cite{Guhr90}; 3) the orbital part is strongly suppressed in spherical
nuclei \cite{Enders99}. Detailed studies of spherical $pf$ shell nuclei
performed within the shell model reproduce the M1 data obtained by inelastic electron scattering very well
\cite{Langanke04}, including the strong dominance of isovector over isoscalar excitations. We note that
results calculated from the shell model GT$_0$ strength
or the experimental or shell model M1 data agreed quite well \cite{Langanke04}, justifying the use of the shell model for the calculation of the allowed contributions to the inelastic neutrino-nucleus cross sections.

In Refs.~\cite{Kolbe99,Toivanen01} a hybrid model, in which the allowed transitions were derived within the shell model and the forbidden within the QRPA approach, has been proposed to calculate inelastic neutrino-nucleus scattering. This hybrid model was then extensively used
to calculate such cross sections for many mid-mass (pf shell) nuclei \cite{Juodagalvis05}.
Calculations of inelastic neutrino-nucleus scattering for selected nuclei, in which all multipole contributions are derived within the QRPA, were reported in Refs.~\cite{Dzhioev14,Dapo12,Paar15,Chasioti07}. 

During the collapse 
the neutrino-nucleus reactions occur in a finite-temperature environment in which nuclei exist as a thermal
ensemble rather than in the ground state. For small neutrino energies this has a significant effect on the cross sections which are strongly enhanced compared to those at $T=0$ \cite{Langanke04,Juodagalvis05}.
Finite temperature effects are relatively unimportant once the neutrino energy is large enough to excite the main GT$_0$ strength at 7--9~MeV.

\begin{figure}
  \begin{center}
    \includegraphics[width=0.7\linewidth]{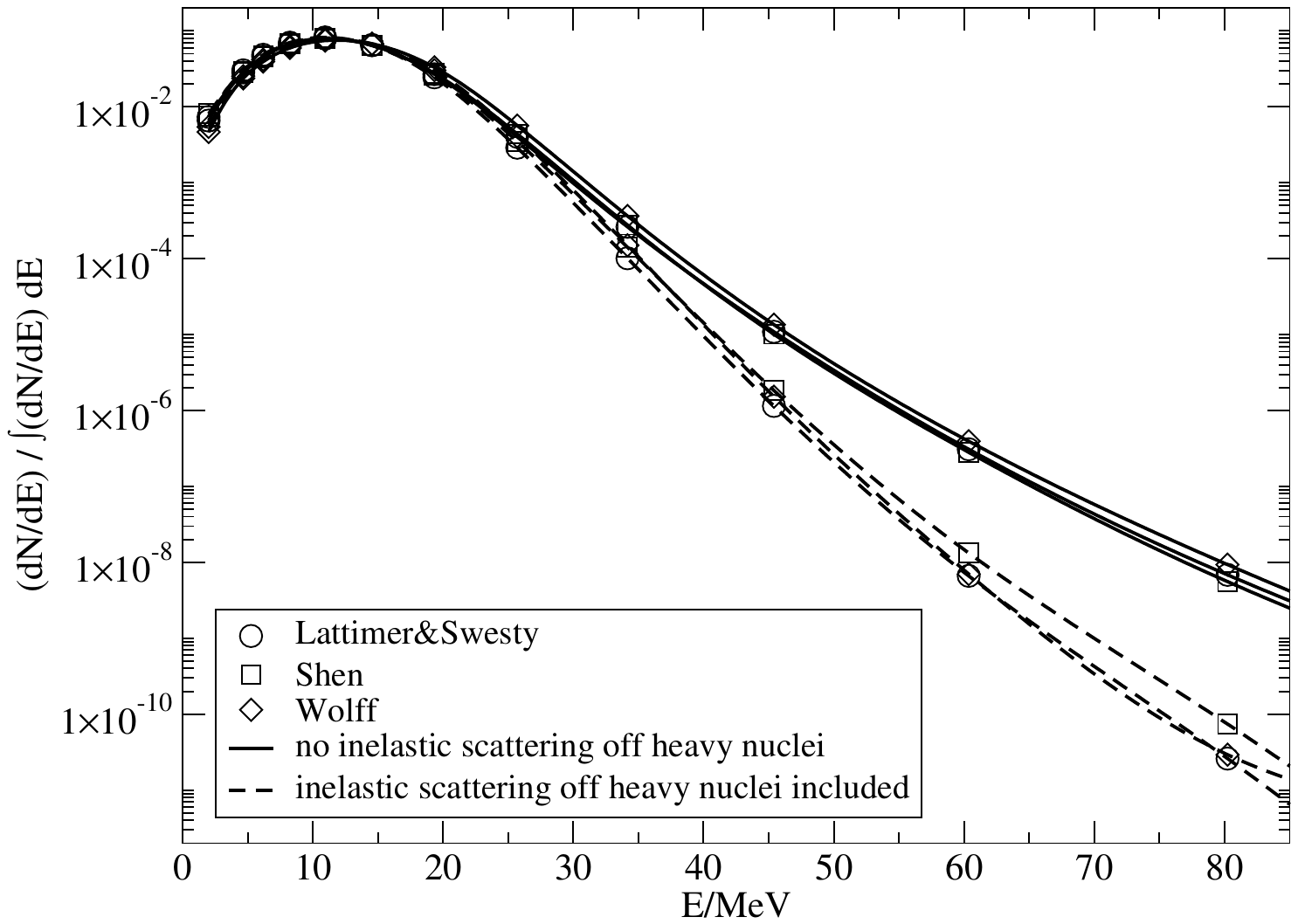}
    \caption{Normalized $\nu_{\mathrm{e}}$ number spectra radiated
      during the shock breakout burst as seen by a distant observer at
      rest. Results are shown for simulations with three different
      nuclear EoSs.  For better comparison of
      the strongly time-dependent spectra during this evolution phase,
      integration in a window of 8$\,$ms around the peak luminosity
      was performed.  Inelastic neutrino scattering off nuclei (dashed
      lines) leads mostly to energy losses of high-energy neutrinos
      and thus reduces the high-energy tails of the spectra. 
      (Figure adapted from
      \cite{Langanke08}). \label{fig:nuespectra}}
  \end{center}
\end{figure}

It has been pointed out \cite{Haxton88} that inelastic neutrino-nucleus scattering can contribute to the shock revival and the thermalization of neutrinos with matter.
A first study performed with relatively simplified neutrino-nucleus cross sections suggested that neutrino-nucleus scattering can compete with inelastic neutrino-electron scattering (reaction~8 in Table~\ref{tab:nu-reactions}) for moderate and high neutrino energies as mechanism to transfer energy from neutrinos to electrons and nucleus to thermalize neutrinos with matter, while it is rather inefficient in reviving the stalled shock \cite{Bruenn91}. A detailed SN simulations employing the hybrid model cross sections of Ref.~\cite{Juodagalvis05} showed that inelastic neutrino-nucleus scattering has a rather negligible effect
on the neutrino thermalization during collapse \cite{Langanke08}. However, the study found that these reactions modify the spectra of the $\nu_e$ burst neutrinos
as they have to traverse a region ahead of the shock where heavy nuclei still exist. Due to the energy dependence of the cross section, scattering occurs more often for high-energy neutrinos, shifting the spectra of $\nu_e$ burst neutrinos slightly to lower energies, in particular reducing its high-energy tail \cite{Langanke08} (see Fig. \ref{fig:nuespectra}). This change of the spectra
reduces in turn the detection event rates for the $\nu_e$ burst neutrinos where the reduction factor depends noticeably on the detection material and its threshold for neutrino observation \cite{Langanke08}.

Neutrino-induced reactions on nuclei contribute to the SN nucleosynthesis after bounce. This can proceed by charged-current $(\nu_e,e^-)$ and $(\bar\nu_e,e^+$) reactions, where the initial and final nuclei are different, but also by neutral-current $(\nu,\nu')$ reactions when the neutrino energy is large enough to excite the nucleus above particle thresholds such that the excited state decays by particle emission. Neutrino-induced neutral-current reactions are possible for all six neutrino types. They are most relevant for muon and tau neutrinos and antineutrinos due to their slightly higher average energies. Also the charged-current reactions can leave the final nucleus in an excited state which decays by particle emission.

Neutrino-induced spallation reactions can be described in a two-step procedure
\cite{Kolbe92,Kolbe01}. In the first step the neutrino-induced excitation cross section to a given final state is calculated on the basis of a chosen nuclear model.
The decay is then evaluated within the statistical model
considering $\gamma$-decay and emission by light particles
(proton, neutron, $\alpha$). For high-energy neutrinos or for nuclei far from stability, for which proton or neutron thresholds are small, the decay can also proceed by multiple particle emissions. The neutrino-induced partial reaction cross sections are then obtained
by summing the excitation cross sections over all final states weighted with the appropriate decay probabilities.

There exists a comprehensive tabulation of partial cross sections for neutrino-induced charged- and neutral-current
reactions on nuclei (up to charge $Z=83$). This tabulation is based on the two-step procedure using
the QRPA model to calculate the nuclear transition strengths. The cross sections are given as functions of neutrino energies so that the results can be used for 
any SN neutrino spectra \cite{Sieverding18}. This is an improvement over previous similar work which was, however, restricted to neutron-rich $r$-process nuclei \cite{Langanke01a,Langanke01b}. We note again that the QRPA gives a good account of the total transition strengths, but underestimates its fragmentation. This can affect partial spallation cross sections if, for example, the GT strength mainly resides around a particle threshold. For specific nuclei like
$^{12}$C, $^{16}$O, $^{20}$Ne or $^{56}$Fe the excitation cross sections have therefore been calculated on the basis of the hybrid model or partially even on GT data if available from charge-exchange experiments (see Refs.~\cite{Yuan12,Balasi15,Langanke21} and references therein).  

For a more detailed review of the calculation of neutrino-nucleus reactions and the involved nuclear structure challenge and its validation the reader is referred to Refs.~\cite{Balasi15,Kolbe03,Suzuki22}.
Note that the framework of neutrino-nucleus interactions introduced above mainly concerns heavy nuclei. 
Light nuclei such as deuteron, triton, and helium isotopes can be present in regions where neutrinos decouple from the PNS (see the bottom panel in Fig.~\ref{fig:acc}).
The nuclear physics properties of these light clusters in dense medium differ from those in vacuum so that their weak rates with neutrinos need to be treated differently~\cite{O'Connor07,Fischer20c}. 
However, studies investigating the impact due to the presence of light nuclear clusters on SN neutrino emission and the dynamical evolution generally found relatively minor impact~\cite{Arcones:2008kv,Fischer20c}.

\subsection{Impact of neutrino-nucleus interaction on neutrino nucleosynthesis in supernovae}\label{sec:nu-nucleo}

Neutrino nucleosynthesis has been suggested as at least part of the Galactic origin of some selected nuclei like $^{11}$B, $^{15}$N,
$^{19}$F, $^{138}$La and $^{180}$Ta \cite{Woosley90}. The mechanism of the process is that neutrinos, emitted from the cooling PNS, strike nuclei in the outer shells of the star and change them by charged-current reactions or by
neutral-current induced spallation of protons or neutrons. For the process to contribute to the abundance
of a nuclide the struck nucleus must be quite abundant and, due to the small neutrino cross sections, the abundance ratio between struck and produced nuclei 
must be of order 1000 or larger.

It was traditionally conjectured that the nucleosynthesis is induced by the neutrinos released from the cooling of the PNS. The luminosities were assumed to be equal for the various neutrino types and to decrease
exponentially with time \cite{Woosley90}. The neutrino spectra were described by a Fermi-Dirac distribution with a characteristic time-independent temperature. The
average neutrino energies related to the temperature parameters by $\langle E_{\nu}\rangle \approx 3.15 T_{\nu}$ were expected to obey the hierarchy $\langle E_{\nu_e}\rangle <\langle E_{\bar\nu_e} \rangle<\langle E_{\nu_x}\rangle$, where $\nu_x$ stands for muon and tau neutrinos and their antiparticles. 

Various studies of neutrino nucleosynthesis indicated that $^{11}$B, $^{15}$N and $^{19}$F are made by neutral-current reactions on $^{12}$C, $^{16}$O and $^{20}$Ne, respectively, when the neutrinos pass through the carbon and neon burning shells of the star. The process is mainly initiated by muon and tau neutrinos and antineutrinos which have sufficiently high energies to excite the struck nuclei above  
particle thresholds where they decay by emissions of $\alpha$ particles, protons and neutrons.
The $\alpha$ branch is unimportant for nucleosynthesis
producing $\alpha$ particles, $^{12}$C and $^{16}$O, which are 
already quite abundant. Proton and neutron decays, however, 
synthesize $^{11}$B (directly or via the $\beta$ decay of $^{11}$C), $^{15}$N (also via $\beta$-unstable $^{15}$O) and $^{19}$F (also via $\beta$ decay of $^{19}$Ne) \cite{Heger05a}. The other two nuclides, synthesized by neutrino nucleosynthesis, $^{138}$La and $^{180}$Ta, are produced by charged-current $(\nu_e,e^-)$ reactions
on $^{138}$Ba and $^{180}$Hf which have been both previously made by $s$-process nucleosynthesis \cite{Heger05a}.
It is intriguing to note that neutrino nucleosynthesis
is sensitive to the neutrino types which have not been observed from SN 1987A. It has even been suggested that neutrino nucleosynthesis might serve as a tool to determine the neutrino mixing angle $\Theta_{13}$ and mass ordering~\cite{Yoshida05,Yoshida08,Mathews12} which, however, requires a complete, far beyond currently available description of stellar evolution and SN modeling as well as the involved neutrino-nucleus reaction cross sections.

\begin{figure}
\includegraphics[width=\linewidth]{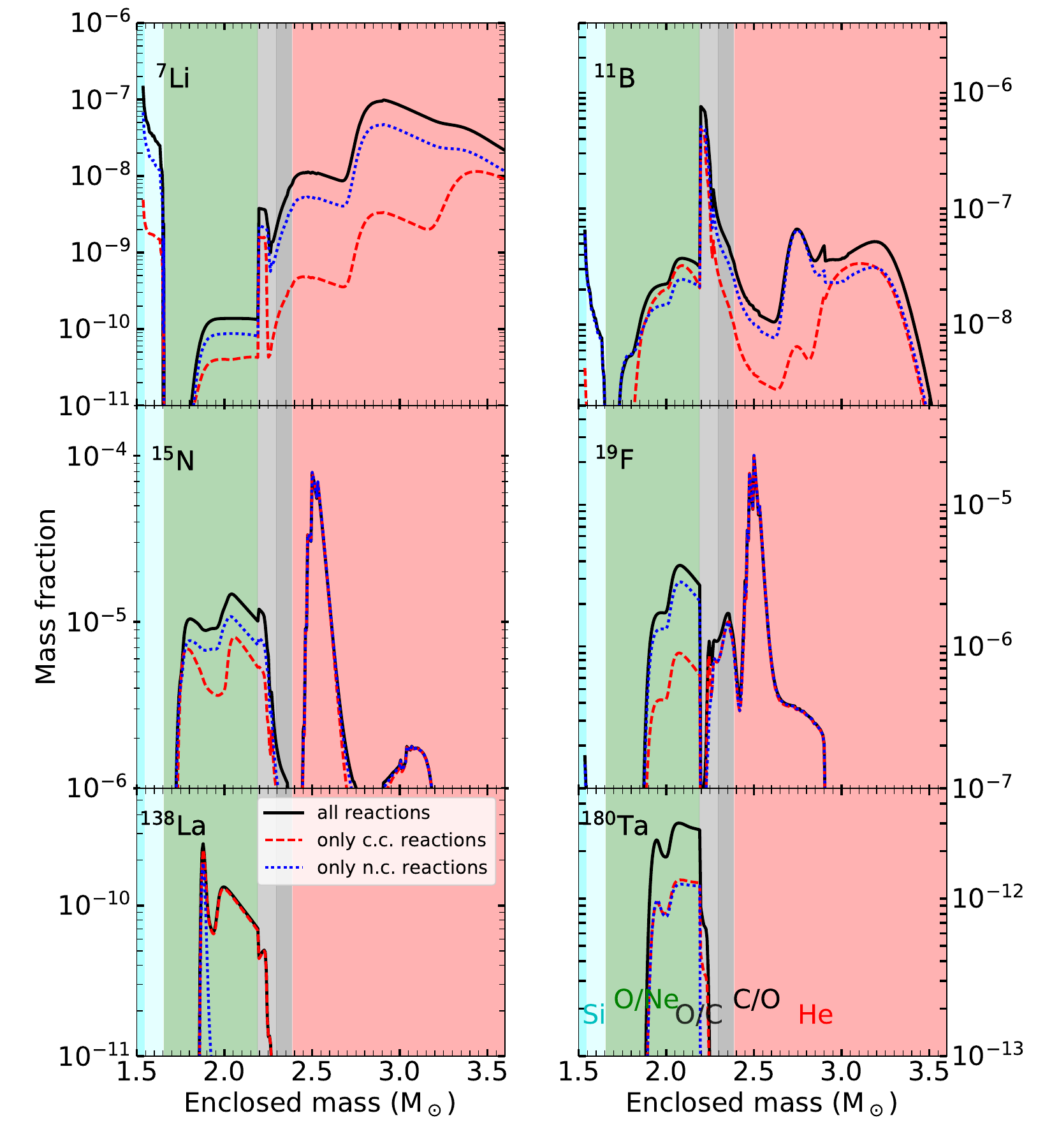}
\caption{Mass fraction profiles for the $15\,\mathrm{M}_\odot$ progenitor model with the low neutrino energies as chosen in 
Ref.~\cite{Sieverding18}
for the six isotopes most affected by the $\nu$~process. Red dashed and blue
dotted lines indicate results with only charged current (c.c.) and only neutral
current (n.c.) reactions. Note that the scale is different in each panel. The background colors 
indicate the different composition layers of the stellar model, as labeled in the bottom right panel.}
\label{fig:all_profile}
\end{figure}

The neutrino nucleosynthesis yields depend sensitively on stellar structure. First of all, the yields depend on the abundances of the primary nuclei from which they are produced. This is most relevant for $^{138}$La and $^{180}$Ta as their primaries have to be made by $s$-process nucleosynthesis. The yields of these two nuclides depend also on a competition between the neutrino-induced reactions which produce them mainly by ($\nu_e,e^-$) reactions and the subsequent fast burning triggered by the shock wave which destroys them
\cite{Sieverding19}. The outcome is further complicated
by the fact that both nuclides can also be synthesized
by the $\gamma$-process which mainly operates in the bottom layers of the O/Ne shells where the temperatures are sufficiently high to dissociate neutrons from $^{139}$La and $^{181}$Ta, respectively. Furthermore,
charged-current reactions on $^{138}$Ba can also excite
$^{138}$La above the neutron threshold, so that this leads to the production of $^{137}$La. In this context it is important that the GT$_+$ distributions on $^{138}$Ba, and also on $^{180}$Hf, have been determined experimentally \cite{Byelikov07}. For $^{180}$Ta a complication arises from the fact the it can be produced
in the stellar environment in two states: the ground state with a half life of about 8 hours (which is noticeably longer than the duration of the nucleosynthesis processes) and a long-lived isomeric state at an excitation energy of 75 keV with a half-life exceeding $10^{15}$ years. The estimate of the contribution which survives in the isomeric state is facilitated by the fact that both states are in equilibrium in the hot stellar environment  from which follows that about 35-39 $\%$ of $^{180}$Ta survives in the isomer \cite{Mohr07,Hayakawa10}.

Obviously a decisive factor for the yields of neutrino nucleosynthesis is the various neutrino spectra and luminosities. Due to more sophisticated and improved SN modeling the spectra predicted for the various neutrino types have changed over the years. This has in particular affected the spectra of muon and tau neutrinos which are now found to still have  higher energies than electron neutrinos, but the differences have decreased significantly. Assuming Fermi-Dirac forms for the various neutrinos, the temperature parameters for $\nu_{\mu,\tau}$,
$\bar\nu_e$ and $\nu_e$ have decreased from $T_\nu=8$, 6, 5 MeV, appropriate at the time of the pioneering work of Woosley et al. \cite{Woosley90}, to $T_\nu=6$, 5, 4 MeV, the accepted values at the time of Ref.~\cite{Heger05a}, and to $T_\nu=4$, 4, 2.8 MeV, the values used in \cite{Sieverding18}. These reductions in the average neutrino energies have significant effects on the yields of $^{11}$B, $^{15}$N and $^{19}$Ne, which are dominantly produced by neutrino-induced neutral-current
spallation. As the primary nuclei in these processes,
$^{12}$C, $^{16}$O and $^{20}$Ne, have rather high thresholds for proton and neutron spallation, the partial spallation reaction rates, and in turn the $^{11}$B, $^{15}$N and $^{19}$F
yields,
are significantly reduced by the changes in the neutrino spectra. Table \ref{tab:prodfac} compares
the calculated neutrino nucleosynthesis production factors
for selected nuclides obtained for a 15 $M_\odot$ progenitor star assuming the high average neutrino energies, as used in Ref.~\cite{Heger05a}, and the low average energies obtained in modern SN simulations and adopted in Ref.~\cite{Sieverding18}. While the studies with higher average neutrino energies suggested that $^{11}$B, $^{138}$La and $^{180}$Ta are produced close to solar abundances by neutrino nucleosynthesis, and $^{15}$N and $^{19}$F to a large extent, this result does not hold for the study with the modern neutrino spectra. We mention that $^7$Li had originally been discussed as another nucleus produced by this process, but this no longer holds, either. Its production, triggered by neutrino-induced spallation of $^4$He followed by a $^4$He fusion reaction, is rather tiny. These findings were confirmed by a large study 
of neutrino nucleosynthesis performed with a complete set of partial neutrino-induced reaction rates and for progenitor stars in the mass range between 13 and 40 $M_\odot$ \cite{Sieverding18}. These studies indicate a strong dependence of the $^{19}$F neutrino nucleosynthesis yield on the stellar mass, as the nuclide is produced in significant fractions only in stars with masses larger than 20 $M_\odot$. This is caused by the fact that $^{19}$F is synthesized by neutrinos in the O/Ne layer of the stars, which is significantly more massive in stars with $M > 20 M_\odot$ than in low-mass progenitors.
The study of Ref.~\cite{Sieverding18} also confirms that neutrino-induced nucleosynthesis contributes noticeably to the yields of the important radionuclides 
$^{22}$Na and $^{26}$Al, as already conjectured in \cite{Woosley90}.

 \begin{table}
  \begin{tabular}{lccc}
   Nucleus & High $\langle E_\nu\rangle$ & Low $\langle E_\nu\rangle$ & $t$-dependent emission  \\
   \hline
           & Ref.~\cite{Heger05a} & Ref.~\cite{Sieverding18} &    Ref.~\cite{Sieverding19}  \\ \hline
 $^{7}$Li         &    ---   &  0.083 &  0.187  \\
 $^{11}$B         &   1.884  &  0.280 &  0.516  \\
 $^{15}$N         &   0.487  &  0.116 &  0.141  \\
 $^{19}$F         &   0.602  &  0.180 &  0.209  \\
 $^{138}$La       &   0.974  &  0.487 &  0.842  \\
 $^{180}$Ta$^{m}$ &   0.964  &  0.484 &  0.636  \\
   \hline
  \end{tabular}
 \caption{Production factors normalized to $^{16}\mathrm{O}$ of $\nu$ process
 isotopes for a $15\,\mathrm{M}_\odot$ progenitor model for a range of model assumptions about the
 neutrino emission spectra in recent studies. The reduction of the expected average 
 neutrino energies in modern SN simulations has reduced the predicted
 production factors, whereas taking into account the full emission information (i.e., time dependence of luminosities and spectra)
 increases the yields. Note that the nuclear and neutrino reaction cross sections have changed 
 from Ref.~\cite{Heger05a} to Ref.~\cite{Sieverding19}.}
 \label{tab:prodfac}
  
 \end{table}

Recently there have been two decisive steps taken to connect neutrino nucleosynthesis studies with modern advances in SN models. First of all, Ref.~\cite{Sieverding19} performed a study of neutrino nucleosynthesis considering the time-dependent
neutrino luminosities, energies and spectral forms as calculated
in a 1D SN simulation of a 27 $M_\odot$ progenitor which followed the post-bounce evolution for about 15 seconds. That nucleosynthesis study
considered for the first time also the neutronization neutrino burst phase which lasts only for about 10 ms, but carries about $10\%$ of the total $\nu_e$ luminosity. The study also accounted for the so-called accretion phase, during which matter falls through the stalled shock and which is accompanied by the emission of neutrinos with typically slightly larger average energies than during the cooling phase.
The duration of the accretion phase is still quite uncertain, but is expected to last for a few 100 ms.
Upon averaging over all three phases (burst, accretion, cooling) one finds
average energies for $\bar\nu_e$ and muon and tau neutrinos that are quite similar to the low values adopted in Ref.~\cite{Sieverding18}. However, the early phases boost the average energy of $\nu_e$ noticeably to $T_\nu=3.46$ MeV. This has a strong effect on those nuclides that are mainly produced
by charged-current $(\nu_e,e^-)$ reactions
($^{138}$La, $^{180}$Ta, and $^{11}$B, the last of which for the modern neutrino spectra is also made partly by charged-current reactions).
However, the average neutrino energies are not constant
during the cooling phase but decrease with time by nearly a factor of 2 during the first 10 seconds. This has two consequences for the neutrino nucleosynthesis yields
compared to studies which adopt time-independent average energies. First,
the energy dependence of the neutrino-nucleus cross sections gives stronger weight to the neutrinos emitted at early times. Second, the fluence of neutrinos through
a particular stellar mass shell is larger at late times, when the neutrinos have smaller average energies but similar luminosities compared to those studies with time-independent spectra. The last column in Table \ref{tab:prodfac} shows the production factors
obtained by considering the changes of neutrino luminosities and average energies with time. Ref.~\cite{Sieverding19} found that the slight deviations of the neutrino spectra from the Fermi-Dirac distribution have a negligible effect on the neutrino nucleosynthesis. 

Steps towards studying neutrino nucleosynthesis in multi-dimensional SN simulations have been taken in
Refs.~\cite{Wanajo.Mueller.ea:2018,Sieverding20}. The challenge of such simulations is the need to follow the calculations for long times ($\sim 10$ seconds) which is computationally very expensive. Ref.~\cite{Sieverding20} performed a 3D simulation
of an 11.8 $M_\odot$ SN (with simplified neutrino transport as in \cite{BMuller15a}) for the innermost ejecta and matched the results to an appropriately chosen 1D-model for the outer regions of the star. Compared to pure 1D models, the multidimensional treatment showed some important consequences for nucleosynthesis studies: the inner ejecta got heated to higher temperatures at similar radii and the ejected material had a noticeably larger spread in $Y_e$. The focus of Ref. \cite{Sieverding20} was not on neutrino nucleosynthesis, but on the production of short-lived radionuclides motivated by
the suggestion that an 11.8$M_\odot$ SN might have triggered the formation of the solar system
\cite{Banerjee16}. The calculation found significantly larger yields for selected radionuclides like
$^{26}$Al, $^{36}$Cl, $^{41}$Ca and $^{53}$Mn than the previous study, to which neutrino-nucleus reactions also contribute.
In \cite{Sieverding20} special attention was paid to the production of $^{10}$Be the 
cosmic origin of which is usually attributed to high-energy nuclear collisions. This had been challenged in
\cite{Banerjee16}, which pointed out that this nucleus could also be made in low-mass CCSNe
initiated by neutrino spallation reactions on $^{12}$C. The calculation of Ref.~\cite{Sieverding20} found that
the $^{10}$Be yield is significantly lowered by modifications in the neutrino spectra and in the explosion energy
obtained in the multidimensional treatment and by additional nuclear reactions ignored in Ref.~\cite{Banerjee16}. The
$^{10}$Be production rates have been further lowered by improved resonant data for the $^{10}$Be(p,$\alpha$)$^7$Li reaction
which makes low-mass CCSNe unlikely to produce enough $^{10}$Be to explain the $^{10}$Be/$^9$Be ratio
observed in meteorites \cite{Sieverding22}.
Regarding short-lived radionuclides, another study \cite{Hayakawa:2018ekx} showed that SN neutrino nucleosynthesis might have contributed $^{98}$Tc to the early solar system. If confirmed by meteoritic studies in the future, $^{98}$Tc can potentially be used to probe the last SN event that polluted the early solar system.

We note that a pioneering study on neutrino nucleosynthesis \cite{ech88} investigated the possibility of an $r$-process with neutrons provided by neutrino reactions on $^4$He in the SN He shell. This proposal was reexamined with modern SN progenitor models in Refs.~\cite{banerjee2011,banerjee2016}, which showed that some type of neutron-capture process resembling an $r$-process can indeed occur in low-mass SNe with low initial metallicities. An intriguing feature of this neutrino-induced nucleosynthesis is that in addition to the heavy nuclei produced by neutron capture on seed nuclei such as $^{56}$Fe, neutron capture on $^7$Li resulting from neutrino interaction with $^4$He can produce the rare nuclide $^9$Be \cite{banerjee2016,banerjee2013}.
We note also that there is a new proposal on the production of a wide range of $p$-nuclei aided by intense neutrino-nucleus interaction in outflows that are moderately neutron-rich and exposed to large neutrino fluxes~\cite{Xiong:2023uyb}.
Although the required conditions have not yet been found in current CCSN or BNSM models, it remains to be seen whether this novel process can be realized in nature.

In summary, recent studies confirmed that neutrino nucleosynthesis contributes to the cosmic origin of selected nuclides (particularly to $^{11}$B, $^{138}$La and $^{180}$Ta, and less significant to $^7$Li, $^{15}$N and $^{19}$F) and is also important for the production of short-lived radionuclides. However, no nuclide seems to be produced entirely by the neutrino process. This is not unexpected as alternative production mechanisms have been discussed for all nuclides.

\newpage
\section{Neutrino flavor conversions and nucleosynthesis}\label{sec:osc}

In what we have discussed in previous sections, a fundamental ingredient -- the flavor oscillations of neutrinos -- has been ignored completely. 
Based on various experiments for solar, atmospheric, reactor, and accelerator neutrinos, it has been well established that a single neutrino can oscillate from one flavor eigenstate (weak interaction eigenstate) to another during their propagation in vacuum and in medium. 
However, in the densest core of CCSNe or BNSM remnants where large amount of neutrinos are produced, trapped, and gradually decouple from matter, their flavor conversions remain poorly understood. 
Although it is not yet possible to consistently include neutrino flavor oscillations in simulations of CCSNe or BNSMs, significant improvements were made in recent years toward this goal. 
Pioneer studies that delineated the potential impact of flavor conversion physics in CCSNe and BNSMs were also performed extensively to obtain useful insights. 
In this section, we first discuss different types of flavor conversion of neutrinos that can happen in environments relevant to CCSNe and BNSMs in Sec.~\ref{sec:osc_types}. 
In Sec.~\ref{sec:osc_nucleo}, we review recent findings on the potential impact of neutrino flavor conversions in CCSNe and in BNSMs.
Table~\ref{tab:osc} summarizes the current understanding of different types of flavor oscillation mechanisms and their impact on relevant physical processes in SNe and in BNSMs\footnote{Notice that in this article we restrict ourselves to neutrino flavor conversions within the Standard Model and do not discuss impacts due to the potential but speculative existence of light sterile neutrinos, which may strongly affect the dynamics and nucleosynthesis in CCSNe and BNSMs~\cite{Tamborra:2011is,Wu:2013gxa,Warren:2014qza,Xiong:2019nvw,Ko:2019asm,Suliga:2019bsq,Syvolap:2019dat,Suliga:2020vpz,Tang:2020pkp,Sigurdarson:2022mcm,Ray:2023gtu}.}.

\begin{table*}[htp]
\caption{Known types of neutrino flavor oscillations that can occur in CCSNe and in BNSMs (first column). 
The second column labels whether a given type is of collective nature or not. 
The third to sixth columns denotes whether they affect the physical processes and/or nucleosynthesis outcome.
The symbols $\checkmark$, \xmark, and $?$ stand for ``yes'', ``no'', and ``not explored yet'' respectively.
}
\begin{tabular}{lccccc}
\hline
\hline
Type & Collective? & SN explosion & SN $\nu$ wind nucleosynthesis & $\nu$ process & BNSM $r$-process \\
\hline 
Slow mode & $\checkmark$ & \xmark & maybe & $\checkmark$ & \xmark \\ 
Fast mode & $\checkmark$ & $\checkmark$ & $\checkmark$ & $\checkmark$ & $\checkmark$ \\ 
Synchronized MSW & $\checkmark$ & \xmark & \xmark & \xmark & \xmark \\  
Matter neutrino resonance & $\checkmark$ & \xmark & \xmark & \xmark & maybe \\ 
Collisional induced & $\checkmark$ & ? & ? & ? & likely \\ 
MSW transformation & \xmark & \xmark & \xmark & $\checkmark$ & \xmark \\
Parametric resonance & \xmark & \xmark & \xmark & ? & \xmark \\
\hline
\hline
\end{tabular}
\label{tab:osc}
\end{table*}

\subsection{Flavor conversions of neutrinos in astrophysical environments}\label{sec:osc_types}

The flavor evolution of an ensemble of neutrinos propagating in flat space-time on the mean-field level\footnote{In this review, we do not discuss potential correction due to the many-body nature of this problem, which can also be viewed as effects beyond the mean field. For interested readers, please refer to Ref.~\cite{Patwardhan:2022mxg} for a recent review as well as Refs.~\cite{Shalgar:2023ooi} and \cite{Johns:2023ewj}.} can be described by \cite{Vlasenko:2013fja,Richers:2019grc,Nagakura:2022qko}
\begin{equation}\label{eq:osc-master}
    \left ( u^\mu \partial_\mu + \frac{dp_i}{dt} 
    \frac{\partial}{\partial p_i}\right ) \varrho(t,\mathbf{x},\mathbf{p})
    = -i[H,\varrho(t,\mathbf{x},\mathbf{p})]
    + C,
\end{equation}
where $u^\mu=(1,\mathbf{v})$ is the four-velocity of an ultra-relativistic neutrino, 
$\mathbf{p}$ is the corresponding three momentum with $p_i$ the corresponding three components, 
$\varrho$ ($\bar\varrho$) is the 3 by 3 Wigner-transformed density matrix for (anti)neutrinos in flavor basis, whose diagonal elements $\varrho_{\alpha\alpha}$ ($\bar\varrho_{\alpha\alpha}$) denote the phase-space occupation numbers of $\nu_\alpha$ ($\bar\nu_\alpha$) with $\alpha\in \{e,\mu,\tau\}$. 
The off-diagonal complex elements $\varrho_{\alpha\beta}$ ($\alpha\neq\beta$) characterize flavor mixing of neutrinos. 

In Eq.~\eqref{eq:osc-master}, the Hamiltonian $H=H_{\rm vac}+H_{\rm mat}+H_{\rm neu}$ includes the contributions from the neutrino mixing in vacuum and from forward scattering of neutrinos with ordinary matter ($e^\pm$, $p$, $n$,...) as well as with neutrino themselves. The vacuum term $H_{\rm vac}=U H_{\rm vac}^{(m)} U^\dagger$, where $U$ is the Pontecorvo-Maki-Nakagawa-Sakata matrix and $H_{\rm vac}^{(m)}$ the vacuum Hamiltonian in mass basis with only non-zero diagonal entries $H_{{\rm vac},ii}^{(m)}\simeq m_i^2/(2p)$, and $m_i$ the mass of neutrino mass eigenstate $i$.
For the matter term, assuming charge neutrality and neglecting the small amount of $\mu^\pm$ and $\tau^\pm$, $H_{\rm mat}=\sqrt{2}G_Fu_\mu j_e^\mu$, where $j_e^\mu=u_{\rm bulk}^\mu\int d^3q/(2\pi)^3 (f_{e^-}-f_{e^+})$, with $u_{\rm bulk}$ the four bulk velocity of fluid and $f_{e^\pm}(q)$ the occupation number of thermal $e^\pm$ with $q$ being their 4-momentum. 
The last term $H_{\rm neu}$ takes a similar form $H_{\rm neu}=\sqrt{2}G_F u_\mu j_{\rm neu}^\mu$, with 
$j_{\rm neu}^\mu=\int d^3p^\prime/(2\pi)^3 u^{\prime\mu}[\varrho(t,\mathbf{x},\mathbf{p}^\prime)-\bar\varrho^*(t,\mathbf{x},\mathbf{p}^\prime)]$.
Clearly, $H_{\rm neu}$ contains the density matrices $\varrho$ and $\bar\varrho$ themselves, which makes Eq.~\eqref{eq:osc-master} nonlinear. 
It is also easy to see that the typical size of the components in $H_{\rm neu}$ is of the order $\mathcal{O}(\sqrt{2}G_F n_\nu)\sim \mathcal{O}(10^{-2}-10^1)$~cm$^{-1}$ for typical $n_\nu\sim \mathcal{O}(10^{30}-10^{32})$~cm$^{-3}$, much larger than that in $H_{\rm vac}\sim\mathcal{O}(1)$~km$^{-1}$ for $\delta m^2_{31}\equiv |m_3^2-m_1^2|$ and $p\sim \mathcal{O}(10)$~MeV, i.e., neutrinos in flavor space are strongly coupled than they are in vacuum. 
The strongly-coupled and nonlinear nature of the problem give rise to various collective effects discussed below. 
For the collision term $C$, we do not explicitly write down its full form here as it requires specifying all different contributions from non-forward scattering of neutrinos. 
We refer interested readers to Ref.~\cite{Blaschke:2016xxt} for detailed expressions of these terms.
For antineutrinos, their equation of motion takes a similar form as Eq.~\eqref{eq:osc-master} by replacing $H$ and $C$ by $\bar H=H_{\rm vac}+{\bar H}_{\rm mat}+{\bar H}_{\rm neu}$ and $\bar C$, respectively, where ${\bar H}_{\rm mat}=-H_{\rm mat}$ and ${\bar H}_{\rm neu}=-H_{\rm neu}^*$.

\subsubsection{Collective flavor oscillations}\label{sec:osc-type-coll}
Like in many other fields, the studies of collective flavor oscillations of neutrinos began with lots of approximations that originate from certain imposed symmetries of the system. 
The earliest studies that assumed homogeneity, isotropy, and two-flavor simplification identified a synchronized collective mode such that all neutrinos with different energy effectively oscillate as in vacuum, with a uniform collective frequency $\omega_{\rm sync}\sim \left. \int dp \omega_p( \varrho_{ee}-\varrho_{xx}+\bar\varrho_{ee}-\bar\varrho_{xx})\right/\int dp ( \varrho_{ee}-\varrho_{xx}-\bar\varrho_{ee}+\bar\varrho_{xx})$ and a matter-suppressed effective mixing angle $\Theta_{\rm eff}\ll 1$ when $\mu\equiv \sqrt{2}G_Fn_{\nu_e}\gg \omega_p\equiv\delta m^2_{31}/(2p)$ for all relevant $p$~\cite{Pastor:2001iu,Duan:2005cp}. 
With large constant matter and neutrino densities, the synchronized mode has little physical relevance because $\Theta_{\rm eff}\ll 1$. 
Meanwhile, studies also found that another collective mode, often dubbed ``bipolar mode'', can cause significant amount of flavor conversions even if $\Theta_{\rm eff}\ll 1$ given that the neutrino mass ordering is inverted~\cite{Duan:2005cp,Hannestad:2006nj}, which may have important implications to CCSN physics. 

Ref.~\cite{Duan:2006an} performed a first set of ``realistic'' simulations that break the assumption of homogeneity and isotropy. 
Assuming spherically symmetric background profiles mimicking the phase of SN neutrino-driven wind, taking certain parametrized neutrino energy spectra, and considering two-flavor approximation, it showed that flavor transitions can possibly happen immediately above the neutrinosphere at $\mathcal{O}(50-100)$~km, later confirmed by~\cite{Fogli:2007bk}. 
Due to the potentially large impact on CCSN physics, its related nucleosynthesis, and the neutrino signals, these important findings opened a long and yet-ended journey of pursuing better understanding of collective flavor oscillations. 
We summarize presently known modes and mechanisms that can cause collective flavor transformation as follows.

\begin{itemize}
\item \textbf{Slow mode:}\\ 
The aforementioned bipolar flavor transformation belongs to this type of collective flavor oscillations. 
This type of collective flavor conversion is triggered by flavor instabilities associated with the ``crossing(s)'' in the energy distribution of neutrinos and antineutrinos~\cite{Dasgupta:2009mg,Banerjee:2011fj,Dasgupta:2021gfs}\footnote{Antineutrinos can be effectively viewed as neutrinos with ``negative energy'' that contributed to $H_{\rm neu}$ with different signs~\cite{Duan:2006an}.}.
The collective frequency is of the order $\omega_{\rm slow}\sim \mathcal{O}(\sqrt{\mu\omega_p})$. 
Studies found that during the SN accretion phase, the slow mode instabilities are usually suppressed by the large matter density for regions inside the SN shock for typical iron-core progenitors~\cite{Chakraborty:2011nf,Chakraborty:2016yeg}\footnote{If a fast time-varying background exists, it may permits slow instabilities despite of the large matter density~\cite{Abbar:2015fwa,Dasgupta:2015iia}. However, the presence of such background in SN environment has not been identified.}. 
However, its occurrence at $r\sim 300-1000$~km should still affect the expected neutrino signals from the next galactic SN as well as the diffuse supernova neutrino background. 
For the subsequent PNS cooling phase, the onset radius of slow instabilities typically decreases with time and may affect the nucleosynthesis in the neutrino-driven wind~\cite{Wu:2014kaa}, which will be discussed in Sec.~\ref{sec:oscnucleo_snenuwind}.

\begin{figure}[htb]
\centering
\subfigure[Fast instability in CCSNe]{
\includegraphics[width=0.6\columnwidth]{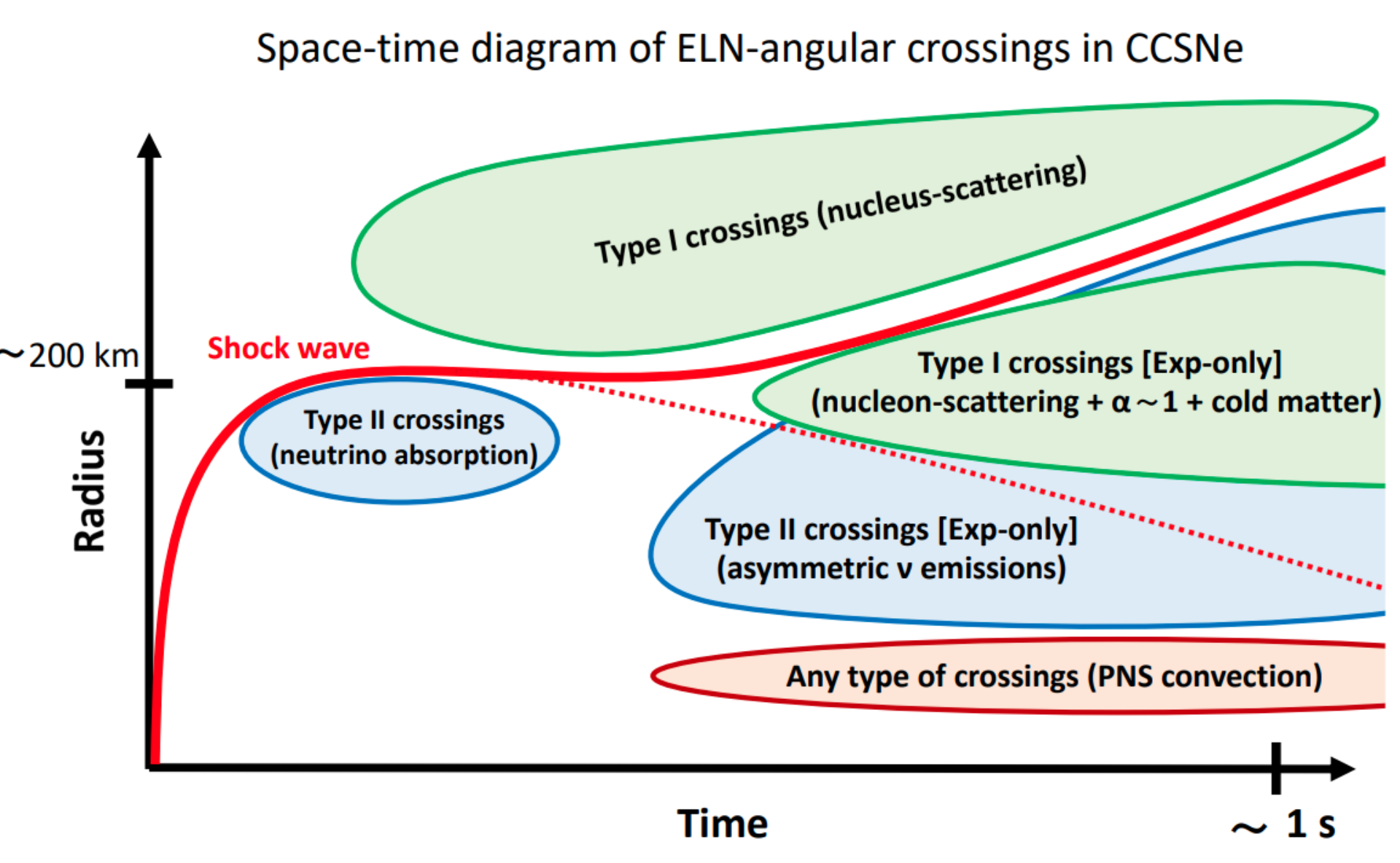}
\label{fig:fast_ins_SNe}}
\hfill
\subfigure[Fast instability in post merger disks]{
\includegraphics[width=0.35\columnwidth]{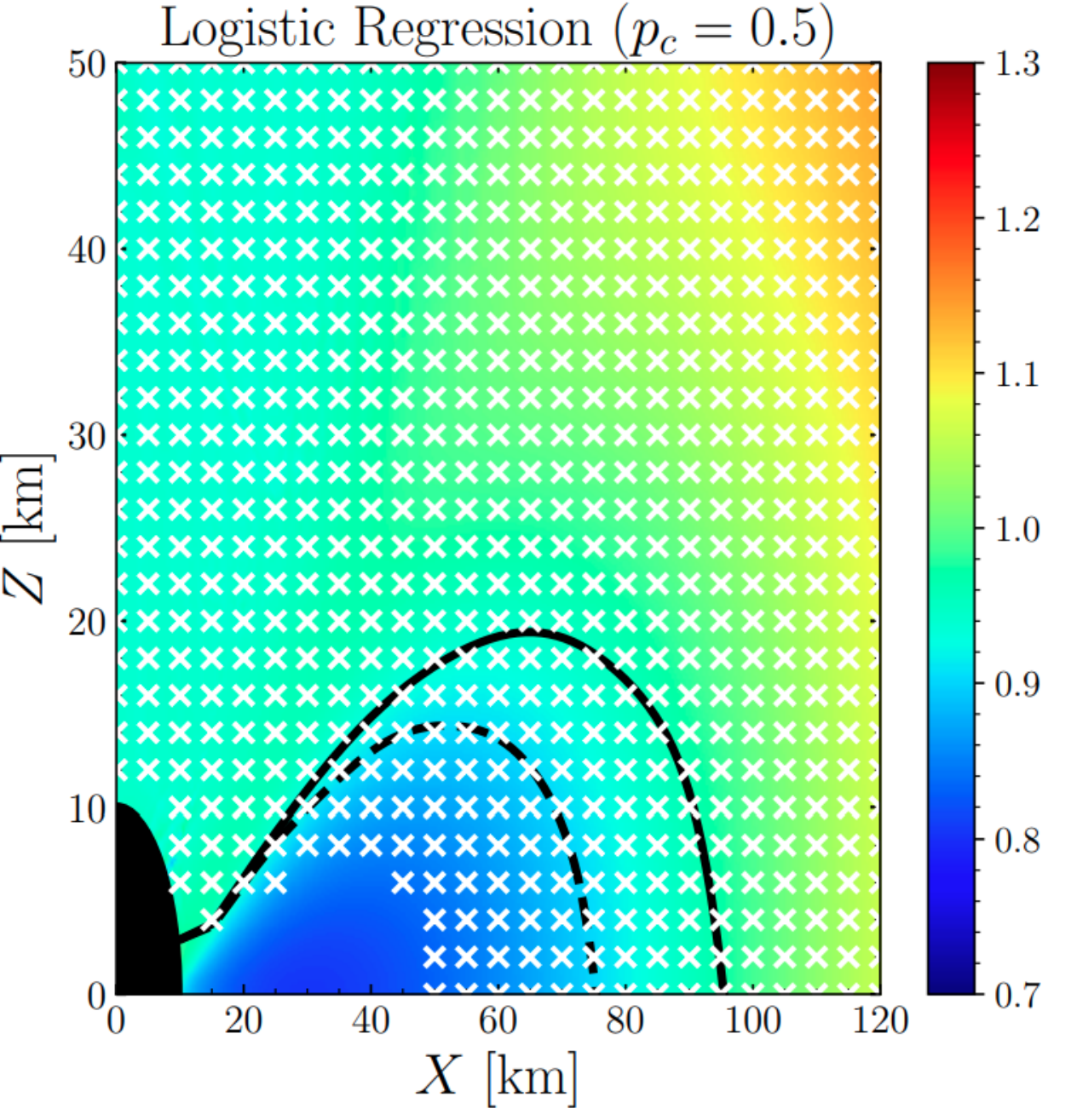}
\label{fig:fast_ins_BHdisk}}
\caption{Panel~(a) (Reprinted figure with permission from \cite{Nagakura:2021hyb} Copyright (2021) by the
American Physical Society): Schematic plot showing the regions where fast flavor instabilities in SN interior may be present. 
For the definition of the Type I and II crossings, please refer to Ref.~\cite{Nagakura:2021hyb} for details. 
Panel~(b): Contour of the number density ratio of $\bar\nu_e$ and $\nu_e$, $n_{\bar\nu_e}/n_{\nu_e}$ indicated by colors, at $t=50$~ms during the BNSM BH accretion disk evolution. The outer (inner) black curve is the decoupling sphere of $\nu_e$ ($\bar\nu_e$). 
Locations where fast instabilities are present are indicated by white crosses.
Figure reprinted from Ref.~\cite{Abbar:2023kta}).}
\label{fig:fast_ins}
\end{figure}

\item \textbf{Fast mode:}\\
The ``re-discovery'' of the fast mode instability, which has a characteristic frequency of $\omega_{\rm fast}\sim \mathcal{O}(\mu)$, in 
the seminal paper~\cite{Sawyer:2015dsa} has triggered extremely intensive studies during the past few years on the onset condition and its potential impact in SNe and in BNSMs. 
Analytical work proved that crossings in the angular distribution of electron-minus-muon neutrino number density is needed for fast flavor oscillations to occur~\cite{Morinaga:2021vmc,Dasgupta:2021gfs}, while numerical works exploring the outcome of fast flavor oscillations that included the advection of neutrinos have been extensively preformed and offered intriguing insights to the outcome of fast flavor conversions~\cite{Bhattacharyya:2020jpj,Wu:2021uvt,Richers:2021xtf,Bhattacharyya:2022eed,Capozzi:2022dtr,Richers:2022bkd,Grohs:2022fyq,Nagakura:2022xwe,Nagakura:2022kic,Shalgar:2022rjj,Shalgar:2022lvv,Zaizen:2023ihz,Xiong:2023vcm,Abbar:2023ltx,Cornelius:2023eop,Nagakura:2023jfi}. 
By analyzing profile snapshots from simulations that do not include flavor oscillation of neutrinos, 
it was found that fast instabilities can generally exist in various time and spatial domain during different SN evolution stages; see Ref.~\cite{Nagakura:2021hyb} and references therein), including regions below the shock during the accretion phase as well as possibly inside the PNS; see Fig.~\ref{fig:fast_ins}(a) for an illustration.  
Based on spherically symmetric SN models, Refs.~\cite{Ehring:2023lcd,Nagakura:2023mhr} recently studied the potential impact of fast flavor conversion on SN shock revival and found that it may in fact lead to a net reduction of neutrino heating behind the shock.  
On the other hand, using 2D models, Ref.~\cite{Ehring:2023abs} found that fast flavor conversions may further facilitate (weaken) the explosion for a progenitor model wherein a successful explosion can (not) be obtained without including parametrized flavor conversion outcome. 
A definitive answer still awaits for studies that couple the flavor evolution equations with hydrodynamics in multi-dimensional simulations in a more consistent fashion.  
The impact of fast flavor conversions on nucleosynthesis in the neutrino-driven wind were studied in Ref.~\cite{Xiong:2020ntn} and will also be discussed in  Sec.~\ref{sec:oscnucleo_snenuwind}.

For BNSMs, fast mode instabilities were found to be even more commonly present inside and above the post-merger accretion disks, due to the overall protonization of the system as well as the mild electron degeneracy and relative low density of the disk~\cite{Wu:2017qpc,Wu:2017drk,George:2020veu,Just:2022flt,Richers:2022dqa,Nagakura:2023wbf,Froustey:2023skf} (see e.g., Fig.~\ref{fig:fast_ins}(b)).
Based on the argument of the separation of scales, (magneto-)hydrodynamic simulations taking into account parametrized flavor oscillation outcomes were performed~\cite{Li:2021vqj,Just:2022flt,Fernandez:2022yyv} and will be further discussed in Sec.~\ref{sec:oscnucleo_bnsm}. 

\item \textbf{Synchronized Mikheyev-Smirnov-Wolfenstein (MSW):}\\
For neutrinos that do not encounter any flavor instabilities on its way propagating outward, the presence of the synchronized collective mode can lead to new resonance conditions at large radii where the size of components in $H_{\rm neu}$ becomes similar to those in $H_{\rm vac}$ (for cases where the components in $H_{\rm mat}$ decreases faster than $H_{\rm neu}$)~\cite{Pastor:2002we}.
Such a condition may exist for SNe with an O-Ne-Mg core that has a very steep density profile above the core during the phase of neutronization burst~\cite{Duan:2007bt,Duan:2007sh}. 
Signature of the synchronized MSW flavor transformation can be encoded in the neutrino signals and may be revealed by the detection of neutrinos from a galactic SN of such kind. 

\item \textbf{Matter neutrino resonance:}\\
For BNSM remnants that overall protonize, the number density of $\bar\nu_e$ can be larger than that of $\nu_e$ outside the remnant. 
As a result, the diagonal components in $H_{\rm mat}$ and $H_{\rm neu}$ can take different signs and therefore cancel with each other at some locations. 
This fact can lead to a new resonance condition and introduces another interesting collective flavor conversion mechanism dubbed the matter neutrino resonance~\cite{Malkus:2012ts,Wu:2015fga,Shalgar:2017pzd}. 
However, the flavor evolution of neutrinos in a realistic multidimensional system that breaks the assumed symmetry has not yet been carried out. 

\item \textbf{Collisional induced:} 
Ref.~\cite{Johns:2021qby} recently found that the emission and absorption of neutrinos, which are part of the $C$ term in Eq.~\eqref{eq:osc-master}, can surprisingly lead to new flavor instabilities, despite the fact that these terms alone tend to destroy flavor coherence. 
This novel instability requires asymmetric rates of $\nu_e$ and $\bar\nu_e$ being larger than the asymmetry in their number density ratio~\cite{Johns:2021qby,Xiong:2022zqz,Liu:2023pjw}. 
A recent study suggested the general presence of collisional flavor instability in the decoupling region of neutrinos in SNe~\cite{Xiong:2022vsy} in the absence of fast mode instability, which may lead to the excessive production of $\nu_x$ from the decoupling region, consistent with what found in Refs.~\cite{Liu:2023vtz,Akaho:2023brj} later. 
On the other hand, another analysis which examined different SN models and took different treatment on neutrino transport claimed subdominant effects of collisional instability~\cite{Shalgar:2023aca}. 
For the BNSM remnants, Ref.~\cite{Xiong:2022zqz} found that collisional flavor instabilities can exist even more ubiquitously inside the remnant disks than the fast modes, and identified that the characteristic collective frequency $\omega_{\rm coll}\sim \mathcal{O}(\Gamma)-\mathcal{O}(\sqrt{\mu\Gamma})$, where $\Gamma$ is $\sim$ the averaged rate for neutrinos. 
As all these findings are relatively preliminary, its interplay with the fast modes~\cite{Johns:2021qby,Johns:2022yqy,Padilla-Gay:2022wck,Kato:2023dcw} and the potential impact on SN explosion and on nucleosynthesis in SNe and BNSMs remain to be explored. 

\end{itemize}

\begin{figure}[htb]
\centering
\includegraphics[width=0.6\columnwidth]{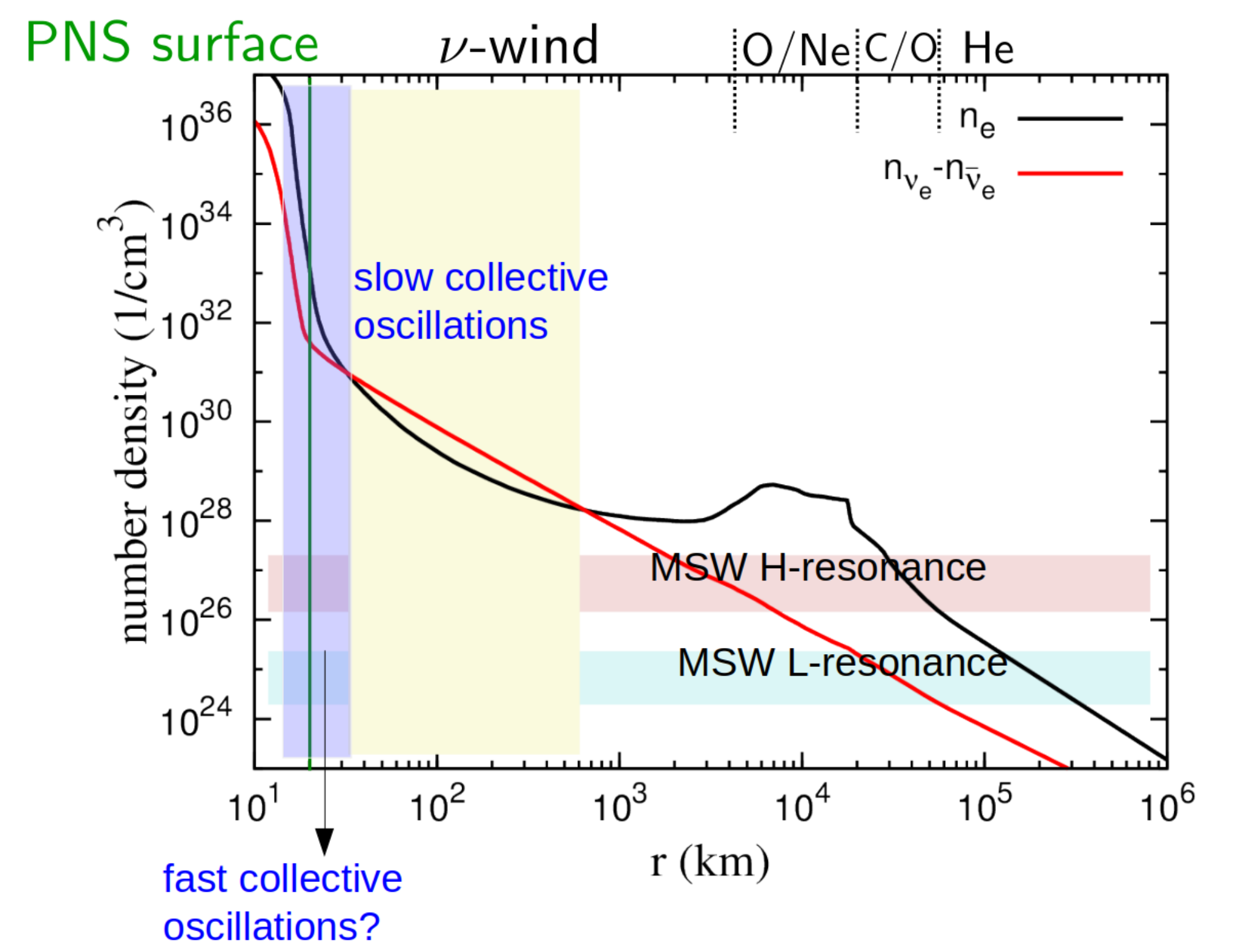}
\caption{Indication of regions where different flavor transformation mechanisms occur in the PNS cooling phase. 
Black and red lines note the profiles $n_e$ and $n_{\nu_e}-n_{\bar\nu_e}$ from 1D SN simulation with an $18$~$M_\odot$ progenitor at $3$~s post the core bounce.
Vertical bands indicate the regions where the fast and slow collective oscillations may occur, while the horizontal bands indicate the corresponding $n_e$ of the two MSW resonances with H and L denoting values corresponding to $\Delta m_{31}^2$ and $\Delta m_{21}^2$.
The short dotted lines separate the layers with different compositions in the progenitor. The green solid line marks the radius of the PNS surface.  
}
\label{fig:SN_wind_osci}
\end{figure}

\subsubsection{Flavor transformation due to varying matter profile}\label{sec:msw}
Outside the regime of collective oscillations, neutrinos can undergo further flavor transformation due to the varying matter density profile. 
Although matter-induced flavor transformation typically occur far outside the central compact object, it can still play an important role in affecting the neutrino nucleosynthesis in SNe as well as shaping the SN neutrino signals. 
In what follows, we summarize known mechanisms. 
\begin{itemize}
\item \textbf{MSW flavor transformation}:\\
For a neutrino with momentum $p$, the well-known MSW resonances occur when the matter potential $\Lambda=\sqrt{2}G_Fn_e \simeq \Delta m_{ij}^2 / (2p)$ 
where the subscripts ${ij}={31}$ and $21$ for the atmospheric and solar neutrino mass squared difference, respectively. 
For $p\sim \mathcal{O}(10)$~MeV, these resonances typically sit at locations where the matter mass density $\rho\sim \mathcal{O}(10^3)$~g~cm$^{-3}$ and $\rho\sim \mathcal{O}(10^2)$~g~cm$^{-3}$ for two different $\Delta m_{ij}^2$, which are located at the C/O shell and the He shell of SNe with an iron core progenitor (see Fig.~\ref{fig:SN_wind_osci}), before the SN shock arrives those layers. 
Since the matter density in the stellar envelopes are often varying slowly enough, the flavor transformation through the MSW resonances are mostly adiabatic.
In this case, the flavor conversion probabilities after going through the resonance only depend on the neutrino mixing angles and can be written down easily (see e.g., Ref.~\cite{Dighe:1999bi}).
For SNe with iron-core progenitors, it is generally expected that the only the adiabatic MSW flavor transformation affects the flavor evolution of neutrinos during the neutronization burst phase (in contrast to the synchronized MSW effect discussed in Sec.~\ref{sec:osc-type-coll}).
If this is indeed true, tPHYSSCR-122426.R1he detection of the neutronization burst neutrinos from the next galactic SN (with an iron core) may provide important information regarding the unknown properties of neutrinos such as their mass ordering; see e.g., Refs.~\cite{Serpico:2011ir,Scholberg:2017czd,Brdar:2022vfr}. 
The passage of SN shock can lead to nonadiabacity and affect the conversion history of neutrinos~\cite{Takahashi:2002yj,Fogli:2006xy,Gava:2009pj,Friedland:2020ecy}. 
We will also discuss the impact of MSW flavor transformation on nucleosynthesis in SNe in Sec.~\ref{sec:oscnucleo_snenuproc}.

\item \textbf{Parametric resonance}:\\
For the matter potential $\Lambda$ 
that contains fluctuations in space or time that are periodic, resonance conditions beyond the simple MSW effect discussed above may also be achieved when the   fluctuation period is exactly an integer multiple of the averaged in-medium oscillation period -- the so-called parametric resonance~\cite{Krastev:1989ix}. 
The parametric resonance can potentially trigger flavor conversions at locations where the density is far from the MSW resonance density. 
Since the SN environment naturally hosts convection and turbulence that may lead to fluctuations in space and time, several studies examined the potential consequences of parametric resonance for shock revival and for neutrino signals~\cite{Kneller:2010sc,Borriello:2013tha,Patton:2014lza}. 

\item \textbf{Turbulence coupling}:
Another interesting effect due to matter fluctuations revealed recently is that they might affect how collective oscillations behave.  
A pioneer work Ref.~\cite{Abbar:2020ror} recently showed that when matter fluctuations exist, it can lead to a ``leakage'' effect that can potentially change how collective oscillations initially develop when the off-diagonal entries of $\varrho$ are still small. 
However, the impact on the full evolution of collective modes remain unclear and needs to be further studied.

\end{itemize}

\subsection{Impact on nucleosynthesis in supernovae and in neutron star mergers}\label{sec:osc_nucleo}

As the flavor conversions of neutrinos alter the energy spectra of different flavors, they can directly affect the neutrino interaction rates with nucleons or nuclei that are critical to different nucleosynthesis processes discussed in Sec.~\ref{sec:third}. 
Besides the direct impact, for flavor conversions that can potentially happen deeply inside the neutrinosphere, they can change the evolution of the system and thus indirectly affect the relevant nucleosynthesis conditions. 

\subsubsection{Neutrino-driven wind in supernovae}\label{sec:oscnucleo_snenuwind}
Earlier works that adopted parametrized wind trajectories and nucleosynthesis conditions as well as simplified treatment in modeling collective oscillations suggested that collective flavor conversions can substantially alter the charged current reaction rates before the recombination of nucleons into $^4$He, which in turn, lead to large effect on the assumed $r$-process nucleosynthesis yields in the neutrino-driven wind~\cite{Balantekin:2004ug,Martinez-Pinedo:2011yhi,Duan:2010af}. 
Later, the long-term evolution of 1D SN models which took into account accurate Boltzmann neutrino transport 
suggested that more likely the neutrino-driven wind hosts mildly neutron-rich initially and transitions to  proton-rich conditions in later phases (see Sec.~\ref{subsec:ndw_nucl}). 
With these long-term simulations becoming available, Ref.~\cite{Wu:2014kaa} performed a comprehensive study that computed the slow mode collective oscillations based on time-dependent wind profiles as well as neutrino distribution functions at neutrinosphere obtained for an $18$~$M_\odot$ progenitor. 
It was found that although slow collective oscillations indeed occur during the entire cooling phase and alter the $\nu_e$ and/or $\bar\nu_e$ absorption rates by nucleons, their onsets generally sit at too large radii to significantly affect $Y_e$ of the wind. 
Moreover, the neutrino spectra evolve in a way that the energy crossings only appear in the neutrino sector but not the antineutrino sector at later times when $\nu p$ process operates. 
As a result, its impact on the $\bar\nu_e$ capture rates on protons, relevant to $\nu p$ process, was found to be negligible. 
In contrast, Ref.~\cite{Sasaki:2017jry} adopted two wind trajectories obtained from SN simulation with a 40~$M_\odot$ progenitor, which permits $\nu p$ process during the early cooling phase.  
Taking parametrized neutrino spectra and perform similar calculations for slow collective oscillations, they found significant impact of slow oscillations on $\nu p$ process yields.  
We note that, however, neither the medium modification on the neutrino-nucleon interaction rates nor the inelastic scattering of neutrinos with nucleons were included in the SN models used in~\cite{Sasaki:2017jry}. 
These missing factors might account for the different conclusions reached in these two studies. 
Nevertheless, it reminds the importance of accurate neutrino transport modeling being needed as the emerging neutrino spectra from the neutrinosphere plays a critical role in determining the outcome of slow collective oscillations. 

\begin{figure}[htb]
\centering
\includegraphics[width=0.6\columnwidth]{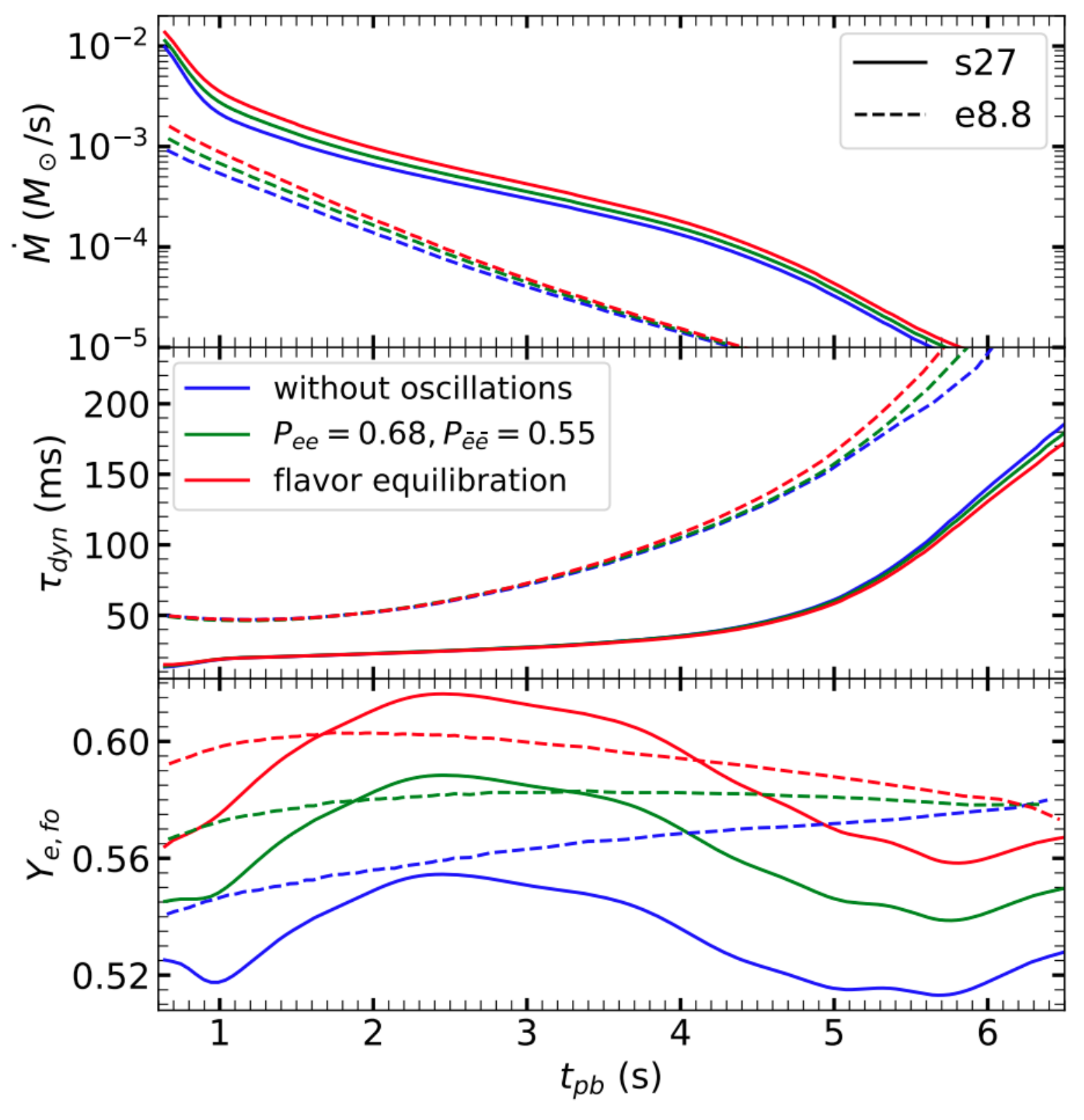}
\caption{Possible impact of fast flavor conversion on the outflow property, including the mass outflow rate $\dot M$ (top panel), the dynamical expansion timescale $\tau_{\rm dyn}$ (middle panel), and $Y_e$ (bottom panel) of the neutrino-driven wind models based on inputs from two SN explosion simulations of different progenitors labeled by $s27$ and $e8.8$. 
Different colors assume different parametrization of flavor oscillation outcome due to fast conversions. 
Figure adapted from Ref.~\cite{Xiong:2020ntn}.
}
\label{fig:fast_nuwind_condi}
\end{figure}

For fast flavor oscillations, although it is yet unclear in terms of when and where they would occur in the PNS cooling phase, Ref.~\cite{Xiong:2020ntn} assumed different parametrized flavor conversion probabilities for neutrinos as potential outcomes of fast flavor conversions and performed steady-state neutrino-driven wind calculations to investigate the potential impact of fast conversions on neutrino-driven wind nucleosynthesis.
Their finding suggests that if fast flavor conversions occur close to the neutrinosphere during the wind phase, it may lead to reduction of the outflow mass and substantially facilitate the $\nu p$ process therein (see Fig.~\ref{fig:fast_nuwind_condi}). 
The reduction of the ejecta mass is simply due to that the flavor swapping between $\nu_e\leftrightarrow\nu_x$ and  $\bar\nu_e\leftrightarrow\bar\nu_x$ reduce the neutrino heating rates responsible for driving the outflow.
The effect on the $\nu p$ process is mainly originated from the increased spectral similarity, i.e., smaller $\langle E_{\bar\nu_e}\rangle - \langle E_{\nu_e}\rangle$ due to flavor conversions, which leads to more proton-rich conditions. 
As a result, enhanced light $p$-nuclides $^{92,94}$Mo and $^{96,98}$Ru may be obtained.

\subsubsection{SN explosive nucleosynthesis}
Works that analyzed the occurrence condition of fast mode instabilities found that they likely appear in regions close to the neutrinosphere where the asymmetry between the number density of $\nu_e$ and $\bar\nu_e$ is small, i.e., $n_{\bar\nu_e}/n_{\nu_e}\sim 1$~\cite{Abbar:2018shq,Nagakura:2021hyb}. 
Based on this observation, Ref.~\cite{Fujimoto:2022njj} conducted a pioneer study on the potential impact of fast conversions on explosive nucleosynthesis by implementing angle-dependence criteria for fast instabilities for mass ejection that immediately follows the shock revival in 2D SN models with approximate neutrino transport scheme. 
It was found that if the emission of $\nu_e$ and $\bar\nu_e$ is non-axisymmetric to a degree of $\sim 10\%$, fast flavor conversions may potentially affect the outcome of explosive nucleosynthesis for nuclei above the iron peak, once again, due to the change of $\nu_e$ and $\bar\nu_e$ capture rates on nucleons that lead to different $Y_e$ relevant for heavy element nucleosynthesis.

\subsubsection{Neutrino nucleosynthesis}\label{sec:oscnucleo_snenuproc}
As discussed in Sec.~\ref{sec:nu-nucleo}, the energy spectra of neutrinos of different flavors are the key to determine the outcome of neutrino nucleosynthesis for several isotopes listed in Table~\ref{tab:prodfac}. 
Obviously, any flavor oscillations that change neutrino energy spectra relevant for charged or neutral current interactions can affect the outcome of neutrino nucleosynthesis that operate outside where flavor oscillations occur. 
Comparing Fig.~\ref{fig:all_profile} and \ref{fig:SN_wind_osci}, one readily sees that collective flavor oscillations which can happen in the deepest layers can possibly affect all of the neutrino nucleosynthesis processes~\cite{Wu:2014kaa,Ko:2019xxm,Kusakabe:2019znq,Ko:2022uqv}, while the MSW flavor transformation can influence those relevant to the production of light nuclei (e.g., $^7$~Li, $^{11}$B, $^{15}$N, and $^{19}$F)~\cite{Yoshida05,Wu:2014kaa,Ko:2019xxm,Kusakabe:2019znq,Ko:2022uqv}.
Effects purely due to the adiabatic MSW flavor transformation on reaction rates relevant to neutrino nucleosynthesis is qualitatively clear: it generally enhances the charged current $\nu_e$ and $\bar\nu_e$ reaction rates due to the swapping of softer $\nu_e$ ($\bar\nu_e$) spectra with the harder spectra of $\nu_x$ ($\bar\nu_x$).
As a result, it tends to enhance both the production of $^7$Li and $^{11}$B. 
However, collective flavor oscillations can change the neutrino spectra prior to the MSW resonances and can thus alter this conclusion.

\subsubsection{$r$-process in neutron star mergers}\label{sec:oscnucleo_bnsm}
The ubiquitous presence of fast mode instabilities in BNSM remnants~\cite{Wu:2017qpc} triggered a number of studies in recent years to access its impact on the $r$-process nucleosynthesis in BNSMs, given the potential relevance to the kilonova observations associated with BNSMs.
Refs.~\cite{Wu:2017drk} and \cite{George:2020veu} assumed total flavor equipartition among all flavors due to fast flavor conversions outside the neutrinosphere and analyzed post-processingly how it affects the nucleosynthesis yields for neutrino-driven wind from BH accretion disk and for dynamical ejecta launched right after the merger with the presence of a HMNS. 
For BH-disk outflow driven by neutrinos, it was found that total flavor equipartition leads to more neutron-rich condition and lead to enhanced production of lanthanides.
The main reason is that since the disk emits little $\nu_x$, flavor conversions reduce both the absorption rates of $\nu_e$ and $\bar\nu_e$, which in turn lead to lower $Y_e$ in the outflow. 
However, for early post-merger ejecta where the HMNS is present, the effect due to total flavor equipartition is the opposite. 
Due to the large amount of $\nu_e$ with higher average energy before flavor oscillations, swapping $\nu_e\leftrightarrow\nu_x$ and  $\bar\nu_e\leftrightarrow\bar\nu_x$ enhances more the absorption rate of $\nu_e$ than that of $\bar\nu_e$. 
Thus, fast oscillations likely make outflow in the polar region more proton rich. 
This results in enhanced (suppressed) production in iron (first) peak nuclei in the polar outflow.

\begin{figure*} [htb!]
  \centering
  \raisebox{+0.6cm}{\includegraphics*[width=.49\textwidth]{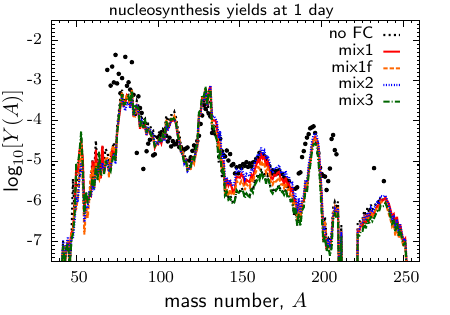}}
  \includegraphics*[width=.46\textwidth]{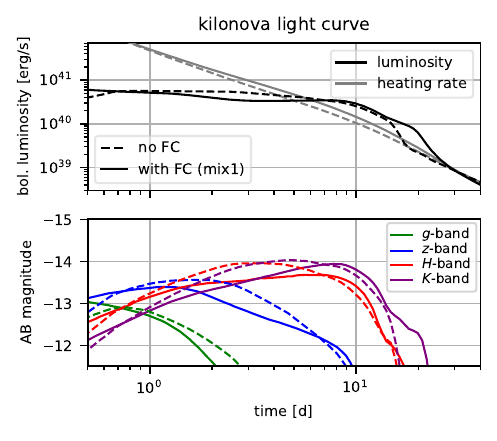}
  \caption{
  Left panel: Impact of fast flavor conversions on the $r$-process abundance yields in BH accretion disk outflows at $t=1\,$d after the birth of the disk. 
  Different models m1, m1mix1, m1mix1f, m1mix2, and m1mix3 assume different parametrization of fast flavor conversions, while ``no FC'' denote the case without including any flavor conversions. 
  Right panel: The kilonova bolometric luminosity (top) and the AB magnitudes in selected bands (bottom) for models m1 (dashed lines) and m1mix1 (solid lines) computed using spherically averaged ejecta properties. 
  Figure reprinted from Ref.~\cite{Just:2022flt}.}
  \label{fig:nuckilo}
\end{figure*}

Due to its potential implication on $r$-process nucleosynthesis and kilonova observable, efforts were made to directly identify fast mode instabilities and implement parametrized flavor conversion outcome in the evolution of post-merger disk simulations with neutrino transport~\cite{Li:2021vqj,Just:2022flt,Fernandez:2022yyv}.
These studies suggested that fast mode can reside deeply inside the neutrinosphere and also affect the dynamical evolution of the disk. 
In terms of impact on nucleosynthesis outcome, they found qualitatively similar conclusions as above: fast flavor conversion enhances the neutron-richness of the outflow launched from BH disk remnants~\cite{Li:2021vqj,Just:2022flt} or remnant with a shortly lived HMNS that collapses to a BH within $\sim \mathcal{O}(10)$~ms~\cite{Fernandez:2022yyv}.
For cases where the HMNS is longer-lived, it tends to make the outflow more proton rich~\cite{Fernandez:2022yyv}. 
Overall, these studies suggest that fast oscillations may change the average $Y_e$ of the outflow by $\sim 0.02-0.05$. 
This amounts to the change of the lanthanide mass fraction by $\sim$ a factor of $2-3$, and can moderately affect the kilonova emission of BNSMs; see e.g., Fig.~\ref{fig:nuckilo}. 

\newpage
\section{Summary and outlook}\label{sec:sum}
This review summarized the current understanding of the role of neutrinos in the astrophysical processes associated with CCSNe and BNSMs, with particular focus on how they affect the dynamics and nucleosynthesis in these explosive events. 
For CCSNe, neutrinos play essential parts in nearly every phase, including the stellar core collapse and bounce, the shock revival through the neutrino heating mechanism, and the launch of the neutrino-driven wind where the heavy element nucleosynthesis can take place.  
In particular, the $\nu p$ process that can produce certain $p$-isotopes with the aid of $\bar\nu_e$ captures on protons can potentially occur in the wind. 
Although state-of-the-art SN models do not find very efficient $\nu p$-process and only obtain the formation of heavy elements up to $A\simeq 90$ (see Sec.~\ref{subsec:ndw_nucl}), it requires improved modeling to include multidimensional effect, detailed neutrino transport, and flavor oscillations of neutrinos to draw a definitive answer.  

In BNSMs, neutrinos are also responsible for determining the neutron richness of the ejecta and hence the likelihood of whether $r$-process nucleosynthesis takes place. Neutrinos are responsible of producing broad distributions in $Y_e$ as required to account for the solar $r$-process abundance pattern. Furthermore, the relevant nuclear physics depends on the specific value of $Y_e$ with masses and beta-decay rates determining the nucleosynthesis yields for about $Y_e \gtrsim 0.1$ and fission for smaller $Y_e$ values. 

We also reviewed recent advances in improving the theoretical modeling of neutrino-matter interactions as well as neutrino-nucleus interactions in Sec.~\ref{sec:third} and their impacts on nucleosynthesis in SNe. 
Neutrino-matter interaction rates are now evaluated taking into account the fully relativistic kinematics of nucleons consistent with the mean-field properties of the underlying nuclear EoS (Sec.~\ref{sec:CC}). 
Long-term and spherically-symmetric simulations of the PNS cooling phase that included the updated set of neutrino-matter interactions as well as the (inverse) neutron decays have found noticeable changes in $Y_e$ by $\Delta Y_e\sim 0.02-0.03$ in the neutrino-driven wind, which can lead to a substantial impact on the yields of wind nucleosynthesis, in particular for nuclei around $Z=40$ (Sec.~\ref{subsec:ndw_nucl}). 
Efforts were also done to improve the modeling of neutrino-matter rates taking into account nucleon-nucleon correlation effects beyond the mean-field level, as well as the nucleon--nucleon bremsstrahlung (Secs.~\ref{subsec:nncorre} and \ref{subsec:nnbrems}). 

For neutrino-nucleus interactions that are keys to the production of several isotopes through the neutrino nucleosynthesis, we outlined the approach based on nuclear shell model and QRPA allowing for computing these inelastic rates including allowed and forbidden transitions at finite temperature for a large set of nuclei (Sec.~\ref{Sec:v-nucleus}).
With these rates, although recent works using updated neutrino properties suggested generally reduced yields of these isotopes compared to earlier works (Sec.~\ref{sec:nu-nucleo}), neutrino-nucleosynthesis in SNe should still contribute a significant fraction to the observed yields in the solar system for specific isotopes, and can also be responsible for certain short-lived radioactive nuclei present in the early solar system. 

Finally, we reviewed our present understanding of neutrino flavor oscillations in CCSNe and in BNSMs, and their possible impact on the dynamics and nucleosynthesis in these systems (Sec.~\ref{sec:osc}). 
Studies in recent years discovered that the collective fast flavor conversions and the collision-induced collective flavor oscillations can both occur around or even inside regions where neutrinos decouple (Sec.~\ref{sec:osc_types}). 
Exploratory works demonstrated their potential importance to SN explosions as well as nucleosynthesis in the neutrino-driven wind and in $r$-process in BNSM outflows (Sec.~\ref{sec:osc_nucleo}).  
Moreover, they can affect the neutrino energy spectra of different flavors before the onset of the potentially slow collective conversions and the MSW flavor transformation, both of which were also found to affect (possibly) the neutrino-driven wind nucleosynthesis and neutrino nucleosynthesis. 

Obviously, much work beyond the progress discussed in this review remains to be done in the future in order to fully elucidate the role of neutrinos in CCSNe and BNSMs.  
For instance, long-term and multidimensional simulations that include the improved neutrino interaction rates described in this review need to be carried out to assess their detailed impacts.
Methods that can effectively take into account the collective neutrino flavor oscillations in the transport of neutrinos need to be developed and applied to simulations. 
Ultimately, given the profound implications provided by the detection of two dozen events from the SN1987A discussed in Sec.~\ref{sec:nu-det}, we fully expect that when the next galactic SN goes off in the coming decades, the unprecedented amount of neutrino events will further shed new lights on our understanding of neutrinos in CCSNe, leading also to improved treatments of the role of neutrinos in BNSMs. 
These, together with the continuous improvement in modeling, will hopefully help clarify the role of neutrinos in various nucleosynthesis processes that are responsible for the creation of elements in the universe.

\newpage
\section*{Acknowledgements}

We thank Projjwal Banerjee, Oliver Just, Andre Sieverding, and Zewei Xiong for fruitful collaborations and discussions. 
TF acknowledges support from the Polish National Science Centre (NCN) under grant number 2020/37/B/ST9/00691. 
GG acknowledges support by the National Natural Science Foundation of China under Grant No. 12205258 and the Natural Science Foundation of Shandong Province, China under Grant No. ZR2022JQ04.
GMP acknowledges support by the European Research Council (ERC) under the European Union's Horizon 2020 research and innovation programme (ERC Advanced Grant KILONOVA No. 885281), by the Deutsche Forschungsgemeinschaft (DFG, German Research Foundation) -- Project-ID 279384907, SFB 1245 and  MA 4248/3-1,  the state of Hesse within the Cluster Project ELEMENTS, and the Helmholtz Forschungsakademie Hessen f\"ur FAIR. 
YZQ acknowledges support by the U.S. Department of Energy under Grant No. DE-FG02-87ER40328.
MRW acknowledges supports from the 
National Science and Technology Council, Taiwan under Grant No.~111-2628-M-001-003-MY4, the Academia Sinica under Project No.~AS-CDA-109-M11, and Physics Division, National Center for Theoretical Sciences, Taiwan.


\begin{thebibliography}{100}
\expandafter\ifx\csname url\endcsname\relax
  \def\url#1{\texttt{#1}}\fi
\expandafter\ifx\csname urlprefix\endcsname\relax\def\urlprefix{URL }\fi
\expandafter\ifx\csname href\endcsname\relax
  \def\href#1#2{#2} \def\path#1{#1}\fi

\bibitem{B2FH}
E.~M. Burbidge, G.~R. Burbidge, W.~A. Fowler, F.~Hoyle, Synthesis of the
  elements in stars, Rev. Mod. Phys. 29 (1957) 547--650.
\newblock \href {https://doi.org/10.1103/RevModPhys.29.547}
  {\path{doi:10.1103/RevModPhys.29.547}}.

\bibitem{Cameron:1957}
A.~G.~W. Cameron, Stellar evolution, nuclear astrophysics, and nucleogenesis,
  Report CRL-41, Chalk River, {reprinted in D. M. Kahl, 2013, Ed.,
  \emph{Stellar Evolution, Nuclear Astrophysics, and Nucleogenesis} (Dover, New
  York).} (1957).

\bibitem{Kappeler:2011xx}
F.~{K{\"a}ppeler}, R.~{Gallino}, S.~{Bisterzo}, W.~{Aoki}, {The s process:
  Nuclear physics, stellar models, and observations}, Reviews of Modern Physics
  83~(1) (2011) 157--194.
\newblock \href {http://arxiv.org/abs/1012.5218} {\path{arXiv:1012.5218}},
  \href {https://doi.org/10.1103/RevModPhys.83.157}
  {\path{doi:10.1103/RevModPhys.83.157}}.

\bibitem{Rauscher.Dauphas.ea:2013}
T.~Rauscher, N.~Dauphas, I.~Dillmann, et~al., Constraining the astrophysical
  origin of the p-nuclei through nuclear physics and meteoritic data, Rep.
  Prog. Phys. 76~(6) (2013) 066201.
\newblock \href {http://arxiv.org/abs/1303.2666} {\path{arXiv:1303.2666}},
  \href {https://doi.org/10.1088/0034-4885/76/6/066201}
  {\path{doi:10.1088/0034-4885/76/6/066201}}.

\bibitem{Pignatari:2016xx}
M.~{Pignatari}, K.~{G{\"o}bel}, R.~{Reifarth}, C.~{Travaglio}, {The production
  of proton-rich isotopes beyond iron: The {\ensuremath{\gamma}}-process in
  stars}, International Journal of Modern Physics E 25~(4) (2016) 1630003--232.
\newblock \href {http://arxiv.org/abs/1605.03690} {\path{arXiv:1605.03690}},
  \href {https://doi.org/10.1142/S0218301316300034}
  {\path{doi:10.1142/S0218301316300034}}.

\bibitem{Cowan:2019pkx}
J.~J. Cowan, C.~Sneden, J.~E. Lawler, et~al., {Origin of the heaviest elements:
  The rapid neutron-capture process}, Rev. Mod. Phys. 93~(1) (2021) 15002.
\newblock \href {http://arxiv.org/abs/1901.01410} {\path{arXiv:1901.01410}},
  \href {https://doi.org/10.1103/RevModPhys.93.015002}
  {\path{doi:10.1103/RevModPhys.93.015002}}.

\bibitem{Arcones:2022jer}
A.~Arcones, F.-K. Thielemann, {Origin of the elements}, Astron. Astrophys. Rev.
  31~(1) (2023) 1.
\newblock \href {https://doi.org/10.1007/s00159-022-00146-x}
  {\path{doi:10.1007/s00159-022-00146-x}}.

\bibitem{Woosley90}
S.~E. Woosley, D.~H. Hartmann, R.~D. Hoffman, W.~C. Haxton, The $\nu$-process,
  Astrophys. J. 356 (1990) 272.

\bibitem{Heger05a}
A.~Heger, E.~Kolbe, W.~C. Haxton, et~al., Neutrino nucleosynthesis, Phys. Lett.
  B 606 (2005) 258.

\bibitem{Froehlich06b}
C.~Frohlich, G.~Martinez-Pinedo, M.~Liebendorfer, et~al., {Neutrino-induced
  nucleosynthesis of a\ensuremath{>}64 nuclei: the nu p-process}, Phys. Rev.
  Lett. 96 (2006) 142502.
\newblock \href {http://arxiv.org/abs/astro-ph/0511376}
  {\path{arXiv:astro-ph/0511376}}, \href
  {https://doi.org/10.1103/PhysRevLett.96.142502}
  {\path{doi:10.1103/PhysRevLett.96.142502}}.

\bibitem{Pruet06}
J.~Pruet, R.~D. Hofmann, S.~E. Woosley, et~al., Nucleosynthesis in early
  supernova winds. ii. the role of neutrinos, Astrophysical Journal 644 (2006)
  1028.

\bibitem{Cowan:1977xx}
J.~J. {Cowan}, W.~K. {Rose}, {Production of $^{14}$C and neutrons in red
  giants.}, Astrophys. J. 212 (1977) 149--158.
\newblock \href {https://doi.org/10.1086/155030} {\path{doi:10.1086/155030}}.

\bibitem{Johns2016xx}
S.~{Jones}, C.~{Ritter}, F.~{Herwig}, et~al., {H ingestion into He-burning
  convection zones in super-AGB stellar models as a potential site for
  intermediate neutron-density nucleosynthesis}, Mon. Not. R. Astron. Soc.
  455~(4) (2016) 3848--3863.
\newblock \href {http://arxiv.org/abs/1510.07417} {\path{arXiv:1510.07417}},
  \href {https://doi.org/10.1093/mnras/stv2488}
  {\path{doi:10.1093/mnras/stv2488}}.

\bibitem{Prantzos:2012}
N.~{Prantzos}, {Production and evolution of Li, Be, and B isotopes in the
  Galaxy}, Astron. Astrophys. 542 (2012) A67.
\newblock \href {http://arxiv.org/abs/1203.5662} {\path{arXiv:1203.5662}},
  \href {https://doi.org/10.1051/0004-6361/201219043}
  {\path{doi:10.1051/0004-6361/201219043}}.

\bibitem{Phillips1999phst.book}
A.~C. {Phillips}, {The Physics of Stars, 2nd Edition}, 1999.

\bibitem{Nomoto87}
K.~{Nomoto}, {Evolution of 8--10 M$_{sun}$ Stars toward Electron Capture
  Supernovae. II. Collapse of an O + NE + MG Core}, Astrophys. J. 322 (1987)
  206.
\newblock \href {https://doi.org/10.1086/165716} {\path{doi:10.1086/165716}}.

\bibitem{Jones13}
S.~{Jones}, R.~{Hirschi}, K.~{Nomoto}, et.al., {Advanced Burning Stages and
  Fate of 8-10 M $_{{\ensuremath{\odot}}}$ Stars}, Astrophys. J. 772~(2) (2013)
  150.
\newblock \href {http://arxiv.org/abs/1306.2030} {\path{arXiv:1306.2030}},
  \href {https://doi.org/10.1088/0004-637X/772/2/150}
  {\path{doi:10.1088/0004-637X/772/2/150}}.

\bibitem{Limongi2023arXiv231200107L}
M.~{Limongi}, L.~{Roberti}, A.~{Chieffi}, K.~{Nomoto}, {Evolution and final
  fate of solar metallicity stars in the mass range 7-15 Msun. I. The
  transition from AGB to SAGB stars, Electron Capture and Core Collapse
  Supernovae progenitors} (Nov. 2023).
\newblock \href {http://arxiv.org/abs/2312.00107} {\path{arXiv:2312.00107}},
  \href {https://doi.org/10.48550/arXiv.2312.00107}
  {\path{doi:10.48550/arXiv.2312.00107}}.

\bibitem{LimongiChieffi2000ApJS129}
M.~{Limongi}, O.~{Straniero}, A.~{Chieffi}, {Massive Stars in the Range 13-25
  M$_{solar}$: Evolution and Nucleosynthesis. II. The Solar Metallicity
  Models}, Astrophys. J., Suppl. 129~(2) (2000) 625--664.
\newblock \href {http://arxiv.org/abs/astro-ph/0003401}
  {\path{arXiv:astro-ph/0003401}}, \href {https://doi.org/10.1086/313424}
  {\path{doi:10.1086/313424}}.

\bibitem{Woosley2002RvMP74}
S.~E. {Woosley}, A.~{Heger}, T.~A. {Weaver}, {The evolution and explosion of
  massive stars}, Reviews of Modern Physics 74~(4) (2002) 1015--1071.
\newblock \href {https://doi.org/10.1103/RevModPhys.74.1015}
  {\path{doi:10.1103/RevModPhys.74.1015}}.

\bibitem{Heger03}
A.~{Heger}, C.~L. {Fryer}, S.~E. {Woosley}, et~al., {How Massive Single Stars
  End Their Life}, Astrophys. J. 591 (2003) 288--300.
\newblock \href {http://arxiv.org/abs/astro-ph/0212469}
  {\path{arXiv:astro-ph/0212469}}, \href {https://doi.org/10.1086/375341}
  {\path{doi:10.1086/375341}}.

\bibitem{UmedaNomoto2008ApJ673}
H.~{Umeda}, K.~{Nomoto}, {How Much $^{56}$Ni Can Be Produced in Core-Collapse
  Supernovae? Evolution and Explosions of 30-100 M$_{{\ensuremath{\odot}}}$
  Stars}, Astrophys. J. 673~(2) (2008) 1014--1022.
\newblock \href {http://arxiv.org/abs/0707.2598} {\path{arXiv:0707.2598}},
  \href {https://doi.org/10.1086/524767} {\path{doi:10.1086/524767}}.

\bibitem{Eggenberger2008Ap&SS316}
P.~{Eggenberger}, G.~{Meynet}, A.~{Maeder}, et~al., {The Geneva stellar
  evolution code}, Astrophys. Space Sci. 316~(1-4) (2008) 43--54.
\newblock \href {https://doi.org/10.1007/s10509-007-9511-y}
  {\path{doi:10.1007/s10509-007-9511-y}}.

\bibitem{Bohm-Vitense:1958}
E.~{B{\"o}hm-Vitense}, {{\"U}ber die Wasserstoffkonvektionszone in Sternen
  verschiedener Effektivtemperaturen und Leuchtkr{\"a}fte. Mit 5
  Textabbildungen}, Zeitschrift f\"ur Astrophysik 46 (1958) 108.

\bibitem{Kippenhahn:2013}
R.~{Kippenhahn}, A.~{Weigert}, A.~{Weiss}, {Stellar Structure and Evolution},
  2013.
\newblock \href {https://doi.org/10.1007/978-3-642-30304-3}
  {\path{doi:10.1007/978-3-642-30304-3}}.

\bibitem{Heger2000ApJ528}
A.~{Heger}, N.~{Langer}, S.~E. {Woosley}, {Presupernova Evolution of Rotating
  Massive Stars. I. Numerical Method and Evolution of the Internal Stellar
  Structure}, Astrophys. J. 528~(1) (2000) 368--396.
\newblock \href {http://arxiv.org/abs/astro-ph/9904132}
  {\path{arXiv:astro-ph/9904132}}, \href {https://doi.org/10.1086/308158}
  {\path{doi:10.1086/308158}}.

\bibitem{UmedaNomoto2003Natur422}
H.~{Umeda}, K.~{Nomoto}, {First-generation black-hole-forming supernovae and
  the metal abundance pattern of a very iron-poor star}, Nature 422~(6934) (2003)
  871--873.
\newblock \href {http://arxiv.org/abs/astro-ph/0301315}
  {\path{arXiv:astro-ph/0301315}}, \href {https://doi.org/10.1038/nature01571}
  {\path{doi:10.1038/nature01571}}.

\bibitem{Hirschi2004A&A425}
R.~{Hirschi}, G.~{Meynet}, A.~{Maeder}, {Stellar evolution with rotation. XII.
  Pre-supernova models}, Astron. Astrophys. 425 (2004) 649--670.
\newblock \href {http://arxiv.org/abs/astro-ph/0406552}
  {\path{arXiv:astro-ph/0406552}}, \href
  {https://doi.org/10.1051/0004-6361:20041095}
  {\path{doi:10.1051/0004-6361:20041095}}.

\bibitem{Iwamoto2005Sci309}
N.~{Iwamoto}, H.~{Umeda}, N.~{Tominaga}, et~al., {The First Chemical Enrichment
  in the Universe and the Formation of Hyper Metal-Poor Stars}, Science
  309~(5733) (2005) 451--453.
\newblock \href {http://arxiv.org/abs/astro-ph/0505524}
  {\path{arXiv:astro-ph/0505524}}, \href
  {https://doi.org/10.1126/science.1112997}
  {\path{doi:10.1126/science.1112997}}.

\bibitem{Tominaga2007ApJ660}
N.~{Tominaga}, H.~{Umeda}, K.~{Nomoto}, {Supernova Nucleosynthesis in
  Population III 13-50 M$_{solar}$ Stars and Abundance Patterns of Extremely
  Metal-poor Stars}, Astrophys. J. 660~(1) (2007) 516--540.
\newblock \href {http://arxiv.org/abs/astro-ph/0701381}
  {\path{arXiv:astro-ph/0701381}}, \href {https://doi.org/10.1086/513063}
  {\path{doi:10.1086/513063}}.

\bibitem{Hirschi2012A&A546A}
L.~{Muijres}, J.~S. {Vink}, A.~{de Koter}, et~al., {Mass-loss predictions for
  evolved very metal-poor massive stars}, Astron. Astrophys. 546 (2012) A42.
\newblock \href {http://arxiv.org/abs/1209.5934} {\path{arXiv:1209.5934}},
  \href {https://doi.org/10.1051/0004-6361/201118666}
  {\path{doi:10.1051/0004-6361/201118666}}.

\bibitem{Langer2012A&A542}
S.~C. {Yoon}, A.~{Dierks}, N.~{Langer}, {Evolution of massive Population III
  stars with rotation and magnetic fields}, Astron. Astrophys. 542 (2012) A113.
\newblock \href {http://arxiv.org/abs/1201.2364} {\path{arXiv:1201.2364}},
  \href {https://doi.org/10.1051/0004-6361/201117769}
  {\path{doi:10.1051/0004-6361/201117769}}.

\bibitem{ChieffiLimongi2013ApJ764}
A.~{Chieffi}, M.~{Limongi}, {Pre-supernova Evolution of Rotating Solar
  Metallicity Stars in the Mass Range 13-120 M $_{{\ensuremath{\odot}}}$ and
  their Explosive Yields}, Astrophys. J. 764~(1) (2013) 21.
\newblock \href {https://doi.org/10.1088/0004-637X/764/1/21}
  {\path{doi:10.1088/0004-637X/764/1/21}}.

\bibitem{O'Connor11}
E.~{O'Connor}, C.~D. {Ott}, {Black Hole Formation in Failing Core-Collapse
  Supernovae}, Astrophys. J. 730 (2011) 70.
\newblock \href {http://arxiv.org/abs/1010.5550} {\path{arXiv:1010.5550}},
  \href {https://doi.org/10.1088/0004-637X/730/2/70}
  {\path{doi:10.1088/0004-637X/730/2/70}}.

\bibitem{Langer:2012jz}
N.~Langer, {Pre-Supernova Evolution of Massive Single and Binary Stars}, Ann.
  Rev. Astron. Astrophys. 50 (2012) 107--164.
\newblock \href {http://arxiv.org/abs/1206.5443} {\path{arXiv:1206.5443}},
  \href {https://doi.org/10.1146/annurev-astro-081811-125534}
  {\path{doi:10.1146/annurev-astro-081811-125534}}.

\bibitem{Yoon:2017nuh}
S.-C. Yoon, L.~Dessart, A.~Clocchiatti, {Type Ib and IIb supernova progenitors
  in interacting binary systems}, Astrophys. J. 840~(1) (2017) 10.
\newblock \href {http://arxiv.org/abs/1701.02089} {\path{arXiv:1701.02089}},
  \href {https://doi.org/10.3847/1538-4357/aa6afe}
  {\path{doi:10.3847/1538-4357/aa6afe}}.

\bibitem{Schneider:2020vvh}
F.~R.~N. Schneider, P.~Podsiadlowski, B.~M\"uller, {Pre-supernova evolution,
  compact object masses and explosion properties of stripped binary stars},
  Astron. Astrophys. 645 (2021) A5.
\newblock \href {http://arxiv.org/abs/2008.08599} {\path{arXiv:2008.08599}},
  \href {https://doi.org/10.1051/0004-6361/202039219}
  {\path{doi:10.1051/0004-6361/202039219}}.

\bibitem{Laplace:2021vre}
E.~Laplace, S.~Justham, M.~Renzo, et~al., {Different to the core: The
  pre-supernova structures of massive single and binary-stripped stars},
  Astron. Astrophys. 656 (2021) A58.
\newblock \href {http://arxiv.org/abs/2102.05036} {\path{arXiv:2102.05036}},
  \href {https://doi.org/10.1051/0004-6361/202140506}
  {\path{doi:10.1051/0004-6361/202140506}}.

\bibitem{Kinugawa:2023hdo}
T.~Kinugawa, S.~Horiuchi, T.~Takiwaki, K.~Kotake, {Fate of supernova
  progenitors in massive binary systems} (11 2023).
\newblock \href {http://arxiv.org/abs/2311.14341} {\path{arXiv:2311.14341}}.

\bibitem{Wang:2024dwq}
H.-S. Wang, K.-C. Pan, {The Influence of Stellar Rotation in Binary Systems on
  Core-Collapse Supernova Progenitors and Multi-messenger Signals} (1 2024).
\newblock \href {http://arxiv.org/abs/2401.08985} {\path{arXiv:2401.08985}}.

\bibitem{KupkaMuthsam:2017}
F.~{Kupka}, H.~J. {Muthsam}, {Modelling of stellar convection}, Living Reviews
  in Computational Astrophysics 3~(1) (2017) 1.
\newblock \href {https://doi.org/10.1007/s41115-017-0001-9}
  {\path{doi:10.1007/s41115-017-0001-9}}.

\bibitem{Arnett:1998}
G.~{Baz{\'a}n}, D.~{Arnett}, {Two-dimensional Hydrodynamics of Pre--Core
  Collapse: Oxygen Shell Burning}, Astrophys. J. 496~(1) (1998) 316--332.
\newblock \href {http://arxiv.org/abs/astro-ph/9702239}
  {\path{arXiv:astro-ph/9702239}}, \href {https://doi.org/10.1086/305346}
  {\path{doi:10.1086/305346}}.

\bibitem{Couch2015ApJ}
S.~M. {Couch}, E.~{Chatzopoulos}, W.~D. {Arnett}, F.~X. {Timmes}, {The
  Three-dimensional Evolution to Core Collapse of a Massive Star}, Astrophys.
  J. Lett. 808~(1) (2015) L21.
\newblock \href {http://arxiv.org/abs/1503.02199} {\path{arXiv:1503.02199}},
  \href {https://doi.org/10.1088/2041-8205/808/1/L21}
  {\path{doi:10.1088/2041-8205/808/1/L21}}.

\bibitem{BMuller16}
B.~{M{\"u}ller}, M.~{Viallet}, A.~{Heger}, H.-T. {Janka}, {The Last Minutes of
  Oxygen Shell Burning in a Massive Star}, Astrophys. J. 833~(1) (2016) 124.
\newblock \href {http://arxiv.org/abs/1605.01393} {\path{arXiv:1605.01393}},
  \href {https://doi.org/10.3847/1538-4357/833/1/124}
  {\path{doi:10.3847/1538-4357/833/1/124}}.

\bibitem{Janka2020ApJ}
N.~{Yadav}, B.~{M{\"u}ller}, H.~T. {Janka}, et~al., {Large-scale Mixing in a
  Violent Oxygen-Neon Shell Merger Prior to a Core-collapse Supernova},
  Astrophys. J. 890~(2) (2020) 94.
\newblock \href {http://arxiv.org/abs/1905.04378} {\path{arXiv:1905.04378}},
  \href {https://doi.org/10.3847/1538-4357/ab66bb}
  {\path{doi:10.3847/1538-4357/ab66bb}}.

\bibitem{Takashi2021MNRAS}
T.~{Yoshida}, T.~{Takiwaki}, D.~R. {Aguilera-Dena}, et~al., {A
  three-dimensional hydrodynamics simulation of oxygen-shell burning in the
  final evolution of a fast-rotating massive star}, Mon. Not. R. Astro. Soc.
  506~(1) (2021) L20--L25.
\newblock \href {http://arxiv.org/abs/2106.09909} {\path{arXiv:2106.09909}},
  \href {https://doi.org/10.1093/mnrasl/slab067}
  {\path{doi:10.1093/mnrasl/slab067}}.

\bibitem{FieldsCouch2021ApJ921}
C.~E. {Fields}, S.~M. {Couch}, {Three-dimensional Hydrodynamic Simulations of
  Convective Nuclear Burning in Massive Stars Near Iron Core Collapse}, Astrophys. J.
  921~(1) (2021) 28.
\newblock \href {http://arxiv.org/abs/2107.04617} {\path{arXiv:2107.04617}},
  \href {https://doi.org/10.3847/1538-4357/ac24fb}
  {\path{doi:10.3847/1538-4357/ac24fb}}.

\bibitem{juoda}
A.~{Juodagalvis}, K.~{Langanke}, W.~R. {Hix}, et~al., {Improved estimate of
  electron capture rates on nuclei during stellar core collapse}, Nuclear
  Physics A 848 (2010) 454--478.
\newblock \href {http://arxiv.org/abs/0909.0179} {\path{arXiv:0909.0179}},
  \href {https://doi.org/10.1016/j.nuclphysa.2010.09.012}
  {\path{doi:10.1016/j.nuclphysa.2010.09.012}}.

\bibitem{langanke03}
K.~{Langanke}, G.~{Mart{\'{\i}}nez-Pinedo}, J.~M. {Sampaio}, et~al., {Electron
  Capture Rates on Nuclei and Implications for Stellar Core Collapse}, Physical
  Review Letters 90~(24) (2003) 241102.
\newblock \href {http://arxiv.org/abs/astro-ph/0302459}
  {\path{arXiv:astro-ph/0302459}}, \href
  {https://doi.org/10.1103/PhysRevLett.90.241102}
  {\path{doi:10.1103/PhysRevLett.90.241102}}.

\bibitem{Fischer13}
T.~{Fischer}, K.~{Langanke}, G.~{Mart{\'{\i}}nez-Pinedo}, {Neutrino-pair
  emission from nuclear de-excitation in core-collapse supernova simulations},
  Phys. Rev. C 88~(6) (2013) 065804.
\newblock \href {http://arxiv.org/abs/1309.4271} {\path{arXiv:1309.4271}},
  \href {https://doi.org/10.1103/PhysRevC.88.065804}
  {\path{doi:10.1103/PhysRevC.88.065804}}.

\bibitem{Fuller:1991}
G.~M. {Fuller}, B.~S. {Meyer}, {High-temperature neutrino-nucleus processes in
  stellar collapse}, Astrophys. J. 376 (1991) 701--716.
\newblock \href {https://doi.org/10.1086/170317} {\path{doi:10.1086/170317}}.

\bibitem{LSEOS}
J.~M. {Lattimer}, F.~{Swesty}, {A generalized equation of state for hot, dense
  matter}, Nuclear Physics A 535 (1991) 331--376.
\newblock \href {https://doi.org/10.1016/0375-9474(91)90452-C}
  {\path{doi:10.1016/0375-9474(91)90452-C}}.

\bibitem{Typel10}
S.~{Typel}, G.~{R{\"o}pke}, T.~{Kl{\"a}hn}, et~al., {Composition and
  thermodynamics of nuclear matter with light clusters}, Phys. Rev. C 81~(1)
  (2010) 015803.
\newblock \href {http://arxiv.org/abs/0908.2344} {\path{arXiv:0908.2344}},
  \href {https://doi.org/10.1103/PhysRevC.81.015803}
  {\path{doi:10.1103/PhysRevC.81.015803}}.

\bibitem{HS}
M.~{Hempel}, J.~{Schaffner-Bielich}, {A statistical model for a complete
  supernova equation of state}, Nuclear Physics A 837 (2010) 210--254.
\newblock \href {http://arxiv.org/abs/0911.4073} {\path{arXiv:0911.4073}},
  \href {https://doi.org/10.1016/j.nuclphysa.2010.02.010}
  {\path{doi:10.1016/j.nuclphysa.2010.02.010}}.

\bibitem{Fischer2016EPJA}
T.~{Fischer}, {Constraining the supersaturation density equation of state from
  core-collapse supernova simulations?. Excluded volume extension of the
  baryons}, European Physical Journal A 52 (2016) 54.
\newblock \href {http://arxiv.org/abs/1604.01629} {\path{arXiv:1604.01629}},
  \href {https://doi.org/10.1140/epja/i2016-16054-9}
  {\path{doi:10.1140/epja/i2016-16054-9}}.

\bibitem{Fischer20c}
T.~{Fischer}, S.~{Typel}, G.~{R{\"o}pke}, et~al., {Medium modifications for
  light and heavy nuclear clusters in simulations of core collapse supernovae:
  Impact on equation of state and weak interactions}, Phys. Rev. C 102~(5)
  (2020) 055807.
\newblock \href {http://arxiv.org/abs/2008.13608} {\path{arXiv:2008.13608}},
  \href {https://doi.org/10.1103/PhysRevC.102.055807}
  {\path{doi:10.1103/PhysRevC.102.055807}}.

\bibitem{Fischer20d}
T.~{Fischer}, G.~{Guo}, G.~{Mart{\'\i}nez-Pinedo}, et~al., {Muonization of
  supernova matter}, Phys. Rev. D 102~(12) (2020) 123001.
\newblock \href {https://doi.org/10.1103/PhysRevD.102.123001}
  {\path{doi:10.1103/PhysRevD.102.123001}}.

\bibitem{Wu:2014kaa}
M.-R. Wu, Y.-Z. Qian, G.~Martinez-Pinedo, et~al., {Effects of neutrino
  oscillations on nucleosynthesis and neutrino signals for an 18 M supernova
  model}, Phys. Rev. D 91~(6) (2015) 065016.
\newblock \href {http://arxiv.org/abs/1412.8587} {\path{arXiv:1412.8587}},
  \href {https://doi.org/10.1103/PhysRevD.91.065016}
  {\path{doi:10.1103/PhysRevD.91.065016}}.

\bibitem{Fischer09}
T.~{Fischer}, S.~C. {Whitehouse}, A.~{Mezzacappa}, et~al., {The neutrino signal
  from protoneutron star accretion and black hole formation}, Astron.
  Astrophys. 499 (2009) 1--15.
\newblock \href {http://arxiv.org/abs/0809.5129} {\path{arXiv:0809.5129}},
  \href {https://doi.org/10.1051/0004-6361/200811055}
  {\path{doi:10.1051/0004-6361/200811055}}.

\bibitem{Janka12}
H.-T. {Janka}, {Explosion Mechanisms of Core-Collapse Supernovae}, Annual
  Review of Nuclear and Particle Science 62 (2012) 407--451.
\newblock \href {http://arxiv.org/abs/1206.2503} {\path{arXiv:1206.2503}},
  \href {https://doi.org/10.1146/annurev-nucl-102711-094901}
  {\path{doi:10.1146/annurev-nucl-102711-094901}}.

\bibitem{Fischer14}
T.~{Fischer}, M.~{Hempel}, I.~{Sagert}, et~al., {Symmetry energy impact in
  simulations of core-collapse supernovae}, European Physical Journal A 50
  (2014) 46.
\newblock \href {http://arxiv.org/abs/1307.6190} {\path{arXiv:1307.6190}},
  \href {https://doi.org/10.1140/epja/i2014-14046-5}
  {\path{doi:10.1140/epja/i2014-14046-5}}.

\bibitem{Fischer12}
T.~{Fischer}, G.~{Mart{\'\i}nez-Pinedo}, M.~{Hempel}, M.~{Liebend{\"o}rfer},
  {Neutrino spectra evolution during protoneutron star deleptonization}, Phys.
  Rev. D 85~(8) (2012) 083003.
\newblock \href {http://arxiv.org/abs/1112.3842} {\path{arXiv:1112.3842}},
  \href {https://doi.org/10.1103/PhysRevD.85.083003}
  {\path{doi:10.1103/PhysRevD.85.083003}}.

\bibitem{Raffelt01}
G.~G. {Raffelt}, {Mu- and Tau-Neutrino Spectra Formation in Supernovae},
  Astrophys. J. 561~(2) (2001) 890--914.
\newblock \href {http://arxiv.org/abs/astro-ph/0105250}
  {\path{arXiv:astro-ph/0105250}}, \href {https://doi.org/10.1086/323379}
  {\path{doi:10.1086/323379}}.

\bibitem{Colgate66}
S.~A. {Colgate}, R.~H. {White}, {The Hydrodynamic Behavior of Supernovae
  Explosions}, Astrophys. J. 143 (1966) 626.
\newblock \href {https://doi.org/10.1086/148549} {\path{doi:10.1086/148549}}.

\bibitem{Bethe85}
H.~A. {Bethe}, J.~R. {Wilson}, {Revival of a stalled supernova shock by
  neutrino heating}, Astrophys. J. 295 (1985) 14--23.
\newblock \href {https://doi.org/10.1086/163343} {\path{doi:10.1086/163343}}.

\bibitem{kitaura06}
F.~S. {Kitaura}, H.-T. {Janka}, W.~{Hillebrandt}, {Explosions of O-Ne-Mg cores,
  the Crab supernova, and subluminous type II-P supernovae}, Astron. Astrophys.
  450 (2006) 345--350.
\newblock \href {http://arxiv.org/abs/arXiv:astro-ph/0512065}
  {\path{arXiv:arXiv:astro-ph/0512065}}, \href
  {https://doi.org/10.1051/0004-6361:20054703}
  {\path{doi:10.1051/0004-6361:20054703}}.

\bibitem{Wanajo09}
S.~Wanajo, K.~Nomoto, H.~T. Janka, et~al., {Nucleosynthesis in Electron Capture
  Supernovae of AGB Stars}, Astrophys. J. 695 (2009) 208--220.
\newblock \href {http://arxiv.org/abs/0810.3999} {\path{arXiv:0810.3999}},
  \href {https://doi.org/10.1088/0004-637X/695/1/208}
  {\path{doi:10.1088/0004-637X/695/1/208}}.

\bibitem{Wanajo.Mueller.ea:2018}
S.~Wanajo, B.~M\"uller, H.-T. Janka, A.~Heger, {Nucleosynthesis in the
  Innermost Ejecta of Neutrino-driven Supernova Explosions in Two Dimensions},
  Astrophys. J. 852~(1) (2018) 40.
\newblock \href {http://arxiv.org/abs/1701.06786} {\path{arXiv:1701.06786}},
  \href {https://doi.org/10.3847/1538-4357/aa9d97}
  {\path{doi:10.3847/1538-4357/aa9d97}}.

\bibitem{Burrows:2019rtd}
A.~Burrows, D.~Radice, D.~Vartanyan, {Three-dimensional supernova explosion
  simulations of 9-, 10-, 11-, 12-, and 13-M\ensuremath{\odot} stars}, Mon.
  Not. Roy. Astron. Soc. 485~(3) (2019) 3153--3168.
\newblock \href {http://arxiv.org/abs/1902.00547} {\path{arXiv:1902.00547}},
  \href {https://doi.org/10.1093/mnras/stz543}
  {\path{doi:10.1093/mnras/stz543}}.

\bibitem{Zha22}
S.~{Zha}, E.~P. {O'Connor}, S.~M. {Couch}, et~al., {Hydrodynamic simulations of
  electron-capture supernovae: progenitor and dimension dependence}, MNRAS
  513~(1) (2022) 1317--1328.
\newblock \href {http://arxiv.org/abs/2112.15257} {\path{arXiv:2112.15257}},
  \href {https://doi.org/10.1093/mnras/stac1035}
  {\path{doi:10.1093/mnras/stac1035}}.

\bibitem{Wanajo11}
S.~{Wanajo}, H.-T. {Janka}, B.~{M{\"u}ller}, {Electron-capturei Supernovae as
  The Origin of Elements Beyond Iron}, Astrophys. J. Lett. 726~(2) (2011) L15.
\newblock \href {http://arxiv.org/abs/1009.1000} {\path{arXiv:1009.1000}},
  \href {https://doi.org/10.1088/2041-8205/726/2/L15}
  {\path{doi:10.1088/2041-8205/726/2/L15}}.

\bibitem{Wang:2023vkk}
T.~Wang, A.~Burrows, {Neutrino-driven Winds in Three-dimensional Core-collapse
  Supernova Simulations}, Astrophys. J. 954~(2) (2023) 114.
\newblock \href {http://arxiv.org/abs/2306.13712} {\path{arXiv:2306.13712}},
  \href {https://doi.org/10.3847/1538-4357/ace7b2}
  {\path{doi:10.3847/1538-4357/ace7b2}}.

\bibitem{Bollig.Yadav.ea:2021}
R.~Bollig, N.~Yadav, D.~Kresse, et~al., {Self-consistent 3D Supernova Models
  From \ensuremath{-}7 Minutes to +7 s: A 1-bethe Explosion of a
  \ensuremath{\sim}19 $M_\odot$ Progenitor}, Astrophys. J. 915~(1) (2021) 28.
\newblock \href {http://arxiv.org/abs/2010.10506} {\path{arXiv:2010.10506}},
  \href {https://doi.org/10.3847/1538-4357/abf82e}
  {\path{doi:10.3847/1538-4357/abf82e}}.

\bibitem{Fogli09}
C.~Fogli, E.~Lisi, A.~Marrone, I.~Tamborra, Supernova neutrino three-flavor
  evolution with dominant collective effects, Cosmol. Astropart. Phys. 10
  (2009) 002.
\newblock \href {http://arxiv.org/abs/0812.3031} {\path{arXiv:0812.3031}},
  \href {https://doi.org/10.1088/1475-7516/2009/04/030}
  {\path{doi:10.1088/1475-7516/2009/04/030}}.

\bibitem{Matsumoto2024MNRAS528}
J.~{Matsumoto}, T.~{Takiwaki}, K.~{Kotake}, {Neutrino-driven massive stellar explosions in 3D fostered by magnetic fields via turbulent alpha-effect}, MNRAS 528 (2024) L96.
\newblock \href {http://arxiv.org/abs/2307.03400} {\path{arXiv:2307.03400}},
  \href {https://doi.org/10.1093/mnrasl/slad173}
  {\path{doi:10.1093/mnrasl/slad173}}.

\bibitem{Liebendorfer09}
M.~{Liebend{\"o}rfer}, S.~C. {Whitehouse}, T.~{Fischer}, {The Isotropic
  Diffusion Source Approximation for Supernova Neutrino Transport}, Astrophys.
  J. 698 (2009) 1174--1190.
\newblock \href {http://arxiv.org/abs/0711.2929} {\path{arXiv:0711.2929}},
  \href {https://doi.org/10.1088/0004-637X/698/2/1174}
  {\path{doi:10.1088/0004-637X/698/2/1174}}.

\bibitem{Kotake18}
K.~{Kotake}, T.~{Takiwaki}, T.~{Fischer}, et~al., {Impact of Neutrino Opacities
  on Core-collapse Supernova Simulations}, Astrophys. J. 853~(2) (2018) 170.
\newblock \href {http://arxiv.org/abs/1801.02703} {\path{arXiv:1801.02703}},
  \href {https://doi.org/10.3847/1538-4357/aaa716}
  {\path{doi:10.3847/1538-4357/aaa716}}.

\bibitem{Takiwaki14}
T.~{Takiwaki}, K.~{Kotake}, Y.~{Suwa}, {A Comparison of Two- and
  Three-dimensional Neutrino-hydrodynamics Simulations of Core-collapse
  Supernovae}, Astrophys. J. 786 (2014) 83.
\newblock \href {http://arxiv.org/abs/1308.5755} {\path{arXiv:1308.5755}},
  \href {https://doi.org/10.1088/0004-637X/786/2/83}
  {\path{doi:10.1088/0004-637X/786/2/83}}.

\bibitem{KurodaT16}
T.~{Kuroda}, T.~{Takiwaki}, K.~{Kotake}, {A New Multi-energy Neutrino
  Radiation-Hydrodynamics Code in Full General Relativity and Its Application
  to the Gravitational Collapse of Massive Stars}, Astrophys. J., Suppl. 222
  (2016) 20.
\newblock \href {http://arxiv.org/abs/1501.06330} {\path{arXiv:1501.06330}},
  \href {https://doi.org/10.3847/0067-0049/222/2/20}
  {\path{doi:10.3847/0067-0049/222/2/20}}.

\bibitem{KurodaT20}
T.~{Kuroda}, A.~{Arcones}, T.~{Takiwaki}, K.~{Kotake}, {Magnetorotational
  Explosion of a Massive Star Supported by Neutrino Heating in General
  Relativistic Three-dimensional Simulations}, Astrophys. J. 896~(2) (2020)
  102.
\newblock \href {https://doi.org/10.3847/1538-4357/ab9308}
  {\path{doi:10.3847/1538-4357/ab9308}}.

\bibitem{Shibata11}
M.~{Shibata}, K.~{Kiuchi}, Y.~{Sekiguchi}, Y.~{Suwa}, {Truncated Moment
  Formalism for Radiation Hydrodynamics in Numerical Relativity}, Progress of
  Theoretical Physics 125 (2011) 1255--1287.
\newblock \href {http://arxiv.org/abs/1104.3937} {\path{arXiv:1104.3937}},
  \href {https://doi.org/10.1143/PTP.125.1255}
  {\path{doi:10.1143/PTP.125.1255}}.

\bibitem{Shibagaki2023arXiv230905161S}
S.~{Shibagaki}, T.~{Kuroda}, K.~{Kotake}, et~al., {Three-dimensional GRMHD
  Simulations of Rapidly Rotating Stellar Core-Collapse}, arXiv e-prints (2023)
  arXiv:2309.05161\href {http://arxiv.org/abs/2309.05161}
  {\path{arXiv:2309.05161}}, \href {https://doi.org/10.48550/arXiv.2309.05161}
  {\path{doi:10.48550/arXiv.2309.05161}}.

\bibitem{Bruenn85}
S.~W. {Bruenn}, {Stellar core collapse - Numerical model and infall epoch},
  Astrophys. J., Suppl. 58 (1985) 771--841.
\newblock \href {https://doi.org/10.1086/191056} {\path{doi:10.1086/191056}}.

\bibitem{Bruenn20}
S.~W. {Bruenn}, J.~M. {Blondin}, W.~R. {Hix}, et~al., {CHIMERA: A Massively
  Parallel Code for Core-collapse Supernova Simulations}, Astrophys. J., Suppl. 248~(1) (2020)
  11.
\newblock \href {http://arxiv.org/abs/1809.05608} {\path{arXiv:1809.05608}},
  \href {https://doi.org/10.3847/1538-4365/ab7aff}
  {\path{doi:10.3847/1538-4365/ab7aff}}.

\bibitem{Bruenn16}
S.~W. {Bruenn}, E.~J. {Lentz}, W.~R. {Hix}, et~al., {The Development of
  Explosions in Axisymmetric Ab Initio Core-collapse Supernova Simulations of
  12-25 M Stars}, Astrophys. J. 818~(2) (2016) 123.
\newblock \href {http://arxiv.org/abs/1409.5779} {\path{arXiv:1409.5779}},
  \href {https://doi.org/10.3847/0004-637X/818/2/123}
  {\path{doi:10.3847/0004-637X/818/2/123}}.

\bibitem{lentz15}
E.~J. {Lentz}, S.~W. {Bruenn}, W.~R. {Hix}, et~al., {Three-dimensional
  Core-collapse Supernova Simulated Using a 15 $M_{sun}$ Progenitor},
  Astrophys. J. Lett. 807 (2015) L31.
\newblock \href {http://arxiv.org/abs/1505.05110} {\path{arXiv:1505.05110}},
  \href {https://doi.org/10.1088/2041-8205/807/2/L31}
  {\path{doi:10.1088/2041-8205/807/2/L31}}.

\bibitem{O'Connor14}
E.~{O'Connor}, {An Open-source Neutrino Radiation Hydrodynamics Code for
  Core-collapse Supernovae}, Astrophys. J., Suppl. 219~(2) (2015) 24.
\newblock \href {http://arxiv.org/abs/1411.7058} {\path{arXiv:1411.7058}},
  \href {https://doi.org/10.1088/0067-0049/219/2/24}
  {\path{doi:10.1088/0067-0049/219/2/24}}.

\bibitem{Skinner19}
M.~A. {Skinner}, J.~C. {Dolence}, A.~{Burrows}, et~al., {FORNAX: A Flexible
  Code for Multiphysics Astrophysical Simulations}, Astrophys. J., Suppl. 241~(1) (2019) 7.
\newblock \href {http://arxiv.org/abs/1806.07390} {\path{arXiv:1806.07390}},
  \href {https://doi.org/10.3847/1538-4365/ab007f}
  {\path{doi:10.3847/1538-4365/ab007f}}.

\bibitem{Skinner16}
M.~A. {Skinner}, A.~{Burrows}, J.~C. {Dolence}, {Should One Use the Ray-by-Ray
  Approximation in Core-collapse Supernova Simulations?}, Astrophys. J. 831~(1) (2016)
  81.
\newblock \href {http://arxiv.org/abs/1512.00113} {\path{arXiv:1512.00113}},
  \href {https://doi.org/10.3847/0004-637X/831/1/81}
  {\path{doi:10.3847/0004-637X/831/1/81}}.

\bibitem{Burrows2021Natur}
A.~{Burrows}, D.~{Vartanyan}, {Core-collapse supernova explosion theory}, Nature
  589~(7840) (2021) 29--39.
\newblock \href {http://arxiv.org/abs/2009.14157} {\path{arXiv:2009.14157}},
  \href {https://doi.org/10.1038/s41586-020-03059-w}
  {\path{doi:10.1038/s41586-020-03059-w}}.

\bibitem{Vartanyan2022MNRAS510}
D.~{Vartanyan}, M.~S.~B. {Coleman}, A.~{Burrows}, {The collapse and
  three-dimensional explosion of three-dimensional massive-star supernova
  progenitor models}, MNRAS 510~(4) (2022) 4689--4705.
\newblock \href {http://arxiv.org/abs/2109.10920} {\path{arXiv:2109.10920}},
  \href {https://doi.org/10.1093/mnras/stab3702}
  {\path{doi:10.1093/mnras/stab3702}}.

\bibitem{Vartanyan2023MNRAS526}
D.~{Vartanyan}, A.~{Burrows}, {Neutrino signatures of 100 2D Axisymmetric
  Core-Collapse Supernova Simulations}, MNRAS 526~(4) (2023) 5900--5910.
\newblock \href {http://arxiv.org/abs/2307.08735} {\path{arXiv:2307.08735}},
  \href {https://doi.org/10.1093/mnras/stad2887}
  {\path{doi:10.1093/mnras/stad2887}}.

\bibitem{Buras06b}
R.~{Buras}, M.~{Rampp}, H.-T. {Janka}, K.~{Kifonidis}, {Two-dimensional
  hydrodynamic core-collapse supernova simulations with spectral neutrino
  transport. I. Numerical method and results for a 15 $M_{sun}$ star}, Astron.
  Astrophys. 447 (2006) 1049--1092.
\newblock \href {http://arxiv.org/abs/arXiv:astro-ph/0507135}
  {\path{arXiv:arXiv:astro-ph/0507135}}, \href
  {https://doi.org/10.1051/0004-6361:20053783}
  {\path{doi:10.1051/0004-6361:20053783}}.

\bibitem{Summa16}
A.~{Summa}, F.~{Hanke}, H.-T. {Janka}, et~al., {Progenitor-dependent Explosion
  Dynamics in Self-consistent, Axisymmetric Simulations of Neutrino-driven
  Core-collapse Supernovae}, Astrophys. J. 825~(1) (2016) 6.
\newblock \href {http://arxiv.org/abs/1511.07871} {\path{arXiv:1511.07871}},
  \href {https://doi.org/10.3847/0004-637X/825/1/6}
  {\path{doi:10.3847/0004-637X/825/1/6}}.

\bibitem{Bollig21}
R.~{Bollig}, N.~{Yadav}, D.~{Kresse}, et~al., {Self-consistent 3D Supernova
  Models From -7 Minutes to +7 s: A 1-bethe Explosion of a 19
  M$_{{\ensuremath{\odot}}}$ Progenitor}, Astrophys. J. 915~(1) (2021) 28.
\newblock \href {http://arxiv.org/abs/2010.10506} {\path{arXiv:2010.10506}},
  \href {https://doi.org/10.3847/1538-4357/abf82e}
  {\path{doi:10.3847/1538-4357/abf82e}}.

\bibitem{Summa18}
A.~{Summa}, H.-T. {Janka}, T.~{Melson}, A.~{Marek}, {Rotation-supported
  Neutrino-driven Supernova Explosions in Three Dimensions and the Critical
  Luminosity Condition}, Astrophys. J. 852~(1) (2018) 28.
\newblock \href {http://arxiv.org/abs/1708.04154} {\path{arXiv:1708.04154}},
  \href {https://doi.org/10.3847/1538-4357/aa9ce8}
  {\path{doi:10.3847/1538-4357/aa9ce8}}.

\bibitem{Glas2019ApJ873}
R.~{Glas}, O.~{Just}, H.~T. {Janka}, M.~{Obergaulinger}, {Three-dimensional
  Core-collapse Supernova Simulations with Multidimensional Neutrino Transport
  Compared to the Ray-by-ray-plus Approximation}, Astrophys. J. 873~(1) (2019) 45.
\newblock \href {http://arxiv.org/abs/1809.10146} {\path{arXiv:1809.10146}},
  \href {https://doi.org/10.3847/1538-4357/ab0423}
  {\path{doi:10.3847/1538-4357/ab0423}}.

\bibitem{Marek06}
A.~{Marek}, H.~{Dimmelmeier}, H.-T. {Janka}, et~al., {Exploring the
  relativistic regime with Newtonian hydrodynamics: an improved effective
  gravitational potential for supernova simulations}, Astron. Astrophys. 445
  (2006) 273--289.
\newblock \href {http://arxiv.org/abs/astro-ph/0502161}
  {\path{arXiv:astro-ph/0502161}}, \href
  {https://doi.org/10.1051/0004-6361:20052840}
  {\path{doi:10.1051/0004-6361:20052840}}.

\bibitem{Mezzacappa93a}
A.~{Mezzacappa}, S.~W. {Bruenn}, {Type II supernovae and Boltzmann neutrino
  transport - The infall phase}, Astrophys. J. 405 (1993) 637--668.
\newblock \href {https://doi.org/10.1086/172394} {\path{doi:10.1086/172394}}.

\bibitem{Mezzacappa93b}
A.~{Mezzacappa}, S.~W. {Bruenn}, {A numerical method for solving the neutrino
  Boltzmann equation coupled to spherically symmetric stellar core collapse},
  Astrophys. J. 405 (1993) 669--684.
\newblock \href {https://doi.org/10.1086/172395} {\path{doi:10.1086/172395}}.

\bibitem{Mezzacappa93c}
A.~{Mezzacappa}, S.~W. {Bruenn}, {Stellar core collapse - A Boltzmann treatment
  of neutrino-electron scattering}, Astrophys. J. 410 (1993) 740--760.
\newblock \href {https://doi.org/10.1086/172791} {\path{doi:10.1086/172791}}.

\bibitem{Liebendorfer04}
M.~{Liebend{\"o}rfer}, O.~E.~B. {Messer}, A.~{Mezzacappa}, et~al., {A Finite
  Difference Representation of Neutrino Radiation Hydrodynamics in Spherically
  Symmetric General Relativistic Spacetime}, Astrophys. J., Suppl. 150 (2004)
  263--316.
\newblock \href {http://arxiv.org/abs/astro-ph/0207036}
  {\path{arXiv:astro-ph/0207036}}, \href {https://doi.org/10.1086/380191}
  {\path{doi:10.1086/380191}}.

\bibitem{O'Connor18_GlobalComparison}
E.~{O'Connor}, R.~{Bollig}, A.~{Burrows}, et~al., {Global comparison of
  core-collapse supernova simulations in spherical symmetry}, Journal of
  Physics G Nuclear Physics 45~(10) (2018) 104001.
\newblock \href {http://arxiv.org/abs/1806.04175} {\path{arXiv:1806.04175}},
  \href {https://doi.org/10.1088/1361-6471/aadeae}
  {\path{doi:10.1088/1361-6471/aadeae}}.

\bibitem{Nagakura14}
H.~{Nagakura}, K.~{Sumiyoshi}, S.~{Yamada}, {Three-dimensional Boltzmann Hydro
  Code for Core Collapse in Massive Stars. I. Special Relativistic Treatments},
  Astrophys. J., Suppl. 214 (2014) 16.
\newblock \href {http://arxiv.org/abs/1407.5632} {\path{arXiv:1407.5632}},
  \href {https://doi.org/10.1088/0067-0049/214/2/16}
  {\path{doi:10.1088/0067-0049/214/2/16}}.

\bibitem{Nagakura17}
H.~{Nagakura}, W.~{Iwakami}, S.~{Furusawa}, et~al., {Three-dimensional
  Boltzmann-Hydro Code for Core-collapse in Massive Stars. II. The
  Implementation of Moving-mesh for Neutron Star Kicks}, Astrophys. J., Suppl. 229~(2) (2017)
  42.
\newblock \href {http://arxiv.org/abs/1605.00666} {\path{arXiv:1605.00666}},
  \href {https://doi.org/10.3847/1538-4365/aa69ea}
  {\path{doi:10.3847/1538-4365/aa69ea}}.

\bibitem{Nagakura18}
H.~{Nagakura}, W.~{Iwakami}, S.~{Furusawa}, et~al., {Simulations of
  Core-collapse Supernovae in Spatial Axisymmetry with Full Boltzmann Neutrino
  Transport}, Astrophys. J. 854~(2) (2018) 136.
\newblock \href {http://arxiv.org/abs/1702.01752} {\path{arXiv:1702.01752}},
  \href {https://doi.org/10.3847/1538-4357/aaac29}
  {\path{doi:10.3847/1538-4357/aaac29}}.

\bibitem{Richers:2017}
S.~{Richers}, H.~{Nagakura}, C.~D. {Ott}, et~al., {A Detailed Comparison of
  Multidimensional Boltzmann Neutrino Transport Methods in Core-collapse
  Supernovae}, Astrophys. J. 847~(2) (2017) 133.
\newblock \href {http://arxiv.org/abs/1706.06187} {\path{arXiv:1706.06187}},
  \href {https://doi.org/10.3847/1538-4357/aa8bb2}
  {\path{doi:10.3847/1538-4357/aa8bb2}}.

\bibitem{Fischer10}
T.~{Fischer}, S.~C. {Whitehouse}, A.~{Mezzacappa}, et~al., {Protoneutron star
  evolution and the neutrino-driven wind in general relativistic neutrino
  radiation hydrodynamics simulations}, Astron. Astrophys. 517 (2010) A80.
\newblock \href {http://arxiv.org/abs/0908.1871} {\path{arXiv:0908.1871}},
  \href {https://doi.org/10.1051/0004-6361/200913106}
  {\path{doi:10.1051/0004-6361/200913106}}.

\bibitem{Nomoto06}
K.~{Nomoto}, N.~{Tominaga}, H.~{Umeda}, et~al., {Nucleosynthesis yields of
  core-collapse supernovae and hypernovae, and galactic chemical evolution},
  Nucl. Phys. A 777 (2006) 424--458.
\newblock \href {http://arxiv.org/abs/astro-ph/0605725}
  {\path{arXiv:astro-ph/0605725}}, \href
  {https://doi.org/10.1016/j.nuclphysa.2006.05.008}
  {\path{doi:10.1016/j.nuclphysa.2006.05.008}}.

\bibitem{bollig17}
R.~{Bollig}, H.-T. {Janka}, A.~{Lohs}, et~al., {Muon Creation in Supernova
  Matter Facilitates Neutrino-driven Explosions}, Phys. Rev. Lett. 119~(24)
  (2017) 242702.
\newblock \href {http://arxiv.org/abs/1706.04630} {\path{arXiv:1706.04630}},
  \href {https://doi.org/10.1103/PhysRevLett.119.242702}
  {\path{doi:10.1103/PhysRevLett.119.242702}}.

\bibitem{Sumiyoshi07}
K.~{Sumiyoshi}, S.~{Yamada}, H.~{Suzuki}, {Dynamics and Neutrino Signal of
  Black Hole Formation in Nonrotating Failed Supernovae. I. Equation of State
  Dependence}, Astrophys. J. 667 (2007) 382--394.
\newblock \href {http://arxiv.org/abs/0706.3762} {\path{arXiv:0706.3762}},
  \href {https://doi.org/10.1086/520876} {\path{doi:10.1086/520876}}.

\bibitem{Rahman22}
N.~{Rahman}, H.~T. {Janka}, G.~{Stockinger}, S.~E. {Woosley}, {Pulsational
  pair-instability supernovae: gravitational collapse, black hole formation,
  and beyond}, Mon. Not. R. Astro. Soc. 512~(3) (2022) 4503--4540.
\newblock \href {http://arxiv.org/abs/2112.09707} {\path{arXiv:2112.09707}},
  \href {https://doi.org/10.1093/mnras/stac758}
  {\path{doi:10.1093/mnras/stac758}}.

\bibitem{Bisnovatyi-Kogan70}
G.~S. {Bisnovatyi-Kogan}, {The Explosion of a Rotating Star As a Supernova
  Mechanism.}, Soviet Astronomy 14 (1971) 652.

\bibitem{LeBlancWilson70}
J.~M. {LeBlanc}, J.~R. {Wilson}, {A Numerical Example of the Collapse of a
  Rotating Magnetized Star}, Astrophys. J. 161 (1970) 541.
\newblock \href {https://doi.org/10.1086/150558} {\path{doi:10.1086/150558}}.

\bibitem{Walder12}
R.~{Walder}, D.~{Folini}, G.~{Meynet}, {Magnetic Fields in Massive Stars, Their
  Winds, and Their Nebulae}, Space Science Reviews 166~(1-4) (2012) 145--185.
\newblock \href {http://arxiv.org/abs/1103.3777} {\path{arXiv:1103.3777}},
  \href {https://doi.org/10.1007/s11214-011-9771-2}
  {\path{doi:10.1007/s11214-011-9771-2}}.

\bibitem{Wada2015}
G.~A. {Wade}, {MiMeS Collaboration}, {Review: Magnetic Fields of O-Type Stars},
  in: Y.~Y. {Balega}, I.~I. {Romanyuk}, D.~O. {Kudryavtsev} (Eds.), Physics and
  Evolution of Magnetic and Related Stars, Vol. 494 of Astronomical Society of
  the Pacific Conference Series, 2015, p.~30.
\newblock \href {http://arxiv.org/abs/1411.3604} {\path{arXiv:1411.3604}},
  \href {https://doi.org/10.48550/arXiv.1411.3604}
  {\path{doi:10.48550/arXiv.1411.3604}}.

\bibitem{Grunhut17}
J.~H. {Grunhut}, G.~A. {Wade}, C.~{Neiner}, et~al., {The MiMeS survey of
  Magnetism in Massive Stars: magnetic analysis of the O-type stars}, Mon. Not.
  R. Astro. Soc. 465~(2) (2017) 2432--2470.
\newblock \href {http://arxiv.org/abs/1610.07895} {\path{arXiv:1610.07895}},
  \href {https://doi.org/10.1093/mnras/stw2743}
  {\path{doi:10.1093/mnras/stw2743}}.

\bibitem{Winteler12}
C.~{Winteler}, R.~{K{\"a}ppeli}, A.~{Perego}, et~al., {Magnetorotationally
  Driven Supernovae as the Origin of Early Galaxy r-process Elements?},
  Astrophys. J. Lett. 750~(1) (2012) L22.
\newblock \href {http://arxiv.org/abs/1203.0616} {\path{arXiv:1203.0616}},
  \href {https://doi.org/10.1088/2041-8205/750/1/L22}
  {\path{doi:10.1088/2041-8205/750/1/L22}}.

\bibitem{Sagert09}
I.~{Sagert}, T.~{Fischer}, M.~{Hempel}, et~al., {Signals of the QCD Phase
  Transition in Core-Collapse Supernovae}, Phys. Rev. Lett. 102~(8) (2009)
  081101.
\newblock \href {http://arxiv.org/abs/0809.4225} {\path{arXiv:0809.4225}},
  \href {https://doi.org/10.1103/PhysRevLett.102.081101}
  {\path{doi:10.1103/PhysRevLett.102.081101}}.

\bibitem{Fischer11}
T.~{Fischer}, I.~{Sagert}, G.~{Pagliara}, et~al., {Core-collapse Supernova
  Explosions Triggered by a Quark-Hadron Phase Transition During the Early
  Post-bounce Phase}, Astrophys. J., Suppl. 194~(2) (2011) 39.
\newblock \href {http://arxiv.org/abs/1011.3409} {\path{arXiv:1011.3409}},
  \href {https://doi.org/10.1088/0067-0049/194/2/39}
  {\path{doi:10.1088/0067-0049/194/2/39}}.

\bibitem{Fischer18}
T.~{Fischer}, N.-U.~F. {Bastian}, M.-R. {Wu}, et~al., {Quark deconfinement as a
  supernova explosion engine for massive blue supergiant stars}, Nature
  Astronomy 2 (2018) 980--986.
\newblock \href {http://arxiv.org/abs/1712.08788} {\path{arXiv:1712.08788}},
  \href {https://doi.org/10.1038/s41550-018-0583-0}
  {\path{doi:10.1038/s41550-018-0583-0}}.

\bibitem{Zha20}
S.~{Zha}, E.~P. {O'Connor}, M.-c. {Chu}, et~al., {Gravitational-wave Signature
  of a First-order Quantum Chromodynamics Phase Transition in Core-Collapse
  Supernovae}, Phys. Rev. Lett. 125~(5) (2020) 051102.
\newblock \href {http://arxiv.org/abs/2007.04716} {\path{arXiv:2007.04716}},
  \href {https://doi.org/10.1103/PhysRevLett.125.051102}
  {\path{doi:10.1103/PhysRevLett.125.051102}}.

\bibitem{KurodaT21}
T.~{Kuroda}, {Impact of a Magnetic Field on Neutrino-Matter Interactions in
  Core-collapse Supernovae}, Astrophys. J. 906~(2) (2021) 128.
\newblock \href {http://arxiv.org/abs/2009.07733} {\path{arXiv:2009.07733}},
  \href {https://doi.org/10.3847/1538-4357/abce61}
  {\path{doi:10.3847/1538-4357/abce61}}.

\bibitem{Jakobus:2022ucs}
P.~Jakobus, B.~Mueller, A.~Heger, et~al., {The role of the hadron-quark phase
  transition in core-collapse supernovae}, Mon. Not. Roy. Astron. Soc. 516~(2)
  (2022) 2554--2574.
\newblock \href {http://arxiv.org/abs/2204.10397} {\path{arXiv:2204.10397}},
  \href {https://doi.org/10.1093/mnras/stac2352}
  {\path{doi:10.1093/mnras/stac2352}}.

\bibitem{KhosraviLargani2023arXiv230412316K}
N.~{Khosravi Largani}, T.~{Fischer}, N.~U.~F. {Bastian}, {Constraining the
  onset density for the QCD phase transition with the neutrino signal from
  core-collapse supernovae}, arXiv e-prints (2023) arXiv:2304.12316\href
  {http://arxiv.org/abs/2304.12316} {\path{arXiv:2304.12316}}, \href
  {https://doi.org/10.48550/arXiv.2304.12316}
  {\path{doi:10.48550/arXiv.2304.12316}}.

\bibitem{Woosley:1993wj}
S.~E. Woosley, {Gamma-ray bursts from stellar mass accretion disks around black
  holes}, Astrophys. J. 405 (1993) 273.
\newblock \href {https://doi.org/10.1086/172359} {\path{doi:10.1086/172359}}.

\bibitem{MacFadyen:1998vz}
A.~MacFadyen, S.~E. Woosley, {Collapsars: Gamma-ray bursts and explosions in
  'failed supernovae'}, Astrophys. J. 524 (1999) 262.
\newblock \href {http://arxiv.org/abs/astro-ph/9810274}
  {\path{arXiv:astro-ph/9810274}}, \href {https://doi.org/10.1086/307790}
  {\path{doi:10.1086/307790}}.

\bibitem{Woosley:2006fn}
S.~E. Woosley, J.~S. Bloom, {The Supernova Gamma-Ray Burst Connection}, Ann.
  Rev. Astron. Astrophys. 44 (2006) 507--556.
\newblock \href {http://arxiv.org/abs/astro-ph/0609142}
  {\path{arXiv:astro-ph/0609142}}, \href
  {https://doi.org/10.1146/annurev.astro.43.072103.150558}
  {\path{doi:10.1146/annurev.astro.43.072103.150558}}.

\bibitem{Kumar:2014upa}
P.~Kumar, B.~Zhang, {The physics of gamma-ray bursts $\textbackslash{\&}$
  relativistic jets}, Phys. Rept. 561 (2014) 1--109.
\newblock \href {http://arxiv.org/abs/1410.0679} {\path{arXiv:1410.0679}},
  \href {https://doi.org/10.1016/j.physrep.2014.09.008}
  {\path{doi:10.1016/j.physrep.2014.09.008}}.

\bibitem{Soker2010}
N.~{Soker}, {Applying the jet feedback mechanism to core-collapse supernova
  explosions}, MNRAS 401~(4) (2010) 2793--2798.
\newblock \href {http://arxiv.org/abs/0909.5276} {\path{arXiv:0909.5276}},
  \href {https://doi.org/10.1111/j.1365-2966.2009.15862.x}
  {\path{doi:10.1111/j.1365-2966.2009.15862.x}}.

\bibitem{Soker:2023mbr}
N.~Soker, {Supernovae in 2023 (review): breakthroughs by late observations} (11
  2023).
\newblock \href {http://arxiv.org/abs/2311.17732} {\path{arXiv:2311.17732}}.

\bibitem{Kushnir:2015mca}
D.~Kushnir, {Thermonuclear explosion of rotating massive stars could explain
  core-collapse supernovae} (2 2015).
\newblock \href {http://arxiv.org/abs/1502.03111} {\path{arXiv:1502.03111}}.

\bibitem{Bugli21}
M.~Bugli, J.~Guilet, M.~Obergaulinger, {Three-dimensional core-collapse
  supernovae with complex magnetic structures \textendash{} I. Explosion
  dynamics}, Mon. Not. Roy. Astron. Soc. 507~(1) (2021) 443--454.
\newblock \href {http://arxiv.org/abs/2105.00665} {\path{arXiv:2105.00665}},
  \href {https://doi.org/10.1093/mnras/stab2161}
  {\path{doi:10.1093/mnras/stab2161}}.

\bibitem{Nishimura.Sawai.ea:2017}
N.~Nishimura, H.~Sawai, T.~Takiwaki, et~al., {The intermediate r-process in
  core-collapse supernovae driven by the magneto-rotational instability},
  Astrophys. J. Lett. 836~(2) (2017) L21.
\newblock \href {http://arxiv.org/abs/1611.02280} {\path{arXiv:1611.02280}},
  \href {https://doi.org/10.3847/2041-8213/aa5dee}
  {\path{doi:10.3847/2041-8213/aa5dee}}.

\bibitem{Reichert23}
M.~{Reichert}, M.~{Obergaulinger}, M.~{\'A}. {Aloy}, et~al., {Magnetorotational
  supernovae: a nucleosynthetic analysis of sophisticated 3D models}, Mon. Not.
  R. Astro. Soc. 518~(1) (2023) 1557--1583.
\newblock \href {http://arxiv.org/abs/2206.11914} {\path{arXiv:2206.11914}},
  \href {https://doi.org/10.1093/mnras/stac3185}
  {\path{doi:10.1093/mnras/stac3185}}.

\bibitem{fischer20b}
T.~{Fischer}, M.-R. {Wu}, B.~{Wehmeyer}, et~al., {Core-collapse Supernova
  Explosions Driven by the Hadron-quark Phase Transition as a Rare r-process
  Site}, Astrophys. J. 894~(1) (2020) 9.
\newblock \href {https://doi.org/10.3847/1538-4357/ab86b0}
  {\path{doi:10.3847/1538-4357/ab86b0}}.

\bibitem{Siegel:2017nub}
D.~M. Siegel, B.~D. Metzger, {Three-Dimensional General-Relativistic
  Magnetohydrodynamic Simulations of Remnant Accretion Disks from Neutron Star
  Mergers: Outflows and $r$-Process Nucleosynthesis}, Phys. Rev. Lett. 119~(23)
  (2017) 231102.
\newblock \href {http://arxiv.org/abs/1705.05473} {\path{arXiv:1705.05473}},
  \href {https://doi.org/10.1103/PhysRevLett.119.231102}
  {\path{doi:10.1103/PhysRevLett.119.231102}}.

\bibitem{Obergaulinger.Reichert:2023}
M.~Obergaulinger, M.~Reichert, {Nucleosynthesis in Jet-Driven and
  Jet-Associated Supernovae} (2023).
\newblock \href {http://arxiv.org/abs/2303.12458} {\path{arXiv:2303.12458}}.

\bibitem{hirata87}
K.~{Hirata}, T.~{Kajita}, M.~{Koshiba}, et~al., {Observation of a neutrino
  burst from the supernova SN1987A}, Physical Review Letters 58 (1987)
  1490--1493.
\newblock \href {https://doi.org/10.1103/PhysRevLett.58.1490}
  {\path{doi:10.1103/PhysRevLett.58.1490}}.

\bibitem{Duncan86}
R.~C. {Duncan}, S.~L. {Shapiro}, I.~{Wasserman}, {Neutrino-driven Winds from
  Young, Hot Neutron Stars}, Astrophys. J. 309 (1986) 141.
\newblock \href {https://doi.org/10.1086/164587} {\path{doi:10.1086/164587}}.

\bibitem{Woosley94}
S.~E. Woosley, J.~R. Wilson, G.~J. Mathews, et~al., The r-process and
  neutrino-heated supernova ejecta, Astrophys. J. 433 (1994) 229.
\newblock \href {https://doi.org/10.1086/174638} {\path{doi:10.1086/174638}}.

\bibitem{Witti94}
J.~{Witti}, H.~T. {Janka}, K.~{Takahashi}, {Nucleosynthesis in neutrino-driven
  winds from protoneutron stars I. The {\ensuremath{\alpha}}-process}, Astron.
  Astrophys. 286 (1994) 841--856.

\bibitem{Qian96}
Y.~Z. {Qian}, S.~E. {Woosley}, {Nucleosynthesis in Neutrino-driven Winds. I.
  The Physical Conditions}, Astrophys. J. 471 (1996) 331.
\newblock \href {http://arxiv.org/abs/astro-ph/9611094}
  {\path{arXiv:astro-ph/9611094}}, \href {https://doi.org/10.1086/177973}
  {\path{doi:10.1086/177973}}.

\bibitem{Thompson01b}
T.~A. {Thompson}, A.~{Burrows}, B.~S. {Meyer}, {The Physics of Proto-Neutron
  Star Winds: Implications for r-Process Nucleosynthesis}, Astrophys. J.
  562~(2) (2001) 887--908.
\newblock \href {http://arxiv.org/abs/astro-ph/0105004}
  {\path{arXiv:astro-ph/0105004}}, \href {https://doi.org/10.1086/323861}
  {\path{doi:10.1086/323861}}.

\bibitem{Janka95}
H.~T. {Janka}, E.~{Mueller}, {The First Second of a Type II Supernova:
  Convection, Accretion, and Shock Propagation}, Astrophys. J. Lett. 448 (1995)
  L109.
\newblock \href {http://arxiv.org/abs/astro-ph/9503015}
  {\path{arXiv:astro-ph/9503015}}, \href {https://doi.org/10.1086/309604}
  {\path{doi:10.1086/309604}}.

\bibitem{Burrows95}
A.~{Burrows}, J.~{Hayes}, B.~A. {Fryxell}, {On the Nature of Core-Collapse
  Supernova Explosions}, Astrophys. J. 450 (1995) 830.
\newblock \href {http://arxiv.org/abs/astro-ph/9506061}
  {\path{arXiv:astro-ph/9506061}}, \href {https://doi.org/10.1086/176188}
  {\path{doi:10.1086/176188}}.

\bibitem{Arcones07}
A.~{Arcones}, H.~T. {Janka}, L.~{Scheck}, {Nucleosynthesis-relevant conditions
  in neutrino-driven supernova outflows. I. Spherically symmetric hydrodynamic
  simulations}, Astron. Astrophys. 467~(3) (2007) 1227--1248.
\newblock \href {http://arxiv.org/abs/astro-ph/0612582}
  {\path{arXiv:astro-ph/0612582}}, \href
  {https://doi.org/10.1051/0004-6361:20066983}
  {\path{doi:10.1051/0004-6361:20066983}}.

\bibitem{Huedepohl10}
L.~{H{\"u}depohl}, B.~{M{\"u}ller}, H.~T. {Janka}, et~al., {Neutrino Signal of
  Electron-Capture Supernovae from Core Collapse to Cooling}, Phys. Rev. Lett.
  104~(25) (2010) 251101.
\newblock \href {http://arxiv.org/abs/0912.0260} {\path{arXiv:0912.0260}},
  \href {https://doi.org/10.1103/PhysRevLett.104.251101}
  {\path{doi:10.1103/PhysRevLett.104.251101}}.

\bibitem{Stockinger:2020hse}
G.~Stockinger, et~al., {Three-dimensional Models of Core-collapse Supernovae
  From Low-mass Progenitors With Implications for Crab}, Mon. Not. Roy. Astron.
  Soc. 496~(2) (2020) 2039--2084.
\newblock \href {http://arxiv.org/abs/2005.02420} {\path{arXiv:2005.02420}},
  \href {https://doi.org/10.1093/mnras/staa1691}
  {\path{doi:10.1093/mnras/staa1691}}.

\bibitem{Witt21}
M.~{Witt}, A.~{Psaltis}, H.~{Yasin}, et~al., {Post-explosion Evolution of
  Core-collapse Supernovae}, Astrophys. J. 921~(1) (2021) 19.
\newblock \href {http://arxiv.org/abs/2107.00687} {\path{arXiv:2107.00687}},
  \href {https://doi.org/10.3847/1538-4357/ac1a6d}
  {\path{doi:10.3847/1538-4357/ac1a6d}}.

\bibitem{Keil03}
M.~T. {Keil}, G.~G. {Raffelt}, H.-T. {Janka}, {Monte Carlo Study of Supernova
  Neutrino Spectra Formation}, Astrophys. J. 590~(2) (2003) 971--991.
\newblock \href {http://arxiv.org/abs/astro-ph/0208035}
  {\path{arXiv:astro-ph/0208035}}, \href {https://doi.org/10.1086/375130}
  {\path{doi:10.1086/375130}}.

\bibitem{Nagakura:2021MNRAS}
H.~{Nagakura}, A.~{Burrows}, D.~{Vartanyan}, {Supernova neutrino signals based
  on long-term axisymmetric simulations}, Mon. Not. R. Astro. Soc. 506~(1)
  (2021) 1462--1479.
\newblock \href {http://arxiv.org/abs/2102.11283} {\path{arXiv:2102.11283}},
  \href {https://doi.org/10.1093/mnras/stab1785}
  {\path{doi:10.1093/mnras/stab1785}}.

\bibitem{Janka08}
H.~T. {Janka}, B.~{M{\"u}ller}, F.~S. {Kitaura}, R.~{Buras}, {Dynamics of shock
  propagation and nucleosynthesis conditions in O-Ne-Mg core supernovae},
  Astron. Astrophys. 485~(1) (2008) 199--208.
\newblock \href {http://arxiv.org/abs/0712.4237} {\path{arXiv:0712.4237}},
  \href {https://doi.org/10.1051/0004-6361:20079334}
  {\path{doi:10.1051/0004-6361:20079334}}.

\bibitem{Sandoval:2021hnk}
M.~A. {Sandoval}, W.~R. {Hix}, O.~E.~B. {Messer}, et~al., {Three-dimensional
  Core-collapse Supernova Simulations with 160 Isotopic Species Evolved to
  Shock Breakout}, Astrophys. J. 921~(2) (2021) 113.
\newblock \href {http://arxiv.org/abs/2106.01389} {\path{arXiv:2106.01389}},
  \href {https://doi.org/10.3847/1538-4357/ac1d49}
  {\path{doi:10.3847/1538-4357/ac1d49}}.

\bibitem{Nagakura:2021MNRAS500}
H.~{Nagakura}, A.~{Burrows}, D.~{Vartanyan}, D.~{Radice}, {Core-collapse
  supernova neutrino emission and detection informed by state-of-the-art
  three-dimensional numerical models}, MNRAS 500~(1) (2021) 696--717.
\newblock \href {http://arxiv.org/abs/2007.05000} {\path{arXiv:2007.05000}},
  \href {https://doi.org/10.1093/mnras/staa2691}
  {\path{doi:10.1093/mnras/staa2691}}.

\bibitem{Wongwathanarat:2014yda}
A.~Wongwathanarat, E.~M\"uller, H.~T. Janka, {Three-Dimensional Simulations of
  Core-Collapse Supernovae: From Shock Revival to Shock Breakout}, Astron.
  Astrophys. 577 (2015) A48.
\newblock \href {http://arxiv.org/abs/1409.5431} {\path{arXiv:1409.5431}},
  \href {https://doi.org/10.1051/0004-6361/201425025}
  {\path{doi:10.1051/0004-6361/201425025}}.

\bibitem{Orlando:2016jxx}
S.~Orlando, M.~Miceli, M.~L. Pumo, F.~Bocchino, {Modeling SNR Cassiopeia A from
  the Supernova Explosion to its Current Age: The role of post-explosion
  anisotropies of ejecta}, Astrophys. J. 822~(1) (2016) 22.
\newblock \href {http://arxiv.org/abs/1603.03690} {\path{arXiv:1603.03690}},
  \href {https://doi.org/10.3847/0004-637X/822/1/22}
  {\path{doi:10.3847/0004-637X/822/1/22}}.

\bibitem{Wongwathanarat:2016jvy}
A.~Wongwathanarat, H.-T. Janka, E.~M\"uller, et~al., {Production and
  Distribution of $^{44}$Ti and $^{56}$Ni in a Three-dimensional Supernova
  Model Resembling Cassiopeia A}, Astrophys. J. 842~(1) (2017) 13.
\newblock \href {http://arxiv.org/abs/1610.05643} {\path{arXiv:1610.05643}},
  \href {https://doi.org/10.3847/1538-4357/aa72de}
  {\path{doi:10.3847/1538-4357/aa72de}}.

\bibitem{Utrobin:2018mjr}
V.~P. Utrobin, A.~Wongwathanarat, H.~T. Janka, et~al., {Three-dimensional
  mixing and light curves: constraints on the progenitor of supernova 1987A},
  Astron. Astrophys. 624 (2019) A116.
\newblock \href {http://arxiv.org/abs/1812.11083} {\path{arXiv:1812.11083}},
  \href {https://doi.org/10.1051/0004-6361/201834976}
  {\path{doi:10.1051/0004-6361/201834976}}.

\bibitem{Muller:2018gok}
B.~M\"uller, D.~Gay, A.~Heger, et~al., {Multidimensional simulations of
  ultrastripped supernovae to shock breakout}, Mon. Not. Roy. Astron. Soc.
  479~(3) (2018) 3675--3689.
\newblock \href {http://arxiv.org/abs/1803.03388} {\path{arXiv:1803.03388}},
  \href {https://doi.org/10.1093/mnras/sty1683}
  {\path{doi:10.1093/mnras/sty1683}}.

\bibitem{Orlando:2019vdf}
S.~Orlando, et~al., {Hydrodynamic simulations unravel the
  progenitor-supernova-remnant connection in SN 1987A}, Astron. Astrophys. 636
  (2020) A22.
\newblock \href {http://arxiv.org/abs/1912.03070} {\path{arXiv:1912.03070}},
  \href {https://doi.org/10.1051/0004-6361/201936718}
  {\path{doi:10.1051/0004-6361/201936718}}.

\bibitem{Ono:2019zhr}
M.~Ono, S.~Nagataki, G.~Ferrand, et~al., {Matter Mixing in Aspherical
  Core-collapse Supernovae: Three-dimensional Simulations with Single Star and
  Binary Merger Progenitor Models for SN 1987A} (12 2019).
\newblock \href {http://arxiv.org/abs/1912.02234} {\path{arXiv:1912.02234}},
  \href {https://doi.org/10.3847/1538-4357/ab5dba}
  {\path{doi:10.3847/1538-4357/ab5dba}}.

\bibitem{Jerkstrand:2020hlf}
A.~Jerkstrand, et~al., {Properties of gamma-ray decay lines in 3D core-collapse
  supernova models, with application to SN 1987A and Cas A}, Mon. Not. Roy.
  Astron. Soc. 494~(2) (2020) 2471--2497.
\newblock \href {http://arxiv.org/abs/2003.05156} {\path{arXiv:2003.05156}},
  \href {https://doi.org/10.1093/mnras/staa736}
  {\path{doi:10.1093/mnras/staa736}}.

\bibitem{Orlando:2020igr}
S.~Orlando, A.~Wongwathanarat, H.~T. Janka, et~al., {The fully developed
  remnant of a neutrino-driven supernova: Evolution of ejecta structure and
  asymmetries in SNR Cassiopeia A}, Astron. Astrophys. 645 (2021) A66.
\newblock \href {http://arxiv.org/abs/2009.01789} {\path{arXiv:2009.01789}},
  \href {https://doi.org/10.1051/0004-6361/202039335}
  {\path{doi:10.1051/0004-6361/202039335}}.

\bibitem{Itoh96}
N.~{Itoh}, A.~{Nishikawa}, Y.~{Kohyama}, {Neutrino Energy Loss in Stellar
  Interiors. VIII. Braaten-Segel Approximation for the Plasma Neutrino
  Process}, Astrophys. J. 470 (1996) 1015.
\newblock \href {https://doi.org/10.1086/177926} {\path{doi:10.1086/177926}}.

\bibitem{Guo16}
G.~{Guo}, Y.-Z. {Qian}, {Spectra and rates of bremsstrahlung neutrino emission
  in stars}, Phys. Rev. D 94~(4) (2016) 043005.
\newblock \href {http://arxiv.org/abs/1608.02852} {\path{arXiv:1608.02852}},
  \href {https://doi.org/10.1103/PhysRevD.94.043005}
  {\path{doi:10.1103/PhysRevD.94.043005}}.

\bibitem{SK:2022}
L.~N. {Machado}, K.~{Abe}, Y.~{Hayato}, et~al., {Pre-supernova Alert System for
  Super-Kamiokande}, Astrophys. J. 935~(1) (2022) 40.
\newblock \href {http://arxiv.org/abs/2205.09881} {\path{arXiv:2205.09881}},
  \href {https://doi.org/10.3847/1538-4357/ac7f9c}
  {\path{doi:10.3847/1538-4357/ac7f9c}}.

\bibitem{juno}
F.~{An}, et~al., {Neutrino physics with JUNO}, Journal of Physics G Nuclear
  Physics 43~(3) (2016) 030401.
\newblock \href {http://arxiv.org/abs/1507.05613} {\path{arXiv:1507.05613}},
  \href {https://doi.org/10.1088/0954-3899/43/3/030401}
  {\path{doi:10.1088/0954-3899/43/3/030401}}.

\bibitem{Kato20}
C.~{Kato}, K.~{Ishidoshiro}, T.~{Yoshida}, {Theoretical Prediction of
  Presupernova Neutrinos and Their Detection}, Annual Review of Nuclear and
  Particle Science 70 (2020) 121--145.
\newblock \href {http://arxiv.org/abs/2006.02519} {\path{arXiv:2006.02519}},
  \href {https://doi.org/10.1146/annurev-nucl-040620-021320}
  {\path{doi:10.1146/annurev-nucl-040620-021320}}.

\bibitem{Odrzywolek:2004a}
A.~{Odrzywolek}, M.~{Misiaszek}, M.~{Kutschera}, {Neutrinos from Pre-Supernova
  Star}, Acta Physica Polonica B 35~(6) (2004) 1981.
\newblock \href {http://arxiv.org/abs/astro-ph/0405006}
  {\path{arXiv:astro-ph/0405006}}, \href
  {https://doi.org/10.48550/arXiv.astro-ph/0405006}
  {\path{doi:10.48550/arXiv.astro-ph/0405006}}.

\bibitem{Odrzywolek:2004b}
A.~Odrzywolek, M.~Misiaszek, M.~Kutschera, {Detection possibility of the pair -
  annihilation neutrinos from the neutrino - cooled pre-supernova star},
  Astropart. Phys. 21 (2004) 303--313.
\newblock \href {http://arxiv.org/abs/astro-ph/0311012}
  {\path{arXiv:astro-ph/0311012}}, \href
  {https://doi.org/10.1016/j.astropartphys.2004.02.002}
  {\path{doi:10.1016/j.astropartphys.2004.02.002}}.

\bibitem{Kutschera:2009}
M.~{Kutschera}, A.~{Odrzywo{\l}ek}, M.~{Misiaszek}, {Presupernovae as Powerful
  Neutrino Sources}, Acta Physica Polonica B 40~(11) (2009) 3063.

\bibitem{Odrzywolek:2010}
A.~{Odrzywo{\l}ek}, A.~{Heger}, {NEUTRINO SIGNATURES OF DYING MASSIVE STARS:
  FROM MAIN SEQUENCE TO THE NEUTRON STAR}, Acta Physica Polonica B 41~(7)
  (2010) 1611.

\bibitem{kato15}
C.~{Kato}, M.~{Delfan Azari}, S.~{Yamada}, et~al., {Pre-supernova Neutrino
  Emissions from ONe Cores in the Progenitors of Core-collapse Supernovae: Are
  They Distinguishable from Those of Fe Cores?}, Astrophys. J. 808~(2) (2015)
  168.
\newblock \href {http://arxiv.org/abs/1506.02358} {\path{arXiv:1506.02358}},
  \href {https://doi.org/10.1088/0004-637X/808/2/168}
  {\path{doi:10.1088/0004-637X/808/2/168}}.

\bibitem{asakura16}
K.~{Asakura}, A.~{Gando}, Y.~{Gando}, et~al., {KamLAND Sensitivity to Neutrinos
  from Pre-supernova Stars}, Astrophys. J. 818~(1) (2016) 91.
\newblock \href {http://arxiv.org/abs/1506.01175} {\path{arXiv:1506.01175}},
  \href {https://doi.org/10.3847/0004-637X/818/1/91}
  {\path{doi:10.3847/0004-637X/818/1/91}}.

\bibitem{kato17}
C.~{Kato}, H.~{Nagakura}, S.~{Furusawa}, et~al., {Neutrino Emissions in All
  Flavors up to the Pre-bounce of Massive Stars and the Possibility of Their
  Detections}, Astrophys. J. 848~(1) (2017) 48.
\newblock \href {http://arxiv.org/abs/1704.05480} {\path{arXiv:1704.05480}},
  \href {https://doi.org/10.3847/1538-4357/aa8b72}
  {\path{doi:10.3847/1538-4357/aa8b72}}.

\bibitem{Patton:2017}
K.~M. Patton, C.~Lunardini, R.~J. Farmer, {Presupernova neutrinos: realistic
  emissivities from stellar evolution}, Astrophys. J. 840~(1) (2017) 2.
\newblock \href {http://arxiv.org/abs/1511.02820} {\path{arXiv:1511.02820}},
  \href {https://doi.org/10.3847/1538-4357/aa6ba8}
  {\path{doi:10.3847/1538-4357/aa6ba8}}.

\bibitem{distance}
G.~M. {Harper}, A.~{Brown}, E.~F. {Guinan}, et~al., {An Updated 2017
  Astrometric Solution for Betelgeuse}, Astrophys. J. 154~(1) (2017) 11.
\newblock \href {http://arxiv.org/abs/1706.06020} {\path{arXiv:1706.06020}},
  \href {https://doi.org/10.3847/1538-3881/aa6ff9}
  {\path{doi:10.3847/1538-3881/aa6ff9}}.

\bibitem{mass16}
M.~M. {Dolan}, G.~J. {Mathews}, D.~D. {Lam}, et~al., {Evolutionary Tracks for
  Betelgeuse}, Astrophys. J. 819~(1) (2016) 7.
\newblock \href {http://arxiv.org/abs/1406.3143} {\path{arXiv:1406.3143}},
  \href {https://doi.org/10.3847/0004-637X/819/1/7}
  {\path{doi:10.3847/0004-637X/819/1/7}}.

\bibitem{Joyce2020}
M.~{Joyce}, S.-C. {Leung}, L.~{Moln{\'a}r}, et~al., {Standing on the Shoulders
  of Giants: New Mass and Distance Estimates for Betelgeuse through Combined
  Evolutionary, Asteroseismic, and Hydrodynamic Simulations with MESA},
  Astrophys. J. 902~(1) (2020) 63.
\newblock \href {http://arxiv.org/abs/2006.09837} {\path{arXiv:2006.09837}},
  \href {https://doi.org/10.3847/1538-4357/abb8db}
  {\path{doi:10.3847/1538-4357/abb8db}}.

\bibitem{Wheeler2023}
J.~C. {Wheeler}, E.~{Chatzopoulos}, {Betelgeuse: a review}, Astronomy and
  Geophysics 64~(3) (2023) 3.11--3.27.
\newblock \href {http://arxiv.org/abs/2306.09449} {\path{arXiv:2306.09449}},
  \href {https://doi.org/10.1093/astrogeo/atad020}
  {\path{doi:10.1093/astrogeo/atad020}}.

\bibitem{Saio2023}
H.~{Saio}, D.~{Nandal}, G.~{Meynet}, S.~{Ekst{\"o}m}, {The evolutionary stage
  of Betelgeuse inferred from its pulsation periods} (2023).
\newblock \href {http://arxiv.org/abs/2306.00287} {\path{arXiv:2306.00287}}.

\bibitem{Guo19}
G.~{Guo}, Y.-Z. {Qian}, A.~{Heger}, {Presupernova neutrino signals as potential
  probes of neutrino mass hierarchy}, Physics Letters B 796 (2019) 126--130.
\newblock \href {http://arxiv.org/abs/1906.06839} {\path{arXiv:1906.06839}},
  \href {https://doi.org/10.1016/j.physletb.2019.07.030}
  {\path{doi:10.1016/j.physletb.2019.07.030}}.

\bibitem{MSW78}
L.~{Wolfenstein}, {Neutrino oscillations in matter}, Phys. Rev. D 17 (1978)
  2369--2374.
\newblock \href {https://doi.org/10.1103/PhysRevD.17.2369}
  {\path{doi:10.1103/PhysRevD.17.2369}}.

\bibitem{MSW85}
S.~P. {Mikheyev}, A.~Y. {Smirnov}, {Resonance enhancement of oscillations in
  matter and solar neutrino spectroscopy}, Yadernaya Fizika 42 (1985)
  1441--1448.

\bibitem{PhysRevC.71.055805}
C.~Galbiati, A.~Pocar, D.~Franco, et~al., Cosmogenic $^{11}\mathrm{C}$
  production and sensitivity of organic scintillator detectors to
  $\mathit{pep}$ and cno neutrinos, Phys. Rev. C 71 (2005) 055805.
\newblock \href {https://doi.org/10.1103/PhysRevC.71.055805}
  {\path{doi:10.1103/PhysRevC.71.055805}}.

\bibitem{PhysRevLett.58.1490}
K.~Hirata, et~al., Observation of a neutrino burst from the supernova sn1987a,
  Phys. Rev. Lett. 58 (1987) 1490--1493.
\newblock \href {https://doi.org/10.1103/PhysRevLett.58.1490}
  {\path{doi:10.1103/PhysRevLett.58.1490}}.

\bibitem{PhysRevD.38.448}
K.~S. Hirata, et~al., Observation in the kamiokande-ii detector of the neutrino
  burst from supernova sn1987a, Phys. Rev. D 38 (1988) 448--458.
\newblock \href {https://doi.org/10.1103/PhysRevD.38.448}
  {\path{doi:10.1103/PhysRevD.38.448}}.

\bibitem{PhysRevLett.58.1494}
R.~M. Bionta, , et~al., Observation of a neutrino burst in coincidence with
  supernova 1987a in the large magellanic cloud, Phys. Rev. Lett. 58 (1987)
  1494--1496.
\newblock \href {https://doi.org/10.1103/PhysRevLett.58.1494}
  {\path{doi:10.1103/PhysRevLett.58.1494}}.

\bibitem{ALEXEYEV1988209}
E.~Alexeyev, et~al., Detection of the neutrino signal from sn 1987a in the lmc
  using the inr baksan underground scintillation telescope, Phys. Lett. B
  205~(2) (1988) 209 -- 214.
\newblock \href {https://doi.org/10.1016/0370-2693(88)91651-6}
  {\path{doi:10.1016/0370-2693(88)91651-6}}.

\bibitem{arnett1989}
W.~D. Arnett, J.~N. Bahcall, R.~P. Kirshner, S.~E. Woosley, Supernova 1987a,
  Annu. Rev. Astron. Astrophys. 27~(1) (1989) 629--700.
\newblock \href {https://doi.org/10.1146/annurev.aa.27.090189.003213}
  {\path{doi:10.1146/annurev.aa.27.090189.003213}}.

\bibitem{Loredo:2001rx}
T.~J. Loredo, D.~Q. Lamb, {Bayesian analysis of neutrinos observed from
  supernova SN-1987A}, Phys. Rev. D 65 (2002) 063002.
\newblock \href {http://arxiv.org/abs/astro-ph/0107260}
  {\path{arXiv:astro-ph/0107260}}, \href
  {https://doi.org/10.1103/PhysRevD.65.063002}
  {\path{doi:10.1103/PhysRevD.65.063002}}.

\bibitem{Costantini_2007}
M.~L. Costantini, A.~Ianni, G.~Pagliaroli, F.~Vissani, Is there a problem with
  low energy {SN}1987a neutrinos?, J. Cosmo. Astropart. Phys. 2007~(05) (2007)
  014--014.
\newblock \href {http://arxiv.org/abs/astro-ph/0608399}
  {\path{arXiv:astro-ph/0608399}}, \href
  {https://doi.org/10.1088/1475-7516/2007/05/014}
  {\path{doi:10.1088/1475-7516/2007/05/014}}.

\bibitem{Olsen21}
J.~{Olsen}, Y.-Z. {Qian}, {Comparison of simulated neutrino emission models
  with data on Supernova 1987A}, Phys. Rev. D 104~(12) (2021) 123020.
\newblock \href {http://arxiv.org/abs/2108.08463} {\path{arXiv:2108.08463}},
  \href {https://doi.org/10.1103/PhysRevD.104.123020}
  {\path{doi:10.1103/PhysRevD.104.123020}}.

\bibitem{Olsen22a}
J.~{Olsen}, Y.-Z. {Qian}, {Erratum: Comparison of simulated neutrino emission
  models with data on Supernova 1987A [Phys. Rev. D 104, 123020 (2021)]}, Phys.
  Rev. D 106~(10) (2022) 109904.
\newblock \href {https://doi.org/10.1103/PhysRevD.106.109904}
  {\path{doi:10.1103/PhysRevD.106.109904}}.

\bibitem{Garching}
https://wwwmpa.mpa-garching.mpg.de/ccsnarchive/.

\bibitem{librc2023}
S.~{Weishi Li}, J.~F. {Beacom}, L.~F. {Roberts}, F.~{Capozzi}, {Old Data, New
  Forensics: The First Second of SN 1987A Neutrino Emission} (2023).
\newblock \href {http://arxiv.org/abs/2306.08024} {\path{arXiv:2306.08024}}.

\bibitem{fiorillo2023supernova}
D.~F.~G. Fiorillo, M.~Heinlein, H.-T. Janka, et~al., Supernova simulations
  confront sn 1987a neutrinos (2023).
\newblock \href {http://arxiv.org/abs/2308.01403} {\path{arXiv:2308.01403}},
  \href {https://doi.org/10.1103/PhysRevD.108.083040}
  {\path{doi:10.1103/PhysRevD.108.083040}}.

\bibitem{Scholberg}
K.~Scholberg, Supernova neutrino detection, Annu. Rev. Nucl. Part. Sci. 62~(1)
  (2012) 81--103.
\newblock \href {http://arxiv.org/abs/1205.6003} {\path{arXiv:1205.6003}},
  \href {https://doi.org/10.1146/annurev-nucl-102711-095006}
  {\path{doi:10.1146/annurev-nucl-102711-095006}}.

\bibitem{Hyper-Kamiokande:2021frf}
K.~Abe, et~al., {Supernova Model Discrimination with Hyper-Kamiokande},
  Astrophys. J. 916~(1) (2021) 15.
\newblock \href {http://arxiv.org/abs/2101.05269} {\path{arXiv:2101.05269}},
  \href {https://doi.org/10.3847/1538-4357/abf7c4}
  {\path{doi:10.3847/1538-4357/abf7c4}}.

\bibitem{totani98}
T.~{Totani}, K.~{Sato}, H.~E. {Dalhed}, J.~R. {Wilson}, {Future Detection of
  Supernova Neutrino Burst and Explosion Mechanism}, Astrophys. J. 496~(1)
  (1998) 216--225.
\newblock \href {http://arxiv.org/abs/astro-ph/9710203}
  {\path{arXiv:astro-ph/9710203}}, \href {https://doi.org/10.1086/305364}
  {\path{doi:10.1086/305364}}.

\bibitem{nakazato13b}
K.~{Nakazato}, K.~{Sumiyoshi}, H.~{Suzuki}, et~al., {Supernova Neutrino Light
  Curves and Spectra for Various Progenitor Stars: From Core Collapse to
  Proto-neutron Star Cooling}, Astrophys. J., Suppl. 205~(1) (2013) 2.
\newblock \href {http://arxiv.org/abs/1210.6841} {\path{arXiv:1210.6841}},
  \href {https://doi.org/10.1088/0067-0049/205/1/2}
  {\path{doi:10.1088/0067-0049/205/1/2}}.

\bibitem{couch20}
S.~M. {Couch}, M.~L. {Warren}, E.~P. {O'Connor}, {Simulating Turbulence-aided
  Neutrino-driven Core-collapse Supernova Explosions in One Dimension},
  Astrophys. J. 890~(2) (2020) 127.
\newblock \href {http://arxiv.org/abs/1902.01340} {\path{arXiv:1902.01340}},
  \href {https://doi.org/10.3847/1538-4357/ab609e}
  {\path{doi:10.3847/1538-4357/ab609e}}.

\bibitem{vartanyan19}
D.~{Vartanyan}, A.~{Burrows}, D.~{Radice}, {Temporal and angular variations of
  3D core-collapse supernova emissions and their physical correlations}, Mon.
  Not. R. Astro. Soc. 489~(2) (2019) 2227--2246.
\newblock \href {http://arxiv.org/abs/1906.08787} {\path{arXiv:1906.08787}},
  \href {https://doi.org/10.1093/mnras/stz2307}
  {\path{doi:10.1093/mnras/stz2307}}.

\bibitem{tamborra14}
I.~{Tamborra}, G.~{Raffelt}, F.~{Hanke}, et~al., {Neutrino emission
  characteristics and detection opportunities based on three-dimensional
  supernova simulations}, Phys. Rev. D 90~(4) (2014) 045032.
\newblock \href {http://arxiv.org/abs/1406.0006} {\path{arXiv:1406.0006}},
  \href {https://doi.org/10.1103/PhysRevD.90.045032}
  {\path{doi:10.1103/PhysRevD.90.045032}}.

\bibitem{Hyper-Kamiokande:2018ofw}
K.~Abe, et~al., {Hyper-Kamiokande Design Report} (2018).
\newblock \href {http://arxiv.org/abs/1805.04163} {\path{arXiv:1805.04163}}.

\bibitem{Migenda:2019xbm}
J.~Migenda, {Supernova Model Discrimination with Hyper-Kamiokande}, Ph.D.
  thesis, University of Sheffield (2019).

\bibitem{Olsen22b}
J.~{Olsen}, Y.-Z. {Qian}, {Prospects for distinguishing supernova models using
  a future neutrino signal}, Phys. Rev. D 105~(8) (2022) 083017.
\newblock \href {http://arxiv.org/abs/2202.09975} {\path{arXiv:2202.09975}},
  \href {https://doi.org/10.1103/PhysRevD.105.083017}
  {\path{doi:10.1103/PhysRevD.105.083017}}.

\bibitem{Lattimier1974}
J.~M. {Lattimer}, D.~N. {Schramm}, {Black-Hole-Neutron-Star Collisions},
  Astrophys. J. Lett. 192 (1974) L145.
\newblock \href {https://doi.org/10.1086/181612} {\path{doi:10.1086/181612}}.

\bibitem{Symbalisty1982}
E.~{Symbalisty}, D.~N. {Schramm}, {Neutron Star Collisions and the r-Process},
  Astrophysical Letters 22 (1982) 143.

\bibitem{Eichler1989}
D.~{Eichler}, M.~{Livio}, T.~{Piran}, D.~N. {Schramm}, {Nucleosynthesis,
  neutrino bursts and {\ensuremath{\gamma}}-rays from coalescing neutron
  stars}, Nature 340~(6229) (1989) 126--128.
\newblock \href {https://doi.org/10.1038/340126a0}
  {\path{doi:10.1038/340126a0}}.

\bibitem{Freiburghaus1999}
C.~{Freiburghaus}, S.~{Rosswog}, F.~K. {Thielemann}, {R-Process in Neutron Star
  Mergers}, Astrophys. J. Lett. 525~(2) (1999) L121--L124.
\newblock \href {https://doi.org/10.1086/312343} {\path{doi:10.1086/312343}}.

\bibitem{Rosswog:1998hy}
S.~Rosswog, M.~Liebendoerfer, F.~K. Thielemann, et~al., {Mass ejection in
  neutron star mergers}, Astron. Astrophys. 341 (1999) 499--526.
\newblock \href {http://arxiv.org/abs/astro-ph/9811367}
  {\path{arXiv:astro-ph/9811367}}.

\bibitem{Goriely:2011vg}
S.~Goriely, A.~Bauswein, H.~T. Janka, {R-Process Nucleosynthesis in Dynamically
  Ejected Matter of Neutron Star Mergers}, Astrophys. J. Lett. 738 (2011) L32.
\newblock \href {http://arxiv.org/abs/1107.0899} {\path{arXiv:1107.0899}},
  \href {https://doi.org/10.1088/2041-8205/738/2/L32}
  {\path{doi:10.1088/2041-8205/738/2/L32}}.

\bibitem{Fernandez:2013tya}
R.~Fern\'andez, B.~D. Metzger, {Delayed outflows from black hole accretion tori
  following neutron star binary coalescence}, Mon. Not. Roy. Astron. Soc. 435
  (2013) 502.
\newblock \href {http://arxiv.org/abs/1304.6720} {\path{arXiv:1304.6720}},
  \href {https://doi.org/10.1093/mnras/stt1312}
  {\path{doi:10.1093/mnras/stt1312}}.

\bibitem{Just:2014fka}
O.~Just, A.~Bauswein, R.~A. Pulpillo, et~al., {Comprehensive nucleosynthesis
  analysis for ejecta of compact binary mergers}, Mon. Not. Roy. Astron. Soc.
  448~(1) (2015) 541--567.
\newblock \href {http://arxiv.org/abs/1406.2687} {\path{arXiv:1406.2687}},
  \href {https://doi.org/10.1093/mnras/stv009}
  {\path{doi:10.1093/mnras/stv009}}.

\bibitem{Wanajo2014}
S.~{Wanajo}, Y.~{Sekiguchi}, N.~{Nishimura}, et~al., {Production of All the
  r-process Nuclides in the Dynamical Ejecta of Neutron Star Mergers},
  Astrophys. J. Lett. 789~(2) (2014) L39.
\newblock \href {http://arxiv.org/abs/1402.7317} {\path{arXiv:1402.7317}},
  \href {https://doi.org/10.1088/2041-8205/789/2/L39}
  {\path{doi:10.1088/2041-8205/789/2/L39}}.

\bibitem{Metzger10}
B.~D. Metzger, G.~Martinez-Pinedo, S.~Darbha, et~al., Electromagnetic
  counterparts of compact object mergers powered by the radioactive decay of
  r-process nuclei, Mon. Not. R. Astron. Soc. 406 (2010) 2650.
\newblock \href {http://arxiv.org/abs/1001.5029} {\path{arXiv:1001.5029}},
  \href {https://doi.org/10.1111/j.1365-2966.2010.16864.x}
  {\path{doi:10.1111/j.1365-2966.2010.16864.x}}.

\bibitem{Barnes:2013wka}
J.~Barnes, D.~Kasen, {Effect of a High Opacity on the Light Curves of
  Radioactively Powered Transients from Compact Object Mergers}, Astrophys. J.
  775 (2013) 18.
\newblock \href {http://arxiv.org/abs/1303.5787} {\path{arXiv:1303.5787}},
  \href {https://doi.org/10.1088/0004-637X/775/1/18}
  {\path{doi:10.1088/0004-637X/775/1/18}}.

\bibitem{Tanaka:2013ana}
M.~Tanaka, K.~Hotokezaka, {Radiative Transfer Simulations of Neutron Star
  Merger Ejecta}, Astrophys. J. 775 (2013) 113.
\newblock \href {http://arxiv.org/abs/1306.3742} {\path{arXiv:1306.3742}},
  \href {https://doi.org/10.1088/0004-637X/775/2/113}
  {\path{doi:10.1088/0004-637X/775/2/113}}.

\bibitem{Tanvir:2013pia}
N.~R. Tanvir, A.~J. Levan, et~al., {A ''kilonova'' associated with
  short-duration gamma-ray burst 130603B}, Nature 500 (2013) 547.
\newblock \href {http://arxiv.org/abs/1306.4971} {\path{arXiv:1306.4971}},
  \href {https://doi.org/10.1038/nature12505} {\path{doi:10.1038/nature12505}}.

\bibitem{Yang:2015pha}
B.~Yang, Z.-P. Jin, X.~Li, et~al., {A possible Macronova in the late afterglow
  of the `long-short' burst GRB 060614}, Nature Commun. 6 (2015) 7323.
\newblock \href {http://arxiv.org/abs/1503.07761} {\path{arXiv:1503.07761}},
  \href {https://doi.org/10.1038/ncomms8323} {\path{doi:10.1038/ncomms8323}}.

\bibitem{Jin:2016pnm}
Z.-P. Jin, K.~Hotokezaka, X.~Li, et~al., {The Macronova in GRB 050709 and the
  GRB/macronova connection}, Nature Commun. 7 (2016) 12898.
\newblock \href {http://arxiv.org/abs/1603.07869} {\path{arXiv:1603.07869}},
  \href {https://doi.org/10.1038/ncomms12898} {\path{doi:10.1038/ncomms12898}}.

\bibitem{LIGOScientific:2017vwq}
B.~P. Abbott, et~al., {GW170817: Observation of Gravitational Waves from a
  Binary Neutron Star Inspiral}, Phys. Rev. Lett. 119~(16) (2017) 161101.
\newblock \href {http://arxiv.org/abs/1710.05832} {\path{arXiv:1710.05832}},
  \href {https://doi.org/10.1103/PhysRevLett.119.161101}
  {\path{doi:10.1103/PhysRevLett.119.161101}}.

\bibitem{LIGOScientific:2017ync}
B.~P. Abbott, et~al., {Multi-messenger Observations of a Binary Neutron Star
  Merger}, Astrophys. J. Lett. 848~(2) (2017) L12.
\newblock \href {http://arxiv.org/abs/1710.05833} {\path{arXiv:1710.05833}},
  \href {https://doi.org/10.3847/2041-8213/aa91c9}
  {\path{doi:10.3847/2041-8213/aa91c9}}.

\bibitem{Shibata:2019wef}
M.~Shibata, K.~Hotokezaka, {Merger and Mass Ejection of Neutron-Star Binaries},
  Ann. Rev. Nucl. Part. Sci. 69 (2019) 41--64.
\newblock \href {http://arxiv.org/abs/1908.02350} {\path{arXiv:1908.02350}},
  \href {https://doi.org/10.1146/annurev-nucl-101918-023625}
  {\path{doi:10.1146/annurev-nucl-101918-023625}}.

\bibitem{Metzger:2019zeh}
B.~D. Metzger, {Kilonovae}, Living Rev. Rel. 23~(1) (2020) 1.
\newblock \href {http://arxiv.org/abs/1910.01617} {\path{arXiv:1910.01617}},
  \href {https://doi.org/10.1007/s41114-019-0024-0}
  {\path{doi:10.1007/s41114-019-0024-0}}.

\bibitem{Baiotti:2019sew}
L.~Baiotti, {Gravitational waves from neutron star mergers and their relation
  to the nuclear equation of state}, Prog. Part. Nucl. Phys. 109 (2019) 103714.
\newblock \href {http://arxiv.org/abs/1907.08534} {\path{arXiv:1907.08534}},
  \href {https://doi.org/10.1016/j.ppnp.2019.103714}
  {\path{doi:10.1016/j.ppnp.2019.103714}}.

\bibitem{Nakar:2019fza}
E.~Nakar, {The electromagnetic counterparts of compact binary mergers}, Phys.
  Rept. 886 (2020) 1--84.
\newblock \href {http://arxiv.org/abs/1912.05659} {\path{arXiv:1912.05659}},
  \href {https://doi.org/10.1016/j.physrep.2020.08.008}
  {\path{doi:10.1016/j.physrep.2020.08.008}}.

\bibitem{George:2020veu}
M.~George, M.-R. Wu, I.~Tamborra, et~al., {Fast neutrino flavor conversion,
  ejecta properties, and nucleosynthesis in newly-formed hypermassive remnants
  of neutron-star mergers}, Phys. Rev. D 102~(10) (2020) 103015.
\newblock \href {http://arxiv.org/abs/2009.04046} {\path{arXiv:2009.04046}},
  \href {https://doi.org/10.1103/PhysRevD.102.103015}
  {\path{doi:10.1103/PhysRevD.102.103015}}.

\bibitem{Ardevol-Pulpillo:2018btx}
R.~Ardevol-Pulpillo, H.~T. Janka, O.~Just, A.~Bauswein, {Improved
  Leakage-Equilibration-Absorption Scheme (ILEAS) for Neutrino Physics in
  Compact Object Mergers}, Mon. Not. Roy. Astron. Soc. 485~(4) (2019)
  4754--4789.
\newblock \href {http://arxiv.org/abs/1808.00006} {\path{arXiv:1808.00006}},
  \href {https://doi.org/10.1093/mnras/stz613}
  {\path{doi:10.1093/mnras/stz613}}.

\bibitem{Hempel12}
M.~{Hempel}, T.~{Fischer}, J.~{Schaffner-Bielich}, M.~{Liebend{\"o}rfer}, {New
  Equations of State in Simulations of Core-collapse Supernovae}, Astrophys. J.
  748 (2012) 70.
\newblock \href {http://arxiv.org/abs/1108.0848} {\path{arXiv:1108.0848}},
  \href {https://doi.org/10.1088/0004-637X/748/1/70}
  {\path{doi:10.1088/0004-637X/748/1/70}}.

\bibitem{SFH}
A.~W. {Steiner}, M.~{Hempel}, T.~{Fischer}, {Core-collapse Supernova Equations
  of State Based on Neutron Star Observations}, Astrophys. J. 774 (2013) 17.
\newblock \href {http://arxiv.org/abs/1207.2184} {\path{arXiv:1207.2184}},
  \href {https://doi.org/10.1088/0004-637X/774/1/17}
  {\path{doi:10.1088/0004-637X/774/1/17}}.

\bibitem{Just:2023wtj}
O.~Just, V.~Vijayan, Z.~Xiong, et~al., {End-to-end Kilonova Models of Neutron
  Star Mergers with Delayed Black Hole Formation}, Astrophys. J. Lett. 951~(1)
  (2023) L12.
\newblock \href {http://arxiv.org/abs/2302.10928} {\path{arXiv:2302.10928}},
  \href {https://doi.org/10.3847/2041-8213/acdad2}
  {\path{doi:10.3847/2041-8213/acdad2}}.

\bibitem{Cusinato:2021zin}
M.~Cusinato, F.~M. Guercilena, A.~Perego, et~al., {Neutrino emission from
  binary neutron star mergers: characterizing light curves and mean energies},
  The European Physical Journal A 58 (2021) 99.
\newblock \href {http://arxiv.org/abs/2111.13005} {\path{arXiv:2111.13005}},
  \href {https://doi.org/10.1140/epja/s10050-022-00743-5}
  {\path{doi:10.1140/epja/s10050-022-00743-5}}.

\bibitem{Radice:2018pdn}
D.~Radice, A.~Perego, K.~Hotokezaka, et~al., {Binary Neutron Star Mergers: Mass
  Ejection, Electromagnetic Counterparts and Nucleosynthesis}, Astrophys. J.
  869~(2) (2018) 130.
\newblock \href {http://arxiv.org/abs/1809.11161} {\path{arXiv:1809.11161}},
  \href {https://doi.org/10.3847/1538-4357/aaf054}
  {\path{doi:10.3847/1538-4357/aaf054}}.

\bibitem{Fujibayashi:2020dvr}
S.~Fujibayashi, S.~Wanajo, K.~Kiuchi, et~al., {Postmerger Mass Ejection of
  Low-mass Binary Neutron Stars}, Astrophys. J. 901~(2) (2020) 122.
\newblock \href {http://arxiv.org/abs/2007.00474} {\path{arXiv:2007.00474}},
  \href {https://doi.org/10.3847/1538-4357/abafc2}
  {\path{doi:10.3847/1538-4357/abafc2}}.

\bibitem{Kullmann:2021gvo}
I.~Kullmann, S.~Goriely, O.~Just, et~al., {Dynamical ejecta of neutron star
  mergers with nucleonic weak processes I: nucleosynthesis}, Mon. Not. Roy.
  Astron. Soc. 510~(2) (2022) 2804--2819.
\newblock \href {http://arxiv.org/abs/2109.02509} {\path{arXiv:2109.02509}},
  \href {https://doi.org/10.1093/mnras/stab3393}
  {\path{doi:10.1093/mnras/stab3393}}.

\bibitem{Foucart:2022bth}
F.~Foucart, {Neutrino transport in general relativistic neutron star merger
  simulations}, Living Reviews in Computational Astrophysics 9 (2022) 1.
\newblock \href {http://arxiv.org/abs/2209.02538} {\path{arXiv:2209.02538}},
  \href {https://doi.org/10.1007/s41115-023-00016-y}
  {\path{doi:10.1007/s41115-023-00016-y}}.

\bibitem{Janka:2022krt}
H.-T. Janka, A.~Bauswein, Dynamics and Equation of State Dependencies of
  Relevance for Nucleosynthesis in Supernovae and Neutron Star Mergers,
  Springer Nature Singapore, Singapore, 2023, pp. 4005--4102.
\newblock \href {http://arxiv.org/abs/2212.07498} {\path{arXiv:2212.07498}},
  \href {https://doi.org/10.1007/978-981-19-6345-2_93}
  {\path{doi:10.1007/978-981-19-6345-2_93}}.

\bibitem{Loffredo:2022prq}
E.~Loffredo, A.~Perego, D.~Logoteta, M.~Branchesi, {Muons in the aftermath of
  neutron star mergers and their impact on trapped neutrinos}, Astron.
  Astrophys. 672 (2023) A124.
\newblock \href {http://arxiv.org/abs/2209.04458} {\path{arXiv:2209.04458}},
  \href {https://doi.org/10.1051/0004-6361/202244927}
  {\path{doi:10.1051/0004-6361/202244927}}.

\bibitem{Fore:2019wib}
B.~Fore, S.~Reddy, {Pions in hot dense matter and their astrophysical
  implications}, Phys. Rev. C 101~(3) (2020) 035809.
\newblock \href {http://arxiv.org/abs/1911.02632} {\path{arXiv:1911.02632}},
  \href {https://doi.org/10.1103/PhysRevC.101.035809}
  {\path{doi:10.1103/PhysRevC.101.035809}}.

\bibitem{Vijayan:2023qrt}
V.~Vijayan, N.~Rahman, A.~Bauswein, et~al., {Impact of pions on binary neutron
  star mergers}, Phys. Rev. D 108~(2) (2023) 023020.
\newblock \href {http://arxiv.org/abs/2302.12055} {\path{arXiv:2302.12055}},
  \href {https://doi.org/10.1103/PhysRevD.108.023020}
  {\path{doi:10.1103/PhysRevD.108.023020}}.

\bibitem{Most:2018eaw}
E.~R. Most, L.~J. Papenfort, V.~Dexheimer, et~al., {Signatures of quark-hadron
  phase transitions in general-relativistic neutron-star mergers}, Phys. Rev.
  Lett. 122~(6) (2019) 061101.
\newblock \href {http://arxiv.org/abs/1807.03684} {\path{arXiv:1807.03684}},
  \href {https://doi.org/10.1103/PhysRevLett.122.061101}
  {\path{doi:10.1103/PhysRevLett.122.061101}}.

\bibitem{Bauswein:2018bma}
A.~Bauswein, N.-U.~F. Bastian, D.~B. Blaschke, et~al., {Identifying a
  first-order phase transition in neutron star mergers through gravitational
  waves}, Phys. Rev. Lett. 122~(6) (2019) 061102.
\newblock \href {http://arxiv.org/abs/1809.01116} {\path{arXiv:1809.01116}},
  \href {https://doi.org/10.1103/PhysRevLett.122.061102}
  {\path{doi:10.1103/PhysRevLett.122.061102}}.

\bibitem{Prakash:2021wpz}
A.~Prakash, D.~Radice, D.~Logoteta, et~al., {Signatures of deconfined quark
  phases in binary neutron star mergers}, Phys. Rev. D 104~(8) (2021) 083029.
\newblock \href {http://arxiv.org/abs/2106.07885} {\path{arXiv:2106.07885}},
  \href {https://doi.org/10.1103/PhysRevD.104.083029}
  {\path{doi:10.1103/PhysRevD.104.083029}}.

\bibitem{Bauswein:2019skm}
A.~Bauswein, N.-U. Friedrich~Bastian, D.~Blaschke, et~al., {Equation-of-state
  Constraints and the QCD Phase Transition in the Era of Gravitational-Wave
  Astronomy}, AIP Conf. Proc. 2127~(1) (2019) 020013.
\newblock \href {http://arxiv.org/abs/1904.01306} {\path{arXiv:1904.01306}},
  \href {https://doi.org/10.1063/1.5117803} {\path{doi:10.1063/1.5117803}}.

\bibitem{Foucart:2020ats}
F.~Foucart, {A brief overview of black hole-neutron star mergers}, Front.
  Astron. Space Sci. 7 (2020) 46.
\newblock \href {http://arxiv.org/abs/2006.10570} {\path{arXiv:2006.10570}},
  \href {https://doi.org/10.3389/fspas.2020.00046}
  {\path{doi:10.3389/fspas.2020.00046}}.

\bibitem{Kyutoku:2021icp}
K.~Kyutoku, M.~Shibata, K.~Taniguchi, {Coalescence of black
  hole\textendash{}neutron star binaries}, Living Rev. Rel. 24~(1) (2021) 5.
\newblock \href {http://arxiv.org/abs/2110.06218} {\path{arXiv:2110.06218}},
  \href {https://doi.org/10.1007/s41114-021-00033-4}
  {\path{doi:10.1007/s41114-021-00033-4}}.

\bibitem{Fujibayashi:2022ftg}
S.~Fujibayashi, K.~Kiuchi, S.~Wanajo, et~al., {Comprehensive Study of Mass
  Ejection and Nucleosynthesis in Binary Neutron Star Mergers Leaving
  Short-lived Massive Neutron Stars}, Astrophys. J. 942~(1) (2023) 39.
\newblock \href {http://arxiv.org/abs/2205.05557} {\path{arXiv:2205.05557}},
  \href {https://doi.org/10.3847/1538-4357/ac9ce0}
  {\path{doi:10.3847/1538-4357/ac9ce0}}.

\bibitem{Fahlman:2022jkh}
S.~Fahlman, R.~Fern\'andez, {Long-term 3D MHD simulations of black hole
  accretion discs formed in neutron star mergers}, Mon. Not. Roy. Astron. Soc.
  513~(2) (2022) 2689--2707.
\newblock \href {http://arxiv.org/abs/2204.03005} {\path{arXiv:2204.03005}},
  \href {https://doi.org/10.1093/mnras/stac948}
  {\path{doi:10.1093/mnras/stac948}}.

\bibitem{Just:2022flt}
O.~Just, S.~Abbar, M.-R. Wu, et~al., {Fast neutrino conversion in hydrodynamic
  simulations of neutrino-cooled accretion disks}, Phys. Rev. D 105~(8) (2022)
  083024.
\newblock \href {http://arxiv.org/abs/2203.16559} {\path{arXiv:2203.16559}},
  \href {https://doi.org/10.1103/PhysRevD.105.083024}
  {\path{doi:10.1103/PhysRevD.105.083024}}.

\bibitem{Perego:2014fma}
A.~Perego, S.~Rosswog, R.~M. Cabez\'on, et~al., {Neutrino-driven winds from
  neutron star merger remnants}, Mon. Not. Roy. Astron. Soc. 443~(4) (2014)
  3134--3156.
\newblock \href {http://arxiv.org/abs/1405.6730} {\path{arXiv:1405.6730}},
  \href {https://doi.org/10.1093/mnras/stu1352}
  {\path{doi:10.1093/mnras/stu1352}}.

\bibitem{Lippuner:2017bfm}
J.~Lippuner, R.~Fern\'andez, L.~F. Roberts, et~al., {Signatures of hypermassive
  neutron star lifetimes on r-process nucleosynthesis in the disc ejecta from
  neutron star mergers}, Mon. Not. Roy. Astron. Soc. 472~(1) (2017) 904--918.
\newblock \href {http://arxiv.org/abs/1703.06216} {\path{arXiv:1703.06216}},
  \href {https://doi.org/10.1093/mnras/stx1987}
  {\path{doi:10.1093/mnras/stx1987}}.

\bibitem{Kiuchi:2022nin}
K.~Kiuchi, S.~Fujibayashi, K.~Hayashi, et~al., {Self-Consistent Picture of the
  Mass Ejection from a One Second Long Binary Neutron Star Merger Leaving a
  Short-Lived Remnant in a General-Relativistic Neutrino-Radiation
  Magnetohydrodynamic Simulation}, Phys. Rev. Lett. 131~(1) (2023) 011401.
\newblock \href {http://arxiv.org/abs/2211.07637} {\path{arXiv:2211.07637}},
  \href {https://doi.org/10.1103/PhysRevLett.131.011401}
  {\path{doi:10.1103/PhysRevLett.131.011401}}.

\bibitem{Just:2015dba}
O.~Just, M.~Obergaulinger, H.~T. Janka, et~al., {Neutron-star merger ejecta as
  obstacles to neutrino-powered jets of gamma-ray bursts}, Astrophys. J. Lett.
  816~(2) (2016) L30.
\newblock \href {http://arxiv.org/abs/1510.04288} {\path{arXiv:1510.04288}},
  \href {https://doi.org/10.3847/2041-8205/816/2/L30}
  {\path{doi:10.3847/2041-8205/816/2/L30}}.

\bibitem{Sun:2022vri}
L.~Sun, M.~Ruiz, S.~L. Shapiro, A.~Tsokaros, {Jet launching from binary neutron
  star mergers: Incorporating neutrino transport and magnetic fields}, Phys.
  Rev. D 105~(10) (2022) 104028.
\newblock \href {http://arxiv.org/abs/2202.12901} {\path{arXiv:2202.12901}},
  \href {https://doi.org/10.1103/PhysRevD.105.104028}
  {\path{doi:10.1103/PhysRevD.105.104028}}.

\bibitem{Reddy98}
S.~Reddy, M.~Prakash, J.~M. Lattimer, {Neutrino interactions in hot and dense
  matter}, Phys. Rev. D 58 (1998) 013009.
\newblock \href {http://arxiv.org/abs/astro-ph/9710115}
  {\path{arXiv:astro-ph/9710115}}, \href
  {https://doi.org/10.1103/PhysRevD.58.013009}
  {\path{doi:10.1103/PhysRevD.58.013009}}.

\bibitem{Guo20a}
G.~Guo, G.~Mart\'\i{}nez-Pinedo, A.~Lohs, T.~Fischer, {Charged-Current Muonic
  Reactions in Core-Collapse Supernovae}, Phys. Rev. D 102~(2) (2020) 023037.
\newblock \href {http://arxiv.org/abs/2006.12051} {\path{arXiv:2006.12051}},
  \href {https://doi.org/10.1103/PhysRevD.102.023037}
  {\path{doi:10.1103/PhysRevD.102.023037}}.

\bibitem{Fischer20a}
T.~{Fischer}, G.~{Guo}, A.~A. {Dzhioev}, et~al., {Neutrino signal from
  proto-neutron star evolution: Effects of opacities from
  charged-current-neutrino interactions and inverse neutron decay}, Phys. Rev.
  C 101~(2) (2020) 025804.
\newblock \href {https://doi.org/10.1103/PhysRevC.101.025804}
  {\path{doi:10.1103/PhysRevC.101.025804}}.

\bibitem{hannestad98}
S.~{Hannestad}, G.~{Raffelt}, {Supernova Neutrino Opacity from Nucleon-Nucleon
  Bremsstrahlung and Related Processes}, Astrophys. J. 507 (1998) 339--352.
\newblock \href {http://arxiv.org/abs/astro-ph/9711132}
  {\path{arXiv:astro-ph/9711132}}, \href {https://doi.org/10.1086/306303}
  {\path{doi:10.1086/306303}}.

\bibitem{Fischer2016AA}
T.~{Fischer}, {The role of medium modifications for neutrino-pair processes
  from nucleon-nucleon bremsstrahlung. Impact on the protoneutron star
  deleptonization}, Astron. Astrophys. 593 (2016) A103.
\newblock \href {http://arxiv.org/abs/1608.05004} {\path{arXiv:1608.05004}},
  \href {https://doi.org/10.1051/0004-6361/201628991}
  {\path{doi:10.1051/0004-6361/201628991}}.

\bibitem{Guo:2019cvs}
G.~Guo, G.~Mart\'\i{}nez-Pinedo, {Chiral effective field theory description of
  neutrino nucleon-nucleon Bremsstrahlung in supernova matter}, Astrophys. J.
  887 (2019) 58.
\newblock \href {http://arxiv.org/abs/1905.13634} {\path{arXiv:1905.13634}},
  \href {https://doi.org/10.3847/1538-4357/ab536d}
  {\path{doi:10.3847/1538-4357/ab536d}}.

\bibitem{Buras02}
R.~Buras, H.-T. Janka, M.~T. Keil, et~al., {Electron neutrino pair
  annihilation: A New source for muon and tau neutrinos in supernovae},
  Astrophys.J. 587 (2003) 320--326.
\newblock \href {http://arxiv.org/abs/astro-ph/0205006}
  {\path{arXiv:astro-ph/0205006}}, \href {https://doi.org/10.1086/368015}
  {\path{doi:10.1086/368015}}.

\bibitem{Formaggio:2012cpf}
J.~A. Formaggio, G.~P. Zeller, {From eV to EeV: Neutrino Cross Sections Across
  Energy Scales}, Rev. Mod. Phys. 84 (2012) 1307--1341.
\newblock \href {http://arxiv.org/abs/1305.7513} {\path{arXiv:1305.7513}},
  \href {https://doi.org/10.1103/RevModPhys.84.1307}
  {\path{doi:10.1103/RevModPhys.84.1307}}.

\bibitem{Burrows:2002jv}
A.~Burrows, T.~A. Thompson, {Neutrino - matter interaction rates in supernovae:
  The Essential microphysics of core collapse}, Springer Netherlands,
  Dordrecht, 2004, pp. 133--174.
\newblock \href {http://arxiv.org/abs/astro-ph/0211404}
  {\path{arXiv:astro-ph/0211404}}, \href
  {https://doi.org/10.1007/978-0-306-48599-2_5}
  {\path{doi:10.1007/978-0-306-48599-2_5}}.

\bibitem{Burrows:2004vq}
A.~Burrows, S.~Reddy, T.~A. Thompson, {Neutrino opacities in nuclear matter},
  Nucl. Phys. A 777 (2006) 356--394.
\newblock \href {http://arxiv.org/abs/astro-ph/0404432}
  {\path{arXiv:astro-ph/0404432}}, \href
  {https://doi.org/10.1016/j.nuclphysa.2004.06.012}
  {\path{doi:10.1016/j.nuclphysa.2004.06.012}}.

\bibitem{Shen98}
H.~{Shen}, H.~{Toki}, K.~{Oyamatsu}, K.~{Sumiyoshi}, {Relativistic equation of
  state of nuclear matter for supernova and neutron star}, Nuclear Physics A
  637 (1998) 435--450.
\newblock \href {http://arxiv.org/abs/arXiv:nucl-th/9805035}
  {\path{arXiv:arXiv:nucl-th/9805035}}, \href
  {https://doi.org/10.1016/S0375-9474(98)00236-X}
  {\path{doi:10.1016/S0375-9474(98)00236-X}}.

\bibitem{Shen10}
G.~Shen, C.~J. Horowitz, S.~Teige, {Equation of State of Dense Matter from a
  density dependent relativistic mean field model}, Phys. Rev. C 82 (2010)
  015806.
\newblock \href {https://doi.org/10.1103/PhysRevC.82.015806}
  {\path{doi:10.1103/PhysRevC.82.015806}}.

\bibitem{Furusawa11}
S.~Furusawa, S.~Yamada, K.~Sumiyoshi, H.~Suzuki, {A new baryonic equation of
  state at sub-nuclear densities for core-collapse simulations}, Astrophys. J.
  738 (2011) 178.
\newblock \href {https://doi.org/10.1088/0004-637X/738/2/178}
  {\path{doi:10.1088/0004-637X/738/2/178}}.

\bibitem{Shen11a}
G.~Shen, C.~J. Horowitz, S.~Teige, {A New Equation of State for Astrophysical
  Simulations}, Phys. Rev. C 83 (2011) 035802.
\newblock \href {https://doi.org/10.1103/PhysRevC.83.035802}
  {\path{doi:10.1103/PhysRevC.83.035802}}.

\bibitem{Shen11b}
G.~Shen, C.~J. Horowitz, E.~O'Connor, {A Second Relativistic Mean Field and
  Virial Equation of State for Astrophysical Simulations}, Phys. Rev. C 83
  (2011) 065808.
\newblock \href {https://doi.org/10.1103/PhysRevC.83.065808}
  {\path{doi:10.1103/PhysRevC.83.065808}}.

\bibitem{roberts17}
L.~F. {Roberts}, S.~{Reddy}, {Charged current neutrino interactions in hot and
  dense matter}, Phys. Rev. C 95~(4) (2017) 045807.
\newblock \href {http://arxiv.org/abs/1612.02764} {\path{arXiv:1612.02764}},
  \href {https://doi.org/10.1103/PhysRevC.95.045807}
  {\path{doi:10.1103/PhysRevC.95.045807}}.

\bibitem{Tews13}
I.~{Tews}, T.~{Kr{\"u}ger}, K.~{Hebeler}, A.~{Schwenk}, {Neutron Matter at
  Next-to-Next-to-Next-to-Leading Order in Chiral Effective Field Theory},
  Phys. Rev. Lett. 110~(3) (2013) 032504.
\newblock \href {http://arxiv.org/abs/1206.0025} {\path{arXiv:1206.0025}},
  \href {https://doi.org/10.1103/PhysRevLett.110.032504}
  {\path{doi:10.1103/PhysRevLett.110.032504}}.

\bibitem{Krueger13}
T.~{Kr{\"u}ger}, I.~{Tews}, K.~{Hebeler}, A.~{Schwenk}, {Neutron matter from
  chiral effective field theory interactions}, Phys. Rev. C 88~(2) (2013)
  025802.
\newblock \href {http://arxiv.org/abs/1304.2212} {\path{arXiv:1304.2212}},
  \href {https://doi.org/10.1103/PhysRevC.88.025802}
  {\path{doi:10.1103/PhysRevC.88.025802}}.

\bibitem{Antoniadis13}
J.~{Antoniadis}, P.~C.~C. {Freire}, N.~{Wex}, et~al., {A Massive Pulsar in a
  Compact Relativistic Binary}, Science 340 (2013) 448.
\newblock \href {http://arxiv.org/abs/1304.6875} {\path{arXiv:1304.6875}},
  \href {https://doi.org/10.1126/science.1233232}
  {\path{doi:10.1126/science.1233232}}.

\bibitem{Fonseca:2021}
E.~{Fonseca}, H.~T. {Cromartie}, T.~T. {Pennucci}, et~al., {Refined Mass and
  Geometric Measurements of the High-mass PSR J0740+6620}, Astrophys. J. Lett.
  915~(1) (2021) L12.
\newblock \href {http://arxiv.org/abs/2104.00880} {\path{arXiv:2104.00880}},
  \href {https://doi.org/10.3847/2041-8213/ac03b8}
  {\path{doi:10.3847/2041-8213/ac03b8}}.

\bibitem{Abbott18}
B.~P. Abbott, et~al., {GW170817: Measurements of Neutron Star Radii and
  Equation of State}, Phys. Rev. Lett. 121~(16) (2018) 161101.
\newblock \href {http://arxiv.org/abs/1805.11581} {\path{arXiv:1805.11581}},
  \href {https://doi.org/10.1103/PhysRevLett.121.161101}
  {\path{doi:10.1103/PhysRevLett.121.161101}}.

\bibitem{Lattimer18}
S.~{De}, D.~{Finstad}, J.~M. {Lattimer}, et~al., {Tidal Deformabilities and
  Radii of Neutron Stars from the Observation of GW170817}, Phys. Rev. Lett.
  121~(9) (2018) 091102.
\newblock \href {http://arxiv.org/abs/1804.08583} {\path{arXiv:1804.08583}},
  \href {https://doi.org/10.1103/PhysRevLett.121.091102}
  {\path{doi:10.1103/PhysRevLett.121.091102}}.

\bibitem{NICER_Miller2019}
M.~C. {Miller}, F.~K. {Lamb}, A.~J. {Dittmann}, et~al., {PSR J0030+0451 Mass
  and Radius from NICER Data and Implications for the Properties of Neutron
  Star Matter}, Astrophys. J. Lett. 887~(1) (2019) L24.
\newblock \href {http://arxiv.org/abs/1912.05705} {\path{arXiv:1912.05705}},
  \href {https://doi.org/10.3847/2041-8213/ab50c5}
  {\path{doi:10.3847/2041-8213/ab50c5}}.

\bibitem{NICER_Watts2019}
A.~V. {Bilous}, A.~L. {Watts}, A.~K. {Harding}, et~al., {A NICER View of PSR
  J0030+0451: Evidence for a Global-scale Multipolar Magnetic Field},
  Astrophys. J. Lett. 887~(1) (2019) L23.
\newblock \href {http://arxiv.org/abs/1912.05704} {\path{arXiv:1912.05704}},
  \href {https://doi.org/10.3847/2041-8213/ab53e7}
  {\path{doi:10.3847/2041-8213/ab53e7}}.

\bibitem{NICER_Miller2021}
M.~C. {Miller}, F.~K. {Lamb}, A.~J. {Dittmann}, et~al., {The Radius of PSR
  J0740+6620 from NICER and XMM-Newton Data}, Astrophys. J. Lett. 918~(2)
  (2021) L28.
\newblock \href {http://arxiv.org/abs/2105.06979} {\path{arXiv:2105.06979}},
  \href {https://doi.org/10.3847/2041-8213/ac089b}
  {\path{doi:10.3847/2041-8213/ac089b}}.

\bibitem{NICER_Riley2021}
T.~E. {Riley}, A.~L. {Watts}, P.~S. {Ray}, et~al., {A NICER View of the Massive
  Pulsar PSR J0740+6620 Informed by Radio Timing and XMM-Newton Spectroscopy},
  Astrophys. J. Lett. 918~(2) (2021) L27.
\newblock \href {http://arxiv.org/abs/2105.06980} {\path{arXiv:2105.06980}},
  \href {https://doi.org/10.3847/2041-8213/ac0a81}
  {\path{doi:10.3847/2041-8213/ac0a81}}.

\bibitem{Lattimer13}
J.~M. {Lattimer}, Y.~{Lim}, {Constraining the Symmetry Parameters of the
  Nuclear Interaction}, Astrophys. J. 771 (2013) 51.
\newblock \href {http://arxiv.org/abs/1203.4286} {\path{arXiv:1203.4286}},
  \href {https://doi.org/10.1088/0004-637X/771/1/51}
  {\path{doi:10.1088/0004-637X/771/1/51}}.

\bibitem{Fischer17}
T.~{Fischer}, N.-U. {Bastian}, D.~{Blaschke}, et~al., {The state of matter in
  simulations of core-collapse supernovae -- Reflections and recent
  developments}, Publ. Astron. Soc. Austral. 34 (2017) 67.
\newblock \href {http://arxiv.org/abs/1711.07411} {\path{arXiv:1711.07411}},
  \href {https://doi.org/10.1017/pasa.2017.63}
  {\path{doi:10.1017/pasa.2017.63}}.

\bibitem{horowitz02}
C.~J. {Horowitz}, {Weak magnetism for antineutrinos in supernovae}, Phys. Rev.
  D 65~(4) (2002) 043001.
\newblock \href {http://arxiv.org/abs/astro-ph/0109209}
  {\path{arXiv:astro-ph/0109209}}, \href
  {https://doi.org/10.1103/PhysRevD.65.043001}
  {\path{doi:10.1103/PhysRevD.65.043001}}.

\bibitem{Fischer20}
T.~{Fischer}, M.-R. {Wu}, B.~{Wehmeyer}, et~al., {Core-collapse Supernova
  Explosions Driven by the Hadron-quark Phase Transition as a Rare r-process
  Site}, Astrophys. J. 894~(1) (2020) 9.
\newblock \href {http://arxiv.org/abs/2003.00972} {\path{arXiv:2003.00972}},
  \href {https://doi.org/10.3847/1538-4357/ab86b0}
  {\path{doi:10.3847/1538-4357/ab86b0}}.

\bibitem{MartinezPinedo12}
G.~{Mart{\'\i}nez-Pinedo}, T.~{Fischer}, A.~{Lohs}, L.~{Huther},
  {Charged-Current Weak Interaction Processes in Hot and Dense Matter and its
  Impact on the Spectra of Neutrinos Emitted from Protoneutron Star Cooling},
  Phys. Rev. Lett. 109~(25) (2012) 251104.
\newblock \href {http://arxiv.org/abs/1205.2793} {\path{arXiv:1205.2793}},
  \href {https://doi.org/10.1103/PhysRevLett.109.251104}
  {\path{doi:10.1103/PhysRevLett.109.251104}}.

\bibitem{MartinezPinedo14}
G.~{Mart{\'\i}nez-Pinedo}, T.~{Fischer}, L.~{Huther}, {Supernova neutrinos and
  nucleosynthesis}, Journal of Physics G Nuclear Physics 41~(4) (2014) 044008.
\newblock \href {http://arxiv.org/abs/1309.5477} {\path{arXiv:1309.5477}},
  \href {https://doi.org/10.1088/0954-3899/41/4/044008}
  {\path{doi:10.1088/0954-3899/41/4/044008}}.

\bibitem{Horowitz:2012us}
C.~J. Horowitz, G.~Shen, E.~O'Connor, C.~D. Ott, {Charged current neutrino
  interactions in core-collapse supernovae in a virial expansion}, Phys. Rev. C
  86 (2012) 065806.
\newblock \href {http://arxiv.org/abs/1209.3173} {\path{arXiv:1209.3173}},
  \href {https://doi.org/10.1103/PhysRevC.86.065806}
  {\path{doi:10.1103/PhysRevC.86.065806}}.

\bibitem{Rrapaj:2014yba}
E.~Rrapaj, J.~W. Holt, A.~Bartl, et~al., {Charged-current reactions in the
  supernova neutrino-sphere}, Phys. Rev. C 91~(3) (2015) 035806.
\newblock \href {http://arxiv.org/abs/1408.3368} {\path{arXiv:1408.3368}},
  \href {https://doi.org/10.1103/PhysRevC.91.035806}
  {\path{doi:10.1103/PhysRevC.91.035806}}.

\bibitem{Shen:2003ih}
C.~Shen, U.~Lombardo, N.~Van~Giai, W.~Zuo, {Neutrino mean free path in neutron
  stars}, Phys. Rev. C 68 (2003) 055802.
\newblock \href {http://arxiv.org/abs/nucl-th/0307101}
  {\path{arXiv:nucl-th/0307101}}, \href
  {https://doi.org/10.1103/PhysRevC.68.055802}
  {\path{doi:10.1103/PhysRevC.68.055802}}.

\bibitem{Vidana:2022ket}
I.~Vidana, D.~Logoteta, I.~Bombaci, {Effect of chiral nuclear forces on the
  neutrino mean free path in hot neutron matter}, Phys. Rev. C 106~(3) (2022)
  035804.
\newblock \href {http://arxiv.org/abs/2206.10190} {\path{arXiv:2206.10190}},
  \href {https://doi.org/10.1103/PhysRevC.106.035804}
  {\path{doi:10.1103/PhysRevC.106.035804}}.

\bibitem{Oertel:2020pcg}
M.~Oertel, A.~Pascal, M.~Mancini, J.~Novak, {Improved neutrino-nucleon
  interactions in dense and hot matter for numerical simulations}, Phys. Rev. C
  102~(3) (2020) 035802.
\newblock \href {http://arxiv.org/abs/2003.02152} {\path{arXiv:2003.02152}},
  \href {https://doi.org/10.1103/PhysRevC.102.035802}
  {\path{doi:10.1103/PhysRevC.102.035802}}.

\bibitem{Hutauruk:2022bii}
P.~T.~P. Hutauruk, H.~Gil, S.-i. Nam, C.~H. Hyun, {Effect of nucleon effective
  mass and symmetry energy on the neutrino mean free path in a neutron star},
  Phys. Rev. C 106~(3) (2022) 035802.
\newblock \href {http://arxiv.org/abs/2204.02061} {\path{arXiv:2204.02061}},
  \href {https://doi.org/10.1103/PhysRevC.106.035802}
  {\path{doi:10.1103/PhysRevC.106.035802}}.

\bibitem{Duan:2023amg}
M.~Duan, M.~Urban, {Energy and angle dependence of neutrino scattering rates in
  proto\textendash{}neutron star and supernova matter within Skyrme RPA}, Phys.
  Rev. C 108~(2) (2023) 025813.
\newblock \href {http://arxiv.org/abs/2305.09499} {\path{arXiv:2305.09499}},
  \href {https://doi.org/10.1103/PhysRevC.108.025813}
  {\path{doi:10.1103/PhysRevC.108.025813}}.

\bibitem{Raffelt:1993ix}
G.~Raffelt, D.~Seckel, {A self-consistent approach to neutral current processes
  in supernova cores}, Phys. Rev. D 52 (1995) 1780--1799.
\newblock \href {http://arxiv.org/abs/astro-ph/9312019}
  {\path{arXiv:astro-ph/9312019}}, \href
  {https://doi.org/10.1103/PhysRevD.52.1780}
  {\path{doi:10.1103/PhysRevD.52.1780}}.

\bibitem{Horowitz:2006pj}
C.~J. Horowitz, A.~Schwenk, {The Neutrino response of low-density neutron
  matter from the virial expansion}, Phys. Lett. B 642 (2006) 326--332.
\newblock \href {http://arxiv.org/abs/nucl-th/0605013}
  {\path{arXiv:nucl-th/0605013}}, \href
  {https://doi.org/10.1016/j.physletb.2006.09.042}
  {\path{doi:10.1016/j.physletb.2006.09.042}}.

\bibitem{Burrows98}
A.~{Burrows}, R.~F. {Sawyer}, {Effects of correlations on neutrino opacities in
  nuclear matter}, Phys. Rev. C 58~(1) (1998) 554--571.
\newblock \href {http://arxiv.org/abs/astro-ph/9801082}
  {\path{arXiv:astro-ph/9801082}}, \href
  {https://doi.org/10.1103/PhysRevC.58.554}
  {\path{doi:10.1103/PhysRevC.58.554}}.

\bibitem{Burrows:1998ek}
A.~Burrows, R.~F. Sawyer, {Many body corrections to charged current neutrino
  absorption rates in nuclear matter}, Phys. Rev. C 59 (1999) 510--514.
\newblock \href {http://arxiv.org/abs/astro-ph/9804264}
  {\path{arXiv:astro-ph/9804264}}, \href
  {https://doi.org/10.1103/PhysRevC.59.510}
  {\path{doi:10.1103/PhysRevC.59.510}}.

\bibitem{Reddy99}
S.~{Reddy}, M.~{Prakash}, J.~M. {Lattimer}, J.~A. {Pons}, {Effects of strong
  and electromagnetic correlations on neutrino interactions in dense matter},
  Phys. Rev. C 59~(5) (1999) 2888--2918.
\newblock \href {http://arxiv.org/abs/astro-ph/9811294}
  {\path{arXiv:astro-ph/9811294}}, \href
  {https://doi.org/10.1103/PhysRevC.59.2888}
  {\path{doi:10.1103/PhysRevC.59.2888}}.

\bibitem{Navarro:1999ij}
J.~Navarro, E.~S. Hern\'andez, D.~Vautherin, {Neutrino mean free path in hot
  neutron matter with skyrme interactions}, Nucl. Phys. A 654~(1) (1999)
  912c--915c.
\newblock \href {https://doi.org/10.1016/S0375-9474(00)88571-1}
  {\path{doi:10.1016/S0375-9474(00)88571-1}}.

\bibitem{Hernandez:1999zz}
E.~S. Hernandez, J.~Navarro, A.~Polls, {Response of asymmetric nuclear matter
  to isospin-flip probes}, Nucl. Phys. A 658 (1999) 327--342.
\newblock \href {https://doi.org/10.1016/S0375-9474(99)00363-2}
  {\path{doi:10.1016/S0375-9474(99)00363-2}}.

\bibitem{Pastore:2014aia}
A.~Pastore, D.~Davesne, J.~Navarro, {Linear response of homogeneous nuclear
  matter with energy density functionals}, Phys. Rept. 563 (2014) 1--67.
\newblock \href {http://arxiv.org/abs/1412.2339} {\path{arXiv:1412.2339}},
  \href {https://doi.org/10.1016/j.physrep.2014.11.002}
  {\path{doi:10.1016/j.physrep.2014.11.002}}.

\bibitem{Dzhioev:2018ovi}
A.~A. Dzhioev, G.~Mart\'\i{}nez-Pinedo, {Skyrme-RPA study of charged-current
  neutrino opacity in hot and dense supernova matter}, EPJ Web Conf. 194 (2018)
  02006.
\newblock \href {http://arxiv.org/abs/1809.08812} {\path{arXiv:1809.08812}},
  \href {https://doi.org/10.1051/epjconf/201819402006}
  {\path{doi:10.1051/epjconf/201819402006}}.

\bibitem{Pascal:2022qeg}
A.~Pascal, J.~Novak, M.~Oertel, {Proto-neutron star evolution with improved
  charged-current neutrino\textendash{}nucleon interactions}, Mon. Not. Roy.
  Astron. Soc. 511~(1) (2022) 356--370.
\newblock \href {http://arxiv.org/abs/2201.01955} {\path{arXiv:2201.01955}},
  \href {https://doi.org/10.1093/mnras/stac016}
  {\path{doi:10.1093/mnras/stac016}}.

\bibitem{Mornas:2002ji}
L.~Mornas, {Neutrino nucleon scattering rate in the relativistic random phase
  approximation}, Nucl. Phys. A 721 (2003) 1040--1043.
\newblock \href {http://arxiv.org/abs/nucl-th/0210035}
  {\path{arXiv:nucl-th/0210035}}, \href
  {https://doi.org/10.1016/S0375-9474(03)01280-6}
  {\path{doi:10.1016/S0375-9474(03)01280-6}}.

\bibitem{Shin:2023sei}
E.~Shin, E.~Rrapaj, J.~W. Holt, S.~K. Reddy, {Chiral EFT calculation of
  neutrino reactions in warm neutron-rich matter} (2023).
\newblock \href {http://arxiv.org/abs/2306.05280} {\path{arXiv:2306.05280}}.

\bibitem{Horowitz:2005zv}
C.~J. Horowitz, A.~Schwenk, {The Virial equation of state of low-density
  neutron matter}, Phys. Lett. B 638 (2006) 153--159.
\newblock \href {http://arxiv.org/abs/nucl-th/0507064}
  {\path{arXiv:nucl-th/0507064}}, \href
  {https://doi.org/10.1016/j.physletb.2006.05.055}
  {\path{doi:10.1016/j.physletb.2006.05.055}}.

\bibitem{Horowitz:2005nd}
C.~J. Horowitz, A.~Schwenk, {Cluster formation and the virial equation of state
  of low-density nuclear matter}, Nucl. Phys. A 776 (2006) 55--79.
\newblock \href {http://arxiv.org/abs/nucl-th/0507033}
  {\path{arXiv:nucl-th/0507033}}, \href
  {https://doi.org/10.1016/j.nuclphysa.2006.05.009}
  {\path{doi:10.1016/j.nuclphysa.2006.05.009}}.

\bibitem{Liu2013}
X.-J. Liu, Virial expansion for a strongly correlated fermi system and its
  application to ultracold atomic fermi gases, Physics Reports 524~(2) (2013)
  37--83, virial expansion for a strongly correlated Fermi system and its
  application to ultracold atomic Fermi gases.
\newblock \href {https://doi.org/https://doi.org/10.1016/j.physrep.2012.10.004}
  {\path{doi:https://doi.org/10.1016/j.physrep.2012.10.004}}.

\bibitem{Lin:2017spm}
Z.~Lin, C.~J. Horowitz, {Neutrino scattering in supernovae and spin
  correlations of a unitary gas}, Phys. Rev. C 96~(5) (2017) 055804.
\newblock \href {http://arxiv.org/abs/1708.01788} {\path{arXiv:1708.01788}},
  \href {https://doi.org/10.1103/PhysRevC.96.055804}
  {\path{doi:10.1103/PhysRevC.96.055804}}.

\bibitem{Alexandru:2019gmp}
A.~Alexandru, P.~F. Bedaque, N.~C. Warrington, {Structure Factors of The
  Unitary Gas Under Supernova Conditions}, Phys. Rev. C 101~(4) (2020) 045805.
\newblock \href {http://arxiv.org/abs/1907.03914} {\path{arXiv:1907.03914}},
  \href {https://doi.org/10.1103/PhysRevC.101.045805}
  {\path{doi:10.1103/PhysRevC.101.045805}}.

\bibitem{Alexandru:2020zti}
A.~Alexandru, P.~Bedaque, E.~Berkowitz, N.~C. Warrington, {Structure Factors of
  Neutron Matter at Finite Temperature}, Phys. Rev. Lett. 126~(13) (2021)
  132701.
\newblock \href {http://arxiv.org/abs/2008.02824} {\path{arXiv:2008.02824}},
  \href {https://doi.org/10.1103/PhysRevLett.126.132701}
  {\path{doi:10.1103/PhysRevLett.126.132701}}.

\bibitem{Bedaque:2018wns}
P.~F. Bedaque, S.~Reddy, S.~Sen, N.~C. Warrington, {Neutrino-nucleon scattering
  in the neutrino-sphere}, Phys. Rev. C 98~(1) (2018) 015802.
\newblock \href {http://arxiv.org/abs/1801.07077} {\path{arXiv:1801.07077}},
  \href {https://doi.org/10.1103/PhysRevC.98.015802}
  {\path{doi:10.1103/PhysRevC.98.015802}}.

\bibitem{Friman79}
B.~L. {Friman}, O.~V. {Maxwell}, {Neutrino emissivities of neutron stars.},
  Astrophys. J. 232 (1979) 541--557.
\newblock \href {https://doi.org/10.1086/157313} {\path{doi:10.1086/157313}}.

\bibitem{Migdal78}
A.~B. {Migdal}, {Pion fields in nuclear matter}, Reviews of Modern Physics
  50~(1) (1978) 107--172.
\newblock \href {https://doi.org/10.1103/RevModPhys.50.107}
  {\path{doi:10.1103/RevModPhys.50.107}}.

\bibitem{Migdal90}
A.~B. {Migdal}, E.~E. {Saperstein}, M.~A. {Troitsky}, D.~N. {Voskresensky},
  {Pion degrees of freedom in nuclear matter}, Phys.Rept. 192~(4-6) (1990)
  179--437.
\newblock \href {https://doi.org/10.1016/0370-1573(90)90132-L}
  {\path{doi:10.1016/0370-1573(90)90132-L}}.

\bibitem{Voskresensky01}
D.~N. {Voskresensky}, {Neutrino Cooling of Neutron Stars: Medium Effects}, in:
  D.~{Blaschke}, N.~K. {Glendenning}, A.~{Sedrakian} (Eds.), Physics of Neutron
  Star Interiors, Vol. 578, 2001, p. 467.
\newblock \href {https://doi.org/10.48550/arXiv.astro-ph/0101514}
  {\path{doi:10.48550/arXiv.astro-ph/0101514}}.

\bibitem{Lykasov:2008yz}
G.~I. Lykasov, C.~J. Pethick, A.~Schwenk, {A Unified approach to structure
  factors and neutrino processes in nucleon matter}, Phys. Rev. C 78 (2008)
  045803.
\newblock \href {http://arxiv.org/abs/0808.0330} {\path{arXiv:0808.0330}},
  \href {https://doi.org/10.1103/PhysRevC.78.045803}
  {\path{doi:10.1103/PhysRevC.78.045803}}.

\bibitem{Bacca:2008yr}
S.~Bacca, K.~Hally, C.~J. Pethick, A.~Schwenk, {Chiral effective field theory
  calculations of neutrino processes in dense matter}, Phys. Rev. C 80 (2009)
  032802.
\newblock \href {http://arxiv.org/abs/0812.0102} {\path{arXiv:0812.0102}},
  \href {https://doi.org/10.1103/PhysRevC.80.032802}
  {\path{doi:10.1103/PhysRevC.80.032802}}.

\bibitem{Bacca:2011qd}
S.~Bacca, K.~Hally, M.~Liebendorfer, et~al., {Neutrino processes in partially
  degenerate neutron matter}, Astrophys. J. 758 (2012) 34.
\newblock \href {http://arxiv.org/abs/1112.5185} {\path{arXiv:1112.5185}},
  \href {https://doi.org/10.1088/0004-637X/758/1/34}
  {\path{doi:10.1088/0004-637X/758/1/34}}.

\bibitem{bartl}
A.~{Bartl}, C.~J. {Pethick}, A.~{Schwenk}, {Supernova Matter at Subnuclear
  Densities as a Resonant Fermi Gas: Enhancement of Neutrino Rates}, Physical
  Review Letters 113~(8) (2014) 081101.
\newblock \href {http://arxiv.org/abs/1403.4114} {\path{arXiv:1403.4114}},
  \href {https://doi.org/10.1103/PhysRevLett.113.081101}
  {\path{doi:10.1103/PhysRevLett.113.081101}}.

\bibitem{Bartl:2016iok}
A.~Bartl, R.~Bollig, H.-T. Janka, A.~Schwenk, {Impact of Nucleon-Nucleon
  Bremsstrahlung Rates Beyond One-Pion Exchange}, Phys. Rev. D 94 (2016)
  083009.
\newblock \href {http://arxiv.org/abs/1608.05037} {\path{arXiv:1608.05037}},
  \href {https://doi.org/10.1103/PhysRevD.94.083009}
  {\path{doi:10.1103/PhysRevD.94.083009}}.

\bibitem{Betranhandy:2020cdf}
A.~Betranhandy, E.~O'Connor, {Impact of neutrino pair-production rates in
  Core-Collapse Supernovae}, Phys. Rev. D 102 (2020) 123015.
\newblock \href {http://arxiv.org/abs/2010.02261} {\path{arXiv:2010.02261}},
  \href {https://doi.org/10.1103/PhysRevD.102.123015}
  {\path{doi:10.1103/PhysRevD.102.123015}}.

\bibitem{Roberts12}
L.~F. {Roberts}, S.~{Reddy}, G.~{Shen}, {Medium modification of the
  charged-current neutrino opacity and its implications}, Phys. Rev. C 86~(6)
  (2012) 065803.
\newblock \href {http://arxiv.org/abs/1205.4066} {\path{arXiv:1205.4066}},
  \href {https://doi.org/10.1103/PhysRevC.86.065803}
  {\path{doi:10.1103/PhysRevC.86.065803}}.

\bibitem{Pons:1998mm}
J.~A. Pons, S.~Reddy, M.~Prakash, et~al., {Evolution of protoneutron stars},
  Astrophys. J. 513 (1999) 780.
\newblock \href {http://arxiv.org/abs/astro-ph/9807040}
  {\path{arXiv:astro-ph/9807040}}, \href {https://doi.org/10.1086/306889}
  {\path{doi:10.1086/306889}}.

\bibitem{Prakash:1992}
M.~Prakash, M.~Prakash, J.~M. Lattimer, C.~J. Pethick, {Rapid Cooling of
  Neutron Stars by Hyperons and Delta Isobars}, Astrophys. J. Lett. 390 (1992)
  L77.
\newblock \href {https://doi.org/10.1086/186376} {\path{doi:10.1086/186376}}.

\bibitem{Reddy:1996tw}
S.~Reddy, M.~Prakash, {Neutrino scattering in a newly born neutron star},
  Astrophys. J. 478 (1997) 689--700.
\newblock \href {http://arxiv.org/abs/astro-ph/9610115}
  {\path{arXiv:astro-ph/9610115}}, \href {https://doi.org/10.1086/303804}
  {\path{doi:10.1086/303804}}.

\bibitem{Mornas:2004vt}
L.~Mornas, {Neutrino scattering rates in the presence of hyperons from a Skyrme
  model in the RPA approximation}, Eur. Phys. J. A 23 (2005) 365--378.
\newblock \href {http://arxiv.org/abs/nucl-th/0407084}
  {\path{arXiv:nucl-th/0407084}}, \href
  {https://doi.org/10.1140/epja/i2004-10085-9}
  {\path{doi:10.1140/epja/i2004-10085-9}}.

\bibitem{Ravenhall:1983}
D.~G. Ravenhall, C.~J. Pethick, J.~R. Wilson, {Structure of matter below
  nuclear saturation density}, Phys. Rev. Lett. 50 (1983) 2066--2069.
\newblock \href {https://doi.org/10.1103/PhysRevLett.50.2066}
  {\path{doi:10.1103/PhysRevLett.50.2066}}.

\bibitem{Hashimoto:1984}
M.~{Hashimoto}, H.~{Seki}, M.~{Yamada}, {Shape of nuclei in the crust of a
  neutron star.}, Progress of Theoretical Physics 71~(2) (1984) 320--326.
\newblock \href {https://doi.org/10.1143/PTP.71.320}
  {\path{doi:10.1143/PTP.71.320}}.

\bibitem{Williams:1985prf}
R.~D. Williams, S.~E. Koonin, {Sub-saturation phases of nuclear matter}, Nucl.
  Phys. A 435 (1985) 844--858.
\newblock \href {https://doi.org/10.1016/0375-9474(85)90191-5}
  {\path{doi:10.1016/0375-9474(85)90191-5}}.

\bibitem{Oyamatsu:1993zz}
K.~Oyamatsu, {Nuclear shapes in the inner crust of a neutron star}, Nucl. Phys.
  A 561 (1993) 431--452.
\newblock \href {https://doi.org/10.1016/0375-9474(93)90020-X}
  {\path{doi:10.1016/0375-9474(93)90020-X}}.

\bibitem{Lorenz:1992zz}
C.~P. Lorenz, D.~G. Ravenhall, C.~J. Pethick, {Neutron star crusts}, Phys. Rev.
  Lett. 70 (1993) 379--382.
\newblock \href {https://doi.org/10.1103/PhysRevLett.70.379}
  {\path{doi:10.1103/PhysRevLett.70.379}}.

\bibitem{Sumiyoshi:1995np}
K.~Sumiyoshi, K.~Oyamatsu, H.~Toki, {Neutron star profiles in the relativistic
  Bruckner-Hartree-Fock theory}, Nucl. Phys. A 595 (1995) 327--345.
\newblock \href {https://doi.org/10.1016/0375-9474(95)00388-5}
  {\path{doi:10.1016/0375-9474(95)00388-5}}.

\bibitem{Pethick:1995di}
C.~J. Pethick, D.~G. Ravenhall, {Matter at large neutron excess and the physics
  of neutron-star crusts}, Ann. Rev. Nucl. Part. Sci. 45 (1995) 429--484.
\newblock \href {https://doi.org/10.1146/annurev.ns.45.120195.002241}
  {\path{doi:10.1146/annurev.ns.45.120195.002241}}.

\bibitem{KIDO:2000}
T.~Kido, T.~Maruyama, K.~Niita, S.~Chiba, Md simulation study for nuclear
  matter, Nuclear Physics A 663-664 (2000) 877c--880c.
\newblock \href {https://doi.org/https://doi.org/10.1016/S0375-9474(99)00736-8}
  {\path{doi:https://doi.org/10.1016/S0375-9474(99)00736-8}}.

\bibitem{Watanabe:2000rj}
G.~Watanabe, K.~Iida, K.~Sato, {Thermodynamic properties of nuclear `pasta' in
  neutron star crusts}, Nucl. Phys. A 676 (2000) 455--473, [Erratum:
  Nucl.Phys.A 726, 357--365 (2003)].
\newblock \href {http://arxiv.org/abs/astro-ph/0001273}
  {\path{arXiv:astro-ph/0001273}}, \href
  {https://doi.org/10.1016/S0375-9474(00)00197-4}
  {\path{doi:10.1016/S0375-9474(00)00197-4}}.

\bibitem{Lopez:2020zne}
J.~A. Lopez, C.~O. Dorso, G.~A. Frank, {Properties of nuclear pastas}, Front.
  Phys. (Beijing) 16~(2) (2021) 24301.
\newblock \href {http://arxiv.org/abs/2007.07417} {\path{arXiv:2007.07417}},
  \href {https://doi.org/10.1007/s11467-020-1004-2}
  {\path{doi:10.1007/s11467-020-1004-2}}.

\bibitem{Watanabe:2000zt}
G.~Watanabe, K.~Iida, K.~Sato, {Effects of neutrino trapping on thermodynamic
  properties of nuclear `pasta'}, Nucl. Phys. A 687 (2001) 512--531.
\newblock \href {http://arxiv.org/abs/astro-ph/0008108}
  {\path{arXiv:astro-ph/0008108}}, \href
  {https://doi.org/10.1016/S0375-9474(00)00585-6}
  {\path{doi:10.1016/S0375-9474(00)00585-6}}.

\bibitem{Horowitz:2004yf}
C.~J. Horowitz, M.~A. Perez-Garcia, J.~Piekarewicz, {Neutrino - pasta
  scattering: The Opacity of nonuniform neutron - rich matter}, Phys. Rev. C 69
  (2004) 045804.
\newblock \href {http://arxiv.org/abs/astro-ph/0401079}
  {\path{arXiv:astro-ph/0401079}}, \href
  {https://doi.org/10.1103/PhysRevC.69.045804}
  {\path{doi:10.1103/PhysRevC.69.045804}}.

\bibitem{Horowitz:2004pv}
C.~J. Horowitz, M.~A. Perez-Garcia, J.~Carriere, D.~K. Berry, J.~Piekarewicz,
  {Nonuniform neutron-rich matter and coherent neutrino scattering}, Phys. Rev.
  C 70 (2004) 065806.
\newblock \href {http://arxiv.org/abs/astro-ph/0409296}
  {\path{arXiv:astro-ph/0409296}}, \href
  {https://doi.org/10.1103/PhysRevC.70.065806}
  {\path{doi:10.1103/PhysRevC.70.065806}}.

\bibitem{Alloy:2010fk}
M.~D. Alloy, D.~P. Menezes, {Nuclear 'pasta phase' and its consequences on
  neutrino opacities}, Phys. Rev. C 83 (2011) 035803.
\newblock \href {http://arxiv.org/abs/1011.0968} {\path{arXiv:1011.0968}},
  \href {https://doi.org/10.1103/PhysRevC.83.035803}
  {\path{doi:10.1103/PhysRevC.83.035803}}.

\bibitem{Alcain:2014cda}
P.~N. Alcain, C.~O. Dorso, {The neutrino opacity of neutron rich matter}, Nucl.
  Phys. A 961 (2017) 183--199.
\newblock \href {http://arxiv.org/abs/1412.6465} {\path{arXiv:1412.6465}},
  \href {https://doi.org/10.1016/j.nuclphysa.2017.02.011}
  {\path{doi:10.1016/j.nuclphysa.2017.02.011}}.

\bibitem{Furtado:2015vga}
U.~J. Furtado, S.~S. Avancini, J.~R. Marinelli, et~al., {Neutrino diffusion in
  the pasta phase matter within the Thomas-Fermi approach}, Eur. Phys. J. A
  52~(9) (2016) 290.
\newblock \href {http://arxiv.org/abs/1505.06776} {\path{arXiv:1505.06776}},
  \href {https://doi.org/10.1140/epja/i2016-16290-y}
  {\path{doi:10.1140/epja/i2016-16290-y}}.

\bibitem{Horowitz:2016fpa}
C.~J. Horowitz, D.~K. Berry, M.~E. Caplan, et~al., {Nuclear pasta and supernova
  neutrinos at late times} (2016).
\newblock \href {http://arxiv.org/abs/1611.10226} {\path{arXiv:1611.10226}}.

\bibitem{Roggero:2017pag}
A.~Roggero, J.~Margueron, L.~F. Roberts, S.~Reddy, {Nuclear pasta in hot dense
  matter and its implications for neutrino scattering}, Phys. Rev. C 97~(4)
  (2018) 045804.
\newblock \href {http://arxiv.org/abs/1710.10206} {\path{arXiv:1710.10206}},
  \href {https://doi.org/10.1103/PhysRevC.97.045804}
  {\path{doi:10.1103/PhysRevC.97.045804}}.

\bibitem{Schuetrumpf:2019hqe}
B.~Schuetrumpf, G.~Mart\'\i{}nez-Pinedo, P.~G. Reinhard, {Survey of nuclear
  pasta in the intermediate-density regime: Structure functions for neutrino
  scattering}, Phys. Rev. C 101~(5) (2020) 055804.
\newblock \href {http://arxiv.org/abs/1912.10510} {\path{arXiv:1912.10510}},
  \href {https://doi.org/10.1103/PhysRevC.101.055804}
  {\path{doi:10.1103/PhysRevC.101.055804}}.

\bibitem{Lin:2020nxy}
Z.~Lin, M.~E. Caplan, C.~J. Horowitz, C.~Lunardini, {Fast neutrino cooling of
  nuclear pasta in neutron stars: molecular dynamics simulations}, Phys. Rev. C
  102~(4) (2020) 045801.
\newblock \href {http://arxiv.org/abs/2006.04963} {\path{arXiv:2006.04963}},
  \href {https://doi.org/10.1103/PhysRevC.102.045801}
  {\path{doi:10.1103/PhysRevC.102.045801}}.

\bibitem{Froehlich06a}
C.~Fr\"ohlich, {\it et al.}, Composition of the innermost supernova ejecta,
  Astrophys. J. 637 (2006) 415.
\newblock \href {http://arxiv.org/abs/astro-ph/0410208}
  {\path{arXiv:astro-ph/0410208}}, \href {https://doi.org/10.1086/498224}
  {\path{doi:10.1086/498224}}.

\bibitem{Wanajo06}
S.~Wanajo, {The rp-process in neutrino-driven winds}, Astrophys. J. 647 (2006)
  1323--1340.
\newblock \href {http://arxiv.org/abs/astro-ph/0602488}
  {\path{arXiv:astro-ph/0602488}}, \href {https://doi.org/10.1086/505483}
  {\path{doi:10.1086/505483}}.

\bibitem{Wanajo:2010mc}
S.~Wanajo, H.-T. Janka, S.~Kubono, {Uncertainties in the $\nu$ p-process:
  supernova dynamics versus nuclear physics}, Astrophys. J. 729 (2011) 46.
\newblock \href {http://arxiv.org/abs/1004.4487} {\path{arXiv:1004.4487}},
  \href {https://doi.org/10.1088/0004-637X/729/1/46}
  {\path{doi:10.1088/0004-637X/729/1/46}}.

\bibitem{Arcones:2012wj}
A.~Arcones, F.~K. Thielemann, {Neutrino-driven wind simulations and
  nucleosynthesis of heavy elements}, J. Phys. G 40 (2013) 013201.
\newblock \href {http://arxiv.org/abs/1207.2527} {\path{arXiv:1207.2527}},
  \href {https://doi.org/10.1088/0954-3899/40/1/013201}
  {\path{doi:10.1088/0954-3899/40/1/013201}}.

\bibitem{Pllumbi:2014saa}
E.~Pllumbi, I.~Tamborra, S.~Wanajo, et~al., {Impact of neutrino flavor
  oscillations on the neutrino-driven wind nucleosynthesis of an
  electron-capture supernova}, Astrophys. J. 808~(2) (2015) 188.
\newblock \href {http://arxiv.org/abs/1406.2596} {\path{arXiv:1406.2596}},
  \href {https://doi.org/10.1088/0004-637X/808/2/188}
  {\path{doi:10.1088/0004-637X/808/2/188}}.

\bibitem{Sieverding18}
A.~Sieverding, L.~Huther, G.~Langanke, K. Martinez-Pinedo, , A.~Heger, Neutrino
  nucleosynthesis of radioactive nuclei in supernovae, Astrophys. J. 865 (2018)
  143.
\newblock \href {http://arxiv.org/abs/1805.10231} {\path{arXiv:1805.10231}},
  \href {https://doi.org/10.3847/1538-4357/aadd48}
  {\path{doi:10.3847/1538-4357/aadd48}}.

\bibitem{Fischer21}
T.~{Fischer}, {QCD phase transition drives supernova explosion of a very
  massive star}, Eur.\ Phys.\ J.\ A 57 (2021) 270.
\newblock \href {http://arxiv.org/abs/2108.00196} {\path{arXiv:2108.00196}},
  \href {https://doi.org/10.1140/epja/s10050-021-00571-z}
  {\path{doi:10.1140/epja/s10050-021-00571-z}}.

\bibitem{mirizzi16}
A.~{Mirizzi}, I.~{Tamborra}, H.-T. {Janka}, N.~{Saviano}, et~al., {Supernova
  neutrinos: production, oscillations and detection}, Nuovo Cimento Rivista
  Serie 39 (2016) 1--112.
\newblock \href {http://arxiv.org/abs/1508.00785} {\path{arXiv:1508.00785}},
  \href {https://doi.org/10.1393/ncr/i2016-10120-8}
  {\path{doi:10.1393/ncr/i2016-10120-8}}.

\bibitem{Nevins:2023tug}
B.~Nevins, L.~F. Roberts, {Proto-neutron star convection and the
  neutrino-driven wind: implications for the r-process}, Mon. Not. Roy. Astron.
  Soc. 520~(3) (2023) 3986--3999.
\newblock \href {http://arxiv.org/abs/2302.01249} {\path{arXiv:2302.01249}},
  \href {https://doi.org/10.1093/mnras/stad372}
  {\path{doi:10.1093/mnras/stad372}}.

\bibitem{Connell72}
J.~S. Connell, T.~W. Donnelly, J.~D. Walecka, Semileptonic weak interactions
  with 12c, Phys. Rev. C 6 (1972) 719.
\newblock \href {https://doi.org/10.1103/PhysRevC.6.719}
  {\path{doi:10.1103/PhysRevC.6.719}}.

\bibitem{Frekers18}
D.~Frekers, M.~Alanssari, Charge-exchange reactions and the quest for
  resolution, Eur. Phys. J. A 54 (2018) 177.
\newblock \href {https://doi.org/10.1140/epja/i2018-12612-5}
  {\path{doi:10.1140/epja/i2018-12612-5}}.

\bibitem{Langanke21}
K.~Langanke, G.~Mart\'\i{}nez-Pinedo, R.~Zegers, {Electron capture in stars},
  Rept. Prog. Phys. 84~(6) (2021) 066301.
\newblock \href {http://arxiv.org/abs/2009.01750} {\path{arXiv:2009.01750}},
  \href {https://doi.org/10.1088/1361-6633/abf207}
  {\path{doi:10.1088/1361-6633/abf207}}.

\bibitem{Fuller82}
G.~M. Fuller, W.~A. Fowler, M.~J. Newman, Stellar weak interaction rates for
  intermediate mass nuclei .3. rate tables for the free nucleons and nuclei
  with a = 21 to a = 60, Astrophys. J. 252 (1982) 715.
\newblock \href {https://doi.org/10.1086/159597} {\path{doi:10.1086/159597}}.

\bibitem{Caurier05}
E.~Caurier, G.~Martinez-Pinedo, F.~Nowacki, others., {The Shell Model as
  Unified View of Nuclear Structure}, Rev. Mod. Phys. 77 (2005) 427--488.
\newblock \href {http://arxiv.org/abs/nucl-th/0402046}
  {\path{arXiv:nucl-th/0402046}}, \href
  {https://doi.org/10.1103/RevModPhys.77.427}
  {\path{doi:10.1103/RevModPhys.77.427}}.

\bibitem{Caurier99}
E.~Caurier, K.~Langanke, G.~Martinez-Pinedo, F.~Nowacki, Shell-model
  calculations of stellar weak interaction rates. i. gamow-teller distributions
  and spectra of nuclei in the mass range a=45-65, Nucl. Phys. A 653 (1999)
  439.
\newblock \href {http://arxiv.org/abs/nucl-th/9903042}
  {\path{arXiv:nucl-th/9903042}}, \href
  {https://doi.org/10.1016/S0375-9474(99)00240-7}
  {\path{doi:10.1016/S0375-9474(99)00240-7}}.

\bibitem{Suzuki22}
T.~Suzuki, {Nuclear weak rates and nuclear weak processes in stars}, Prog.
  Part. Nucl. Phys. 126 (2022) 103974.
\newblock \href {http://arxiv.org/abs/2205.09262} {\path{arXiv:2205.09262}},
  \href {https://doi.org/10.1016/j.ppnp.2022.103974}
  {\path{doi:10.1016/j.ppnp.2022.103974}}.

\bibitem{Brown88}
B.~A. Brown, B.~H. Wildenthal, Status of the nuclear shell model, Ann. Rev.
  Nucl. Part. Sci. 38 (1988) 29.
\newblock \href {https://doi.org/10.1146/annurev.ns.38.120188.000333}
  {\path{doi:10.1146/annurev.ns.38.120188.000333}}.

\bibitem{Langanke95}
K.~Langanke, D.~J. Dean, P.~B. Radha, et~al., {Shell-model Monte Carlo studies
  of fp-shell nuclei}, Phys. Rev. C 52 (1995) 718--725.
\newblock \href {http://arxiv.org/abs/nucl-th/9504019}
  {\path{arXiv:nucl-th/9504019}}, \href
  {https://doi.org/10.1103/PhysRevC.52.718}
  {\path{doi:10.1103/PhysRevC.52.718}}.

\bibitem{Martinez97}
G.~Martinez-Pinedo, A.~P. Zuker, A.~Poves, E.~Caurier, {Full pf shell study of
  A=47 and A=49 nuclei}, Phys. Rev. C 55 (1997) 187--205.
\newblock \href {http://arxiv.org/abs/nucl-th/9608044}
  {\path{arXiv:nucl-th/9608044}}, \href
  {https://doi.org/10.1103/PhysRevC.55.187}
  {\path{doi:10.1103/PhysRevC.55.187}}.

\bibitem{Scott14}
M.~Scott, {\it et al}., Gamow-teller transition strengths from 56fe extracted
  from the 56fe(t,3he) reaction, Phys. Rev. C 90 (2014) 025801.
\newblock \href {https://doi.org/10.1103/PhysRevC.90.025801}
  {\path{doi:10.1103/PhysRevC.90.025801}}.

\bibitem{Langanke00}
K.~{Langanke}, G.~{Mart{\'{\i}}nez-Pinedo}, {Shell-model calculations of
  stellar weak interaction rates: II. Weak rates for nuclei in the mass range
  A=45-65 in supernovae environments}, Nuclear Physics A 673 (2000) 481--508.
\newblock \href {http://arxiv.org/abs/nucl-th/0001018}
  {\path{arXiv:nucl-th/0001018}}, \href
  {https://doi.org/10.1016/S0375-9474(00)00131-7}
  {\path{doi:10.1016/S0375-9474(00)00131-7}}.

\bibitem{Janka07}
H.-T. {Janka}, K.~{Langanke}, A.~{Marek}, et~al., {Theory of core-collapse
  supernovae},  Phys.Rept. 442 (2007) 38--74.
\newblock \href {http://arxiv.org/abs/arXiv:astro-ph/0612072}
  {\path{arXiv:arXiv:astro-ph/0612072}}, \href
  {https://doi.org/10.1016/j.physrep.2007.02.002}
  {\path{doi:10.1016/j.physrep.2007.02.002}}.

\bibitem{Johnson92}
C.~W. Johnson, S.~E. Koonin, G.~H. Lang, W.~E. Ormand, {Monte Carlo methods for
  the nuclear shell model}, Phys. Rev. Lett. 69 (1992) 3157--3160.
\newblock \href {http://arxiv.org/abs/nucl-th/9210014}
  {\path{arXiv:nucl-th/9210014}}, \href
  {https://doi.org/10.1103/PhysRevLett.69.3157}
  {\path{doi:10.1103/PhysRevLett.69.3157}}.

\bibitem{Koonin97}
S.~E. Koonin, D.~J. Dean, K.~Langanke, {Shell model Monte Carlo methods}, Phys.
  Rept. 278 (1997) 1--77.
\newblock \href {http://arxiv.org/abs/nucl-th/9602006}
  {\path{arXiv:nucl-th/9602006}}, \href
  {https://doi.org/10.1016/S0370-1573(96)00017-8}
  {\path{doi:10.1016/S0370-1573(96)00017-8}}.

\bibitem{Dzhioev20}
A.~A. Dzhioev, K.~Langanke, et. al., {Unblocking of stellar electron captures
  for neutron-rich $N=50$ nuclei at finite temperature}, Phys. Rev. C 101~(2)
  (2020) 025805.
\newblock \href {http://arxiv.org/abs/1910.03335} {\path{arXiv:1910.03335}},
  \href {https://doi.org/10.1103/PhysRevC.101.025805}
  {\path{doi:10.1103/PhysRevC.101.025805}}.

\bibitem{Litvinova21}
E.~Litvinova, C.~Robin, Impact of complex many-body correlations on electron
  capture in thermally excited nuclei around 78ni, Phys. Rev. C 103 (2021)
  024326.
\newblock \href {https://doi.org/10.1103/PhysRevC.103.024326}
  {\path{doi:10.1103/PhysRevC.103.024326}}.

\bibitem{Dzhioev22}
A.~A. Dzhioev, A.~I. Vdovin, {Superoperator Approach to the Theory of Hot
  Nuclei and Astrophysical Applications: II\textemdash{}Electron Capture in
  Stars}, Phys. Part. Nucl. 53~(5) (2022) 939--999.
\newblock \href {https://doi.org/10.1134/S1063779622050045}
  {\path{doi:10.1134/S1063779622050045}}.

\bibitem{Giraud2022}
S.~Giraud, R.~G.~T. Zegers, B.~A. Brown, et~al., Finite-temperature electron
  capture rates for neutron-rich nuclei near n=50 and effects on core-collapse
  supernova simulations, Phys. Rev. C 105 (2022) 055801.
\newblock \href {https://doi.org/10.1103/PhysRevC.105.055801}
  {\path{doi:10.1103/PhysRevC.105.055801}}.

\bibitem{Zeitnitz94}
B.~Zeitnitz, {\it et al.}, Karmen - neutrino physics at isis, Prog. Part. Nucl.
  Phys. 32 (1994) 351.
\newblock \href {https://doi.org/10.1016/0146-6410(94)90034-5}
  {\path{doi:10.1016/0146-6410(94)90034-5}}.

\bibitem{Armbruster98}
B.~Armbruster, {\it et al.}, Measurement of the weak neutral current excitation
  c-12(nu(mu)nu(mu))c-12*(1(+),1;15.1 mev) at e-nu mu=29.8, Phys. Lett. B 423
  (1998) 15.
\newblock \href {https://doi.org/10.1016/S0370-2693(98)00087-2}
  {\path{doi:10.1016/S0370-2693(98)00087-2}}.

\bibitem{Guhr90}
T.~{Guhr}, H.~{Diesener}, A.~{Richter}, et~al., {Electroexcitation of magnetic
  dipole and other modes in$^{46}$Ti and$^{48}$Ti}, Zeitschrift fur Physik A
  Hadrons and Nuclei 336~(2) (1990) 159--178.
\newblock \href {https://doi.org/10.1007/BF01290617}
  {\path{doi:10.1007/BF01290617}}.

\bibitem{Enders99}
J.~Enders, {\it et al.}, Comprehensive analysis of the scissors mode in heavy
  even-even nuclei, Phys. Rev. C 59 (1999) R1851.
\newblock \href {https://doi.org/10.1103/PhysRevC.59.R1851}
  {\path{doi:10.1103/PhysRevC.59.R1851}}.

\bibitem{Langanke04}
K.~Langanke, G.~Martinez-Pinedo, von Neumann-Cosel~P., R.~A., Supernova
  inelastic neutrino-nucleus cross sections from precision m1 data and shell
  model calculations, Phys. Rev. Lett. 93 (2004) 202501.
\newblock \href {http://arxiv.org/abs/nucl-th/0402001}
  {\path{arXiv:nucl-th/0402001}}, \href
  {https://doi.org/10.1103/PhysRevLett.93.202501}
  {\path{doi:10.1103/PhysRevLett.93.202501}}.

\bibitem{Kolbe99}
E.~Kolbe, K.~Langanke, G.~Martinez-Pinedo, The inclusive
  $^{56}$fe($\nu_e$,$e^-$)$^{56}$co cross section, Phys. Rev. C 60 (1999)
  052801.
\newblock \href {http://arxiv.org/abs/nucl-th/9905001}
  {\path{arXiv:nucl-th/9905001}}, \href
  {https://doi.org/10.1103/PhysRevC.60.052801}
  {\path{doi:10.1103/PhysRevC.60.052801}}.

\bibitem{Toivanen01}
J.~Toivanen, E.~Kolbe, K.~Langanke, et~al., Supernova neutrino induced
  reactions on iron isotopes, Nucl. Phys. A 694 (2001) 395.
\newblock \href {https://doi.org/10.1016/S0375-9474(01)00992-7}
  {\path{doi:10.1016/S0375-9474(01)00992-7}}.

\bibitem{Juodagalvis05}
A.~Juodagalvis, K.~Langanke, G.~Martinez-Pinedo, et~al., {Neutral-current
  neutrino-nucleus cross-sections for $A \sim$ 50 - 65 nuclei}, Nucl. Phys. A
  747 (2005) 87--108.
\newblock \href {http://arxiv.org/abs/nucl-th/0404078}
  {\path{arXiv:nucl-th/0404078}}, \href
  {https://doi.org/10.1016/j.nuclphysa.2004.09.005}
  {\path{doi:10.1016/j.nuclphysa.2004.09.005}}.

\bibitem{Dzhioev14}
A.~A. Dzhioev, A.~I. Vdovin, J.~Wambach, V.~Y. Ponomarev, {Inelastic neutrino
  scattering off hot nuclei in supernova environments}, Phys. Rev. C 89~(3)
  (2014) 035805.
\newblock \href {http://arxiv.org/abs/1401.4008} {\path{arXiv:1401.4008}},
  \href {https://doi.org/10.1103/PhysRevC.89.035805}
  {\path{doi:10.1103/PhysRevC.89.035805}}.

\bibitem{Dapo12}
H.~Dapo, N.~Paar, {Neutral-current neutrino-nucleus cross sections based on
  relativistic nuclear energy density functional}, Phys. Rev. C 86 (2012)
  035804.
\newblock \href {http://arxiv.org/abs/1203.5224} {\path{arXiv:1203.5224}},
  \href {https://doi.org/10.1103/PhysRevC.86.035804}
  {\path{doi:10.1103/PhysRevC.86.035804}}.

\bibitem{Paar15}
N.~Paar, T.~Marketin, D.~Vale, D.~Vretenar, {Modeling nuclear weak-interaction
  processes with relativistic energy density functionals}, Int. J. Mod. Phys. E
  24~(09) (2015) 1541004.
\newblock \href {http://arxiv.org/abs/1505.07486} {\path{arXiv:1505.07486}},
  \href {https://doi.org/10.1142/S0218301315410049}
  {\path{doi:10.1142/S0218301315410049}}.

\bibitem{Chasioti07}
V.~Chasioti, T.~S. Kosmas, P.~Divari, Realistic calculations for
  neutrino‐nucleus reactions cross sections, Prog. Part. Nucl. Phys. 59
  (2007) 481.
\newblock \href {https://doi.org/10.1016/j.ppnp.2007.01.003}
  {\path{doi:10.1016/j.ppnp.2007.01.003}}.

\bibitem{Langanke08}
K.~Langanke, G.~Martinez-Pinedo, B.~M\"uller, et~al., {Effects of Inelastic
  Neutrino-Nucleus Scattering on Supernova Dynamics and Radiated Neutrino
  Spectra}, Phys. Rev. Lett. 100 (2008) 011101.
\newblock \href {http://arxiv.org/abs/0706.1687} {\path{arXiv:0706.1687}},
  \href {https://doi.org/10.1103/PhysRevLett.100.011101}
  {\path{doi:10.1103/PhysRevLett.100.011101}}.

\bibitem{Haxton88}
W.~C. Haxton, Neutrino heating in supernovae, Phys. Rev. Lett. 60 (1988) 1999.
\newblock \href {https://doi.org/10.1103/PhysRevLett.60.1999}
  {\path{doi:10.1103/PhysRevLett.60.1999}}.

\bibitem{Bruenn91}
S.~W. {Bruenn}, W.~C. {Haxton}, {Neutrino-nucleus interactions in core-collapse
  supernovae}, Astrophys. J. 376 (1991) 678--700.
\newblock \href {https://doi.org/10.1086/170316} {\path{doi:10.1086/170316}}.

\bibitem{Kolbe92}
E.~Kolbe, K.~Langanke, S.~Krewald, F.-K. Thielemann, Inelastic neutrino
  scattering on $^{12}$c and $^{16}$o above the particle emission threshold,
  Nucl. Phys. A 540 (1992) 599.
\newblock \href {https://doi.org/10.1016/0375-9474(92)90175-J}
  {\path{doi:10.1016/0375-9474(92)90175-J}}.

\bibitem{Kolbe01}
E.~Kolbe, K.~Langanke, {The Role of neutrino induced reactions on lead and iron
  in neutrino detectors}, Phys. Rev. C 63 (2001) 025802.
\newblock \href {http://arxiv.org/abs/nucl-th/0003060}
  {\path{arXiv:nucl-th/0003060}}, \href
  {https://doi.org/10.1103/PhysRevC.63.025802}
  {\path{doi:10.1103/PhysRevC.63.025802}}.

\bibitem{Langanke01a}
K.~Langanke, E.~Kolbe, Neutrino-induced charged-current reaction rates for
  r-process nuclei, Atomic Data Nuclear Data Tables 79 (2001) 2003.
\newblock \href {https://doi.org/10.1006/adnd.2001.0872}
  {\path{doi:10.1006/adnd.2001.0872}}.

\bibitem{Langanke01b}
K.~Langanke, E.~Kolbe, Neutrino-induced neutral-current reaction rates for
  r-process nuclei, Atomic Data Nuclear Data Tables 82 (2002) 191.
\newblock \href {https://doi.org/10.1006/adnd.2002.0883}
  {\path{doi:10.1006/adnd.2002.0883}}.

\bibitem{Yuan12}
C.~Yuan, T.~Suzuki, T.~Otsuka, et~al., {Shell-model study of boron, carbon,
  nitrogen, and oxygen isotopes with a monopole-based universal interaction},
  Phys. Rev. C 85 (2012) 064324.
\newblock \href {http://arxiv.org/abs/1209.5587} {\path{arXiv:1209.5587}},
  \href {https://doi.org/10.1103/PhysRevC.85.064324}
  {\path{doi:10.1103/PhysRevC.85.064324}}.

\bibitem{Balasi15}
K.~G. Balasi, K.~Langanke, G.~Mart\'\i{}nez-Pinedo,
  {Neutrino\textendash{}nucleus reactions and their role for supernova dynamics
  and nucleosynthesis}, Prog. Part. Nucl. Phys. 85 (2015) 33--81.
\newblock \href {http://arxiv.org/abs/1503.08095} {\path{arXiv:1503.08095}},
  \href {https://doi.org/10.1016/j.ppnp.2015.08.001}
  {\path{doi:10.1016/j.ppnp.2015.08.001}}.

\bibitem{Kolbe03}
E.~Kolbe, K.~Langanke, G.~Martinez-Pinedo, P.~Vogel, {Neutrino nucleus
  reactions and nuclear structure}, J. Phys. G 29 (2003) 2569--2596.
\newblock \href {http://arxiv.org/abs/nucl-th/0311022}
  {\path{arXiv:nucl-th/0311022}}, \href
  {https://doi.org/10.1088/0954-3899/29/11/010}
  {\path{doi:10.1088/0954-3899/29/11/010}}.

\bibitem{O'Connor07}
E.~{O'Connor}, D.~{Gazit}, C.~J. {Horowitz}, et~al., {Neutrino breakup of A=3
  nuclei in supernovae}, Phys. Rev. C 75~(5) (2007) 055803.
\newblock \href {http://arxiv.org/abs/nucl-th/0702044}
  {\path{arXiv:nucl-th/0702044}}, \href
  {https://doi.org/10.1103/PhysRevC.75.055803}
  {\path{doi:10.1103/PhysRevC.75.055803}}.

\bibitem{Arcones:2008kv}
A.~Arcones, G.~Martinez-Pinedo, O'Connor, et~al., {Influence of light nuclei on
  neutrino-driven supernova outflows}, Phys. Rev. C 78 (2008) 015806.
\newblock \href {http://arxiv.org/abs/0805.3752} {\path{arXiv:0805.3752}},
  \href {https://doi.org/10.1103/PhysRevC.78.015806}
  {\path{doi:10.1103/PhysRevC.78.015806}}.

\bibitem{Yoshida05}
T.~Yoshida, T.~Kajino, D.~H. Hartmann, {Constraining the spectrum of supernova
  neutrinos from neutrino-process-induced light-element synthesis}, Phys. Rev.
  Lett. 94 (2005) 231101.
\newblock \href {http://arxiv.org/abs/astro-ph/0505043}
  {\path{arXiv:astro-ph/0505043}}, \href
  {https://doi.org/10.1103/PhysRevLett.94.231101}
  {\path{doi:10.1103/PhysRevLett.94.231101}}.

\bibitem{Yoshida08}
T.~Yoshida, T.~Suzuki, S.~Chiba, et~al., {Neutrino-Nucleus Reaction Cross
  Sections for Light Element Synthesis in Supernova Explosions}, Astrophys. J.
  686 (2008) 448--466.
\newblock \href {http://arxiv.org/abs/0807.2723} {\path{arXiv:0807.2723}},
  \href {https://doi.org/10.1086/591266} {\path{doi:10.1086/591266}}.

\bibitem{Mathews12}
G.~J. Mathews, T.~Kajino, W.~Aoki, et~al., {Exploring the Neutrino Mass
  Hierarchy Probability with Meteoritic Supernova Material, nu-Process
  Nucleosynthesis, and theta(13) Mixing}, Phys. Rev. D 85 (2012) 105023.
\newblock \href {http://arxiv.org/abs/1108.0725} {\path{arXiv:1108.0725}},
  \href {https://doi.org/10.1103/PhysRevD.85.105023}
  {\path{doi:10.1103/PhysRevD.85.105023}}.

\bibitem{Sieverding19}
A.~Sieverding, G.~Mart\'\i{}nez-Pinedo, K.~Langanke, et~al., {The $\nu$-process
  with Fully Time-dependent Supernova Neutrino Emission Spectra}, Astrophys. J.
  876~(2) (2019) 151.
\newblock \href {http://arxiv.org/abs/1902.06643} {\path{arXiv:1902.06643}},
  \href {https://doi.org/10.3847/1538-4357/ab17e2}
  {\path{doi:10.3847/1538-4357/ab17e2}}.

\bibitem{Byelikov07}
A.~Byelikov, et. al., Gamow-teller strength in the exotic odd-odd nuclei
  $^{138}\mathrm{La}$ and $^{180}\mathrm{Ta}$ and its relevance for neutrino
  nucleosynthesis, Phys. Rev. Lett. 98 (2007) 082501.
\newblock \href {https://doi.org/10.1103/PhysRevLett.98.082501}
  {\path{doi:10.1103/PhysRevLett.98.082501}}.

\bibitem{Mohr07}
P.~Mohr, F.~K\"appeler, R.~Gallino, {Survival of Nature's Rarest Isotope Ta-180
  under Stellar Conditions}, Phys. Rev. C 75 (2007) 012802.
\newblock \href {http://arxiv.org/abs/astro-ph/0612427}
  {\path{arXiv:astro-ph/0612427}}, \href
  {https://doi.org/10.1103/PhysRevC.75.012802}
  {\path{doi:10.1103/PhysRevC.75.012802}}.

\bibitem{Hayakawa10}
T.~Hayakawa, P.~Mohr, T.~Kajino, et~al., Reanalysis of the ($j=5$) state at
  $592$ kev in $^{180}\mathrm{Ta}$ and its role in the
  $\ensuremath{\nu}$-process nucleosynthesis of $^{180}\mathrm{Ta}$ in
  supernovae, Phys. Rev. C 82 (2010) 058801.
\newblock \href {https://doi.org/10.1103/PhysRevC.82.058801}
  {\path{doi:10.1103/PhysRevC.82.058801}}.

\bibitem{Sieverding20}
A.~{Sieverding}, B.~{M{\"u}ller}, Y.~Z. {Qian}, {Nucleosynthesis of an 11.8
  M$_{{\ensuremath{\odot}}}$ Supernova with 3D Simulation of the Inner Ejecta:
  Overall Yields and Implications for Short-lived Radionuclides in the Early
  Solar System}, Astrophys. J. 904~(2) (2020) 163.
\newblock \href {http://arxiv.org/abs/2008.12831} {\path{arXiv:2008.12831}},
  \href {https://doi.org/10.3847/1538-4357/abc61b}
  {\path{doi:10.3847/1538-4357/abc61b}}.

\bibitem{BMuller15a}
B.~{M{\"u}ller}, H.-T. {Janka}, {Non-radial instabilities and progenitor
  asphericities in core-collapse supernovae}, Mon. Not. R. Astro. Soc. 448
  (2015) 2141--2174.
\newblock \href {http://arxiv.org/abs/1409.4783} {\path{arXiv:1409.4783}},
  \href {https://doi.org/10.1093/mnras/stv101}
  {\path{doi:10.1093/mnras/stv101}}.

\bibitem{Banerjee16}
P.~Banerjee, Y.-Z. Qian, A.~Heger, W.~C. Haxton, {Evidence from stable isotopes
  and Be-10 for solar system formation triggered by a low-mass supernova},
  Nature Commun. 7 (2016) 3639.
\newblock \href {http://arxiv.org/abs/1611.07162} {\path{arXiv:1611.07162}},
  \href {https://doi.org/10.1038/NCOMMS13639} {\path{doi:10.1038/NCOMMS13639}}.

\bibitem{Sieverding22}
A.~Sieverding, J.~S. Randhawa, D.~Zetterberg, et~al., {Role of low-lying
  resonances for the Be10(p,\ensuremath{\alpha})Li7 reaction rate and
  implications for the formation of the Solar System}, Phys. Rev. C 106~(1)
  (2022) 015803.
\newblock \href {http://arxiv.org/abs/2203.06524} {\path{arXiv:2203.06524}},
  \href {https://doi.org/10.1103/PhysRevC.106.015803}
  {\path{doi:10.1103/PhysRevC.106.015803}}.

\bibitem{Hayakawa:2018ekx}
T.~Hayakawa, et~al., {Short-Lived Radioisotope Tc98 Synthesized by the
  Supernova Neutrino Process}, Phys. Rev. Lett. 121~(10) (2018) 102701.
\newblock \href {https://doi.org/10.1103/PhysRevLett.121.102701}
  {\path{doi:10.1103/PhysRevLett.121.102701}}.

\bibitem{ech88}
R.~I. {Epstein}, S.~A. {Colgate}, W.~C. {Haxton}, {Neutrino-induced r-process
  nucleosynthesis}, Phys. Rev. Lett. 61~(18) (1988) 2038--2041.
\newblock \href {https://doi.org/10.1103/PhysRevLett.61.2038}
  {\path{doi:10.1103/PhysRevLett.61.2038}}.

\bibitem{banerjee2011}
P.~{Banerjee}, W.~C. {Haxton}, Y.-Z. {Qian}, {Long, Cold, Early r Process?
  Neutrino-Induced Nucleosynthesis in He Shells Revisited}, Phys. Rev. Lett.
  106~(20) (2011) 201104.
\newblock \href {https://doi.org/10.1103/PhysRevLett.106.201104}
  {\path{doi:10.1103/PhysRevLett.106.201104}}.

\bibitem{banerjee2016}
P.~{Banerjee}, Y.-Z. {Qian}, A.~{Heger}, W.~{Haxton}, {Neutrino-Induced
  Nucleosynthesis in Helium Shells of Early Core-Collapse Supernovae}, in:
  European Physical Journal Web of Conferences, Vol. 109 of European Physical
  Journal Web of Conferences, 2016, p. 06001.
\newblock \href {https://doi.org/10.1051/epjconf/201610906001}
  {\path{doi:10.1051/epjconf/201610906001}}.

\bibitem{banerjee2013}
P.~{Banerjee}, Y.-Z. {Qian}, W.~C. {Haxton}, A.~{Heger}, {New Primary
  Mechanisms for the Synthesis of Rare Be9 in Early Supernovae}, Phys. Rev.
  Lett. 110~(14) (2013) 141101.
\newblock \href {https://doi.org/10.1103/PhysRevLett.110.141101}
  {\path{doi:10.1103/PhysRevLett.110.141101}}.

\bibitem{Xiong:2023uyb}
Z.~Xiong, G.~Mart\'\i{}nez-Pinedo, O.~Just, A.~Sieverding, {Production of
  $p$-nuclei from $r$-process seeds: the $\nu r$-process} (2023).
\newblock \href {http://arxiv.org/abs/2305.11050} {\path{arXiv:2305.11050}}.

\bibitem{Tamborra:2011is}
I.~Tamborra, G.~G. Raffelt, L.~Hudepohl, H.-T. Janka, {Impact of eV-mass
  sterile neutrinos on neutrino-driven supernova outflows}, JCAP 01 (2012) 013.
\newblock \href {http://arxiv.org/abs/1110.2104} {\path{arXiv:1110.2104}},
  \href {https://doi.org/10.1088/1475-7516/2012/01/013}
  {\path{doi:10.1088/1475-7516/2012/01/013}}.

\bibitem{Wu:2013gxa}
M.-R. Wu, T.~Fischer, L.~Huther, et~al., {Impact of active-sterile neutrino
  mixing on supernova explosion and nucleosynthesis}, Phys. Rev. D 89~(6)
  (2014) 061303.
\newblock \href {http://arxiv.org/abs/1305.2382} {\path{arXiv:1305.2382}},
  \href {https://doi.org/10.1103/PhysRevD.89.061303}
  {\path{doi:10.1103/PhysRevD.89.061303}}.

\bibitem{Warren:2014qza}
M.~L. Warren, M.~Meixner, G.~Mathews, et~al., {Sterile neutrino oscillations in
  core-collapse supernovae}, Phys. Rev. D 90~(10) (2014) 103007.
\newblock \href {http://arxiv.org/abs/1405.6101} {\path{arXiv:1405.6101}},
  \href {https://doi.org/10.1103/PhysRevD.90.103007}
  {\path{doi:10.1103/PhysRevD.90.103007}}.

\bibitem{Xiong:2019nvw}
Z.~Xiong, M.-R. Wu, Y.-Z. Qian, {Active-sterile Neutrino Oscillations in
  Neutrino-driven Winds: Implications for Nucleosynthesis}, Astrophys. J.
  880~(2) (2019) 81.
\newblock \href {http://arxiv.org/abs/1904.09371} {\path{arXiv:1904.09371}},
  \href {https://doi.org/10.3847/1538-4357/ab2870}
  {\path{doi:10.3847/1538-4357/ab2870}}.

\bibitem{Ko:2019asm}
H.~Ko, D.~Jang, M.~Kusakabe, M.-K. Cheoun, {The Viability of the 3 + 1 Neutrino
  Model in the Supernova Neutrino Process}, Astrophys. J. 894~(2) (2020) 99.
\newblock \href {http://arxiv.org/abs/1910.04984} {\path{arXiv:1910.04984}},
  \href {https://doi.org/10.3847/1538-4357/ab84e4}
  {\path{doi:10.3847/1538-4357/ab84e4}}.

\bibitem{Suliga:2019bsq}
A.~M. Suliga, I.~Tamborra, M.-R. Wu, {Tau lepton asymmetry by sterile neutrino
  emission -- Moving beyond one-zone supernova models}, JCAP 12 (2019) 019.
\newblock \href {http://arxiv.org/abs/1908.11382} {\path{arXiv:1908.11382}},
  \href {https://doi.org/10.1088/1475-7516/2019/12/019}
  {\path{doi:10.1088/1475-7516/2019/12/019}}.

\bibitem{Syvolap:2019dat}
V.~Syvolap, O.~Ruchayskiy, A.~Boyarsky, {Resonance production of keV sterile
  neutrinos in core-collapse supernovae and lepton number diffusion}, Phys.
  Rev. D 106~(1) (2022) 015017.
\newblock \href {http://arxiv.org/abs/1909.06320} {\path{arXiv:1909.06320}},
  \href {https://doi.org/10.1103/PhysRevD.106.015017}
  {\path{doi:10.1103/PhysRevD.106.015017}}.

\bibitem{Suliga:2020vpz}
A.~M. Suliga, I.~Tamborra, M.-R. Wu, {Lifting the core-collapse supernova
  bounds on keV-mass sterile neutrinos}, JCAP 08 (2020) 018.
\newblock \href {http://arxiv.org/abs/2004.11389} {\path{arXiv:2004.11389}},
  \href {https://doi.org/10.1088/1475-7516/2020/08/018}
  {\path{doi:10.1088/1475-7516/2020/08/018}}.

\bibitem{Tang:2020pkp}
J.~Tang, T.~Wang, M.-R. Wu, {Constraining sterile neutrinos by core-collapse
  supernovae with multiple detectors}, JCAP 10 (2020) 038.
\newblock \href {http://arxiv.org/abs/2005.09168} {\path{arXiv:2005.09168}},
  \href {https://doi.org/10.1088/1475-7516/2020/10/038}
  {\path{doi:10.1088/1475-7516/2020/10/038}}.

\bibitem{Sigurdarson:2022mcm}
G.~Sigur\dh{}arson, I.~Tamborra, M.-R. Wu, {Resonant production of light
  sterile neutrinos in compact binary merger remnants}, Phys. Rev. D 106~(12)
  (2022) 123030.
\newblock \href {http://arxiv.org/abs/2209.07544} {\path{arXiv:2209.07544}},
  \href {https://doi.org/10.1103/PhysRevD.106.123030}
  {\path{doi:10.1103/PhysRevD.106.123030}}.

\bibitem{Ray:2023gtu}
A.~Ray, Y.-Z. Qian, {Evolution of tau-neutrino lepton number in protoneutron
  stars due to active-sterile neutrino mixing}, Phys. Rev. D 108~(6) (2023)
  063025.
\newblock \href {http://arxiv.org/abs/2306.08209} {\path{arXiv:2306.08209}},
  \href {https://doi.org/10.1103/PhysRevD.108.063025}
  {\path{doi:10.1103/PhysRevD.108.063025}}.

\bibitem{Patwardhan:2022mxg}
A.~V. Patwardhan, M.~J. Cervia, E.~Rrapaj, et~al., {Many-body collective
  neutrino oscillations: recent developments} (2022).
\newblock \href {http://arxiv.org/abs/2301.00342} {\path{arXiv:2301.00342}},
  \href {https://doi.org/10.1007/978-981-15-8818-1_126-1}
  {\path{doi:10.1007/978-981-15-8818-1_126-1}}.

\bibitem{Shalgar:2023ooi}
S.~Shalgar, I.~Tamborra, {Do we have enough evidence to invalidate the
  mean-field approximation adopted to model collective neutrino oscillations?},
  Phys. Rev. D 107~(12) (2023) 123004.
\newblock \href {http://arxiv.org/abs/2304.13050} {\path{arXiv:2304.13050}},
  \href {https://doi.org/10.1103/PhysRevD.107.123004}
  {\path{doi:10.1103/PhysRevD.107.123004}}.

\bibitem{Johns:2023ewj}
L.~Johns, {Neutrino many-body correlations} (2023).
\newblock \href {http://arxiv.org/abs/2305.04916} {\path{arXiv:2305.04916}}.

\bibitem{Vlasenko:2013fja}
A.~Vlasenko, G.~M. Fuller, V.~Cirigliano, {Neutrino Quantum Kinetics}, Phys.
  Rev. D 89~(10) (2014) 105004.
\newblock \href {http://arxiv.org/abs/1309.2628} {\path{arXiv:1309.2628}},
  \href {https://doi.org/10.1103/PhysRevD.89.105004}
  {\path{doi:10.1103/PhysRevD.89.105004}}.

\bibitem{Richers:2019grc}
S.~A. Richers, G.~C. McLaughlin, J.~P. Kneller, A.~Vlasenko, {Neutrino Quantum
  Kinetics in Compact Objects}, Phys. Rev. D 99~(12) (2019) 123014.
\newblock \href {http://arxiv.org/abs/1903.00022} {\path{arXiv:1903.00022}},
  \href {https://doi.org/10.1103/PhysRevD.99.123014}
  {\path{doi:10.1103/PhysRevD.99.123014}}.

\bibitem{Nagakura:2022qko}
H.~Nagakura, {General-relativistic quantum-kinetics neutrino transport}, Phys.
  Rev. D 106~(6) (2022) 063011.
\newblock \href {http://arxiv.org/abs/2206.04098} {\path{arXiv:2206.04098}},
  \href {https://doi.org/10.1103/PhysRevD.106.063011}
  {\path{doi:10.1103/PhysRevD.106.063011}}.

\bibitem{Blaschke:2016xxt}
D.~N. Blaschke, V.~Cirigliano, {Neutrino Quantum Kinetic Equations: The
  Collision Term}, Phys. Rev. D 94~(3) (2016) 033009.
\newblock \href {http://arxiv.org/abs/1605.09383} {\path{arXiv:1605.09383}},
  \href {https://doi.org/10.1103/PhysRevD.94.033009}
  {\path{doi:10.1103/PhysRevD.94.033009}}.

\bibitem{Pastor:2001iu}
S.~Pastor, G.~G. Raffelt, D.~V. Semikoz, {Physics of synchronized neutrino
  oscillations caused by selfinteractions}, Phys. Rev. D 65 (2002) 053011.
\newblock \href {http://arxiv.org/abs/hep-ph/0109035}
  {\path{arXiv:hep-ph/0109035}}, \href
  {https://doi.org/10.1103/PhysRevD.65.053011}
  {\path{doi:10.1103/PhysRevD.65.053011}}.

\bibitem{Duan:2005cp}
H.~Duan, G.~M. Fuller, Y.-Z. Qian, {Collective neutrino flavor transformation
  in supernovae}, Phys. Rev. D 74 (2006) 123004.
\newblock \href {http://arxiv.org/abs/astro-ph/0511275}
  {\path{arXiv:astro-ph/0511275}}, \href
  {https://doi.org/10.1103/PhysRevD.74.123004}
  {\path{doi:10.1103/PhysRevD.74.123004}}.

\bibitem{Hannestad:2006nj}
S.~Hannestad, G.~G. Raffelt, G.~Sigl, Y.~Y.~Y. Wong, {Self-induced conversion
  in dense neutrino gases: Pendulum in flavour space}, Phys. Rev. D 74 (2006)
  105010, [Erratum: Phys.Rev.D 76, 029901 (2007)].
\newblock \href {http://arxiv.org/abs/astro-ph/0608695}
  {\path{arXiv:astro-ph/0608695}}, \href
  {https://doi.org/10.1103/PhysRevD.74.105010}
  {\path{doi:10.1103/PhysRevD.74.105010}}.

\bibitem{Duan:2006an}
H.~Duan, G.~M. Fuller, J.~Carlson, Y.-Z. Qian, {Simulation of Coherent
  Non-Linear Neutrino Flavor Transformation in the Supernova Environment. 1.
  Correlated Neutrino Trajectories}, Phys. Rev. D 74 (2006) 105014.
\newblock \href {http://arxiv.org/abs/astro-ph/0606616}
  {\path{arXiv:astro-ph/0606616}}, \href
  {https://doi.org/10.1103/PhysRevD.74.105014}
  {\path{doi:10.1103/PhysRevD.74.105014}}.

\bibitem{Fogli:2007bk}
G.~L. Fogli, E.~Lisi, A.~Marrone, A.~Mirizzi, {Collective neutrino flavor
  transitions in supernovae and the role of trajectory averaging}, JCAP 12
  (2007) 010.
\newblock \href {http://arxiv.org/abs/0707.1998} {\path{arXiv:0707.1998}},
  \href {https://doi.org/10.1088/1475-7516/2007/12/010}
  {\path{doi:10.1088/1475-7516/2007/12/010}}.

\bibitem{Dasgupta:2009mg}
B.~Dasgupta, A.~Dighe, G.~G. Raffelt, A.~Y. Smirnov, {Multiple Spectral Splits
  of Supernova Neutrinos}, Phys. Rev. Lett. 103 (2009) 051105.
\newblock \href {http://arxiv.org/abs/0904.3542} {\path{arXiv:0904.3542}},
  \href {https://doi.org/10.1103/PhysRevLett.103.051105}
  {\path{doi:10.1103/PhysRevLett.103.051105}}.

\bibitem{Banerjee:2011fj}
A.~Banerjee, A.~Dighe, G.~Raffelt, {Linearized flavor-stability analysis of
  dense neutrino streams}, Phys. Rev. D 84 (2011) 053013.
\newblock \href {http://arxiv.org/abs/1107.2308} {\path{arXiv:1107.2308}},
  \href {https://doi.org/10.1103/PhysRevD.84.053013}
  {\path{doi:10.1103/PhysRevD.84.053013}}.

\bibitem{Dasgupta:2021gfs}
B.~Dasgupta, {Collective Neutrino Flavor Instability Requires a Crossing},
  Phys. Rev. Lett. 128~(8) (2022) 081102.
\newblock \href {http://arxiv.org/abs/2110.00192} {\path{arXiv:2110.00192}},
  \href {https://doi.org/10.1103/PhysRevLett.128.081102}
  {\path{doi:10.1103/PhysRevLett.128.081102}}.

\bibitem{Chakraborty:2011nf}
S.~Chakraborty, T.~Fischer, A.~Mirizzi, et~al., {No collective neutrino flavor
  conversions during the supernova accretion phase}, Phys. Rev. Lett. 107
  (2011) 151101.
\newblock \href {http://arxiv.org/abs/1104.4031} {\path{arXiv:1104.4031}},
  \href {https://doi.org/10.1103/PhysRevLett.107.151101}
  {\path{doi:10.1103/PhysRevLett.107.151101}}.

\bibitem{Chakraborty:2016yeg}
S.~Chakraborty, R.~Hansen, I.~Izaguirre, G.~Raffelt, {Collective neutrino
  flavor conversion: Recent developments}, Nucl. Phys. B 908 (2016) 366--381.
\newblock \href {http://arxiv.org/abs/1602.02766} {\path{arXiv:1602.02766}},
  \href {https://doi.org/10.1016/j.nuclphysb.2016.02.012}
  {\path{doi:10.1016/j.nuclphysb.2016.02.012}}.

\bibitem{Abbar:2015fwa}
S.~Abbar, H.~Duan, {Neutrino flavor instabilities in a time-dependent supernova
  model}, Phys. Lett. B 751 (2015) 43--47.
\newblock \href {http://arxiv.org/abs/1509.01538} {\path{arXiv:1509.01538}},
  \href {https://doi.org/10.1016/j.physletb.2015.10.019}
  {\path{doi:10.1016/j.physletb.2015.10.019}}.

\bibitem{Dasgupta:2015iia}
B.~Dasgupta, A.~Mirizzi, {Temporal Instability Enables Neutrino Flavor
  Conversions Deep Inside Supernovae}, Phys. Rev. D 92~(12) (2015) 125030.
\newblock \href {http://arxiv.org/abs/1509.03171} {\path{arXiv:1509.03171}},
  \href {https://doi.org/10.1103/PhysRevD.92.125030}
  {\path{doi:10.1103/PhysRevD.92.125030}}.

\bibitem{Nagakura:2021hyb}
H.~Nagakura, L.~Johns, A.~Burrows, G.~M. Fuller, {Where, when, and why:
  Occurrence of fast-pairwise collective neutrino oscillation in
  three-dimensional core-collapse supernova models}, Phys. Rev. D 104~(8)
  (2021) 083025.
\newblock \href {http://arxiv.org/abs/2108.07281} {\path{arXiv:2108.07281}},
  \href {https://doi.org/10.1103/PhysRevD.104.083025}
  {\path{doi:10.1103/PhysRevD.104.083025}}.

\bibitem{Abbar:2023kta}
S.~Abbar, {Applications of machine learning to detecting fast neutrino flavor
  instabilities in core-collapse supernova and neutron star merger models},
  Phys. Rev. D 107~(10) (2023) 103006.
\newblock \href {http://arxiv.org/abs/2303.05560} {\path{arXiv:2303.05560}},
  \href {https://doi.org/10.1103/PhysRevD.107.103006}
  {\path{doi:10.1103/PhysRevD.107.103006}}.

\bibitem{Sawyer:2015dsa}
R.~F. Sawyer, {Neutrino cloud instabilities just above the neutrino sphere of a
  supernova}, Phys. Rev. Lett. 116~(8) (2016) 081101.
\newblock \href {http://arxiv.org/abs/1509.03323} {\path{arXiv:1509.03323}},
  \href {https://doi.org/10.1103/PhysRevLett.116.081101}
  {\path{doi:10.1103/PhysRevLett.116.081101}}.

\bibitem{Morinaga:2021vmc}
T.~Morinaga, {Fast neutrino flavor instability and neutrino flavor lepton
  number crossings}, Phys. Rev. D 105~(10) (2022) L101301.
\newblock \href {http://arxiv.org/abs/2103.15267} {\path{arXiv:2103.15267}},
  \href {https://doi.org/10.1103/PhysRevD.105.L101301}
  {\path{doi:10.1103/PhysRevD.105.L101301}}.

\bibitem{Bhattacharyya:2020jpj}
S.~Bhattacharyya, B.~Dasgupta, {Fast Flavor Depolarization of Supernova
  Neutrinos}, Phys. Rev. Lett. 126~(6) (2021) 061302.
\newblock \href {http://arxiv.org/abs/2009.03337} {\path{arXiv:2009.03337}},
  \href {https://doi.org/10.1103/PhysRevLett.126.061302}
  {\path{doi:10.1103/PhysRevLett.126.061302}}.

\bibitem{Wu:2021uvt}
M.-R. Wu, M.~George, C.-Y. Lin, Z.~Xiong, {Collective fast neutrino flavor
  conversions in a 1D box: Initial conditions and long-term evolution}, Phys.
  Rev. D 104~(10) (2021) 103003.
\newblock \href {http://arxiv.org/abs/2108.09886} {\path{arXiv:2108.09886}},
  \href {https://doi.org/10.1103/PhysRevD.104.103003}
  {\path{doi:10.1103/PhysRevD.104.103003}}.

\bibitem{Richers:2021xtf}
S.~Richers, D.~Willcox, N.~Ford, {Neutrino fast flavor instability in three
  dimensions}, Phys. Rev. D 104~(10) (2021) 103023.
\newblock \href {http://arxiv.org/abs/2109.08631} {\path{arXiv:2109.08631}},
  \href {https://doi.org/10.1103/PhysRevD.104.103023}
  {\path{doi:10.1103/PhysRevD.104.103023}}.

\bibitem{Bhattacharyya:2022eed}
S.~Bhattacharyya, B.~Dasgupta, {Elaborating the ultimate fate of fast
  collective neutrino flavor oscillations}, Phys. Rev. D 106~(10) (2022)
  103039.
\newblock \href {http://arxiv.org/abs/2205.05129} {\path{arXiv:2205.05129}},
  \href {https://doi.org/10.1103/PhysRevD.106.103039}
  {\path{doi:10.1103/PhysRevD.106.103039}}.

\bibitem{Capozzi:2022dtr}
F.~Capozzi, M.~Chakraborty, S.~Chakraborty, M.~Sen, {Supernova fast flavor
  conversions in 1+1D: Influence of mu-tau neutrinos}, Phys. Rev. D 106~(8)
  (2022) 083011.
\newblock \href {http://arxiv.org/abs/2205.06272} {\path{arXiv:2205.06272}},
  \href {https://doi.org/10.1103/PhysRevD.106.083011}
  {\path{doi:10.1103/PhysRevD.106.083011}}.

\bibitem{Richers:2022bkd}
S.~Richers, H.~Duan, M.-R. Wu, et~al., {Code comparison for fast flavor
  instability simulations}, Phys. Rev. D 106~(4) (2022) 043011.
\newblock \href {http://arxiv.org/abs/2205.06282} {\path{arXiv:2205.06282}},
  \href {https://doi.org/10.1103/PhysRevD.106.043011}
  {\path{doi:10.1103/PhysRevD.106.043011}}.

\bibitem{Grohs:2022fyq}
E.~Grohs, S.~Richers, S.~M. Couch, et~al., {Neutrino fast flavor instability in
  three dimensions for a neutron star merger}, Phys. Lett. B 846 (2023) 138210.
\newblock \href {http://arxiv.org/abs/2207.02214} {\path{arXiv:2207.02214}},
  \href {https://doi.org/10.1016/j.physletb.2023.138210}
  {\path{doi:10.1016/j.physletb.2023.138210}}.

\bibitem{Nagakura:2022xwe}
H.~Nagakura, M.~Zaizen, {Connecting small-scale to large-scale structures of
  fast neutrino-flavor conversion}, Phys. Rev. D 107~(6) (2023) 063033.
\newblock \href {http://arxiv.org/abs/2211.01398} {\path{arXiv:2211.01398}},
  \href {https://doi.org/10.1103/PhysRevD.107.063033}
  {\path{doi:10.1103/PhysRevD.107.063033}}.

\bibitem{Nagakura:2022kic}
H.~Nagakura, M.~Zaizen, {Time-Dependent and Quasisteady Features of Fast
  Neutrino-Flavor Conversion}, Phys. Rev. Lett. 129~(26) (2022) 261101.
\newblock \href {http://arxiv.org/abs/2206.04097} {\path{arXiv:2206.04097}},
  \href {https://doi.org/10.1103/PhysRevLett.129.261101}
  {\path{doi:10.1103/PhysRevLett.129.261101}}.

\bibitem{Shalgar:2022rjj}
S.~Shalgar, I.~Tamborra, {Neutrino decoupling is altered by flavor conversion},
  Phys. Rev. D 108~(4) (2023) 043006.
\newblock \href {http://arxiv.org/abs/2206.00676} {\path{arXiv:2206.00676}},
  \href {https://doi.org/10.1103/PhysRevD.108.043006}
  {\path{doi:10.1103/PhysRevD.108.043006}}.

\bibitem{Shalgar:2022lvv}
S.~Shalgar, I.~Tamborra, {Neutrino flavor conversion, advection, and
  collisions: Toward the full solution}, Phys. Rev. D 107~(6) (2023) 063025.
\newblock \href {http://arxiv.org/abs/2207.04058} {\path{arXiv:2207.04058}},
  \href {https://doi.org/10.1103/PhysRevD.107.063025}
  {\path{doi:10.1103/PhysRevD.107.063025}}.

\bibitem{Zaizen:2023ihz}
M.~Zaizen, H.~Nagakura, {Characterizing quasisteady states of fast
  neutrino-flavor conversion by stability and conservation laws}, Phys. Rev. D
  107~(12) (2023) 123021.
\newblock \href {http://arxiv.org/abs/2304.05044} {\path{arXiv:2304.05044}},
  \href {https://doi.org/10.1103/PhysRevD.107.123021}
  {\path{doi:10.1103/PhysRevD.107.123021}}.

\bibitem{Xiong:2023vcm}
Z.~Xiong, M.-R. Wu, S.~Abbar, et~al., {Evaluating approximate asymptotic
  distributions for fast neutrino flavor conversions in a periodic 1D box},
  Phys. Rev. D 108~(6) (2023) 063003.
\newblock \href {http://arxiv.org/abs/2307.11129} {\path{arXiv:2307.11129}},
  \href {https://doi.org/10.1103/PhysRevD.108.063003}
  {\path{doi:10.1103/PhysRevD.108.063003}}.

\bibitem{Abbar:2023ltx}
S.~Abbar, M.-R. Wu, Z.~Xiong, {Physics-Informed Neural Networks for Predicting
  the Asymptotic Outcome of Fast Neutrino Flavor Conversions} (11 2023).
\newblock \href {http://arxiv.org/abs/2311.15656} {\path{arXiv:2311.15656}}.

\bibitem{Cornelius:2023eop}
M.~Cornelius, S.~Shalgar, I.~Tamborra, {Perturbing Fast Neutrino Flavor
  Conversion} (12 2023).
\newblock \href {http://arxiv.org/abs/2312.03839} {\path{arXiv:2312.03839}}.

\bibitem{Nagakura:2023jfi}
H.~Nagakura, L.~Johns, M.~Zaizen, {BGK subgrid model for neutrino quantum
  kinetics} (12 2023).
\newblock \href {http://arxiv.org/abs/2312.16285} {\path{arXiv:2312.16285}}.

\bibitem{Ehring:2023lcd}
J.~Ehring, S.~Abbar, H.-T. Janka, G.~Raffelt, {Fast neutrino flavor conversion
  in core-collapse supernovae: A parametric study in 1D models}, Phys. Rev. D
  107~(10) (2023) 103034.
\newblock \href {http://arxiv.org/abs/2301.11938} {\path{arXiv:2301.11938}},
  \href {https://doi.org/10.1103/PhysRevD.107.103034}
  {\path{doi:10.1103/PhysRevD.107.103034}}.

\bibitem{Nagakura:2023mhr}
H.~Nagakura, {Roles of Fast Neutrino-Flavor Conversion on the Neutrino-Heating
  Mechanism of Core-Collapse Supernova}, Phys. Rev. Lett. 130~(21) (2023)
  211401.
\newblock \href {http://arxiv.org/abs/2301.10785} {\path{arXiv:2301.10785}},
  \href {https://doi.org/10.1103/PhysRevLett.130.211401}
  {\path{doi:10.1103/PhysRevLett.130.211401}}.

\bibitem{Ehring:2023abs}
J.~Ehring, S.~Abbar, H.-T. Janka, et~al., {Fast Neutrino Flavor Conversions Can
  Help and Hinder Neutrino-Driven Explosions}, Phys. Rev. Lett. 131~(6) (2023)
  061401.
\newblock \href {http://arxiv.org/abs/2305.11207} {\path{arXiv:2305.11207}},
  \href {https://doi.org/10.1103/PhysRevLett.131.061401}
  {\path{doi:10.1103/PhysRevLett.131.061401}}.

\bibitem{Xiong:2020ntn}
Z.~Xiong, A.~Sieverding, M.~Sen, Y.-Z. Qian, {Potential Impact of Fast Flavor
  Oscillations on Neutrino-driven Winds and Their Nucleosynthesis}, Astrophys.
  J. 900~(2) (2020) 144.
\newblock \href {http://arxiv.org/abs/2006.11414} {\path{arXiv:2006.11414}},
  \href {https://doi.org/10.3847/1538-4357/abac5e}
  {\path{doi:10.3847/1538-4357/abac5e}}.

\bibitem{Wu:2017qpc}
M.-R. Wu, I.~Tamborra, {Fast neutrino conversions: Ubiquitous in compact binary
  merger remnants}, Phys. Rev. D 95~(10) (2017) 103007.
\newblock \href {http://arxiv.org/abs/1701.06580} {\path{arXiv:1701.06580}},
  \href {https://doi.org/10.1103/PhysRevD.95.103007}
  {\path{doi:10.1103/PhysRevD.95.103007}}.

\bibitem{Wu:2017drk}
M.-R. Wu, I.~Tamborra, O.~Just, H.-T. Janka, {Imprints of neutrino-pair flavor
  conversions on nucleosynthesis in ejecta from neutron-star merger remnants},
  Phys. Rev. D 96~(12) (2017) 123015.
\newblock \href {http://arxiv.org/abs/1711.00477} {\path{arXiv:1711.00477}},
  \href {https://doi.org/10.1103/PhysRevD.96.123015}
  {\path{doi:10.1103/PhysRevD.96.123015}}.

\bibitem{Richers:2022dqa}
S.~Richers, {Evaluating approximate flavor instability metrics in neutron star
  mergers}, Phys. Rev. D 106~(8) (2022) 083005.
\newblock \href {http://arxiv.org/abs/2206.08444} {\path{arXiv:2206.08444}},
  \href {https://doi.org/10.1103/PhysRevD.106.083005}
  {\path{doi:10.1103/PhysRevD.106.083005}}.

\bibitem{Nagakura:2023wbf}
H.~Nagakura, {Global features of fast neutrino-flavor conversion in binary
  neutron star mergers}, Phys. Rev. D 108~(10) (2023) 103014.
\newblock \href {http://arxiv.org/abs/2306.10108} {\path{arXiv:2306.10108}},
  \href {https://doi.org/10.1103/PhysRevD.108.103014}
  {\path{doi:10.1103/PhysRevD.108.103014}}.

\bibitem{Froustey:2023skf}
J.~Froustey, S.~Richers, E.~Grohs, et~al., {Neutrino fast flavor oscillations
  with moments: linear stability analysis and application to neutron star
  mergers} (11 2023).
\newblock \href {http://arxiv.org/abs/2311.11968} {\path{arXiv:2311.11968}}.

\bibitem{Li:2021vqj}
X.~Li, D.~M. Siegel, {Neutrino Fast Flavor Conversions in Neutron-Star
  Postmerger Accretion Disks}, Phys. Rev. Lett. 126~(25) (2021) 251101.
\newblock \href {http://arxiv.org/abs/2103.02616} {\path{arXiv:2103.02616}},
  \href {https://doi.org/10.1103/PhysRevLett.126.251101}
  {\path{doi:10.1103/PhysRevLett.126.251101}}.

\bibitem{Fernandez:2022yyv}
R.~Fern\'andez, S.~Richers, N.~Mulyk, S.~Fahlman, {Fast flavor instability in
  hypermassive neutron star disk outflows}, Phys. Rev. D 106~(10) (2022)
  103003.
\newblock \href {http://arxiv.org/abs/2207.10680} {\path{arXiv:2207.10680}},
  \href {https://doi.org/10.1103/PhysRevD.106.103003}
  {\path{doi:10.1103/PhysRevD.106.103003}}.

\bibitem{Pastor:2002we}
S.~Pastor, G.~Raffelt, {Flavor oscillations in the supernova hot bubble region:
  Nonlinear effects of neutrino background}, Phys. Rev. Lett. 89 (2002) 191101.
\newblock \href {http://arxiv.org/abs/astro-ph/0207281}
  {\path{arXiv:astro-ph/0207281}}, \href
  {https://doi.org/10.1103/PhysRevLett.89.191101}
  {\path{doi:10.1103/PhysRevLett.89.191101}}.

\bibitem{Duan:2007bt}
H.~Duan, G.~M. Fuller, J.~Carlson, Y.-Z. Qian, {Neutrino Mass Hierarchy and
  Stepwise Spectral Swapping of Supernova Neutrino Flavors}, Phys. Rev. Lett.
  99 (2007) 241802.
\newblock \href {http://arxiv.org/abs/0707.0290} {\path{arXiv:0707.0290}},
  \href {https://doi.org/10.1103/PhysRevLett.99.241802}
  {\path{doi:10.1103/PhysRevLett.99.241802}}.

\bibitem{Duan:2007sh}
H.~Duan, G.~M. Fuller, J.~Carlson, Y.-Z. Qian, {Flavor Evolution of the
  Neutronization Neutrino Burst from an O-Ne-Mg Core-Collapse Supernova}, Phys.
  Rev. Lett. 100 (2008) 021101.
\newblock \href {http://arxiv.org/abs/0710.1271} {\path{arXiv:0710.1271}},
  \href {https://doi.org/10.1103/PhysRevLett.100.021101}
  {\path{doi:10.1103/PhysRevLett.100.021101}}.

\bibitem{Malkus:2012ts}
A.~Malkus, J.~P. Kneller, G.~C. McLaughlin, R.~Surman, {Neutrino oscillations
  above black hole accretion disks: disks with electron-flavor emission}, Phys.
  Rev. D 86 (2012) 085015.
\newblock \href {http://arxiv.org/abs/1207.6648} {\path{arXiv:1207.6648}},
  \href {https://doi.org/10.1103/PhysRevD.86.085015}
  {\path{doi:10.1103/PhysRevD.86.085015}}.

\bibitem{Wu:2015fga}
M.-R. Wu, H.~Duan, Y.-Z. Qian, {Physics of neutrino flavor transformation
  through matter\textendash{}neutrino resonances}, Phys. Lett. B 752 (2016)
  89--94.
\newblock \href {http://arxiv.org/abs/1509.08975} {\path{arXiv:1509.08975}},
  \href {https://doi.org/10.1016/j.physletb.2015.11.027}
  {\path{doi:10.1016/j.physletb.2015.11.027}}.

\bibitem{Shalgar:2017pzd}
S.~Shalgar, {Multi-angle calculation of the matter-neutrino resonance near an
  accretion disk}, JCAP 02 (2018) 010.
\newblock \href {http://arxiv.org/abs/1707.07692} {\path{arXiv:1707.07692}},
  \href {https://doi.org/10.1088/1475-7516/2018/02/010}
  {\path{doi:10.1088/1475-7516/2018/02/010}}.

\bibitem{Johns:2021qby}
L.~Johns, {Collisional Flavor Instabilities of Supernova Neutrinos}, Phys. Rev.
  Lett. 130~(19) (2023) 191001.
\newblock \href {http://arxiv.org/abs/2104.11369} {\path{arXiv:2104.11369}},
  \href {https://doi.org/10.1103/PhysRevLett.130.191001}
  {\path{doi:10.1103/PhysRevLett.130.191001}}.

\bibitem{Xiong:2022zqz}
Z.~Xiong, L.~Johns, M.-R. Wu, H.~Duan, {Collisional flavor instability in dense
  neutrino gases}, Phys. Rev. D 108~(8) (2023) 083002.
\newblock \href {http://arxiv.org/abs/2212.03750} {\path{arXiv:2212.03750}},
  \href {https://doi.org/10.1103/PhysRevD.108.083002}
  {\path{doi:10.1103/PhysRevD.108.083002}}.

\bibitem{Liu:2023pjw}
J.~Liu, M.~Zaizen, S.~Yamada, {Systematic study of the resonancelike structure
  in the collisional flavor instability of neutrinos}, Phys. Rev. D 107~(12)
  (2023) 123011.
\newblock \href {https://doi.org/10.1103/PhysRevD.107.123011}
  {\path{doi:10.1103/PhysRevD.107.123011}}.

\bibitem{Xiong:2022vsy}
Z.~Xiong, M.-R. Wu, G.~Mart\'\i{}nez-Pinedo, et~al., {Evolution of collisional
  neutrino flavor instabilities in spherically symmetric supernova models},
  Phys. Rev. D 107~(8) (2023) 083016.
\newblock \href {http://arxiv.org/abs/2210.08254} {\path{arXiv:2210.08254}},
  \href {https://doi.org/10.1103/PhysRevD.107.083016}
  {\path{doi:10.1103/PhysRevD.107.083016}}.

\bibitem{Liu:2023vtz}
J.~Liu, H.~Nagakura, R.~Akaho, et~al., {Universality of the neutrino
  collisional flavor instability in core-collapse supernovae}, Phys. Rev. D
  108~(12) (2023) 123024.
\newblock \href {http://arxiv.org/abs/2310.05050} {\path{arXiv:2310.05050}},
  \href {https://doi.org/10.1103/PhysRevD.108.123024}
  {\path{doi:10.1103/PhysRevD.108.123024}}.

\bibitem{Akaho:2023brj}
R.~Akaho, J.~Liu, H.~Nagakura, et~al., {Collisional and fast neutrino flavor
  instabilities in two-dimensional core-collapse supernova simulation with
  Boltzmann neutrino transport}, Phys. Rev. D 109~(2) (2024) 023012.
\newblock \href {http://arxiv.org/abs/2311.11272} {\path{arXiv:2311.11272}},
  \href {https://doi.org/10.1103/PhysRevD.109.023012}
  {\path{doi:10.1103/PhysRevD.109.023012}}.

\bibitem{Shalgar:2023aca}
S.~Shalgar, I.~Tamborra, {Do Neutrinos Become Flavor Unstable Due to Collisions
  with Matter in the Supernova Decoupling Region?} (2023).
\newblock \href {http://arxiv.org/abs/2307.10366} {\path{arXiv:2307.10366}}.

\bibitem{Johns:2022yqy}
L.~Johns, Z.~Xiong, {Collisional instabilities of neutrinos and their interplay
  with fast flavor conversion in compact objects}, Phys. Rev. D 106~(10) (2022)
  103029.
\newblock \href {http://arxiv.org/abs/2208.11059} {\path{arXiv:2208.11059}},
  \href {https://doi.org/10.1103/PhysRevD.106.103029}
  {\path{doi:10.1103/PhysRevD.106.103029}}.

\bibitem{Padilla-Gay:2022wck}
I.~Padilla-Gay, I.~Tamborra, G.~G. Raffelt, {Neutrino fast flavor pendulum. II.
  Collisional damping}, Phys. Rev. D 106~(10) (2022) 103031.
\newblock \href {http://arxiv.org/abs/2209.11235} {\path{arXiv:2209.11235}},
  \href {https://doi.org/10.1103/PhysRevD.106.103031}
  {\path{doi:10.1103/PhysRevD.106.103031}}.

\bibitem{Kato:2023dcw}
C.~Kato, H.~Nagakura, M.~Zaizen, {Flavor conversions with energy-dependent
  neutrino emission and absorption}, Phys. Rev. D 108~(2) (2023) 023006.
\newblock \href {http://arxiv.org/abs/2303.16453} {\path{arXiv:2303.16453}},
  \href {https://doi.org/10.1103/PhysRevD.108.023006}
  {\path{doi:10.1103/PhysRevD.108.023006}}.

\bibitem{Dighe:1999bi}
A.~S. Dighe, A.~Y. Smirnov, {Identifying the neutrino mass spectrum from the
  neutrino burst from a supernova}, Phys. Rev. D 62 (2000) 033007.
\newblock \href {http://arxiv.org/abs/hep-ph/9907423}
  {\path{arXiv:hep-ph/9907423}}, \href
  {https://doi.org/10.1103/PhysRevD.62.033007}
  {\path{doi:10.1103/PhysRevD.62.033007}}.

\bibitem{Serpico:2011ir}
P.~D. Serpico, S.~Chakraborty, T.~Fischer, et~al., {Probing the neutrino mass
  hierarchy with the rise time of a supernova burst}, Phys. Rev. D 85 (2012)
  085031.
\newblock \href {http://arxiv.org/abs/1111.4483} {\path{arXiv:1111.4483}},
  \href {https://doi.org/10.1103/PhysRevD.85.085031}
  {\path{doi:10.1103/PhysRevD.85.085031}}.

\bibitem{Scholberg:2017czd}
K.~Scholberg, {Supernova Signatures of Neutrino Mass Ordering}, J. Phys. G
  45~(1) (2018) 014002.
\newblock \href {http://arxiv.org/abs/1707.06384} {\path{arXiv:1707.06384}},
  \href {https://doi.org/10.1088/1361-6471/aa97be}
  {\path{doi:10.1088/1361-6471/aa97be}}.

\bibitem{Brdar:2022vfr}
V.~Brdar, X.-J. Xu, {Timing and multi-channel: novel method for determining the
  neutrino mass ordering from supernovae}, JCAP 08 (2022) 067.
\newblock \href {http://arxiv.org/abs/2204.13135} {\path{arXiv:2204.13135}},
  \href {https://doi.org/10.1088/1475-7516/2022/08/067}
  {\path{doi:10.1088/1475-7516/2022/08/067}}.

\bibitem{Takahashi:2002yj}
K.~Takahashi, K.~Sato, H.~E. Dalhed, J.~R. Wilson, {Shock propagation and
  neutrino oscillation in supernova}, Astropart. Phys. 20 (2003) 189--193.
\newblock \href {http://arxiv.org/abs/astro-ph/0212195}
  {\path{arXiv:astro-ph/0212195}}, \href
  {https://doi.org/10.1016/S0927-6505(03)00175-0}
  {\path{doi:10.1016/S0927-6505(03)00175-0}}.

\bibitem{Fogli:2006xy}
G.~L. Fogli, E.~Lisi, A.~Mirizzi, D.~Montanino, {Damping of supernova neutrino
  transitions in stochastic shock-wave density profiles}, JCAP 06 (2006) 012.
\newblock \href {http://arxiv.org/abs/hep-ph/0603033}
  {\path{arXiv:hep-ph/0603033}}, \href
  {https://doi.org/10.1088/1475-7516/2006/06/012}
  {\path{doi:10.1088/1475-7516/2006/06/012}}.

\bibitem{Gava:2009pj}
J.~Gava, J.~Kneller, C.~Volpe, G.~C. McLaughlin, {A Dynamical collective
  calculation of supernova neutrino signals}, Phys. Rev. Lett. 103 (2009)
  071101.
\newblock \href {http://arxiv.org/abs/0902.0317} {\path{arXiv:0902.0317}},
  \href {https://doi.org/10.1103/PhysRevLett.103.071101}
  {\path{doi:10.1103/PhysRevLett.103.071101}}.

\bibitem{Friedland:2020ecy}
A.~Friedland, P.~Mukhopadhyay, {Near-critical supernova outflows and their
  neutrino signatures}, Phys. Lett. B 834 (2022) 137403.
\newblock \href {http://arxiv.org/abs/2009.10059} {\path{arXiv:2009.10059}},
  \href {https://doi.org/10.1016/j.physletb.2022.137403}
  {\path{doi:10.1016/j.physletb.2022.137403}}.

\bibitem{Krastev:1989ix}
P.~I. Krastev, A.~Y. Smirnov, {Parametric Effects in Neutrino Oscillations},
  Phys. Lett. B 226 (1989) 341--346.
\newblock \href {https://doi.org/10.1016/0370-2693(89)91206-9}
  {\path{doi:10.1016/0370-2693(89)91206-9}}.

\bibitem{Kneller:2010sc}
J.~P. Kneller, C.~Volpe, {Turbulence effects on supernova neutrinos}, Phys.
  Rev. D 82 (2010) 123004.
\newblock \href {http://arxiv.org/abs/1006.0913} {\path{arXiv:1006.0913}},
  \href {https://doi.org/10.1103/PhysRevD.82.123004}
  {\path{doi:10.1103/PhysRevD.82.123004}}.

\bibitem{Borriello:2013tha}
E.~Borriello, S.~Chakraborty, H.-T. Janka, et~al., {Turbulence patterns and
  neutrino flavor transitions in high-resolution supernova models}, JCAP 11
  (2014) 030.
\newblock \href {http://arxiv.org/abs/1310.7488} {\path{arXiv:1310.7488}},
  \href {https://doi.org/10.1088/1475-7516/2014/11/030}
  {\path{doi:10.1088/1475-7516/2014/11/030}}.

\bibitem{Patton:2014lza}
K.~M. Patton, J.~P. Kneller, G.~C. McLaughlin, {Stimulated neutrino
  transformation through turbulence on a changing density profile and
  application to supernovae}, Phys. Rev. D 91~(2) (2015) 025001.
\newblock \href {http://arxiv.org/abs/1407.7835} {\path{arXiv:1407.7835}},
  \href {https://doi.org/10.1103/PhysRevD.91.025001}
  {\path{doi:10.1103/PhysRevD.91.025001}}.

\bibitem{Abbar:2020ror}
S.~Abbar, {Turbulence Fingerprint on Collective Oscillations of Supernova
  Neutrinos}, Phys. Rev. D 103~(4) (2021) 045014.
\newblock \href {http://arxiv.org/abs/2007.13655} {\path{arXiv:2007.13655}},
  \href {https://doi.org/10.1103/PhysRevD.103.045014}
  {\path{doi:10.1103/PhysRevD.103.045014}}.

\bibitem{Balantekin:2004ug}
A.~B. Balantekin, H.~Yuksel, {Neutrino mixing and nucleosynthesis in
  core-collapse supernovae}, New J. Phys. 7 (2005) 51.
\newblock \href {http://arxiv.org/abs/astro-ph/0411159}
  {\path{arXiv:astro-ph/0411159}}, \href
  {https://doi.org/10.1088/1367-2630/7/1/051}
  {\path{doi:10.1088/1367-2630/7/1/051}}.

\bibitem{Martinez-Pinedo:2011yhi}
G.~Martinez-Pinedo, B.~Ziebarth, T.~Fischer, K.~Langanke, {Effect of collective
  neutrino flavor oscillations on vp-process nucleosynthesis}, Eur. Phys. J. A
  47 (2011) 98.
\newblock \href {http://arxiv.org/abs/1105.5304} {\path{arXiv:1105.5304}},
  \href {https://doi.org/10.1140/epja/i2011-11098-y}
  {\path{doi:10.1140/epja/i2011-11098-y}}.

\bibitem{Duan:2010af}
H.~Duan, A.~Friedland, G.~McLaughlin, R.~Surman, {The influence of collective
  neutrino oscillations on a supernova r-process}, J. Phys. G 38 (2011) 035201.
\newblock \href {http://arxiv.org/abs/1012.0532} {\path{arXiv:1012.0532}},
  \href {https://doi.org/10.1088/0954-3899/38/3/035201}
  {\path{doi:10.1088/0954-3899/38/3/035201}}.

\bibitem{Sasaki:2017jry}
H.~Sasaki, T.~Kajino, T.~Takiwaki, et~al., {Possible effects of collective
  neutrino oscillations in three-flavor multiangle simulations of supernova
  $\nu p$ processes}, Phys. Rev. D 96~(4) (2017) 043013.
\newblock \href {http://arxiv.org/abs/1707.09111} {\path{arXiv:1707.09111}},
  \href {https://doi.org/10.1103/PhysRevD.96.043013}
  {\path{doi:10.1103/PhysRevD.96.043013}}.

\bibitem{Abbar:2018shq}
S.~Abbar, H.~Duan, K.~Sumiyoshi, et~al., {On the occurrence of fast neutrino
  flavor conversions in multidimensional supernova models}, Phys. Rev. D
  100~(4) (2019) 043004.
\newblock \href {http://arxiv.org/abs/1812.06883} {\path{arXiv:1812.06883}},
  \href {https://doi.org/10.1103/PhysRevD.100.043004}
  {\path{doi:10.1103/PhysRevD.100.043004}}.

\bibitem{Fujimoto:2022njj}
S.-i. Fujimoto, H.~Nagakura, {Explosive nucleosynthesis with fast
  neutrino-flavour conversion in core-collapse supernovae}, Mon. Not. Roy.
  Astron. Soc. 519~(2) (2022) 2623--2629.
\newblock \href {http://arxiv.org/abs/2210.02106} {\path{arXiv:2210.02106}},
  \href {https://doi.org/10.1093/mnras/stac3763}
  {\path{doi:10.1093/mnras/stac3763}}.

\bibitem{Ko:2019xxm}
H.~Ko, et~al., {Neutrino Process in Core-collapse Supernovae with Neutrino
  Self-interaction and MSW Effects}, Astrophys. J. Lett. 891~(1) (2020) L24.
\newblock \href {http://arxiv.org/abs/1903.02086} {\path{arXiv:1903.02086}},
  \href {https://doi.org/10.3847/2041-8213/ab775b}
  {\path{doi:10.3847/2041-8213/ab775b}}.

\bibitem{Kusakabe:2019znq}
M.~Kusakabe, M.-K. Cheoun, K.~S. Kim, et~al., {Supernova Neutrino Process of Li
  and B Revisited}, Astrophys. J. 872~(2) (2019) 164.
\newblock \href {http://arxiv.org/abs/1901.01715} {\path{arXiv:1901.01715}},
  \href {https://doi.org/10.3847/1538-4357/aafc35}
  {\path{doi:10.3847/1538-4357/aafc35}}.

\bibitem{Ko:2022uqv}
H.~Ko, et~al., {Comprehensive Analysis of the Neutrino Process in
  Core-collapsing Supernovae}, Astrophys. J. 937~(2) (2022) 116.
\newblock \href {http://arxiv.org/abs/2203.13365} {\path{arXiv:2203.13365}},
  \href {https://doi.org/10.3847/1538-4357/ac88cd}
  {\path{doi:10.3847/1538-4357/ac88cd}}.

\end{thebibliography}
\end{document}